\documentclass[a4paper,11pt]{article}
\pdfoutput=1 % if your are submitting a pdflatex (i.e. if you have images in pdf, png or jpg format)

\usepackage{jheppub} % for details on the use of the package, please see the JHEP-author-manual

\usepackage[T1]{fontenc}
\usepackage{lmodern}
\usepackage{booktabs}
\usepackage{textcomp}
\usepackage{xspace}
\usepackage{relsize}
\usepackage{listings}
\usepackage{microtype}
\usepackage{multirow}
\usepackage{tabularx}
\usepackage{array}
\usepackage{placeins}
\usepackage{cuted}
\usepackage{soul} % only for \st; delete if this causes you problems.
\usepackage[dvipsnames]{xcolor}
\usepackage[clockwise,figuresright]{rotating}
\usepackage{siunitx}

\allowdisplaybreaks

\newcolumntype{L}{>{\raggedright\let\newline\\\arraybackslash\hspace{0pt}}X}
\newcolumntype{R}{>{\raggedleft\let\newline\\\arraybackslash\hspace{0pt}}X}
\newcolumntype{C}{>{\centering\let\newline\\\arraybackslash\hspace{0pt}}X}

\newcommand{\imperial}{Department of Physics, Imperial College London, Blackett Laboratory, Prince Consort Road, London SW7 2AZ, UK}

\newcommand{\adelaide}{Department of Physics, University of Adelaide, Adelaide, SA 5005, Australia}

\newcommand{\grappa}{GRAPPA, Institute of Physics, University of Amsterdam, Science Park 904, 1098 XH Amsterdam, Netherlands}

\newcommand{\rwth}{Institute for Theoretical Particle Physics and Cosmology (TTK), RWTH Aachen University, \\ D-52056 Aachen, Germany}

\makeatletter

\newcommand{\preprintnumber}[1]{\gdef\@preprintnumber{\begin{flushright}{#1}\end{flushright}}}

\g@addto@macro\bfseries{\boldmath}
\makeatother

\let\underscore\_
\renewcommand{\_}{\discretionary{\underscore}{}{\underscore}}

\makeatletter
\let\orgdescriptionlabel\descriptionlabel
\renewcommand*{\descriptionlabel}[1]{%
  \let\orglabel\label
  \let\label\@gobble
  \phantomsection
  \protected@edef\@currentlabel{#1}%
  \let\label\orglabel
  \orgdescriptionlabel{#1}%
}
\makeatother

\newcommand\postnewlinemarker{\hbox{\ensuremath{\hookrightarrow}}}
\newcommand\cpp[1]{{\lstinline!#1!}}  % Apparently curly braces are only "experimental"

\newcommand\yaml[1]{{\lstset{style=yaml}\lstinline!#1!\lstset{style=cpp}}}

\newcommand\term[1]{{\lstset{style=terminal}\lstinline!#1!\lstset{style=cpp}}}
\newcommand\fortran[1]{{\lstset{style=fortran}\lstinline!#1!\lstset{style=cpp}}}
\newcommand\py[1]{{\lstset{style=python}\lstinline!#1!\lstset{style=cpp}}}
\newcommand\customtilde{{\raisebox{0.2ex}{\scalebox{0.6}{\boldmath$\sim$}}}}
\newcommand\mathematica[1]{{\lstset{style=Mathematica}\lstinline!#1!\lstset{style=cpp}}}

\lstnewenvironment{lstlistingyaml}{\lstset{style=yaml}}{\lstset{style=cpp}}
\lstnewenvironment{lstlistingterm}{\lstset{style=terminal}}{\lstset{style=cpp}}
\lstnewenvironment{lstlistingfortran}{\lstset{style=fortran}}{\lstset{style=cpp}}
\lstnewenvironment{lstcpp}{\lstset{style=cpp}}{\lstset{style=cpp}}
\lstnewenvironment{lstcppalt}{\lstset{style=cppalt}}{\lstset{style=cpp}}
\lstnewenvironment{lstcppnum}{\lstset{style=cppnum}}{\lstset{style=cpp}}
\lstnewenvironment{lstyaml}{\lstset{style=yaml}}{\lstset{style=cpp}}
\lstnewenvironment{lstterm}{\lstset{style=terminal}}{\lstset{style=cpp}}
\lstnewenvironment{lsttermalt}{\lstset{style=terminalalt}}{\lstset{style=cpp}}
\lstnewenvironment{lsttext}{\lstset{style=text}}{\lstset{style=cpp}}
\lstnewenvironment{lstfortran}{\lstset{style=fortran}}{\lstset{style=cpp}}
\lstnewenvironment{lstpy}{\lstset{style=python}}{\lstset{style=cpp}}

\newcommand{\tmpname}{}
\newcommand{\tmplistingname}{}
\makeatletter
\newif\ifATOlabelname
\lst@Key{labelname}{Listing}{\def\ATOlabelname{#1}\global\ATOlabelnametrue}
\makeatother
\lstnewenvironment{lstcpplabel}[1][]{
  \lstset{style=cpp,#1} % #1 allows to add new options with [] same as for normal lstlistings environment
  \ifATOlabelname
  \renewcommand{\tmpname}{\lstlistingname}
  \renewcommand{\tmplistingname}{\lstlistlistingname}
  \renewcommand{\lstlistingname}{\ATOlabelname}% Listing -> labelname
  \renewcommand{\lstlistlistingname}{List of \lstlistingname s}% List of Listings -> List of labelname
  \fi
}{
  \renewcommand{\lstlistingname}{\tmpname}
  \renewcommand{\lstlistlistingname}{\tmplistingname}
  \lstset{style=cpp}
}
\definecolor{solarized@base03}{HTML}{002B36}
\definecolor{solarized@base02}{HTML}{073642}
\definecolor{solarized@base01}{HTML}{586e75}
\definecolor{solarized@base00}{HTML}{657b83}
\definecolor{solarized@base0}{HTML}{839496}
\definecolor{solarized@base1}{HTML}{93a1a1}
\definecolor{solarized@base2}{HTML}{EEE8D5}
\definecolor{solarized@base3}{HTML}{FDF6E3}
\definecolor{solarized@yellow}{HTML}{B58900}
\definecolor{solarized@orange}{HTML}{CB4B16}
\definecolor{solarized@red}{HTML}{DC322F}
\definecolor{solarized@magenta}{HTML}{D33682}
\definecolor{solarized@violet}{HTML}{6C71C4}
\definecolor{solarized@blue}{HTML}{268BD2}
\definecolor{solarized@cyan}{HTML}{2AA198}
\definecolor{solarized@green}{HTML}{859900}
\definecolor{darkred}{HTML}{550003}
\definecolor{darkgreen}{HTML}{00AA00}

\newcommand\YAMLstringstyle{\footnotesize\color{solarized@green}\mdseries}
\newcommand\YAMLkeystyle{\footnotesize\color{solarized@blue}\ttfamily}
\newcommand\YAMLvaluestyle{\footnotesize\color{blue}\mdseries}
\newcommand\ProcessThreeDashes{\llap{\color{cyan}\mdseries-{-}-}}
\newcommand\CPPcommentstyle{\color{solarized@violet}\footnotesize\ttfamily}
\newcommand\CPPdirectivestyle{\color{solarized@magenta}\footnotesize\ttfamily}
\newcommand\termplainstyle{\footnotesize\ttfamily}

\newcommand\processLongMacroDelimiter
{%
  \CPPdirectivestyle%
  \#define%
}

\lstdefinestyle{cpp}
{
  language=C++,
  basicstyle=\footnotesize\ttfamily,
  basewidth={0.53em,0.44em}, %Ben: experimenting a bit with the fixed-width width (first argument); feels a bit more readable to me with the slightly smaller width (was 0.6em by default)
  numbers=none,
  tabsize=2,
  breaklines=true,
  escapeinside={@}{@},
  showstringspaces=false,
  numberstyle=\tiny\color{solarized@base01},
  keywordstyle=\color{solarized@orange},
  stringstyle=\color{solarized@red}\ttfamily,
  identifierstyle=\color{solarized@blue},
  commentstyle=\CPPcommentstyle,
  directivestyle=\CPPdirectivestyle,
  emphstyle=\color{solarized@green},
  frame=single,
  rulecolor=\color{solarized@base2},
  rulesepcolor=\color{solarized@base2},
  literate={~} {\customtilde}1,
  moredelim=*[directive]\ \ \#,
  moredelim=*[directive]\ \ \ \ \#
}

\lstdefinestyle{cppalt}
{
  language=C++,
  basicstyle=\footnotesize\ttfamily,
  basewidth={0.53em,0.44em}, %Ben: experimenting a bit with the fixed-width width (first argument); feels a bit more readable to me with the slightly smaller width (was 0.6em by default)
  numbers=none,
  tabsize=2,
  breaklines=true,
  escapeinside={*@}{@*},
  showstringspaces=false,
  numberstyle=\tiny\color{solarized@base01},
  keywordstyle=\color{solarized@orange},
  stringstyle=\color{solarized@red}\ttfamily,
  identifierstyle=\color{solarized@blue},
  commentstyle=\CPPcommentstyle,
  directivestyle=\CPPdirectivestyle,
  emphstyle=\color{solarized@green},
  frame=single,
  rulecolor=\color{solarized@base2},
  rulesepcolor=\color{solarized@base2},
  literate={~}{\customtilde}1,
  moredelim=**[is][\processLongMacroDelimiter]{BeginLongMacro}{EndLongMacro} %special delimiter for long macros that go over several lines
}

\lstdefinestyle{cppnum}
{
  language=C++,
  basicstyle=\footnotesize\ttfamily,
  basewidth={0.53em,0.44em}, %Ben: experimenting a bit with the fixed-width width (first argument); feels a bit more readable to me with the slightly smaller width (was 0.6em by default)
  numbers=none,
  tabsize=2,
  breaklines=true,
  escapeinside={@}{@},
  numberstyle=\tiny\color{solarized@base01},
  showstringspaces=false,
  numberstyle=\tiny\color{solarized@base01},
  keywordstyle=\color{solarized@orange},
  stringstyle=\color{solarized@red}\ttfamily,
  identifierstyle=\color{solarized@blue},
  commentstyle=\CPPcommentstyle,
  directivestyle=\CPPdirectivestyle,
  emphstyle=\color{solarized@green},
  frame=single,
  rulecolor=\color{solarized@base2},
  rulesepcolor=\color{solarized@base2},
  literate={~} {\customtilde}1,
  moredelim=*[directive]\ \ \#,
  moredelim=*[directive]\ \ \ \ \#
}

\lstdefinestyle{python}
{
  language=Python,
  basicstyle=\footnotesize\ttfamily,
  basewidth={0.53em,0.44em},
  numbers=none,
  tabsize=2,
  breaklines=true,
  escapeinside={@}{@},
  showstringspaces=false,
  numberstyle=\tiny\color{solarized@base01},
  keywordstyle=\color{blue},
  stringstyle=\color{orange}\ttfamily,
  identifierstyle=\color{darkred},
  commentstyle=\color{purple},
  emphstyle=\color{green},
  frame=single,
  rulecolor=\color{solarized@base2},
  rulesepcolor=\color{solarized@base2},
  literate = {~}{\customtilde}1
  {\ as\ }{{\color{blue}\ as\ \color{black}}}3
}

\lstdefinestyle{fortran}
{
  language=Fortran,
  basicstyle=\footnotesize\ttfamily,
  basewidth={0.53em,0.44em},
  numbers=none,
  tabsize=2,
  breaklines=true,
  escapeinside={@}{@},
  showstringspaces=false,
  numberstyle=\tiny\color{solarized@base01},
  keywordstyle=\color{blue},
  stringstyle=\color{orange}\ttfamily,
  identifierstyle=\color{Periwinkle},
  commentstyle=\color{purple},
  emphstyle=\color{green},
  morekeywords={and, or, true, false},
  frame=single,
  rulecolor=\color{solarized@base2},
  rulesepcolor=\color{solarized@base2},
  literate={~}{\customtilde}1
}

\lstdefinestyle{terminal}
{
  language=bash,
  basicstyle=\termplainstyle,
  numbers=none,
  tabsize=2,
  breaklines=true,
  escapeinside={@}{@},
  frame=single,
  showstringspaces=false,
  numberstyle=\tiny\color{solarized@base01},
  keywordstyle=\color{solarized@orange},
  stringstyle=\color{solarized@red}\ttfamily,
  identifierstyle=\color{black},
  commentstyle=\color{solarized@violet},
  emphstyle=\color{solarized@green},
  frame=single,
  rulecolor=\color{solarized@base2},
  rulesepcolor=\color{solarized@base2},
  morekeywords={gambit, cmake, make, mkdir},
  deletekeywords={test},
  literate = {\ gambit}{{\ }{\color{black}}gambit}7
  {/gambit}{{/}{\color{black}}gambit}6
  {gambit/}{{\color{black}}gambit{/}}6
  {/include}{{/}{\color{black}}include}8
  {cmake/}{{\color{black}}cmake/}6
  {.cmake}{{.}{\color{black}}cmake}6
  {~}{\customtilde}1
}

\lstdefinestyle{terminalalt}
{
  language=bash,
  basicstyle=\footnotesize\ttfamily,
  numbers=none,
  tabsize=2,
  breaklines=true,
  escapeinside={*@}{@*},
  frame=single,
  showstringspaces=false,
  numberstyle=\tiny\color{solarized@base01},
  keywordstyle=\color{solarized@orange},
  stringstyle=\color{solarized@red}\ttfamily,
  identifierstyle=\color{black},
  commentstyle=\color{solarized@violet},
  emphstyle=\color{solarized@green},
  frame=single,
  rulecolor=\color{solarized@base2},
  rulesepcolor=\color{solarized@base2},
  morekeywords={gambit, cmake, make, mkdir},
  deletekeywords={test},
  literate = {\ gambit}{{\ }{\color{black}}gambit}7
  {/gambit}{{/}{\color{black}}gambit}6
  {gambit/}{{\color{black}}gambit{/}}6
  {/include}{{/}{\color{black}}include}8
  {cmake/}{{\color{black}}cmake/}6
  {.cmake}{{.}{\color{black}}cmake}6
  {~}{\customtilde}1
}

\lstdefinestyle{text}
{
  language={},
  basicstyle=\footnotesize\ttfamily,
  identifierstyle=\color{black},
  numbers=none,
  tabsize=2,
  breaklines=true,
  escapeinside={*@}{@*},
  showstringspaces=false,
  frame=single,
  rulecolor=\color{solarized@base2},
  rulesepcolor=\color{solarized@base2},
  literate={~}{\customtilde}1
}

\lstdefinestyle{yaml}
{
  language=bash,
  escapeinside={@}{@},
  keywords={true,false,null},
  otherkeywords={},
  keywordstyle=\color{solarized@base0}\bfseries,
  basicstyle=\footnotesize\color{black}\ttfamily,
  identifierstyle=\YAMLkeystyle,
  sensitive=false,
  commentstyle=\color{solarized@orange}\ttfamily,
  morecomment=[l]{\#},
  morecomment=[s]{/*}{*/},
  stringstyle=\YAMLstringstyle\ttfamily,
  moredelim=**[s][\YAMLkeystyle]{,}{:},   % switch to value style at : but back to key style at,
  moredelim=**[l][\YAMLvaluestyle]{:},    % switch to value style at :
  morestring=[b]',
  morestring=[b]",
  literate =    {---}{{\ProcessThreeDashes}}3
  {>}{{\textcolor{solarized@red}\textgreater}}1
  {|}{{\textcolor{solarized@red}\textbar}}1
  {\ -\ }{{\mdseries\color{black}\ -\ \negmedspace}}3
  {\}}{{{\color{black} \}}}}1
  {\{}{{{\color{black} \{}}}1
  {[}{{{\color{black} [}}}1
  {]}{{{\color{black} ]}}}1
  {~}{\customtilde}1,
  breakindent=0pt,
  breakatwhitespace,
  columns=fullflexible
}

\lstdefinestyle{mathematica}
{
  language={Mathematica},
  basicstyle=\footnotesize\ttfamily,
  basewidth={0.53em,0.44em},
  numbers=none,
  tabsize=2,
  breaklines=true,
  escapeinside={@}{@},
  numberstyle=\tiny\color{black},
  showstringspaces=false,
  numberstyle=\tiny\color{solarized@base01},
  keywordstyle=\color{solarized@orange},
  stringstyle=\color{solarized@red}\ttfamily,
  identifierstyle=\color{solarized@orange}\ttfamily,
  commentstyle=\color{solarized@gray}\ttfamily,
  directivestyle=\color{solarized@orange}\ttfamily,
  emphstyle=\color{solarized@green},
  frame=single,
  rulecolor=\color{solarized@base2},
  rulesepcolor=\color{solarized@base2},
  literate={~} {\customtilde}1,
  moredelim=*[directive]\ \ \#,
  moredelim=*[directive]\ \ \ \ \#,
  mathescape=true
}

\newcommand{\doublecross}[2]{\hyperref[#2]{\textbf{#1}}}
\newcommand{\doublecrosssf}[2]{\hyperref[#2]{\textbf{\textsf{#1}}}}

\newcommand{\startglossary}{\section{Glossary}\label{glossary}Here we explain some terms that have specific technical definitions in \GB.\begin{description}}
  \newcommand{\finishglossary}{\end{description}}

\newcommand{\bcode}{\begin{lstlisting}}
  \newcommand{\ecode}{\end{lstlisting}}
\newcommand{\metavarf}[1]{\textit{\color{darkgreen}\footnotesize\textrm{#1}}}

\newcommand{\metavar}{\metavarf}

\newcommand{\eV}{\ensuremath{\text{e}\mspace{-0.8mu}\text{V}}\xspace}
\newcommand{\MeV}{\text{M\eV}\xspace}
\newcommand{\GeV}{\text{G\eV}\xspace}

\newcommand{\gambit}{\textsf{GAMBIT}\xspace}

\newcommand{\darkbit}{\textsf{DarkBit}\xspace}

\newcommand{\GB}{\gambit}

\newcommand\MultiNest{\textsf{MultiNest}\xspace}
\newcommand\multinest{\MultiNest}

\newcommand\twalk{\textsf{T-Walk}\xspace}
\newcommand\diver{\textsf{Diver}\xspace}

\newcommand\YAML{\textsf{YAML}\xspace}

\newcommand\beq{\begin{equation}}
  \newcommand\eeq{\end{equation}}

\renewcommand{\url}[1]{\href{#1}{#1}}

\usepackage[english]{babel}
\usepackage{xspace}
\usepackage[font=small,tableposition=above]{caption}
\usepackage{makecell}
\usepackage{threeparttable}
\usepackage{tikz}

\usetikzlibrary{trees}
\usetikzlibrary{shapes}
\newcommand{\gambitver}{\gambit\ \textsf{1.3.0}\xspace}
\newcommand{\gambitverupdate}{\gambit\ \textsf{1.3.1}\xspace}
\newcommand{\updated}[2]{#2}
\newcommand{\confirmed}[1]{#1}
\newcommand{\SH}[1]{#1}
\newcommand{\vc}[1]{\mathbf{#1}}
\newcommand{\refeq}[1]{(\ref{#1})}
\newcommand{\reffig}[1]{Fig.~\ref{#1}}
\newcommand{\highl}[1]{{\color{RedOrange}#1}}
\newcommand{\fa}{f_a}
\newcommand{\mazero}{m_{a,0}}
\newcommand{\ma}{m_a}
\newcommand{\gagg}{g_{a\gamma\gamma}}
\newcommand{\cagg}{C_{a\gamma\gamma}}
\newcommand{\caggtilde}{\widetilde{C}_{a\gamma\gamma}}
\newcommand{\gaee}{g_{aee}}
\newcommand{\caee}{C_{aee}}
\newcommand{\OmegaA}{\Omega_a}
\newcommand{\Tcrit}{T_\chi}
\newcommand{\sol}{\ensuremath{\odot}}
\newcommand{\lnL}{\ln \left(\mathcal{L}\right)}
\newcommand{\maxlnL}{\ln (\hat{\mathcal{L}})}
\newcommand{\lnLvar}[1]{\ln \left(\mathcal{L}_#1\right)}
\DeclareMathAlphabet{\pazocal}{OMS}{zplm}{m}{n}
\newcommand{\lagrangian}{\pazocal{L}}
\newcommand{\del}{\partial}

\newcommand{\ee}{\mathrm{e}}
\newcommand{\mrm}[2]{#1_\text{#2}}
\newcommand{\osc}[1]{{#1}_\text{osc}}
\newcommand{\mpl}{\mrm{m}{Pl}}
\newcommand{\ede}{\varepsilon}
\newcommand{\otherpi}{\pi}

\newcommand{\genalp}{\textsf{GeneralALP}\xspace}
\newcommand{\simpalp}{\textsf{ConstantMassALP}\xspace}
\newcommand{\qcdaxion}{\textsf{QCDAxion}\xspace}
\newcommand{\ksvz}{\textsf{KSVZAxion}\xspace}
\newcommand{\dfsz}{\textsf{DFSZAxion}\xspace}
\newcommand{\dfszI}{\textsf{DFSZAxion-I}\xspace}
\newcommand{\dfszII}{\textsf{DFSZAxion-II}\xspace}
\newcommand{\order}{\pazocal{O}}
\newcommand{\prob}[1]{\pazocal{P}\left(#1\right)}
\newcommand{\prrange}[2]{$[\num{#1}, \, \num{#2}]$}
\newcommand{\iuo}[2]{$#1 \, [\si{#2}]$}

\newcommand{\lsim}{\mathrel{\rlap{\lower4pt\hbox{$\sim$}}\raise1pt\hbox{$<$}}}
\newcommand{\gsim}{\mathrel{\rlap{\lower4pt\hbox{$\sim$}}\raise1pt\hbox{$>$}}}
\newcommand{\thetai}{\mrm{\theta}{i}}
\newcommand{\LambdaQCD}{\Lambda_\chi}
\newcommand{\dd}{\mathrm{d}}
\newcommand{\OmegaCDM}{\mrm{\Omega}{DM}}
\newcommand*\widebar[1]{\hbox{%
    \vbox{\hrule height 0.5pt % The actual bar
      \kern0.3ex%         % Distance between bar and symbol
      \hbox{%
        \kern-0.1em%      % Shortening on the left side
        \ensuremath{#1}%
        \kern-0.1em%      % Shortening on the right side
      }
}}}
\newcommand{\mc}[1]{\multicolumn{1}{c}{#1}}
\newcommand{\mr}[1]{\makecell[tl]{#1}}

\preprint{TTK-18-40}

\title{Axion global fits with Peccei-Quinn symmetry breaking before inflation using GAMBIT}

\author[a]{Sebastian Hoof,}
\author[b]{Felix Kahlhoefer,}
\author[a]{Pat Scott,}
\author[c]{Christoph Weniger,}
\author[d]{Martin White}

\affiliation[a]{\imperial}
\affiliation[b]{\rwth}
\affiliation[c]{\grappa}
\affiliation[d]{\adelaide}

\emailAdd{s.hoof15@imperial.ac.uk}
\emailAdd{kahlhoefer@physik.rwth-aachen.de}
\emailAdd{p.scott@imperial.ac.uk}
\emailAdd{c.weniger@uva.nl}
\emailAdd{martin.white@adelaide.edu.au}

\abstract{We present global fits of cosmologically stable axion-like particle and QCD~axion models in the mass range 0.1\,neV to~10\,eV. We focus on the case where the Peccei-Quinn symmetry is broken before the end of inflation, such that the initial value of the axion field can be considered to be homogeneous throughout the visible Universe. We include detailed likelihood functions from light-shining-through-wall experiments, haloscopes, helioscopes, the axion relic density, horizontal branch stars, supernova~1987A, white dwarf cooling, and gamma-ray observations. We carry out both frequentist and Bayesian analyses, with and without the inclusion of white dwarf cooling. We explore the degree of fine-tuning present in different models and identify parameter regions where it is possible for QCD~axion models to account for both the dark matter in the Universe and the cooling hints, comparing them to specific DFSZ- and KSVZ-type models. We find the most credible parameter regions, allowing us to set (prior-dependent) upper and lower bounds on the axion mass. Our analysis also suggests that QCD~axions in this scenario most probably make up a non-negligible but sub-dominant component of the dark matter in the Universe.}

\begin{document}
\maketitle
\flushbottom

\section{Introduction}\label{sec:intro}
QCD~axions~\cite{1977_pq_axion1,1977_pq_axion2,1978_weinberg_axion,1978_wilczek_axion} and axion-like particles~(ALPs) are perhaps among the most intriguing classes of hypothetical particles. They have been extensively studied in the literature and experiments are expected to probe the relevant parameter space for many axion models in the near future (see Ref.~\cite{1801.08127} for an up-to-date review on axion searches). Because QCD~axions can behave as cold dark matter~(DM)~\cite{Preskill:1982cy,Abbott:1982af,Dine:1982ah,1986_turner_axiondensity}, the particularly interesting regions of the parameter space are where they contribute significantly to the DM density of the Universe. If so, they would solve at least two outstanding problems in physics at the same time (the other being the Strong~CP problem~\cite{1977_pq_axion1,1977_pq_axion2}).

There are now many complementary searches for axions\footnote{In this paper, we use the generic term ``axion'' to refer to both QCD~axions and ALPs. We use more specific terms such as ``QCD~axion'' to make more model-specific statements.} underway, and many new search strategies have emerged in recent years~(see Ref.~\cite{1801.08127}). Axions could also explain some apparent anomalies in the cooling of white dwarf stars~\cite{Isern:1992gia,1205.6180,1211.3389,1512.08108,1605.06458,1605.07668,1708.02111}, or the transparency of the Universe to gamma rays~\cite{0707.4312,0712.2825,1001.0972,1106.1132,1201.4711,1302.1208}. It is therefore crucial to combine all available results in order to extract the maximum information from the data. In doing so, we can learn more about the parameter space of different models, help guide the planning of future searches towards the most promising search areas and -- if axions do indeed exist -- find them in the correlated signals of several experiments and determine their properties.

Some of these goals can be satisfactorily achieved by over-plotting exclusion limits from several experiments. However, such simple exclusion plots generally have some shortcomings. They make assumptions about the relative importance of interactions with other matter, and about important inputs such as solar physics, the local density of DM and theoretical uncertainties. Incompatibilities between the assumptions can undermine the validity of a simple combination. They also do not allow for a quantitative model comparison between different theories that include axions.

A solution is to perform a consistent global statistical analysis of all available constraints, accounting for the leading theoretical and experimental uncertainties. Implementing such an analysis requires careful design, along with compromises relating to the availability of data and the total computational runtime required. An example of such a computational framework for BSM physics is {\gambit}~\cite{gambit}. Full details of \gambit's features can be found in the relevant publications~\cite{gambit,ColliderBit,DarkBit,FlavBit,SDPBit,ScannerBit} and physics analyses~\cite{MSSM,SSDM,CMSSM,SSDM2,HiggsPortal,EWMSSM}. The most relevant features for this work are \gambit's modularity, and the options that it offers for carrying out both Bayesian and frequentist analyses. \gambit's structure allows easy integration of new components such as models, theory calculations, likelihoods and sampling algorithms. \gambit contains a variety of advanced samplers for both Bayesian and frequentist analyses, which are particularly useful for including nuisance parameters and assessing fine-tuning.

In the following section, we review some aspects of axion physics, including specific issues that should be accounted for in global fits. In Sec.~\ref{sec:axionmodels}, we turn to axion models, their corresponding effective field theories and the family tree of axion models available in \gambit. Sec.~\ref{sec:obsexplike} describes our observables and likelihood functions, including their implementation, incorporated experimental data, potential caveats and restrictions. Results and discussion of the first global scans of axion models can be found in Sec.~\ref{sec:results}, and Sec.~\ref{sec:outlook} summarises our results.

The axion routines developed for this paper are available in the {\darkbit}~\cite{DarkBit} module of \gambitverupdate,\footnote{\SH{We discovered a mistake in the initial version~(\gambitver) used for this paper, resulting in values of the axion-photon coupling for \qcdaxion models being too large by a factor of~$\sqrt{2\otherpi}$. This mistake was corrected in \gambitverupdate and all affected figures were replaced and marked with the updated version number.}} available at \url{https://gambit.hepforge.org} under the 3-clause BSD license.\footnote{\href{http://opensource.org/licenses/BSD-3-Clause}{http://opensource.org/licenses/BSD-3-Clause}} Likelihood and posterior samples from this study can be downloaded from \textsf{Zenodo}~\cite{Zenodo_axions}.

\section{Axion physics}\label{sec:axionphysics}
Here we present a brief overview of axion physics, highlighting caveats of our current implementation and pointing out opportunities for future extensions. More details can be found in the literature~\cite{1987_kim_lightpseudoscalars,Kuster:2008zz,1201.5902,1301.1123,1510.07633,1801.08127}.

\subsection{The QCD axion}
The Peccei-Quinn~(PQ) mechanism~\cite{1977_pq_axion1,1977_pq_axion2} is a proposed solution to the Strong~CP problem that gives rise to a pseudo-scalar particle: the QCD~axion~\cite{1978_weinberg_axion,1978_wilczek_axion}. The QCD~axion is an excellent DM candidate~\cite{Preskill:1982cy,Abbott:1982af,Dine:1982ah,1986_turner_axiondensity}, as it can account for the entire cosmological abundance of DM via the \textit{vacuum misalignment} or \textit{realignment} mechanism (Sec.~\ref{sec:realignment}). The original QCD~axion model also inspired further archetypical models, such as the KSVZ~\cite{1979_kim_ksvz,1980_shifman_ksvz} and DFSZ~\cite{1980_zhitnitsky_dfsz,1981_dine_dfsz} axion models, which we will introduce in Secs~\ref{sec:mod:KSVZAxion} and~\ref{sec:mod:DFSZAxion}.

The symmetries of the Standard Model (SM) Lagrangian permit a term of the form
\begin{equation}
\mrm{\lagrangian}{QCD} \supset - \frac{\mrm{\alpha}{S}}{8\otherpi} \mrm{\theta}{QCD} G_{\mu \nu}^a \widetilde{G}^{\mu \nu, a} \, , \label{eq:QCDLagrangian}
\end{equation}
where $G_{\mu \nu}^a$ is the gluon field strength, $a$~is the $SU(3)$ gauge index and $\mrm{\alpha}{S}$ is the strong coupling constant. The angle $\mrm{\theta}{QCD} \in [-\otherpi, \otherpi]$ is an unknown parameter and $\widetilde{G}^{\mu \nu, a} = \epsilon^{\mu \nu \kappa \lambda}G_{\kappa \lambda}^a/2$. A contribution is also generated by chiral transformations due to the chiral anomaly, which replaces $\mrm{\theta}{QCD}$ by an effective angle,
\begin{equation}
\mrm{\theta}{eff} \equiv \mrm{\theta}{QCD} - \arg \left [ \det (Y_dY_u) \right ] \, , \label{eq:thetaeff}
\end{equation}
where $Y_d$ and $Y_u$ are the down- and up-type Yukawa matrices, respectively~\cite[Sec.~29.5]{Book_2014_Schwartz}. The $G\widetilde{G}$-term is anti-symmetric under the discrete parity ($P$) and charge-parity ($CP$) transformations. Due to the presence of this term, one would na{\"i}vely expect the strong interaction to show some $CP$-violating effects -- especially because weak interactions are known to violate $P$ maximally and $CP$ mildly. There is no $CP$-violation if and only if $\mrm{\theta}{eff}$ vanishes (modulo the periodicity).

Experiments trying to measure the CP-violating electric dipole moment of the neutron~(nEDM) can place limits on the value of $\mrm{\theta}{eff}$. Within 40--50\% uncertainty, the dipole moment induced by the $G\widetilde{G}$-term is given by~\cite{1979_crewther,1980_crewther,hep-ph/9904483}
\begin{equation}
|d_n| \approx \left( \SI{2.4E-16}{\elementarycharge \centi \metre} \right) \left|\mrm{\theta}{eff}\right| \, .\label{eq:EDMn}
\end{equation}
The current limit on the nEDM is $|d_n| < \SI{3.6E-26}{\elementarycharge \centi \metre}$ at 95\% confidence level~\cite{1509.04411}, resulting in $|\mrm{\theta}{eff}|\lsim \num{e-10}$.

This observation poses a fine-tuning issue in the SM, commonly referred to as the Strong~CP problem. The mechanism suggested by Peccei and Quinn~\cite{1977_pq_axion1,1977_pq_axion2} solves this problem by introducing a new global, axial $U(1)$ symmetry. This so-called PQ~symmetry is spontaneously broken by the vacuum~expectation value~$v$ of a complex scalar field. The resulting Nabu-Goldstone boson is the QCD~axion, a pseudo-scalar field denoted by~$a(x)$. It adds another contribution to~$\mrm{\theta}{eff}$ of the form~$Na(x)/v$, where the non-zero integer~$N$ is the colour anomaly of the PQ symmetry. The associated shift symmetry can then be used to cancel the $\mrm{\theta}{eff}$~term by driving $\mrm{\theta}{eff} + Na(x)/v$ to zero. In fact, it can be shown by the Vafa-Witten theorem~\cite{1984_vafa_vafawitten1,1984_vafa_vafawitten2} that the axion dynamically and asymptotically relaxes $\mrm{\theta}{eff}$ to the $CP$-conserving minimum. The Strong~CP problem is therefore effectively solved by promoting~$\mrm{\theta}{eff}$ to a dynamical degree of freedom.

The continuous shift symmetry of the PQ $U(1)$ phase forbids any mass terms for the axion at the Lagrangian level. In the presence of an $a \, G\widetilde{G}$~term in the Lagrangian, however, this symmetry is broken after the QCD phase transition due to topologically non-trivial fluctuations of the gluon fields, leaving only a discrete shift symmetry of size $2\otherpi N v$.

The resulting effective potential can be written as\footnote{The full potential~\cite{DiVecchia:1980yfw,1511.02867} includes corrections to the simple cosine shape. There is evidence that the potential in~\refeq{eq:axion_eff_pot} is a good approximation at higher temperatures~\cite{1508.06917,1606.07494}, relevant for solving the axion field equation in Sec.~\ref{sec:realignment}.}
\begin{equation}
	V(a) = \fa^2 \, \ma^2 \, \left[1 - \cos (a/\fa) \right] \, , \label{eq:axion_eff_pot}
\end{equation}
where $\ma$ is the temperature-dependent axion mass and $\fa \equiv v/N$. The axion is therefore practically massless until the time of the QCD phase transition, where it picks up a small mass that can be calculated using the chiral Lagrangian formalism.

The zero-temperature mass of the QCD~axion was initially calculated by relating $v$~to the scale of weak interactions~\cite{1978_weinberg_axion}. However, this model was ruled out and the parameter space for $\fa$ was opened up in order to make the axion ``invisible'' -- in the sense that it evaded experimental constraints at the time. Neglecting suppressed quark mass ratios, the QCD~axion mass at zero temperature is given by~\cite{1978_weinberg_axion}
\begin{equation}
	 \mazero \simeq \frac{\sqrt{z_d}}{1+z_d} \, \frac{f_{\pi^0}}{\fa} \, m_{\pi^0} \, ,
\end{equation}
where $z_d=m_u/m_d$ is the up- and down-quark mass ratio, and $f_{\pi^0}$ and $m_{\pi^0}$ are the decay constant and mass of the pion respectively. Using next-to-leading order chiral perturbation theory, a recent study found the QCD~axion mass at zero temperature to be~\cite{1511.02867}
\begin{equation}
	 \mazero = \SI{5.70(7)}{\micro \electronvolt} \left ( \frac{\SI{e12}{\giga \electronvolt}}{\fa} \right )\, . \label{eq:intro:axionmass}
\end{equation}
At temperatures exceeding the QCD scale, the axion becomes increasingly light as the shift symmetry is restored. Numerical estimates of the temperature dependence can be obtained directly from recent lattice QCD simulations~\cite{1606.03145,1606.07494}. These results are well-approximated at higher temperatures by a power law $\ma(T)\propto T^{-\beta/2}$ for some $\beta > 0$, and by a constant axion mass below some transition temperature~$\Tcrit$. This also agrees with the behaviour predicted from the finite-temperature dilute instanton
gas approximation~\cite{1980_Pisarski}.

As we solve all relevant equations numerically in this paper, it would be straightforward to include the full lattice QCD results and calculate the axion mass at every given temperature. However, including the statistical uncertainties\footnote{From lattice QCD results, we can infer that (at zero temperature) the statistical uncertainties in the axion mass are about twice as important as the systematic ones~\cite{1606.07494}.} is most easily achieved by having a parametrised form of the temperature dependence of the QCD~axion mass. This is commonly done by introducing two parameters, $\beta$ and $\Tcrit$, which we fit to lattice QCD results. More details about the implementation can be found in Sec.~\ref{sec:mod:QCDAxion}.

\subsection{Axion-like particles}
Apart from the original QCD~axion, pseudo-Nambu-Goldstone bosons with fundamental shift symmetries appear in many other contexts. These are usually referred to as ALPs (axion-like particles), and are appealing for theorists and model builders because of the shift symmetry, and the ability of their field values to undergo relaxation via the realignment mechanism in the same manner as axions. They do not necessarily solve the Strong~CP problem. However, under the right circumstances, they can be cold DM candidates, or play an important role in solving other physics puzzles. ALPs typically arise from the breaking of a $U(1)$ symmetry at some scale~$\fa$ and generate a mass from explicit breaking of the residual symmetry at scale~$\Lambda$. As a result, they can be fairly light, with masses of the order $\ma \sim \Lambda^2\left/\fa\right.$, and have suppressed couplings to the SM. Due to the lack of a direct relation between the two scales involved ($\fa$ and $\ma$), they occupy a larger parameter space than QCD~axions. More details on the theory and phenomenology of these particles can be found in the literature~\cite{1987_kim_lightpseudoscalars,1002.0329,1801.08127}.

\subsection{Axion creation mechanisms}\label{sec:axioncreation}
QCD~axions (with roughly~\si{\micro\eV} to~\si{\milli\eV} masses) and many ALP models are expected to be lighter than other hypothetical particles, such as WIMPs, which typically have \si{\GeV} to~\si{\tera\eV}-scale masses~\cite{1003.0904}. For thermal DM, this would be problematic, as DM would not be sufficiently cold today to reproduce the observed large-scale structure of the Universe. However, although a small population of axions is produced thermally, the relic abundance is typically dominated by non-thermal mechanisms, namely the realignment mechanism and topological defects. In this work, we focus on the realignment mechanism, which allows axions to be both ultra-cold and very light at the same time.

\subsubsection{Realignment mechanism}\label{sec:realignment}
The equation of motion for a homogeneous QCD~axion or ALP field~$\theta(t)=a(t)/\fa$ with potential~$V(\theta)$ in a Friedmann-Robertson-Walker-{Lema\^itre} universe reads~\cite[app.~B.12]{weinberg_cosmology}
\begin{equation}
	\ddot{\theta}+3H(t)\dot{\theta}+ \frac{1}{\fa^2}\frac{\delta V(\theta)}{\delta \theta} = 0 \, , \label{eq:AxionFieldEqGeneral}
\end{equation}
where $H\equiv\dot{a}/a$ is the Hubble parameter. For the canonical axion potential,
\begin{equation}
  V(\theta) = \fa^2\ma^2\left[1-\cos (\theta)\right], \label{eq:potential}
\end{equation}
the equation of motion becomes
\begin{equation}
  \ddot{\theta} + 3H(t) \, \dot{\theta} + \ma^2(t) \, \sin (\theta) = 0 \, .
\label{eq:AxionFieldEq}
\end{equation}
The general form of this equation does not possess analytic solutions. In the early Universe $\ma \ll H$ and the system is an overdamped oscillator. We can integrate the differential equation with boundary conditions $\theta(\mrm{t}{i}) = \thetai$ and $\dot{\theta}(\mrm{t}{i}) = 0$, where the value~$\thetai$ of the axion field is called the \textit{initial misalignment angle}.

At later times, around the time when $\ma \sim H$, the system becomes critically damped and the field starts to oscillate. In the regime of $\ma \gg H$, the axion field oscillations are adiabatic while their amplitude continuously decreases due to Hubble damping. In other words, at some time~$t_\star$, the axion field evolution is guaranteed to eventually enter an epoch where $\left|\theta(t)\right|\ll 1$ for all $t>t_\star$. In this harmonic limit, (\ref{eq:AxionFieldEq})~is a damped harmonic oscillator, and the field evolution is very well described by the WKB~approximation~\cite{1201.5902}
\begin{equation}
	\theta (t) \propto \left[\ma(t) \; a^3(t)\right]^{-\frac{1}{2}} \cos \left( \int^{t}_{t_\star} \! \ma(\tau) \, \dd \tau + \delta_\star \right) , \label{eq:wkbansatz}
\end{equation}
where we can match the solution at time $t_\star$ with the phase $\delta_\star$. This limiting form of the axion scalar field oscillations can be used to demonstrate that they behave like cold DM at sufficiently late times by averaging~(\ref{eq:wkbansatz}) over an oscillation period to find the effective behaviour of the energy density, $\ede$, such that~\cite{1201.5902}
\begin{equation}
\langle \ede \rangle \propto \ma \; a^{-3} \, . \label{eq:axionscaling}
\end{equation}
Once the mass becomes constant, this is the scaling behaviour of cold matter. We can rephrase this statement for non-relativistic axions to see that the (averaged) \emph{comoving} axion number density is conserved:
\begin{equation}
n^\text{com}_a \equiv \frac{\rho a^3}{\ma} \simeq \frac{\ede a^3}{\ma} \, . \label{eq:conscomnumdens}
\end{equation}
In fact, there is an adiabatic invariant in the harmonic limit of $\sin (\theta) \simeq \theta$, which has been studied in the literature for QCD~axions and ALPs~\cite{Abbott:1982af,1986_turner_axiondensity,Frieman:1991qv}. However, if the initial misalignment angle $\thetai \sim \order(1)$, then the adiabatic invariant is not sufficient to describe the system, and anharmonic effects arising from the full potential must be taken into account. These have been estimated and calculated by several authors~\cite{1986_turner_axiondensity,Lyth:1991ub, astro-ph/9405028,0806.0497,0910.1066}.

The energy density in axions today can hence be parametrised by an overall transfer function $F$, with the properties that $F$ is bounded from below, and becomes constant in the harmonic limit $|\thetai|\ll 1$~\cite{1986_turner_axiondensity,astro-ph/9405028,0806.0497}. This allows us to conveniently write the axion energy density in the Universe today as
\begin{equation}
\rho_a = F\left(|\thetai|\, ;\, \beta,\, \Tcrit,\,  \mazero\right) \,  \thetai^2 \, , \label{eq:axionedetoday}
\end{equation}
where we explicitly denote the dependence of the transfer function~$F$ on the values of the axion parameters. For a given set of these parameters, the transfer function -- and hence the axion relic density -- could be determined once and used in subsequent calculations. For sampling over the parameter space, it is necessary to repeat this process for each set of parameters (or to tabulate the results on a sufficiently dense grid ahead of the scan).

We have thus far only considered the \emph{homogeneous} field equation~(\ref{eq:AxionFieldEq}). It is not obvious that this is sufficient for determining the energy density in axions today. Let us consider the two possible scenarios in a cosmology with inflation: either PQ symmetry breaks \emph{after} inflation, or it breaks \emph{before} inflation ends.

In the first case, the Universe consists of a large number of causally disconnected ``bubbles'' where the initial misalignment angles take random values from a uniform distribution on the interval $- \otherpi$ to $\otherpi$~\cite{1986_turner_axiondensity}. As a consequence, the initial misalignment angle~$\thetai$ effectively becomes a function of space. Nonetheless, the resulting overall energy density in axions is fixed, because it can be calculated as the average over all the misalignment angles in all the patches. As~$\thetai$ is constant within the patches, we may use~(\ref{eq:axionedetoday}) to calculate the result:
\begin{equation}
	\rho_a^\text{post-inf} = \frac{1}{2\otherpi} \, \int_{-\otherpi}^{\otherpi} \! F\left(|\thetai|\right)\thetai^2 \, \dd\thetai = \frac{1}{\otherpi} \, \int_{0}^{\otherpi} \! F\left(|\thetai|\right) \thetai^2 \, \dd\thetai \, . \label{eq:rhoApost}
\end{equation}

In the second case, where the PQ symmetry breaks \emph{before} the end of inflation, one causally connected patch gets blown up to at least the size of the observable Universe. The axion field is therefore homogeneous, with a random initial value in the interval $(-\otherpi,\otherpi]$. The stochastic nature of the initial misalignment angle gives rise to a physically motivated prior probability for~$\thetai$ in a Bayesian analysis, which will be discussed in Sec.~\ref{sec:results}.

A major issue with this scenario is perturbations from inflation, which will affect the initial misalignment angle by some amount that depends on the energy scale of inflation \cite[see e.g.][]{Lyth:1991ub}. It has been argued that this scenario is therefore finely-tuned to a degree that can be considered ``worse'' than the Strong~CP problem itself~\cite{0911.0418,0911.0421}. Since the necessary additional implementation of inflationary models is beyond the scope of this work, we neglect the issue of field fluctuations during inflation.

\subsubsection{Topological defects}
Because axions arise from the breaking of a $U(1)$ symmetry, topological defects known as cosmic strings and domain walls can appear~\cite{Kibble:1976sj,book_kolbturner}. Decay of these defects to axions will increase the axion DM density. They have been studied extensively in the literature, but the relative importance of their contribution is still not firmly established~\cite{Davis:1985pt,Davis:1986xc,Harari:1987ht,Hagmann:1990mj,astro-ph/9311017,astro-ph/9403018,hep-ph/9807428,hep-ph/9811311,hep-ph/0012361,1202.5851,1012.4558,1012.5502,1412.0789,1509.00026,1707.05566,1708.07521,1806.04677}.

Some authors have parametrised the topological defect contribution as $\OmegaA = \OmegaA^\mathrm{realign}(1+\alpha)$, where $\OmegaA^\mathrm{realign}$ is the contribution from realignment. Including $\alpha$ as a nuisance parameter in our work would remove much of the predictability of the model, and would assume similar scaling of the energy density contributions. As we consider PQ symmetry to break before inflation, we assume that contributions from topological defects are diluted away during inflation~\cite{astro-ph/0610440}.

\subsubsection{Thermal creation}
Axions are also thermally produced in the early Universe~\cite{1987_Turner,book_kolbturner}. The resulting abundance is however rather dependent on the additional new field content associated with the axion.

For the QCD~axion, the Boltzmann equations for the least model-dependent processes give useful estimates. The contributions of low-temperature processes such as $\pi +\pi \rightleftharpoons \pi + a$ have been computed~\cite{1987_Turner,book_kolbturner,hep-ph/9306216,hep-ph/0504059}, including all combinations of pionic states $\pi$. Thermal axion production during reheating has also been studied~\cite{hep-ph/0203221,1008.4528,1310.6982}. As hadronisation would not yet have taken place by the time of reheating, these calculations consider processes like axion-gluon interactions, i.e.\ $g + g \rightleftharpoons g + a$. Recent calculations also considered model-dependent processes with quarks~$q$, i.e. $q + \bar{q} \rightleftharpoons g + a$ and $q + g \rightleftharpoons q + a$, which can be dominant in the early Universe~\cite{1801.06090}.

A significant abundance of thermal axions modifies the effective number of relativistic species, which can be used to set limits on the axion mass~\cite{hep-ph/0312154,0705.2695,0706.4198,0803.1585,1004.0695,1307.0615,1403.4852,1503.00911,1507.08665}. Generally speaking, thermally-produced QCD~axions with a mass of more than $\order(\si{\electronvolt})$ are hot DM with a relic abundance comparable to that of neutrinos and photons. Those axion models are therefore excluded, as they would exceed the observational bounds on the fraction of DM that can be hot~\cite[e.g.][]{book_kolbturner, astro-ph/0610440, 1510.07633}. However, such bounds have some dependence on the choice of cosmological datasets~\cite{1503.00911}, and the limits can be relaxed if the Universe had a non-standard thermal history~\cite{0711.1352}.

\section{Axion models and theoretical uncertainties}\label{sec:axionmodels}
Let us now discuss the various axion models that we consider in this work with a focus on the interactions between axions and other particles. We already saw that axions acquire a mass, so that they interact gravitationally. QCD~axions also interact via the strong interaction. For generic ALPs, however, it is not clear \textit{a priori} what the PQ charges of SM particles are, nor whether the axion interacts with a given particle at tree level or only at higher order in perturbation theory.

In order to study axion phenomenology and to identify useful observables, axions have been studied in an effective field theory~(EFT) framework~\cite{Kaplan:1985dv,1985_srednicki_axioneft,1986_georgi_axioneft}, which can be adjusted to a given energy scale and scenario. One can then apply exclusion limits on the effective interactions to specific axion models that establish a relation between the effective couplings and the fundamental axion parameters; examples of such models include the so-called KSVZ model~\cite{1979_kim_ksvz,1980_shifman_ksvz} and the DFSZ model~\cite{1980_zhitnitsky_dfsz,1981_dine_dfsz}.

One important consequence of the fundamental shift symmetry is that axions can only directly interact with matter via derivative couplings. Below the electroweak scale, the effective Lagrangian for axion-SM interactions is
\begin{align}
	\lagrangian_a = -\sum_{f}^{} \frac{g_{aff}}{2m_f} \, \bar{f} \, \gamma^\mu\gamma_5f \, \del_\mu a - \frac{\gagg}{4} \; aF_{\mu\nu}\widetilde{F}^{\mu\nu} - \frac{\mrm{\alpha}{S}}{8\otherpi} \; \frac{a}{\fa}G_{\mu\nu}^b\widetilde{G}^{\mu\nu, \, b} \, , \label{eq:axioneft}
\end{align}
where $a$ is the axion field, $\widetilde{F}$ and $\widetilde{G}$ are the duals of electromagnetic and strong field strengths and $g_{aff}$ and $\gagg$ are effective coupling constants of mass dimension~$-1$. In principle, $f$ runs over all SM fermions with mass~$m_f$, but the couplings most relevant for axion searches are those to electrons and nucleons, the latter arising from matching the EFT to chiral perturbation theory~\cite{1986_georgi_axioneft}.\footnote{The same matching also gives rise to interactions between axions and mesons, which can in particular induce flavour-changing rare decays. While of phenomenological relevance for heavier ALPs~\cite{Dolan:2014ska}, these processes do not lead to relevant constraints on the parameter space considered in the present work.} Also note that the Lagrangian~\refeq{eq:axioneft} is flavour-diagonal, which is not necessarily the case in all models. While we do not consider the possibility of flavour non-diagonal interactions in this paper, the resulting models can be phenomenologically interesting, rather predictive, and within reach of existing and future experimental searches~\cite[e.g.][]{10.1103/PhysRevLett.49.1549,1612.05492,1612.08040}.

\begin{figure}
	\centering
	\begin{tikzpicture}[
	sibling distance=3.25cm,
	level distance=2cm,
	treenode/.style = {draw, rectangle, font=\footnotesize, rounded corners=2pt, align=center, inner sep=3pt, execute at begin node=\setlength{\baselineskip}{1.25em}},
	lvl1/.style={treenode},
	lvl2/.style={treenode, yshift=-0.9cm}]

\node [treenode] (a0) {\highl{\textsf{GeneralALP (7)}}\\ 7~parameter model\\ $\fa$, $\mazero$, $\gagg$, $\gaee$, $\beta$, $\Tcrit$, $\thetai$}
	child { node [lvl1, xshift=-0.2cm] (a1) {\highl{\textsf{QCDAxion (4+4)}}\\ Free parameters:\\ $\fa$, $E/N$, $C_{aee}$, $\thetai$\\ Nuisance parameters:\\ $\LambdaQCD$, $\Tcrit$, $\beta$, $\widetilde{C}_{a\gamma\gamma}$}
		child { node [lvl2, xshift=-0.2cm] (a11) {\highl{\textsf{DFSZAxion-I (3+4)}}\\ Free parameters:\\ $\fa$, $\tan (\beta^\prime)$, $\thetai$\\ Nuisance parameters:\\ $\LambdaQCD$, $\Tcrit$, $\beta$, $\widetilde{C}_{a\gamma\gamma}$\\ Fixed parameters:\\ $E/N=8/3$} }
		child { node [lvl2] (a12) {\highl{\textsf{DFSZAxion-II (3+4)}}\\ Free parameters:\\ $\fa$, $\tan (\beta^\prime)$, $\thetai$\\ Nuisance parameters:\\ $\LambdaQCD$, $\Tcrit$, $\beta$, $\widetilde{C}_{a\gamma\gamma}$\\ Fixed parameters:\\ $E/N=2/3$} }
		child { node [lvl2, xshift=0.2cm] (a13) {\highl{\textsf{KSVZAxion (3+4)}}\\ Free parameters:\\ $\fa$, $E/N$, $\thetai$\\ Nuisance parameters:\\ $\LambdaQCD$, $\Tcrit$, $\beta$, $\widetilde{C}_{a\gamma\gamma}$\\ Fixed parameters:\\ $C_{aee}$} }
	}
	child { node [lvl1, xshift=0.2cm] (a2) {\highl{\textsf{ConstantMassALP (5)}}\\  Free parameters:\\ $\fa$, $\Lambda$, $C_{a\gamma\gamma}$, $C_{aee}$, $\thetai$\\ Fixed parameters:\\ $\Tcrit$ irrelevant, $\beta \equiv 0$}
	}
;

\end{tikzpicture}
	\caption{Family tree of axion models in \gambit. The numbers in brackets refer to the number of model parameters; $(n+m)$ indicates $n$ (largely unconstrained) fundamental parameters of the model and $m$ (typically well-constrained) nuisance parameters.}
	\label{fig:AxionModelTree}
\end{figure}
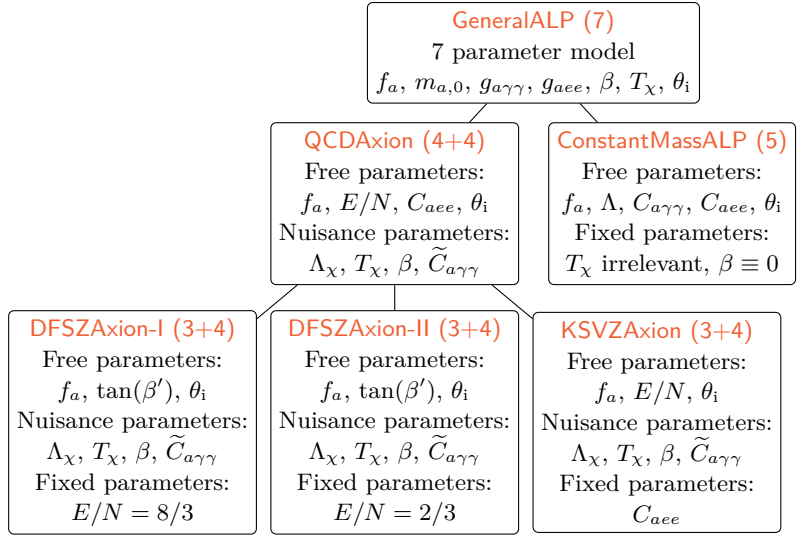
We now turn to the specific axion models implemented in \gambit (see \reffig{fig:AxionModelTree}). We describe these models in the following subsections, while we leave the discussion of the parameter ranges and prior distributions for Sec.~\ref{sec:results}. Note that the models presented here are only a subset of the many general and more specific ALP models studied in the literature; detailed overviews can be found in review articles~\cite{1510.07633,1801.08127}.

Models in \gambit are defined as collections of named parameters. All relevant observables for a model must be computable from these parameters. Models can have relationships to other models, allowing parameter combinations in one model to be translated to equivalent combinations in another model or to an alternative parameterisation of the same model. This is achieved by adding new ``children'' to the family tree of a more general model and defining a translation between the two models. This translation may, e.g., fix the values of some of the ``parent'' parameters or compute them as functions of (possibly new) parameters in the child model. This ensures flexibility in the choice of the independent model parameters to work with in any given theory calculation, allowing the most convenient definitions to always be used for any calculation. More details on the general implementation of models in \gambit can be found in Ref.~\cite{gambit}.

\subsection{General ALP model}
\label{sec:generalalpmodel}
We define a new family of \gambit models for axions and ALPs. On top of the family tree is the \genalp model, with seven parameters. This model describes an effective Lagrangian that is a subset of~(\ref{eq:axioneft}),
\begin{equation}
	\lagrangian_a^\mathrm{int} = -\frac{\gagg}{4} aF_{\mu\nu}\widetilde{F}^{\mu\nu} - \frac{\gaee}{2m_e} \bar{e}\gamma^\mu\gamma_5e\del_\mu a \, .\label{eq:ax:lagrange}
\end{equation}
In this study we do not include couplings of axions to nucleons. However, future versions of \gambit may include additional interactions such as these, subject to the availability of interesting observables and constraints.

We employ the rescaled field value $\theta = a/\fa$, assume the canonical cosine potential~(\ref{eq:potential}) for all types of axions,\footnote{Note that this is a non-trivial assumption, as the potential could be any periodic function of~$\theta$; more general potentials have been invoked to e.g.\ construct models of inflation~\cite{1401.5212}. Allowing a different shape of the potential would also require a modification of the relic density calculator, presented in Appendix~\ref{app:axioneom}.} and take
\begin{equation}
\ma(T) =  \mazero
\begin{cases}
\hfil 1 \hfil & \mathrm{if \; } T \leq \Tcrit \\
\left ( \frac{\Tcrit}{T} \right )^{\beta/2} & \mathrm{otherwise}
\end{cases} \, . \label{eq:axionmass}
\end{equation}
A summary of all the model parameters can be found in Table~\ref{tab:GeneralALP}.

\begin{table}
	\renewcommand{\arraystretch}{1.1}
	\caption{Parameters for the family of axion models. For dimensionful quantities, we also give the units with which they are defined within \gambit. The first section refers to parameters that are part of the \genalp model. The second section lists the parameters used in child models of the \genalp.\label{tab:GeneralALP}}
	\centering
	\small
	\begin{tabular}{@{}cll@{}}
		\toprule
		\multicolumn{1}{@{}l}{\textbf{Parameter}} & \textbf{Description} & \textbf{Comment} \\
		\midrule
		$\gagg$ & Effective axion-photon coupling & Units of $\si{\per\GeV}$\\
		$\gaee$ & Effective axion-electron coupling & \\
		$\fa$ & Axion decay constant & Units of $\si{\giga\electronvolt}$\\
		$ \mazero$ & Axion zero-temperature mass & Units of $\si{\electronvolt}$\\
		$\Tcrit$ & Transition temperature in the broken power law for $\ma$ & Units of $\si{\mega\electronvolt}$ \\
		$\beta$ & Exponent of the broken power law for $\ma$ & $\beta > 0$\\
		$\thetai$ & Initial misalignment angle & $-\otherpi < \thetai \le \otherpi$\\
		\midrule
		$\Lambda$ & Breaking scale of the residual symmetry & Units of $\si{\mega\electronvolt}$\\
		$\caee$ & Axion-electron coupling & \\
		$\LambdaQCD$ & Zero-temperature topological susceptibility, $\LambdaQCD^4 \equiv \chi (T=0)$ & Units of $\si{\mega\electronvolt}$\\
		$E/N$ & Anomaly ratio in QCD axion models & \\
		$\caggtilde$ & Axion-photon coupling contribution from axion-meson mixing & \\
		$\tan (\beta^\prime)$ & Ratio of the two Higgs vacuum expectation values & \\
		\bottomrule
	\end{tabular}
\end{table}

\subsection{QCD axion models}\label{sec:mod:QCDAxion}
\begin{table}
	\renewcommand{\arraystretch}{1.15}
	\caption{Nuisance parameters for the QCD axion model. We only quote best-fit estimates of~$\beta$ and $\Tcrit$, as the likelihood includes correlations between them (cf.\ Sec.~\ref{sec:mod:QCDAxion}). \label{tab:QCDAxion}}
	\centering
	\begin{tabularx}{0.77\textwidth}{@{}clX@{}}
		\toprule
		\multicolumn{1}{@{}l}{\textbf{Parameter}} & \textbf{Value} & \textbf{Comment} \\
		\midrule
		$\LambdaQCD$ & $\SI{75.5(5)}{\mega\electronvolt}$ & Ref.~\cite{1511.02867}\\
		$\caggtilde$ & $\num{1.92(4)}$ & Ref.~\cite{1511.02867}\\
		$\beta$ & $\hat{\beta}\approx 7.94$ & \multirow{2}{*}{$\left\{\begin{tabular}{l}\text{Via lattice QCD nuisance} \\\text{likelihood based on Ref.~\cite{1606.07494}}\end{tabular}~\right.$}\\
		$\Tcrit$ & $\hat{T}_\chi \approx \SI{147.0}{\mega\electronvolt}$ & \\
		\bottomrule
	\end{tabularx}
\end{table}

The \qcdaxion model is a child of the \genalp model. It is inspired by the original QCD~axion models, which solve the Strong~CP problem. The scale of the explicit breaking of the shift symmetry by instanton-like effects is therefore linked to the QCD scale. This connection can be exploited to uniquely determine the parameters $ \mazero$, $\beta$ and $\Tcrit$. However, there are uncertainties from theory, experiment and lattice QCD simulations that should be taken into account. We treat these as nuisance parameters; Table~\ref{tab:QCDAxion} provides an overview.

\begin{figure}
	\centering
	{
		\includegraphics[width=0.46340504409369\linewidth]{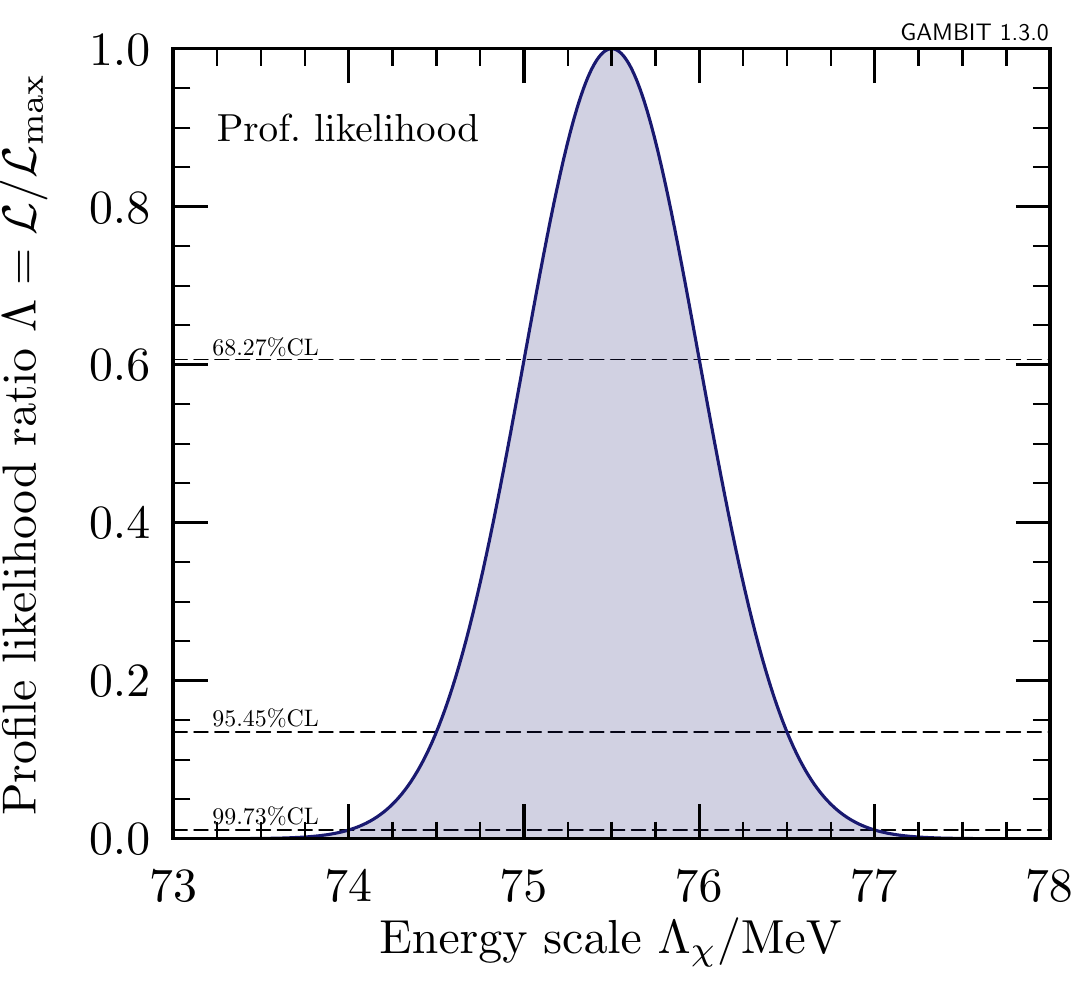}
		\hfil
		\includegraphics[width=0.52659495590631\linewidth]{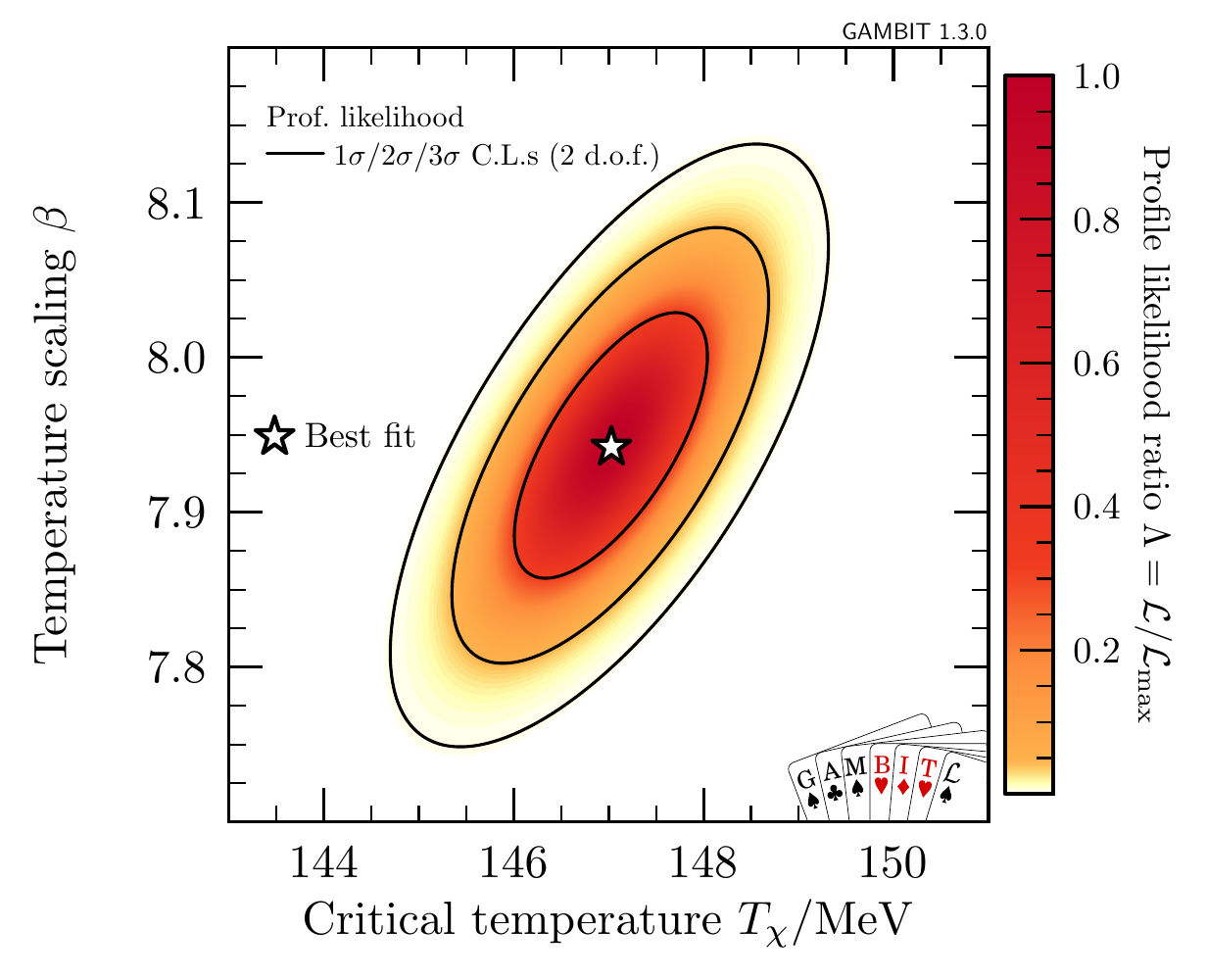}
	}
	\caption{Nuisance likelihoods for the scale $\LambdaQCD$~(\textit{left}, from direct theory calculations~\cite{1511.02867}) and $\beta$ and $\Tcrit$~(\textit{right}, from lattice QCD~\cite{1606.07494}). Note that~$\beta$ is correlated with $\Tcrit$ when the full lattice QCD results are taken into account.\label{fig:QCDAxion:Nuisance}}
\end{figure}

Exploiting the connection to QCD, we can replace the parameter $ \mazero$ of the \genalp model with the energy scale~$\LambdaQCD$, defined via the topological susceptibility at zero temperature, $\LambdaQCD^4 \equiv \chi(T=0)$, such that
\begin{equation}
	\mazero \equiv \frac{\LambdaQCD^2}{\fa} \, .
\end{equation}
To determine~$\LambdaQCD$, we use first principle calculations of the zero-temperature axion mass~\cite{1511.02867}. We include these results via a Gaussian likelihood
\begin{equation}
  \lnL =  - \frac{1}{2} \frac{(\LambdaQCD-\hat{\Lambda}_\chi)^2}{\sigma_{\LambdaQCD}^2} \, ,
\end{equation}
where $\hat{\Lambda}_\chi = \SI{75.5}{\MeV}$ and $\sigma_{\LambdaQCD} = \SI{0.5}{\MeV}$ are the most likely values for $\LambdaQCD$ and its uncertainty from Ref.~\cite{1511.02867}.\footnote{The authors of Ref.~\cite{1606.07494} find a comparable result of $\LambdaQCD = \SI[parse-numbers=false]{75.6 \pm 1.8 (stat) \pm 0.9 (sys)}{\MeV}$ using lattice~QCD methods.} This is shown in the left panel of \reffig{fig:QCDAxion:Nuisance}.

The two parameters $\beta$ and $\Tcrit$ can then be constrained by using, e.g., the full lattice QCD results from Table~S.7 in the supplementary material of Ref.~\cite{1606.07494}. We construct a likelihood function by performing a chi-squared fit to the $N=20$ data points,
\begin{equation}
  \lnL =  - \frac{1}{2} \sum\limits_{i=1}^{N} \frac{(X-\hat{X}_i)^2}{\sigma_{X_i}^2} \, ,
\end{equation}
where $X = \log_{10}\left[\chi(T \, | \, \beta, \, \Tcrit) / \chi(T=0)\right]$ is the logarithm of the normalised topological susceptibility, and $\hat{X}_i$ and $\sigma_{X_i}^2$ are its value and uncertainty for the $i$th data point.\footnote{As the uncertainty quoted in Ref.~\cite{1606.07494} also includes an uncertainty on the value of~$\LambdaQCD$, we divide the topological susceptibility~$\chi=\fa^2\ma^2$ by their best-fit value for $\chi(T=0)$, and remove this uncertainty by assuming simple error propagation.} In \reffig{fig:QCDAxion:Nuisance}, we show the two-dimensional profile likelihood for~$\beta$ and~$\Tcrit$. These parameters show a clear correlation. This is expected, as a higher transition temperature implies a steeper slope in the temperature-dependent branch of the axion mass (corresponding to larger~$\beta$) in order to maintain a good fit to the shape of~$\ma(T)$, as determined by lattice QCD. We provide the best-fit values in Table~\ref{tab:QCDAxion}.

We note that the fit to our functional form for the QCD~axion mass~(\ref{eq:axionmass}) captures the temperature dependence established by lattice~QCD well everywhere except in the region around~$T=\Tcrit$. This is because~(\ref{eq:axionmass}) is not smooth there and the overall fit is poor ($\chi^2 = 55.7$ for 18 d.o.f., which corresponds to a $p$-value of about $10^{-5}$). However, the disagreement stems only from a narrow temperature range, and has no impact on any of our results. Excluding the only data point in that region improves the fit to an acceptable level ($\chi^2 = 21.6$ for 17 degrees of freedom, which corresponds to a $p$-value of about 0.2).

We also replace the axion-electron coupling with the model-dependent factor~$\caee$
\begin{equation}
	\gaee = \frac{m_e}{\fa} \; \caee\, , \label{eq:qcdaxioncouplings1}
\end{equation}
and the axion-photon coupling with the model-dependent ratio of the electromagnetic and colour anomalies~$E/N$
\begin{equation}
	\gagg = \frac{\mrm{\alpha}{EM}}{2\otherpi \fa}\left(\frac{E}{N} - \caggtilde\right) \, ,\label{eq:qcdaxioncouplings2}
\end{equation}
where $\caggtilde$ is the model-independent contribution from axion-pion mixing. We use the value obtained in Ref.~\cite{1511.02867} for~$\caggtilde$ and include it as a simple nuisance likelihood,
\begin{equation}
	\lnL =  - \frac{1}{2} \frac{(X-\hat{X})^2}{\sigma_X^2} \, ,
\end{equation}
with $X=\caggtilde$ and $\hat{X}$ and $\sigma_X$ again the most likely value and its uncertainty (see Table~\ref{tab:QCDAxion}).

Finally, we want to emphasise some subtle considerations about the coupling strengths of the QCD~axion model. The possible ratios $E/N$ are rational numbers arising from group theoretical considerations. In this study, we sample over~$E/N$ as if it were a continuous parameter, as the possible rational numbers that it can take on are quite densely spaced along the real line, at least over the range of values that we consider. The sometimes so-called ``classical axion window'' considers a canonical, small and somewhat arbitrarily-defined range of couplings for the prototypical axion models~\cite{Kaplan:1985dv,hep-ph/9506295,hep-ph/9802220,PDG17}, arising from only quite a small range in $E/N$. It has recently been pointed out that the range of choices can, indeed, be extended to include more possibilities~\cite{1610.07593,1705.05370,1709.06085,1712.04940}. To assess the whole range of various axion models, we use the minimum and maximum values for~$E/N$ from Table~IV in Ref.~\cite{1705.05370}, so that $E/N \in [-4/3,\, 524/3]$. These values arise from a systematic study of DFSZ- and KSVZ-type axion models, where the additional heavy quarks in KSVZ-type models have cosmologically safe lifetimes and do not introduce Landau poles below the Planck scale. Furthermore, in DFSZ-type models, the number of Higgs doublets may go up to the maximum of nine.

Note that~(\ref{eq:qcdaxioncouplings2}) implies the possibility of having~$\gagg < 0$ within the valid range for $E/N$. Note however that all the likelihood functions that we use in this paper only depend on the absolute value of~$\gagg$. We therefore plot only~$|\gagg|$ in our results, even though we do scan over parameter values that lead to negative couplings in the range $-3.25 \lsim E/N - \caggtilde \leq 0$.

\subsubsection{KSVZ models}\label{sec:mod:KSVZAxion}
The archetypical axion model is the KSVZ model~\cite{1979_kim_ksvz,1980_shifman_ksvz}, where the SM is supplemented by one or more electrically neutral, heavy quarks. In our implementation, the \ksvz is a child model of the \qcdaxion, where the anomaly ratio, $E/N$, is a free parameter. In these models, axions have no tree-level interactions with fermions. However, there is a loop-induced coupling to electrons due to the axion-photon interaction, which -- in the absence of a leading order contribution -- must be taken into account~\cite{1985_srednicki_axioneft}:
\begin{equation}
	\gaee \approx \frac{3\mrm{\alpha}{EM}^2}{2\otherpi} \left[\frac{E}{N}\ln\left(\frac{\fa}{m_e}\right) - \caggtilde \ln\left(\frac{\tilde{\Lambda}}{m_e}\right) \right] \frac{m_e}{\fa}\, , \label{eq:ksvzgaee}
\end{equation}
where $\tilde{\Lambda}\sim\SI{1}{\giga\electronvolt}$ is the QCD confinement scale. Several previous works have used this expression with $\tilde{\Lambda}=\SI{1}{\giga\electronvolt}$~\cite{1307.1488,1708.02111}, even though this quantity is not uniquely defined. We too assume $\tilde{\Lambda}=\SI{1}{\giga\electronvolt}$, relying on the fact that any deviations enter only logarithmically. Although the anomaly ratio in the original KSVZ model was $E/N=0$~\cite{1979_kim_ksvz,1980_shifman_ksvz}, other assignments are possible. As in Ref.~\cite{1708.02111}, we will consider four different \ksvz models: $E/N=0,\, 2/3,\, 5/3,$ and $8/3$.

\subsubsection{DFSZ models}\label{sec:mod:DFSZAxion}
In contrast to the KSVZ model, DFSZ models are obtained by adding an additional Higgs doublet to the SM~\cite{1980_zhitnitsky_dfsz,1981_dine_dfsz}. This results in direct axion-electron interactions. One can define two manifestations of this model, often called \dfszI and \dfszII. The couplings in these two models are given by
\begin{align}
	\begin{array}{lll}
		\caee = \sin^2 (\beta^\prime)\left/3\right., \quad \phantom{.} & E/N = 8/3  \quad \phantom{.}& (\dfszI) \, \\
		\caee = \left[1-\sin^2 (\beta^\prime) \right]\left/3\right., \quad \phantom{.} & E/N = 2/3 \quad \phantom{.} & (\dfszII) \,
	\end{array} \label{eq:dfsz:caee}
\end{align}
where $\tan (\beta^\prime)$ is the ratio of the two Higgs vacuum expectation values~\cite{1985_srednicki_axioneft}. Perturbative unitary requires $0.28 < \tan(\beta^\prime) < 140$. It is therefore convenient to replace the parameter~$\caee$ in the \qcdaxion model by $\tan (\beta^\prime)$.

\subsection{ALP models with constant mass}
As a template model for ALPs that arise as pseudo-Nabu-Goldstone bosons but do not have a temperature-dependent mass, we define the \simpalp model. This is mainly for convenience in studies where we want to parametrically explore the coupling space whilst keeping the inverse dependence on~$\fa$, but are not interested in a temperature-dependent mass.

For \simpalp models, $\Tcrit$ is irrelevant because $\beta = 0$, reducing the total number of parameters to five. Similar to \qcdaxion models, we replace the ALP mass with a pseudo-Nambu-Goldstone scale~$\Lambda$ and introduce dimensionless coupling constants $\cagg$ and $\caee$, consistent with the other models:
\begin{equation}
 \mazero = \frac{\Lambda^2}{\fa} \, , \qquad \gagg = \frac{\mrm{\alpha}{EM}}{2\otherpi \fa} \; \cagg \, , \qquad \gaee = \frac{m_e}{\fa} \; \caee \, .
\end{equation}

\section{Observables, experiments and likelihoods}\label{sec:obsexplike}
\begin{table*}
	\caption{Overview of the likelihood functions that we employ in this paper (in the order they are discussed). \label{tab:likelihoods}}
	\renewcommand{\arraystretch}{1.15}
	\centering
	\footnotesize
	\begin{tabularx}{0.99\linewidth}{@{}lXl@{}}
		\toprule
		\textbf{Likelihood/Observable} & \textbf{Comments} & \textbf{References}\\
		\midrule
		QCD nuisance parameters & Table~\ref{tab:QCDAxion} & \cite{1511.02867,1606.07494}; Sec.~\ref{sec:mod:QCDAxion}\\
		\midrule
		ALPS final limits & vacuum and argon data & \cite{1004.1313}; Sec.~\ref{sec:lsw}\\
		CAST 2007 & vacuum data, CCD (2004) & \cite{hep-ex/0702006}; Sec.~\ref{sec:helioscopes:CAST07} \& Appendix~\ref{app:solarmodelintegration}\\
		CAST 2017 & vacuum data, all detectors & \cite{1705.02290}; Sec.~\ref{sec:helioscopes:CAST17} \& Appendix~\ref{app:solarmodelintegration}\\
		RBF &  & based on~\cite{DePanfilis:1987dk,Wuensch:1989sa}; Sec.~\ref{sec:haloscopes:rbf}\\
		UF &  & based on~\cite{Hagmann:1990tj}; Sec.~\ref{sec:haloscopes:uf}\\
		ADMX 1998 -- 2009 &  & based on~\cite{astro-ph/9801286,Asztalos:2001tf,astro-ph/0310042,astro-ph/0603108,0910.5914}; Sec~\ref{sec:haloscopes:admx98}\\
		ADMX 2018 &  & based on~\cite{1804.05750}; Sec.~\ref{sec:haloscopes:admx18}\\
		DM relic density & built-in calculator; Sec.~\ref{sec:cdmconstraints} and Appendix~\ref{app:axioneom} & limits~\cite{Planck15cosmo}\\
		H.E.S.S. & axion-photon conversion in galactic cluster magnetic fields & based on~\cite{1311.3148}; Sec.~\ref{sec:astro:hess}\\
		Supernova limits & axion-photon conversion in magnetic fields of the Milky Way & based on~\cite{1410.3747}; Sec.~\ref{sec:astro:SN1987A}\\
		$R$~parameter & & \cite{1406.6053,1512.08108}; Sec.~\ref{sec:astro:rparameter}\\
		\midrule
		\multicolumn{3}{l}{\textit{Stellar cooling hints (optional likelihood)}}\\
		White dwarf cooling hints & only considered in Sec.~\ref{sec:results:coolinghints} & \cite{1205.6180,1211.3389,1605.06458,1605.07668}; Sec.~\ref{sec:coolinghints}\\
		\bottomrule
	\end{tabularx}
\end{table*}

Table~\ref{tab:likelihoods} gives an overview of the observables and likelihood functions that we use in this paper. In what follows, we give details of the experimental data, computational methods and likelihood implementations that they employ.

\subsection{Light-shining-through-wall experiments}\label{sec:lsw}
Light-shining-through-wall~(LSW) experiments shine laser light through a magnetic field onto an opaque material, and attempt to detect it on the other side. A photon may convert into an axion within the magnetic field before the material, pass through it as an axion, and convert back into a photon in the magnetic field on the other side~\cite{VanBibber:1987rq,Anselm:1986gz,Anselm:1987vj}. Examples of experiments based on this technique are ALPS~\cite{0905.4159,1004.1313} and OSQAR~\cite{0712.3362,1506.08082}.

The predicted number of photons on the opposite side of the wall to the laser source is
\begin{equation}
	s = \mrm{\epsilon}{tot} \; \prob{\gamma\to a\to\gamma}\frac{P}{\omega_\gamma} \, \mrm{t}{obs} \, , \label{eq:lswsig}
\end{equation}
where~$\epsilon_\mathrm{tot}$ is the detector efficiency, $P$ is the laser power, $\omega_\gamma$ the laser energy, and $\mrm{t}{obs}$ the observation period. $\prob{\gamma\to a\to\gamma}$ is the probability of a photon converting into an axion and back, in an appropriately aligned magnetic field of length~$L$ and strength~$B$. It is given by $\prob{\gamma\to a\to\gamma}=\pazocal{P}^2\left(\gamma\to a\right)$ with~\cite{1004.1313}
\begin{equation}
	\prob{\gamma\to a} = \prob{a\to\gamma} = \left(\frac{\gagg B L}{2}\right)^2 \, \mathrm{sinc}^2\left(\frac{M^2 L}{2 \omega_\gamma}\right) \, , \label{eq:convprobability}
\end{equation}
where $M^2 \equiv \ma^2/2 + \omega_\gamma^2(n-1)$, $\mathrm{sinc}(x) \equiv \sin(x)/x$, $\ma$ is the axion mass, $\omega_\gamma$ is the photon energy, and $n$ is the refractive index of the medium in the experimental setup ($n=1$ for vacuum).

Our LSW likelihood is based on the final results of the ALPS-I experiment, using data for both evacuated and gas-filled magnets~\cite{1004.1313}. The ALPS Collaboration took data in ``frames'' of $\mrm{t}{f}=\SI{1}{\hour}$ each, binning physical pixels of the detector into $3\times 3$-pixel blocks. ALPs refer to these simply as ``pixels'' of area $\SI{42}{\micro\metre}\times\SI{42}{\micro\metre}$. The collaboration searched their data frames for cosmic rays and signatures of other systematics, over a wide region around the single pixel where most of the laser light would fall in the absence of a wall, referred to as the ``signal pixel''.

ALPS adjusted the raw \texttt{ADU} values (electron counts) of the signal pixel in an attempt to account for the average background in surrounding pixels. These reduced \texttt{ADU} (\texttt{ADUred}) values are obtained by subtracting the average \texttt{ADU} values across all pixels in the region surrounding the signal pixel, from the \texttt{ADU} value of the signal pixel. Doing this for every frame where the laser was on (``signal frames'') and off (``background frames''), ALPS constructed histograms of \texttt{ADUred} values for both signal and background. By fitting Gaussian functions to these histograms, they were able to estimate $\hat{b}$ and $\sigma_b$, the expected value and standard deviation of \texttt{ADUred} for the background, as well as $\hat{o}$ and $\sigma_o$, the equivalent quantities for signal frames. An example and more details can be found in Fig.~2 of Ref.~\cite{1004.1313} (see also Ref.~\cite{0905.4159}).

From (\ref{eq:lswsig}), the expected signal~$s$ from photon-axion-photon production per frame of ALPS-I data (with $B=\SI{4.98}{\tesla}$, $L=\SI{4.2}{\metre}$, and $\omega_\gamma = \SI{2.33}{\eV}$) is
\begin{equation}
	s \simeq \num{12.1} \; \mrm{\epsilon}{tot} \; \left(\frac{P}{\SI{1096}{\watt}}\right) \left[ \frac{\gagg}{\SI{E-7}{\GeV^{-1}}} \; \mathrm{sinc}\left(\frac{M^2 L}{2 \omega_\gamma} \right) \right]^4 \, .
\end{equation}

\begin{figure}
	\centering
	\includegraphics[width=0.618\linewidth]{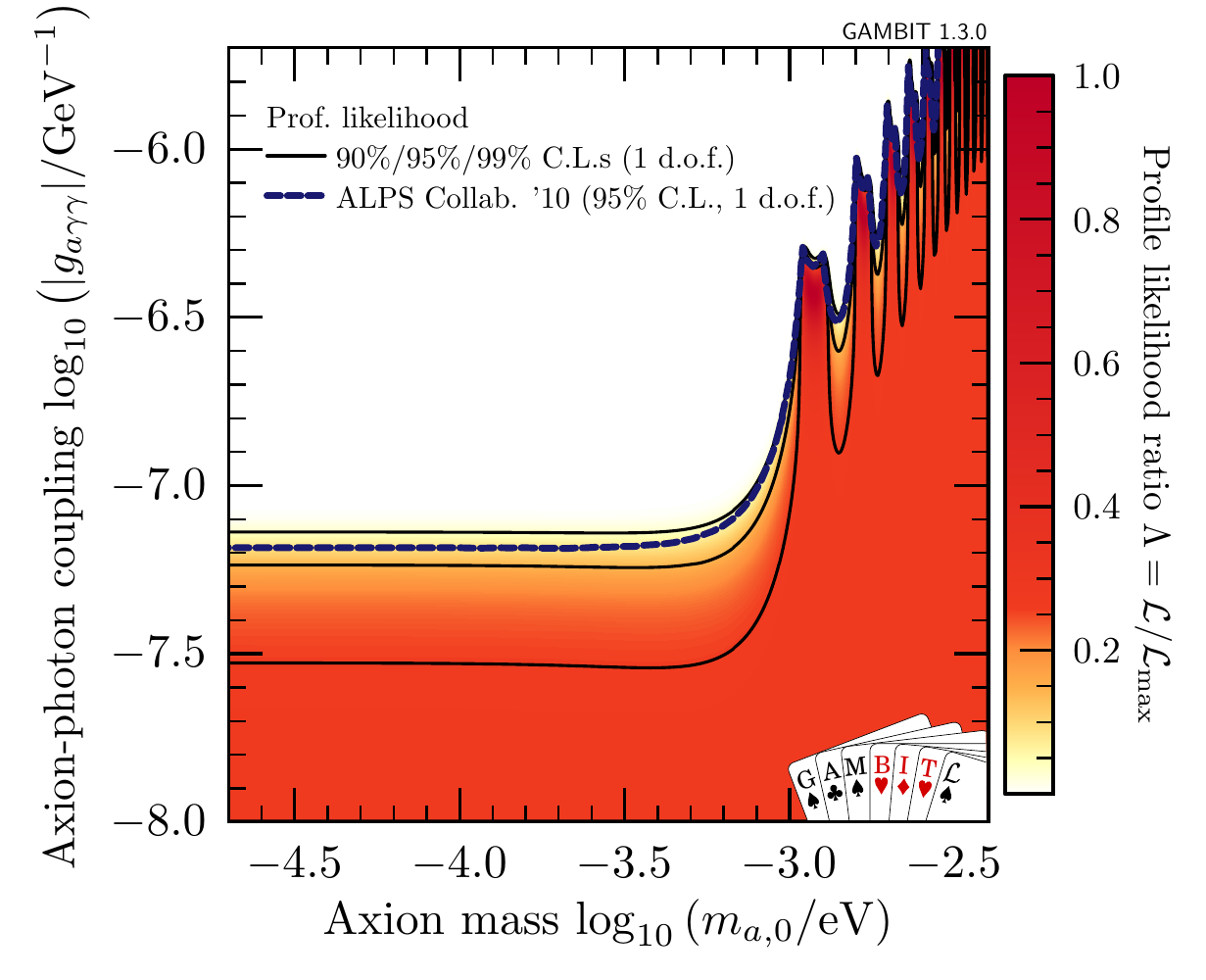}
	\caption{Validation of our implementation of ALPS result for limits on the axion-photon coupling. We show the 90\%, 95\% and 99\% C.L.s compared to the envelope of their strongest vacuum or gas results. The difference between our likelihood and the limit published by ALPS arises mostly due to the fact that we combine both likelihoods, rather than taking the envelope.\label{fig:validation:lsw}}
\end{figure}
Apart from fluctuations in the experimental performance~$\epsilon_\mathrm{tot}$, which amounts to an uncertainty of $\Delta \mrm{\epsilon}{tot}/\mrm{\epsilon}{tot}\approx6\%$, the experimental parameters are known to sufficient precision to be fixed to their reference values. We incorporate the uncertainty on $\epsilon_\mathrm{tot}$ into the estimate of the signal prediction, $\sigma_{s}= s \Delta \mrm{\epsilon}{tot}/\mrm{\epsilon}{tot}$. As a result of the \texttt{ADU} reduction procedure, the measured signal and background estimators are non-integer numbers. We test the signal-plus-background hypothesis with a Gaussian likelihood:
\begin{equation}
	\lnLvar{i} = - \frac{1}{2} \frac{\left[s_i - \left(\hat{o}_i - \hat{b}_i \right) \right]^2}{\sigma_{s,i\phantom{b}}^2+\sigma_{\hat{o}-\hat{b},i}^2} \, .
\end{equation}
We add together the log-likelihoods for data sets $i=1$ to~$3$, where two data sets (five and six frames compared to 122 and 47~background frames, respectively) consist of vacuum data and one data set (eight frames compared to 155~background frames) consists of argon gas data.\footnote{Courtesy of Axel Lindner, private communication.}

We show the resulting exclusion limits in \reffig{fig:validation:lsw} and compare with the envelope of the strongest vacuum or argon gas limits in Ref.~\cite{1004.1313}. The differences between the published results and our implementation are due to fact that we combine the likelihoods instead of just adopting the more constraining of the two, and also because the authors of Ref.~\cite{1004.1313} used the Feldman-Cousins method~\cite{physics/9711021}, assuming Gaussianity and physical signals $s>0$. Considering the vacuum data alone, the method of Ref.~\cite{physics/9711021} gives a slightly stronger limit than our log-likelihood ratio method (\SI{6.5e-8}{\GeV^{-1}} at 90\% confidence limit (C.L.) in the low mass limit, as compared to \SI{6.9e-8}{\GeV^{-1}} in our implementation). By combining the vacuum and argon likelihoods however, our final limit is somewhat stronger: $\gagg <$ \SI{5.8e-8}{\GeV^{-1}} at 90\% C.L. at low masses.

\subsection{Helioscopes}\label{sec:helioscopes}
Axion helioscopes attempt to detect axions produced by interactions in the Sun by observing the solar disc with a ``telescope'' consisting of a long magnet contained in an opaque casing~\cite{1983_Sikivie,1985_Sikivie,1989_Bibber}. Solar-produced axions would pass through the casing, convert into photons in the field of the magnet, and be observed in a detector behind the magnet.

Multiple processes in the Sun can produce axions: resonant production in the oscillating electric field of the solar plasma, non-resonant production in solar magnetic fields, and emission from the interaction of electrons with photons, nuclei or one another. The predicted axion-induced photon flux at Earth therefore depends on the assumed solar model, and on the couplings of the axion to both photons and electrons \cite[see e.g.][]{1310.0823}.

The dominant process for axion-photon interactions is Primakoff production~\cite{10.1103/PhysRev.81.899,10.1103/PhysRev.152.1295}, where photons are resonantly converted into axions in the presence of an atomic nucleus. The rate at which axions of energy~$E$ can be produced in a plasma from photons of the same energy is given by~\cite{Raffelt:1985nk,Raffelt:1987np}
\begin{equation}
	\Gamma_{\gamma\to a} = \gagg^2 \; \frac{\kappa_s^2T}{32\otherpi} \left[ \left( 1 + \frac{\mrm{\kappa}{s}^2}{4E^2} \right) \; \ln \left( 1 + \frac{4E}{\mrm{\kappa}{s}^2} \right) -1 \right] \, . \label{eq:ggA}
\end{equation}
Here, the inverse screening length is given in the Debye-H\"uckel approximation by
\begin{equation}
	\mrm{\kappa}{s}^2 = \frac{4\otherpi \mrm{\alpha}{EM}}{T} \left(n_e + \sum_{j}^{} Z_j^2n_j\right) \, , \label{eq:screeningscale}
\end{equation}
with $n_e$ denoting the electron number density, and $n_j$ and $Z_j$ representing the number density and charge, respectively, of the $j$th nucleus. Note that the number densities and temperature, and hence the conversion rate, vary with the radial position $r$. Using the expression for the axion-photon conversion probability~\refeq{eq:convprobability}, the differential photon flux at the detector is~\cite{hep-ex/0702006}
\begin{align}
	\frac{\dd \Phi (E)}{\dd E} &= 2\otherpi \, \prob{a\to\gamma} \, \int_{0}^{r_s}\! r \; \varphi_{\gamma\to a}(E, \, r) \, \dd r \, , \label{eq:diffsolarphotonflux} \\
	\text{where} \quad \varphi_{\gamma\to a}(E, \, r) &= \frac{R_\sol^3}{2\otherpi^3 D_\sol^2} \int_{r}^{1} \! \frac{\rho}{\sqrt{\rho^2-r^2}} \; \frac{E\sqrt{E^2 - \mrm{\omega}{pl}^2(\rho)}}{\ee^{E/T(\rho)}-1} \; \Gamma_{\gamma\to a}(E, \, \rho) \, \dd \rho \, , \label{eq:solardiscflux}
\end{align}
with~$\rho$ a dimensionless radial co-ordinate in the Sun, and $r$ a dimensionless radial co-ordinate on the solar disc on the sky. The quantity~$D_\sol$ is the (average) Sun-Earth distance, which we take to be one astronomical unit, and $\mrm{\omega}{pl}^2(\rho)=4\otherpi\mrm{\alpha}{EM} n_e(\rho)/m_e$ is the plasma frequency calculated from the electron number density~$n_e(\rho)$ and electron mass~$m_e$. The upper limit of outer integral, $r_s$, controls how much of the inner part of the image of the solar disc on the detector is included in the analysis. This need not always be~1, as the signal-to-noise ratio can be maximised by introducing a cut-off $r_s < 1$ \cite[see e.g.][]{hep-ex/0702006}.

The contribution to the solar axion flux from axion-electron interactions can be taken into account by including the additional interaction rate~$\Gamma_{e\to a}(E, \, \rho)$ in~\refeq{eq:solardiscflux}. However, these contributions are not so straightforward to calculate from first principles. This is due to narrow free-bound transition lines~\cite{Dimopoulos:1986kc,Dimopoulos:1986mi,0807.3279,1007.1833}, axion bremsstrahlung~\cite{Zhitnitsky:1979cn,Krauss:1984gm,Raffelt:1985nk} and Compton scattering~\cite{Mikaelian:1978jg,Fukugita:1982ep,Fukugita:1982gn}. To include these contributions in our signal prediction, we use tabulated data for the axion-electron spectrum provided in Ref.~\cite{1310.0823}.

The spectrum in Ref.~\cite{1310.0823} was computed with the 2009 iteration of the AGSS09met solar model~\cite{0909.2668}, which is based on photospheric abundances for non-refractory species and meteoritic abundances for refractory elements~\cite{AGSS}. The AGSS09met model is thus the default solar model in \gambit, and the one that we use throughout the rest of this paper. We however make use of its latest iteration~\cite{1611.09867} in preference to the earlier version wherever possible, such as when computing the axion flux from axion-photon interactions. In the limit of small axion mass, the predicted flux from axion-photon interations deviates by no more than 4.4\% between solar models, with the greatest difference ocurring between the GS98~\cite{10.1023/A:1005161325181,1611.09867} and most recent AGSS09met models~\cite{1611.09867}.

Full details of our integration routines for the axion-photon and axion-electron contributions, as well as the options available for the inclusion of solar models for axion physics in \gambit, can be found in Appendix~\ref{app:solarmodelintegration}.

\subsubsection{CAST results 2007}\label{sec:helioscopes:CAST07}
The first of our two CAST likelihoods is based on the CCD results published in 2007~\cite{hep-ex/0702006} (CCD detector data from the 2004 vacuum run in Table~1 of Ref.~\cite{hep-ex/0702006}). The other data in Ref.~\cite{hep-ex/0702006} are less constraining; including them only improves the upper limit by~1\%~\cite{hep-ex/0702006}. We therefore do not provide separate likelihoods for the other runs.

Although the 2007 CAST analysis was based on the solar model of Ref.~\cite{astro-ph/0402114}, and a follow-up analysis~\cite{1302.6283} on axion-electron interactions was based on a different model~\cite{TurckChieze:2001ye}, here we use the AGSS09met model of Ref.~\cite{1611.09867}. For both the axion-photon and axion-electron interactions, we integrate the total flux over all 20~energy bins (from \SIrange{0.8}{6.8}{\kilo \electronvolt}), taking into account the observation time, effective area, and detector efficiency (see Appendix~\ref{app:solarmodelintegration}). In this case $r_s \approx 0.23$ for the axion-photon contribution. As we use the interpolated spectrum for the axion-electron contribution (calculated for $r_s = 1$), we can only rescale the resulting flux by an overall factor in order to estimate the flux inside $r_s = 0.23$. We take this number to be~0.826, assuming the findings from the axion-photon interaction also apply in the axion-electron case~\cite{hep-ex/0702006}.

\begin{figure}
	\centering
	{
		\includegraphics[width=0.49\linewidth]{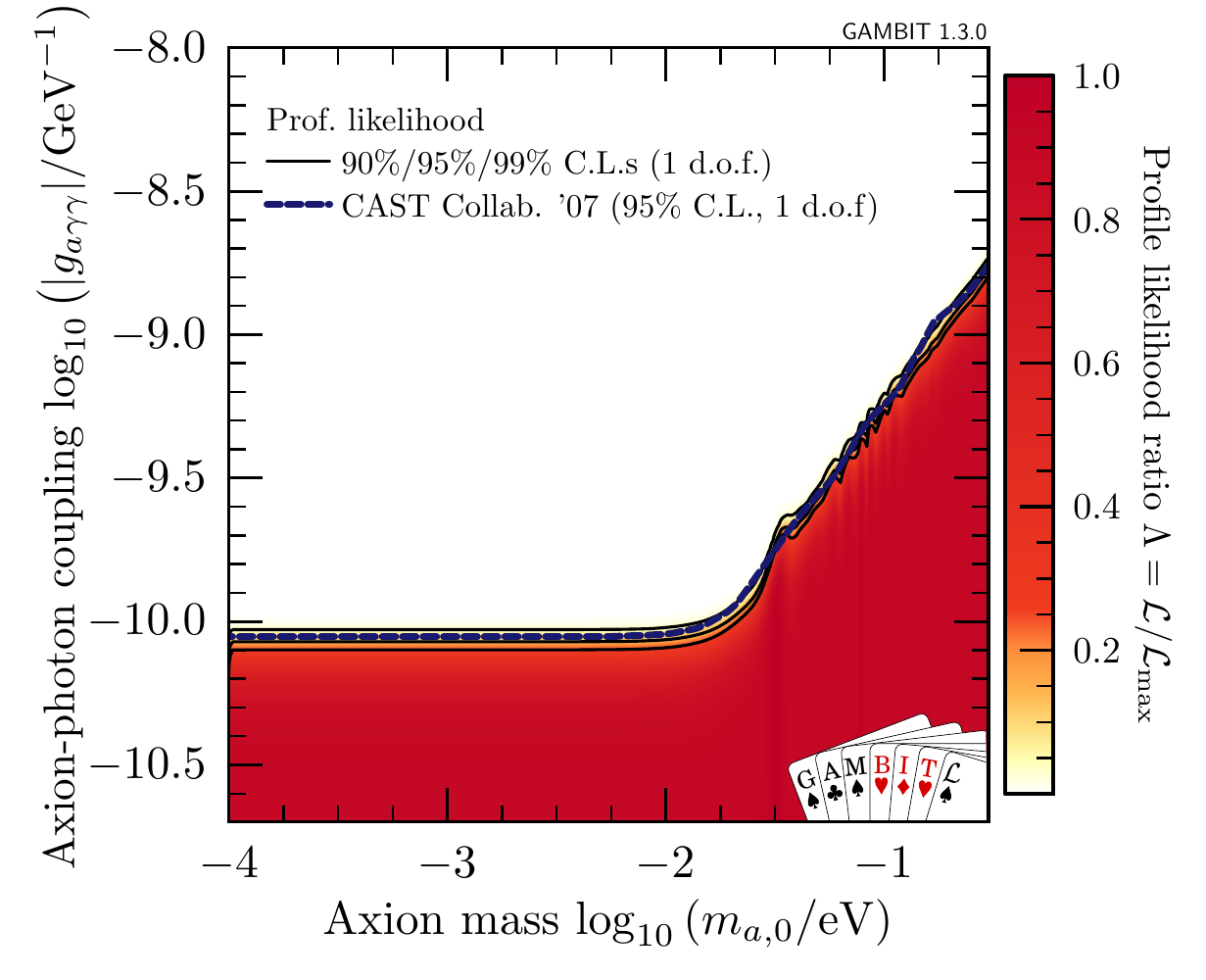}
		\hfill
		\includegraphics[width=0.49\linewidth]{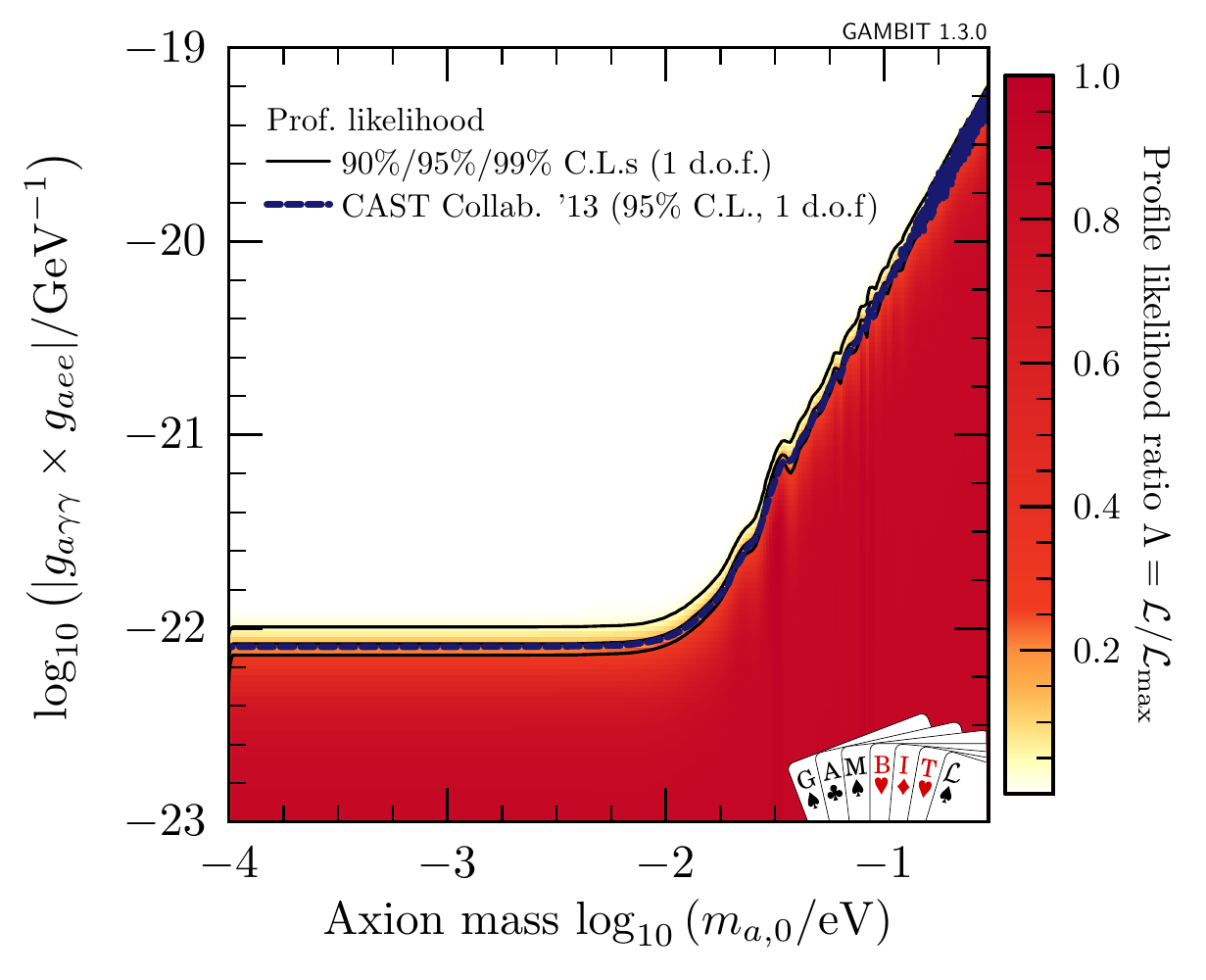}
	}
	\caption{Validation of our implementation of limits arising from the 2007 CAST data \cite{hep-ex/0702006}, including the 2013 re-interpretation in terms of axion-electron couplings \cite{1302.6283}. \textit{Left}: Exclusion limits for the axion-photon coupling~\cite{hep-ex/0702006}. \textit{Right}: Exclusion limits for the product of axion-photon and axion-electron coupling, assuming that axion-electron interactions dominate the axion production inside the Sun. We only make this assumption here to compare our implementation to Ref.~\cite{1302.6283}.\label{fig:validation:cast07}}
\end{figure}

Our implementation follows the CAST analysis~\cite{hep-ex/0702006,1302.6283}, using a Poisson likelihood in each of the 20 energy bins
\begin{equation}
  \lnL = \sum\limits_{i=1}^{20} o_i \ln\left(s_i+b_i\right) - \ln\left(o_i!\right) - (s_i + b_i) \, ,
\end{equation}
where $o_i$, $s_i$ and $b_i$ are respectively the observed number of photons, the number of expected signal photons, and the expected number of background photons based on observations away from the Sun, in the $i$th energy bin. In total, 26~photons were observed in the detector during data-taking compared to 30.9~expected background events. The resulting exclusion limits can be found in \reffig{fig:validation:cast07}.

\subsubsection{CAST results 2017}\label{sec:helioscopes:CAST17}
Our implementation of the latest CAST results~\cite{1705.02290} is analogous to that of the 2007 results, using the signal and expected background counts, exposure and detector efficiencies for the 2017 data.\footnote{I.~Irastorza, private communication.}

To calculate the signal predictions, we integrate the axion-photon and axion-electron fluxes over each of the 10 energy bins from~\SIrange{2}{7}{\kilo\electronvolt}, and then scale the predictions by the effective exposure for each of the ten datasets in Ref.~\cite{1705.02290}.

\begin{figure}
	\centering
	\includegraphics[width=0.618\linewidth]{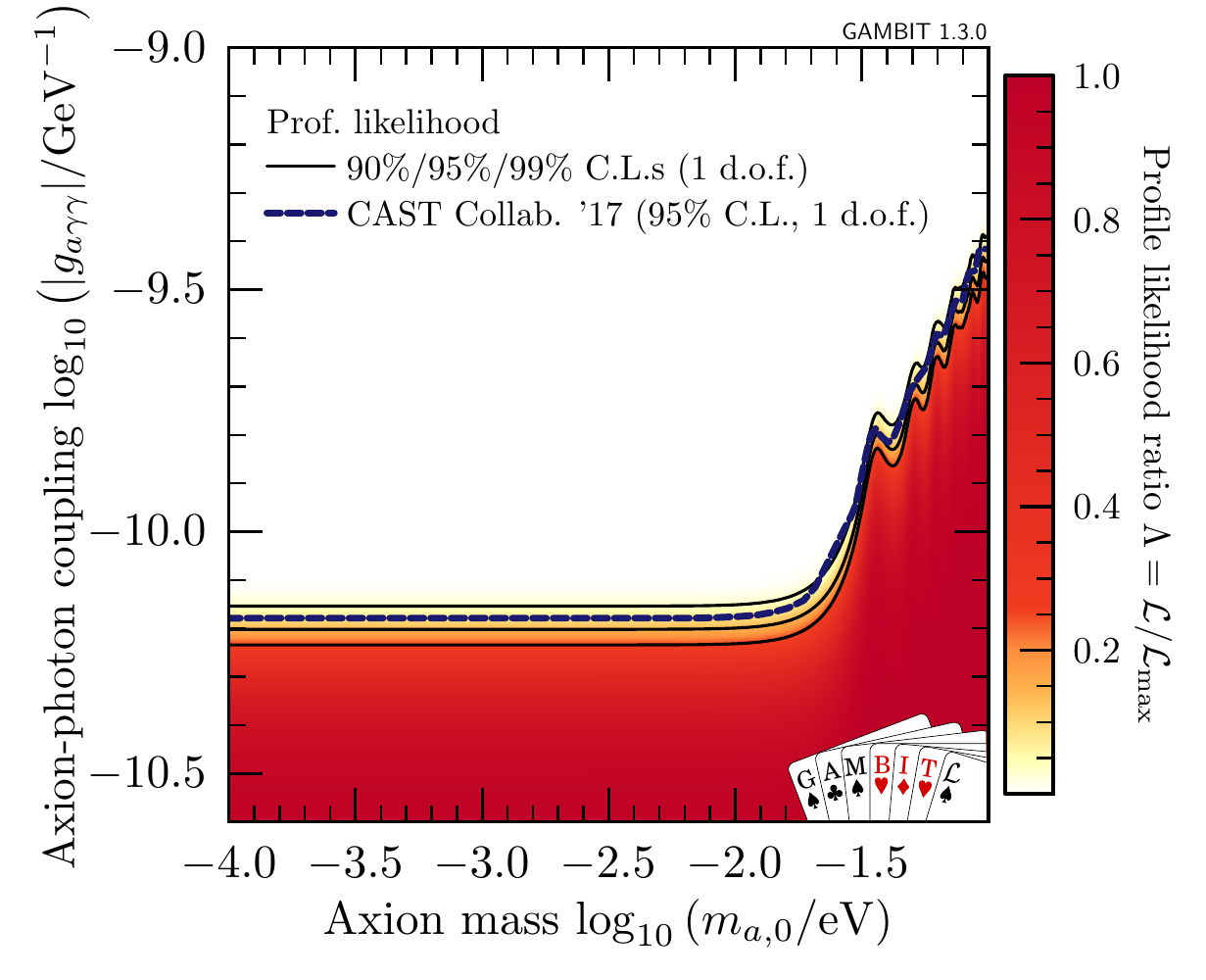}
	\caption{Validation of our implementation of the 2017 CAST limits~\cite{1705.02290}. Differences are mainly due to our simplified implementation of the likelihood function, which does not employ event-level information. \label{fig:validation:cast17}}
\end{figure}

The exclusion limits presented in Ref.~\cite{1705.02290} are based on an unbinned likelihood. Here, we treat each energy bin in each of the ten datasets as a separate counting experiment, combining them into a binned Poisson likelihood
\begin{equation}
	\lnL = \sum\limits_{j=1}^{12}\sum\limits_{i=1}^{10} o_i^{(j)} \ln\left(s_i^{(j)}+b_i^{(j)}\right) + \ln\left(o_i^{(j)}!\right) - (s_i^{(j)} + b_i^{(j)}) \, .
\end{equation}
Here $o_i^{(j)}$, $s_i^{(j)}$ and $b_i^{(j)}$ are respectively the observed number of photons, the expected number of signal events and the expected number of background events in the $i$th energy bin of the $j$th experiment. In total, 226~photons were observed in the detector during data taking compared to 246.6~expected background events.
Our slightly different choice of likelihood function to the original CAST analysis is significantly simpler, because it does not require event-level information -- but it still reproduces the published exclusion limits rather well (see Fig.~\ref{fig:validation:cast17}).

\subsection{Haloscopes (cavity experiments)}\label{sec:haloscopes}
Axion haloscopes are designed to detect DM axions by resonant axion-photon conversion inside a tunable cavity~\cite{1983_Sikivie, 1985_Sikivie}. Microwave cavities are the most sensitive axion experiments in existence, but only cover a small mass range compared to other techniques. The ability of haloscopes to detect axions therefore directly depends on their cosmological abundance, and to a lesser extent, their velocities in the Galactic halo~\cite{2011_Hoskins}. Here we consider only the case where axions are fully virialised within the halo.

The power expected to be converted from axions to photons in a cavity is~\cite{1983_Sikivie,Krauss:1985ub,1985_Sikivie}
\begin{equation}
	P = \gagg^2 \, B_0^2 \, C \, \frac{\rho_{a,\text{ local}} \, V}{\mazero} \, Q \, \min\left(1,\frac{Q_a}{Q}\right) \, , \label{eq:HaloscopeRate}
\end{equation}
where $C$ is the form factor (a dimensionless integral over the $\vc{E}$- and $\vc{B}$-field configuration of the cavity), $B_0$ is magnetic field strength in the cavity, $V$ is its volume, and $Q$ and $Q_a$ are the quality factors of the cavity and axions, respectively. In this context, $Q$~describes the ratio of stored vs dissipated energy of the cavity, while~$Q_a$ is proportional to the axion velocity dispersion (just as~$Q$ effectively characterises the bandwidth of the cavity).

The signal prediction also depends on the local DM density in axions, which we obtain by rescaling the local DM density as
\begin{equation}
  \rho_{a,\text{ local}} = \min\left(\frac{\OmegaA}{\OmegaCDM},1\right) \, \rho_\text{DM, local} \, .
\end{equation}

Obtaining exclusion limits from cavity experiments is often quite involved, generally requiring simulation of the selection procedure of the detector \cite[e.g.][]{Asztalos:2001tf}. Without access to this information, we must approximate the underlying likelihood functions based on the publicly available limits and publications.

\begin{figure}
	\centering
	\includegraphics[width=0.49\linewidth]{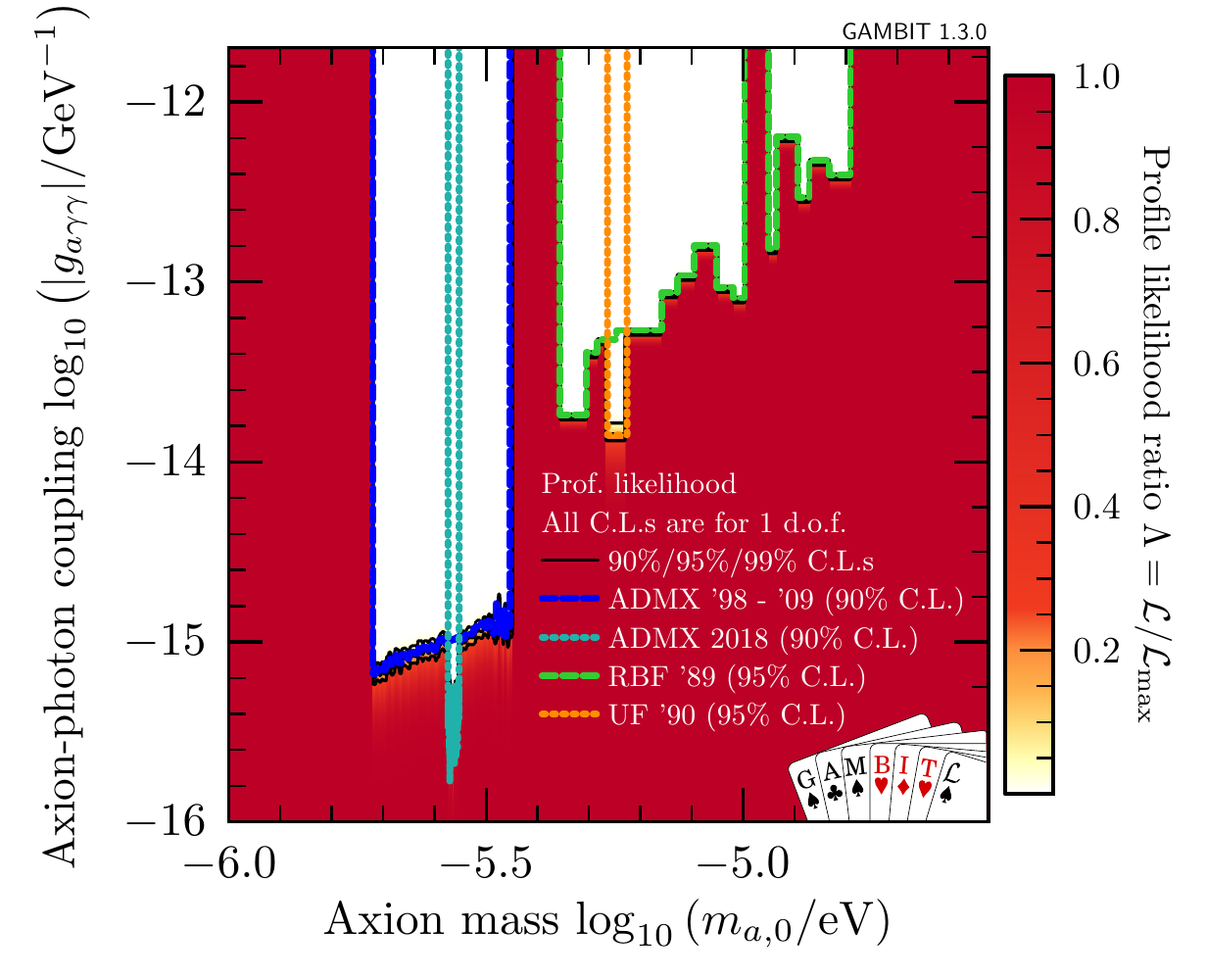}
	\hfill
	\includegraphics[width=0.49\linewidth]{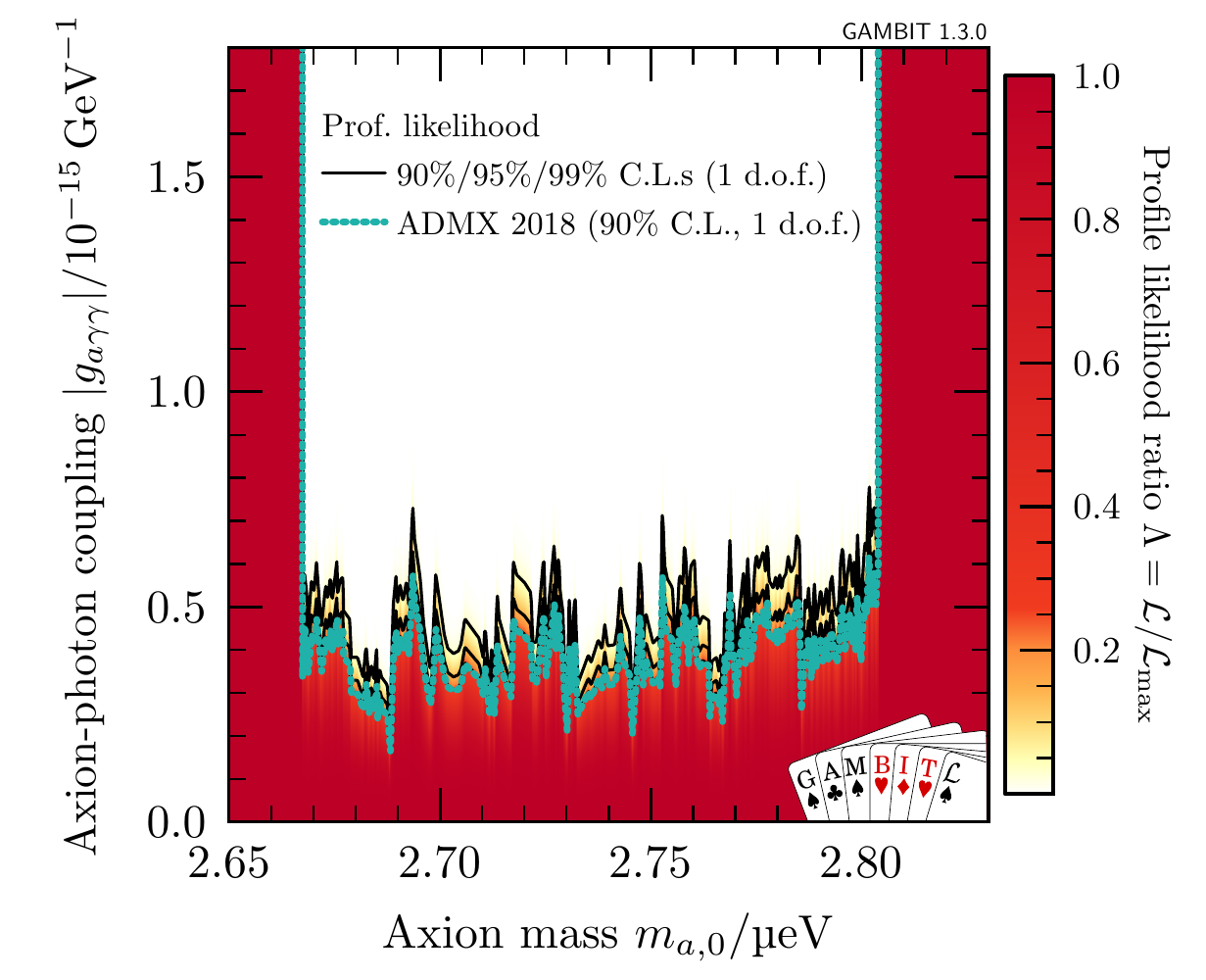}
	\caption{\textit{Left:} Our implementation of haloscope likelihoods compared to the exclusion limits for the RBF~\cite{DePanfilis:1987dk,Wuensch:1989sa}, UF~\cite{Hagmann:1990tj} and ADMX~\cite{astro-ph/9801286,astro-ph/0104200,Asztalos:2001tf,astro-ph/0310042,astro-ph/0603108,0910.5914,1105.4203,1804.05750}  experiments. \textit{Right:} Magnified details of the latest ADMX results~\cite{1804.05750}. \label{fig:validation:haloscope}}
\end{figure}

In the following, we describe our likelihoods for three different haloscope experiments. An overview of the resulting exclusion limits can be found in \reffig	{fig:validation:haloscope}.

\subsubsection{RBF results}\label{sec:haloscopes:rbf}
The Rochester-Brookhaven-Fermi~(RBF) collaboration performed a search for axions using several cavities~\cite{DePanfilis:1987dk,Wuensch:1989sa}. Table~I, Eq.~(26) and Fig.~14 in Ref.~\cite{Wuensch:1989sa} provide useful information for approximating their results.

By~\refeq{eq:HaloscopeRate}, the axion-induced power in a haloscope is proportional to $\rho_{a,\,\text{local}}^{\phantom{2}}\;\gagg^2$, which we define as the ``reduced signal''
\begin{equation}
	s \equiv \rho_{a,\,\text{local}}^{\phantom{2}}\;\gagg^2 \, . \label{eq:genhaloscopesignal}
\end{equation}
The remaining factors from \refeq{eq:HaloscopeRate} are effectively constant across all frequency/mass bins and detectors. The signal is expected to occur in a single frequency bin $i$, which satisfies $\mazero \in [\omega_i, \omega_{i+1})$. Using this definition, we adopt an ansatz for the likelihood function inspired by Eq.~(26) of Ref.~\cite{Wuensch:1989sa}:
\begin{equation}
  \lnL = -\frac{1}{2}\Theta\left(s-a_i\right)\frac{(s-a_i)^2}{b_i^2} \, .
\end{equation}
Here~$a_i$ is a threshold parameter, $b_i$ effectively corresponds to an expected standard deviation of the reduced signal, and~$\Theta$ is the Heaviside step function. The threshold values~$a_i$ arise because RBF manually inspected all candidate frequencies in their data over a certain significance level. The two parameters are related as $a_i=Nb_i$, where $N$ is the number of standard deviations required for manual inspection of a candidate signal.

Although Table~I in Ref.~\cite{Wuensch:1989sa} would allow us to determine~$b_i$, using the central frequency of the bin as well as the 95\% C.L. on the coupling strength, the resulting bins are not quite identical to what is shown in Fig.~14 of the same paper. We therefore determine $a_i$ and $b_i$ for each frequency bin from the limits in Fig.~15 of that paper, assuming $N=4$ in all cases (cf.\ Table~I in the same paper). This leads to limits in 14~bins ranging from $ \mazero = \num{4.4}$~to $\SI{10.1}{\micro\electronvolt}$ and from $ \mazero = \num{11.2}$~to $\SI{16.2}{\micro\electronvolt}$.

\subsubsection{UF results}\label{sec:haloscopes:uf}
While the results from the University of Florida~(UF) collaboration~\cite{Hagmann:1990tj} could be implemented in the same way as the RBF~experiment, the published data do not allow us to infer the threshold parameter~$a_i$ for the one mass bin. However, Eq.~(6) in Ref.~\cite{Hagmann:1990tj} quotes the ``noise power fluctuation'', which we use as a standard deviation~$\sigma_P\approx\SI{2.86e-22}{\watt}$ for the expected power~$P$. We obtain the expected power for each axion model using the information provided in Ref.~\cite{Hagmann:1990tj}, and check that the quoted limit is comparable to the expected sensitivity (which we obtain by assuming that the observed data is equal to the background expectation). The corresponding likelihood function for the single bin from $ \mazero = \num{5.4}$ to $\SI{5.9}{\micro\electronvolt}$ and signal~$s$ from~\refeq{eq:genhaloscopesignal} is given by
\begin{equation}
  \lnL = -\frac{1}{2}\left(\frac{P(s)}{\sigma_P}\right)^2 \, .
\end{equation}

\subsubsection{ADMX results 1998--2009}\label{sec:haloscopes:admx98}
The procedure used by the ADMX Collaboration for setting limits in the absence of a detection is highly customised for the experiment~\cite{Asztalos:2001tf}. Unfortunately, the necessary information for fully implementing their numerous results~\cite{astro-ph/9801286,astro-ph/0104200,Asztalos:2001tf,astro-ph/0310042,astro-ph/0603108,0910.5914} is not available. Similar to the RBF likelihood, we therefore use the reduced signal~\refeq{eq:genhaloscopesignal} and the following ansatz for the likelihood:
\begin{equation}
	\lnL = -\frac{1}{2}\Theta\left(s/s_i-a\right)\frac{(s/s_i-a)^2}{b^2} \, .
\end{equation}
Here $s_i$ is the known limit in the $i$th frequency/mass bin, and $a$ and $b$ are free parameters that have to be determined by a fit to published exclusion curves. We do this using the exclusion limits at three confidence levels published in Ref.~\cite{astro-ph/9801286}, and by assuming that the shape of the likelihood function is representative of the shape in all other ADMX bins over the range from $\mazero = \num{1.90}$ to $\SI{3.65}{\micro\electronvolt}$. Doing so results in $a= 0.0131$ and $b= 0.455$.

\subsubsection{ADMX results 2018}\label{sec:haloscopes:admx18}
In their recent publication, the ADMX collaboration increased the sensitivity of their setup, which is now able to rule out some DFSZ-type models~\cite{1804.05750} in the range $2.66\le \mazero \le\SI{2.81}{\micro\electronvolt}$. We approximate the likelihood of this result using the 90\% C.L. limits in Fig.~4 of Ref.~\cite{1804.05750}, for the Maxwell-Boltzmann velocity distribution (consistent with the model of the halo velocity distribution that we assume for analysing the results of other searches).

Because the experimental setup changed compared to the previous runs, we do not employ the shape parameters from Sec.~\ref{sec:haloscopes:admx98} for the 2018 dataset. Instead, as with the UF experiment we assume that ADMX saw no signal events, approximating the 2018 likelihood as
\begin{equation}
	\lnL = -\frac{1}{2} \frac{s^2}{\mrm{\sigma}{eff}^2(\mazero, \, s)} \, .
\end{equation}
Here, the effective standard deviation is given by $\mrm{\sigma}{eff}^2 = \mrm{\sigma}{stat}^2(\mazero) + \mrm{\sigma}{sys}^2(s)$. We infer the statistical contribution~$\mrm{\sigma}{stat}$ by setting the log-likelihood at the observed values of the limits in Fig.~4 of Ref.~\cite{1804.05750} to that corresponding to a 90\% C.L. for one degree of freedom. We read off $\mrm{\sigma}{stat}$ at 194~different masses, and interpolate between them linearly for intermediate mass values (ignoring the narrow region from \SIrange{2.7302}{2.7307}{\micro\eV} where the ADMX limits do not apply, due to radio interference~\cite{1804.05750}). We add the systematic uncertainty of~$13\%$ of the signal prediction (quoted in Ref.~\cite{1804.05750}) in quadrature with the statistical uncertainty.

\subsection{Dark matter relic density}\label{sec:cdmconstraints}
\begin{figure}
	\centering
	{
		\includegraphics[width=0.49\linewidth]{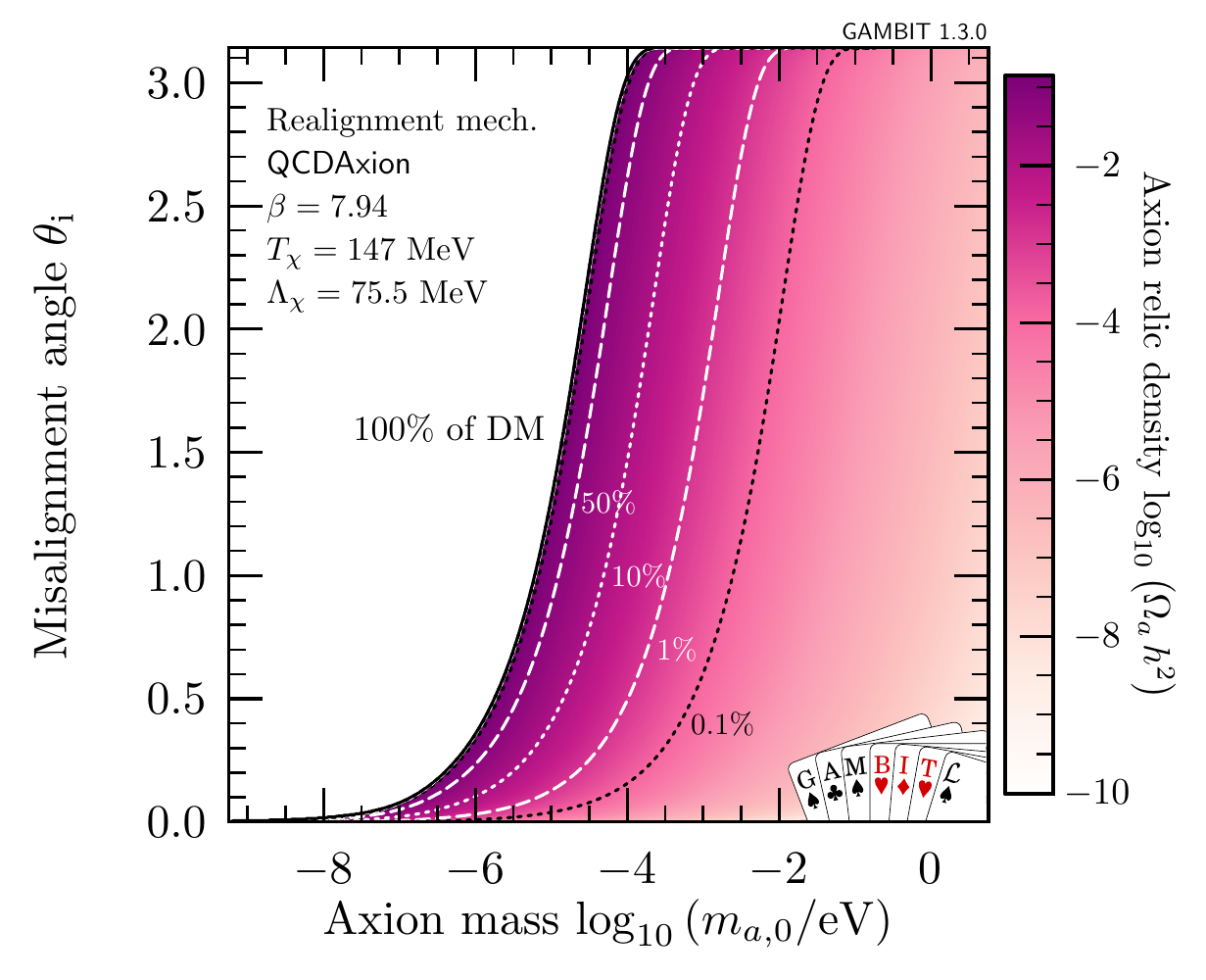}
		\hfil
		\includegraphics[width=0.49\linewidth]{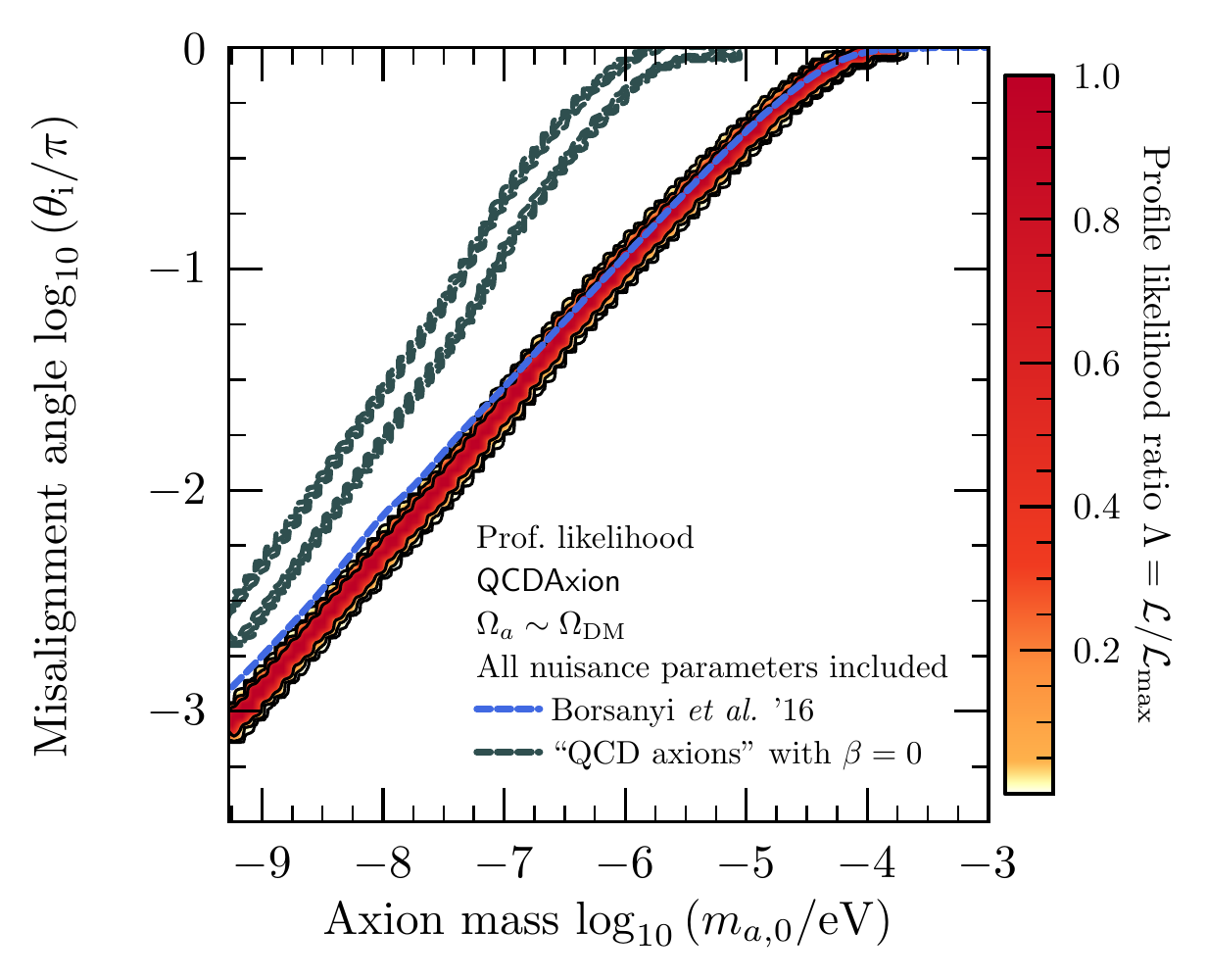}
	}
	\caption{\textit{Left:} Realignment energy density in QCD axions today as a function of mass and initial misalignment angle (fixing $\beta$, $\Tcrit$, and $\LambdaQCD$ to their best-fit values for comparison). To guide the eye, we have included dotted and dashed lines to indicate where axions make up certain fractions of the DM relic density. \textit{Right:} Band of $\mazero$-$\thetai$ combinations (from \diver) to get the correct DM density (including $\beta$, $\Tcrit$, and $\LambdaQCD$ as nuisance parameters). We show the results of Ref.~\cite{1606.07494} for comparison and also include the hypothetical case of a ``temperature-independent QCD axion'' (with $\LambdaQCD$ as a nuisance parameter, but $\beta = 0$ and $\Tcrit$ irrelevant) as an example.\label{fig:validation:realignment}}
\end{figure}

The realignment mechanism gives an axion contribution to the observed cold DM relic density. We solve~(\ref{eq:AxionFieldEq}) numerically as a function of temperature to obtain $\OmegaA$, the fraction of the critical energy density in axions today; details can be found in Appendix~\ref{app:axioneom}. The resulting energy densities for the \qcdaxion are shown in \reffig{fig:validation:realignment}, similar to the presentation in other works~\cite[e.g.]{0903.4377}. We reiterate that in this paper, we do not consider other contributions to the relic density than vacuum realignment.

The relic density of DM is very well constrained by the most recent \textit{Planck} analysis~\cite{Planck15cosmo}. We employ a Gaussian likelihood with the central value and standard deviation from~\cite{Planck15cosmo} ($\OmegaCDM h^2 = 0.1188$, $\sigma_\mathrm{exp} = 0.0010$), combining the experimental uncertainty in quadrature with a further 5\% theory uncertainty,\footnote{We adopt this entirely heuristic number from the default value in \darkbit~\cite{DarkBit}. The estimated sub-percent numerical systematic uncertainty of our code is smaller, but the theoretical uncertainty due to the possibility of non-standard cosmologies can be much larger \cite[e.g.][]{0912.0015}. Several authors have considered scenarios designed to avoid overproducing axions, including entropy dilution~\cite{Dine:1982ah,10.1016/0370-2693(83)90727-X,10.1016/0370-2693(85)90763-4,10.1016/0370-2693(87)90115-8}, inflationary models~\cite{10.1016/0370-2693(83)90148-X,10.1103/PhysRevLett.60.1899,1507.08660,1709.01090}, and hidden magnetic monopoles~\cite{1511.05030}.} $\sigma_\mathrm{theo} = 0.05 \, \OmegaA h^2$, to give
\begin{equation}
	\lnL = -\frac{1}{2} \frac{\left(\OmegaA h^2-\OmegaCDM h^2\right)^2}{\mrm{\sigma}{exp}^2+\mrm{\sigma}{theo}^2} \, . \label{omegalnl}
\end{equation}
\GB offers two options for this likelihood: a detection or an upper limit. These allow us to demand either that axions are responsible for all DM, or only a fraction. For the upper limit, we simply set $\lnL=0$ for $\OmegaA<\OmegaCDM$ in~\eqref{omegalnl}. Except where we state otherwise, we show results based on the upper limit option.

\subsection{Astrophysical probes}
Astrophysical systems can provide significant additional constraints on axions, especially the axion-electron and axion-neutron couplings, which are not well constrained by other probes. Due to their weak interactions with matter, axions can efficiently transport energy across large distances in free space, or through stellar matter, thereby influencing stellar structure and evolution.

Intriguingly, a number of astrophysical systems appear to exhibit an unexplained mechanism of energy transport, which might be due to ALPs: white dwarfs display apparently anomalous cooling rates (Sec.~\ref{sec:coolinghints}) or deviations in the shape of their luminosity function~\cite{1406.7712,1805.00135} and highly-energetic gamma rays seem to experience significantly less attenuation through intergalactic space than might be expected~\cite{0707.4312,0712.2825,1001.0972,1106.1132,1201.4711,1302.1208}. Unfortunately, the systematic uncertainties associated with these potential hints of new physics are quite difficult to quantify. Nonetheless, if the observations and associated theoretical uncertainties turn out to be robust, ALPs can indeed explain the observed deviations from expectation.

\subsubsection{Distortions of gamma-ray spectra}\label{sec:astro:hess}
Axions can be generated in galactic or intergalactic magnetic fields, distorting or dimming the spectra of distant sources~\cite{Raffelt:1987im,hep-ph/0111311,hep-ph/0204216,0704.3044}. Several studies have investigated the effects of ALPs on otherwise featureless spectra~\cite{0706.3203,0707.2695,0708.1144,0911.0015,1205.6428,1305.2114}, and used the resulting limits to constrain ALP properties~\cite{1311.3148,1603.06978,1801.01646,1801.08813,1802.08420}.

The probability of photon-axion conversion in a domain of size~$\ell$ filled with a suitably aligned magnetic field~$\mrm{B}{eff}$ and plasma with electron number density~$n_e$ is given by~\cite{Raffelt:1987im,0911.0015}
\begin{equation}
	p_1(E) \equiv \prob{\gamma\to a} = \frac{1}{1+\left(\frac{\mrm{E}{crit}}{E}\right)^2} \; \sin^2\left( \frac{1}{2} \, \gagg\, \mrm{B}{eff} \, \ell \; \sqrt{1+\left(\frac{\mrm{E}{crit}}{E}\right)^2} \right) \, , \label{eq:axionmagfield:single}
\end{equation}
where $\mrm{\omega}{pl}$ and $\mrm{E}{crit}$ are the plasma frequency and critical energy, respectively:
\begin{equation}
\mrm{\omega}{pl}^2 = \frac{4\otherpi \mrm{\alpha}{EM} \, n_e}{m_e} \, , \quad \mrm{E}{crit} = \frac{1}{2} \frac{\left| \ma^2 - \mrm{\omega}{pl}^2\right|}{\gagg \, \mrm{B}{eff}} \, . \label{eq:axionmagfield:thresholderg}
\end{equation}
The quantity~$\mrm{E}{crit}$ describes the energy scale at which photons will efficiently convert into axions in the extragalactic magnetic field domains. In the absence of dust or any other photon absorber, after traversing $N$~such domains of size~$\ell$, the remaining fraction of photons is~\cite{hep-ph/0111311,hep-ph/0204216,astro-ph/0607415,0911.0015}
\begin{equation}
\prob{\gamma\to\gamma} = \frac{2}{3} + \frac{1}{3}\ee^{-\frac{3Np_1}{2}} \, . \label{eq:axionmagfield:many}
\end{equation}
Equations~\refeq{eq:axionmagfield:single} and~\refeq{eq:axionmagfield:many} reveal that we do not expect to see any effect due to axions for $E \ll \mrm{E}{crit}$, because $p_1 \simeq 0$. For a given photon energy~$E$, this happens for large axion masses~$\ma$ and small axion-photon couplings~$\gagg$.

For $E \gg \mrm{E}{crit}$, on the other hand, conversion is very efficient, but the observed spectrum would simply decrease by a constant factor over the entire energy range. In this case it is also not possible to test the axion hypothesis: the expected spectral normalisation of any source is not well constrained and it therefore has to be a free fitting parameter in the analysis.

\begin{figure}
  \centering
  \includegraphics[width=0.618\linewidth]{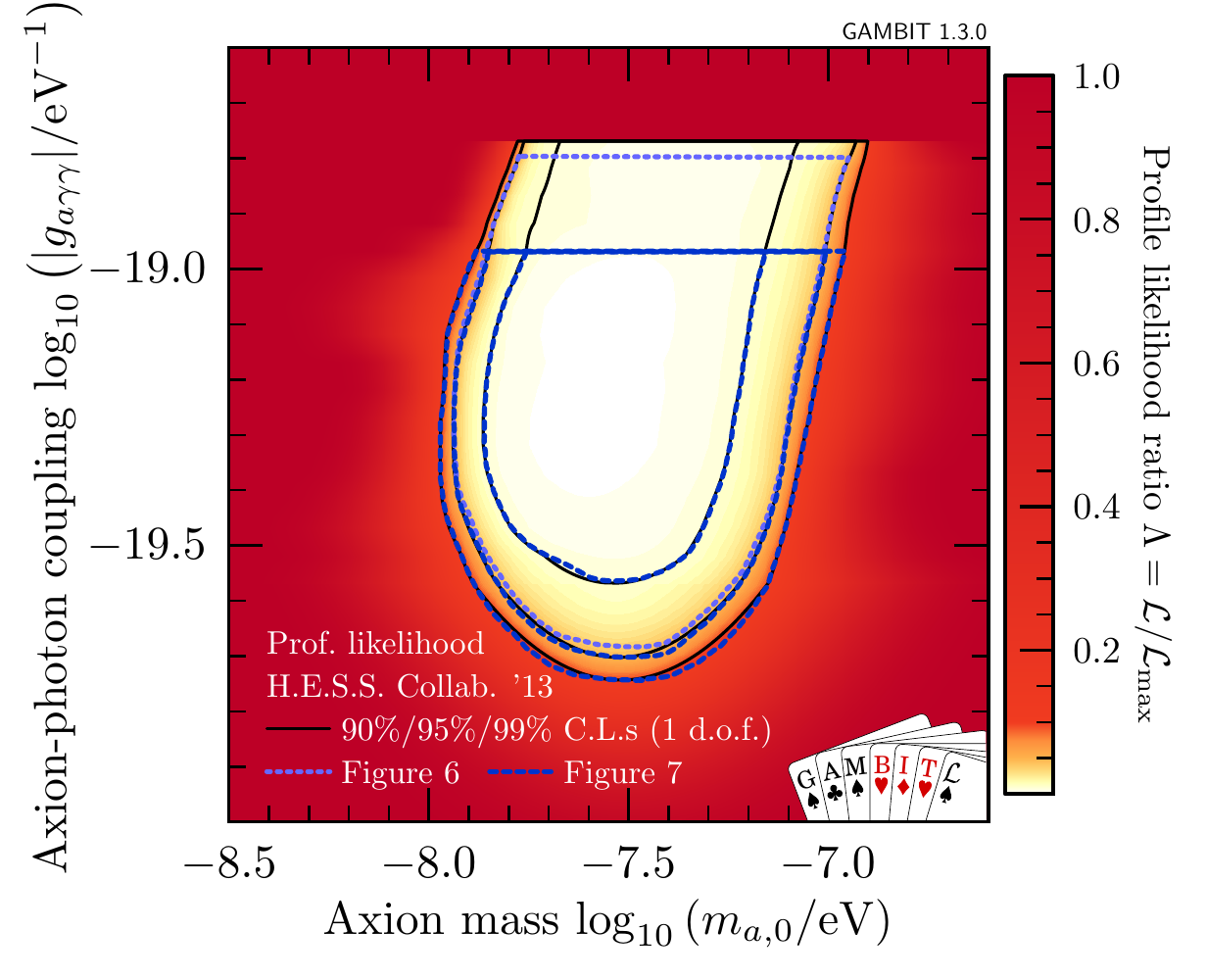}
  \caption{Our implementation of the H.E.S.S. exclusion limits from gamma-ray spectral distortions of PKS 2155-304, based on the limits quoted in Figs~6 and~7 of Ref.~\cite{1311.3148}. Full details of the likelihood construction can be found in Appendix~\ref{app:hessdetails}.\label{fig:validation:h.e.s.s.}}
\end{figure}
It is only possible to constrain models where the critical energy lies within the spectral window of the instrument, such that one end of the spectrum is suppressed, but the other is not~\cite{1205.6428,1305.2114}. Limits from the distortion of gamma-ray spectra are therefore strongest at axion masses that lead to threshold energies similar to the photon energies observed by the experiment. This explains the characteristic shape of the limits (in particular why this method is not sensitive to axion masses below a certain value; see \reffig{fig:validation:h.e.s.s.}).

The H.E.S.S. Collaboration applied this technique to the spectrum of the active galactic nucleus PKS 2155-304, using data obtained with their Cherenkov telescope array~\cite{1311.3148}. Unfortunately, their signal prediction requires Monte Carlo simulation of magnetic field realisations, which are no longer available.\footnote{P.~Brun, private communication.}

We therefore approximate their likelihood function for the galactic cluster magnetic field using Figs~6 and~7 of Ref.~\cite{1311.3148}, based on a scheme that we describe in detail in Appendix~\ref{app:hessdetails}. The main idea is to use common interpolation methods inside the published exclusion contours, and to extrapolate to likelihood values outside the known contours using a method that mimics the shape of the known iso-likelihood contours and preserves the mathematical properties of the likelihood. Our approximation procedure is of course somewhat arbitrary. The advantages of our scheme are that it exactly reproduces the known exclusion curves by construction, and that the general likelihood function is well-behaved. The obvious downsides are that we can neither guarantee that the likelihood in the outermost and innermost regions are completely accurate, nor can we extend it to larger couplings than the values shown in Fig.~7 of Ref.~\cite{1311.3148}.

\subsubsection{Supernova 1987A}\label{sec:astro:SN1987A}
Supernovae are excellent particle laboratories. Supernova~1987A (SN1987A) provides significant constraints on axion parameters, based on the neutrino burst duration~\cite{Ellis:1987pk,Raffelt:1987yt,Turner:1987by,Mayle:1987as} and axion-photon interaction in magnetic fields external to the supernova~\cite{astro-ph/9605197,astro-ph/9606028}.

Our likelihood for SN1987A is based on the results from Ref.~\cite{1410.3747}. The authors of that study derived limits based on the absence of a coincident gamma-ray burst from SN1987A, which should have been observed by instruments on board of the Solar Maximum Mission~\cite{Chupp:1989kx} if axions were produced in the explosion and converted to gamma rays in the Galactic magnetic field.

The gamma-ray spectrum of photons with energy~$E$ expected at Earth per unit time from axions produced in SN1987A is
\begin{equation}
	\frac{\dd\Phi_{a\to\gamma}}{\dd E} = \frac{1}{4\otherpi d^2} \; \frac{\dd \dot{N}}{\dd E} \; \prob{a\to \gamma} \, , \label{eq:snflux}
\end{equation}
where $d$ is the distance to the supernova, $\prob{a\to \gamma}$ is the conversion probability~\refeq{eq:axionmagfield:single}, $N$ is the number of axions created in the supernova and $\dd \dot{N}/\dd E$ is the axion spectrum at the source, as predicted from a supernova model. To obtain the measured photon fluence at Earth (gamma rays per unit area during the observation), $\mathcal{F}_{a\to\gamma}$, equation~(\ref{eq:snflux}) has to be integrated over the energy range and time duration of the observation, while modelling the transport of the axions through the Galactic magnetic field.\footnote{The authors of Ref.~\cite{1410.3747} consider two magnetic field models from Refs~\cite{1103.0814} and~\cite{1204.3662}. The latter yields weaker limits and is their reference model as well as the basis for our implementation.}

From Fig.~6 in Ref.~\cite{1410.3747}, we can see that the photon fluence at Earth for a given axion-photon coupling becomes constant below a certain mass scale~$m_\ast$ and rapidly decreases for bigger masses. This is not surprising, given that most axion experiments lose sensitivity at large masses due to the loss of coherence in axion-photon conversion. We therefore make the following ansatz for the fluence,
\begin{equation}
	\mathcal{F}_{a\to\gamma} = F \, \left(\frac{\gagg}{\SI{5.3e-10}{\per\GeV}}\right)^4 \,
	\begin{cases}
		\hfil 1 & \text{for } \ma \leq m_\ast \\
		\left ( \frac{m_\ast}{\ma} \right )^{b} & \text{for } \ma > m_\ast \, ,
	\end{cases}
\end{equation}
where $F \approx \SI{0.57e-12}{\centi\metre^{-2}}$ is the fluence for small axion masses at the reference value of $\gagg\approx \SI{5.3e-10}{\GeV^{-1}}$; we obtain this value by integrating Eq.~(2.11) in Ref.~\cite{1410.3747}.

We determine the best-fit values for~$m_\ast$ and the exponent~$b$ from the higher-mass region~($ \mazero>\SI{6.0e-10}{\electronvolt}$) via a least-squares fit to the fluence contour in Fig.~6 of Ref.~\cite{1410.3747}, giving $\hat{m}_\ast \approx \SI{5.43e-10}{\electronvolt}$ and $\hat{b}\approx 4.02$. The value of~$\hat{b}$ that we obtain can be qualitatively understood by examining the axion-photon conversion probability given in~(\ref{eq:axionmagfield:single}), which can be written as $\prob{\gamma\to a} = \sin^2 (c \sqrt{x})/x$, where $c=\gagg\mrm{B}{eff}\ell/2$ does not depend on~$\ma$. For axions masses $\ma \gg \mrm{\omega}{pl}$ (with $\mrm{\omega}{pl} \sim \SI{4e-12}{\eV}$), we have $x \propto \ma^4$ and, since the oscillatory part is washed out by the turbulent magnetic fields, the conversion probability is effectively suppressed by a factor of~$\ma^4$.

\begin{figure}
	\centering
	\includegraphics[width=0.618\linewidth]{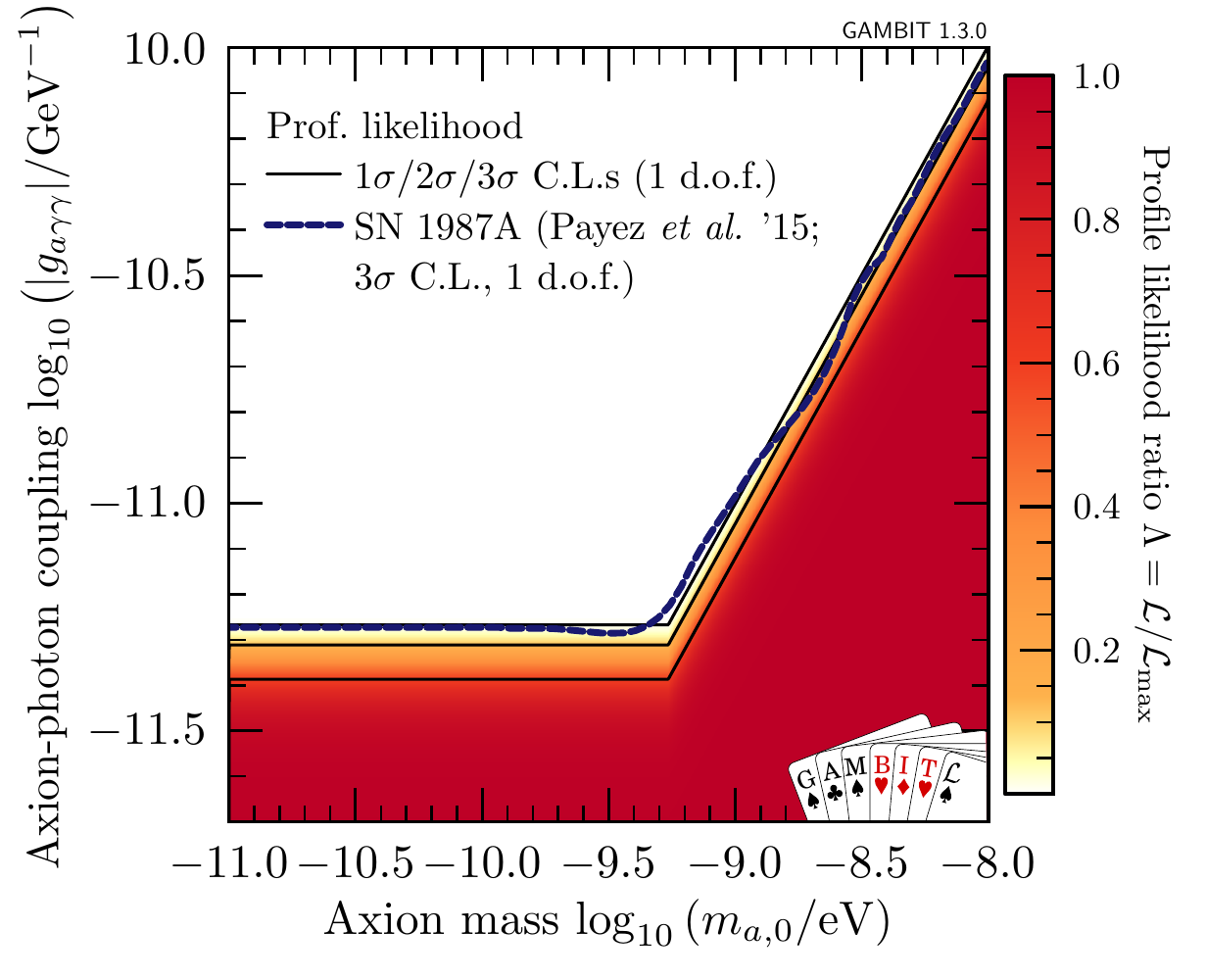}
	\caption{Our implementation of SN1987A limits compared to Ref.~\cite{1410.3747} (dashed line).\label{fig:validation:sn1987a}}
\end{figure}
The likelihood for $s=\mathcal{F}_{a\to\gamma}$ in the absence of a photon signal with background fluctuations~$\sigma_\mathcal{F}^2$ is then given by
\begin{equation}
  \lnL = - \frac{1}{2} \frac{s^2}{\sigma_\mathcal{F}^2} \, .
\end{equation}
Fig.~\ref{fig:validation:sn1987a} shows how the limits obtained from our approximation compare to the original reference.

\subsubsection{Horizontal Branch stars and R parameter}\label{sec:astro:rparameter}
Weakly-interacting particles can influence stellar evolution by providing an additional energy loss mechanism, cooling stars over the course of their evolution~\cite{Sato:1975vy,Raffelt:1990yz,book_raffelt_laboratories}. The so-called $R$~parameter, $R = \mrm{N}{HB}/\mrm{N}{RGB}$, is the ratio between the number of Horizontal Branch~(HB) stars, $\mrm{N}{HB}$, and upper Red Giant Branch~(RGB) stars, $\mrm{N}{RGB}$, in Galactic globular clusters. Its value depends on the relative time that stars spend on each branch, which is sensitive to the details of stellar evolution and cooling. Axions are expected to be produced in the cores of both types of stars, but would remove heat more efficiently from the cores of HB stars, reducing the time that they spend on the HB and leading to a reduction in $R$.

Based on a weighted average of a selection of cluster count observations~\cite{astro-ph/0403600}, the observed value is $\mrm{R}{obs}=1.39 \pm 0.03$~\cite{1406.6053}. The dependence of the predicted value of the $R$~parameter on the properties of axions can be approximated as~\cite{1512.08108,1983A&A...128...94B,Raffelt:1989xu,1311.1669,1406.6053}
\begin{align}
	\mrm{R}{pred} \approx 7.33\, Y &- 0.422 - 0.0949 \left( -4.68 + \sqrt{21.9 + 21.1 \, x_{a\gamma\gamma}} \right) \nonumber\\
	&-0.00533 \, x_{aee}^2 - 0.0387\left( -1.23 - 0.138 \, x_{aee}^{1.5} + \sqrt{1.51 + x_{aee}^2} \right) \, , \label{eq:Rpred}
\end{align}
where $Y$ is the Helium abundance, $x_{a\gamma\gamma}\equiv\gagg/\SI{e-10}{\GeV^{-1}}$ and $x_{aee}\equiv\gaee/\num{e-13}$. The equation above is valid only if axions are sufficiently light compared to the typical temperatures of the stellar interior, which are $T \sim \SI{e8}{K} \approx \SI{10}{\keV}$~\cite{book_raffelt_laboratories}, i.e.\ much higher than the axion masses we consider. Our $R$~parameter likelihood is then simply
\begin{equation}
\lnL = - \frac{1}{2} \frac{\left(\mrm{R}{pred}-\mrm{R}{obs}\right)^2}{ \mrm{\sigma}{pred}^2+\mrm{\sigma}{obs}^2} \, ,
\end{equation}
where $\sigma_{\mathrm{pred}}$ and $\sigma_{\mathrm{obs}}$ are the uncertainties of the predicted and observed values.

\begin{figure}
	\centering
	\includegraphics[width=0.618\linewidth]{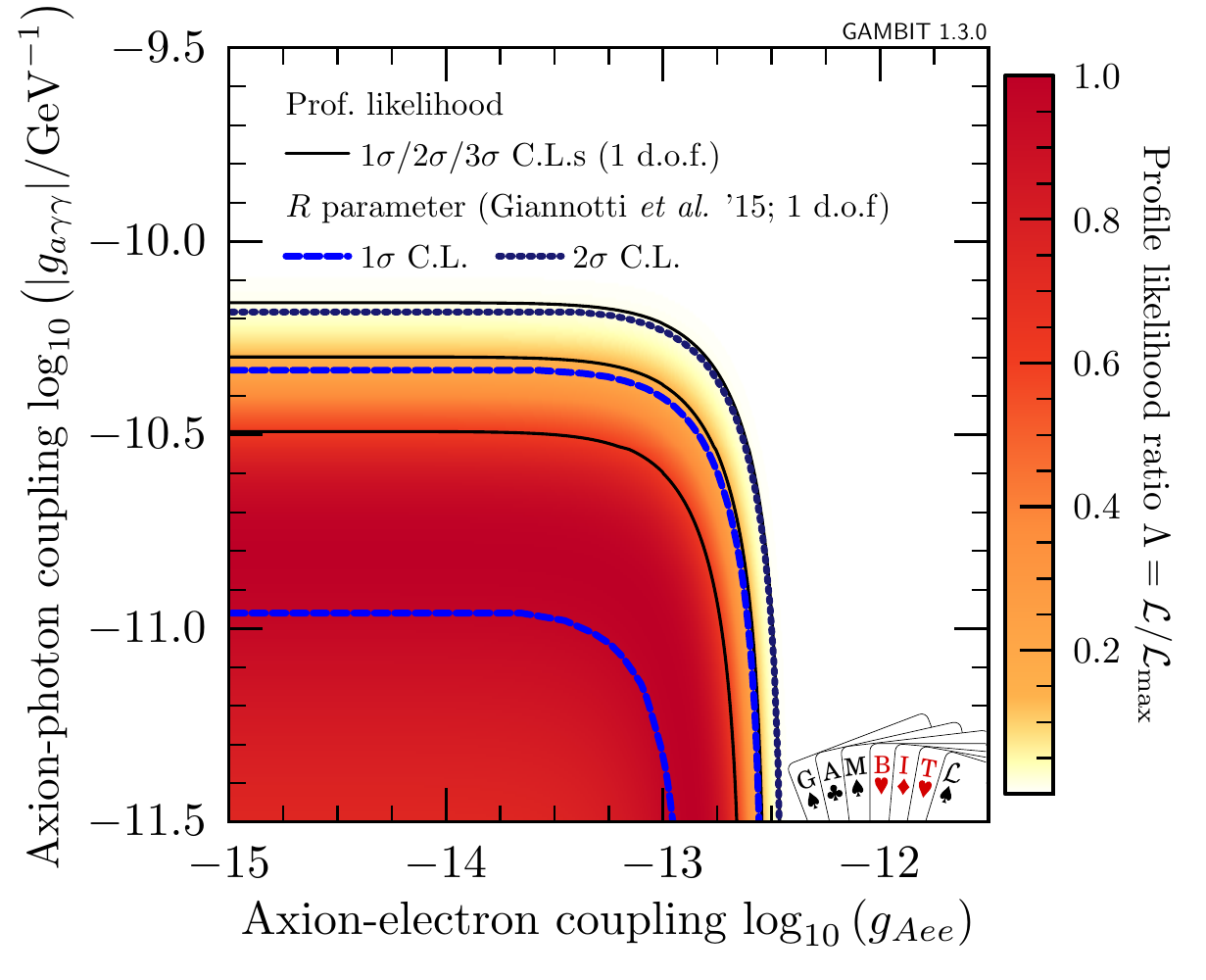}
	\caption{Comparison of our implemented $R$~parameter likelihood with the $1\sigma$ and $2\sigma$ contours (dashed and dotted blue lines) of Ref.~\cite{1512.08108}. Note that our adopted likelihood leads to limits only, whereas the results of Ref.~\cite{1512.08108} indicate an almost $2\sigma$ preference for signal (hence the presence of $1\sigma$ upper \textit{and} lower limits in the results that we plot from Ref.~\cite{1512.08108}). The difference is due to the updated He abundance that we employ here.\label{fig:validation:rparameter}}
\end{figure}
The authors of Ref.~\cite{1512.08108} adopted a Helium abundance of $Y = 0.255 \pm 0.002$~\cite{1408.6953}, leading to a predicted value from standard (axion-free) stellar evolution calculations of $\mrm{R}{theo} = 1.45 \pm 0.01$, almost $2\sigma$ higher than the observed value. We adopt the updated value for such low-metallicity environments of $Y = 0.2515 \pm 0.0017$~\cite{1503.08146}, leading to $\mrm{R}{theo} = 1.42 \pm 0.01$, entirely consistent with the observed $R$~parameter. The effect of the uncertainty of~$Y$ on~$\mrm{R}{pred}$ can be estimated according to~\refeq{eq:Rpred}, as $\mrm{\sigma}{pred} = 7.33\times 0.0017 \approx 0.012$. A~comparison to the exclusion curves in Ref.~\cite{1512.08108} can be found in Fig.~\ref{fig:validation:rparameter}.

\subsubsection{White Dwarf cooling hints}\label{sec:coolinghints}
White dwarfs~(WDs) are a particularly interesting environment in which to study axion-electron interactions, due to their electron-degenerate cores~\cite{Raffelt:1985nj, Nakagawa:1987pga,Nakagawa:1988rhp,Isern:1992gia}. Current observations can be interpreted as indicating a need for additional cooling in WDs compared to standard models. The coupling necessary to explain the cooling with axions has been estimated to be $\gaee \sim \order\left(\num{e-13}\right)$~\cite{1512.08108}. A more recent analysis also considered these cooling hints in a global fitting framework~\cite{1708.02111}. Whilst the systematics of such analyses are still a matter of debate, and alternative explanations (whether involving BSM physics or not) are certainly still possible, it is intriguing to investigate the impact on axion global fits of including the WD cooling hints. Due to the speculative nature of theses hints, we will do so in a separate (alternative) analysis in Sec.~\ref{sec:results:coolinghints}, presented alongside our main global fits.

WDs typically pulsate, allowing the oscillation of their radii and luminosity to be used to probe their internal structure via astroseismology. The periods, $\Pi$, of their pulsations decrease with time, with a rate $X=\dd \Pi/\dd t$, which can be related to the energy loss in the system.

\begin{table}
  \caption{Overview of available cooling hints for WD variables of spectral types~DA and DB. We list the couplings allowed at $2\sigma$ confidence (whether limits or intervals). The numerical values necessary for constructing the corresponding likelihoods were kindly provided by A.~H.~C{\'o}rsico and T.~Battich. In the case of R548, we have two more data points compared to what is shown in Fig.~1 of Ref.~\cite{1211.3389}.\label{tab:coolinghints}}
  \centering
  \begin{tabular}{lcccll}
    \toprule
    \textbf{Object} & \textbf{Type} & \multicolumn{2}{l}{\textbf{Mode}} & \textbf{$2\sigma$ C.L.} & \textbf{References} \\
    & & \textbf{$k$} & \textbf{$\ell$} & {$\gaee/\num{e-13}$} & \\
    \midrule
     G117-B15A & DA variable & 2 & 1 & $\left[3.4,\, 6.0\right]$ & Fig.~5 \cite{1205.6180}; \cite{hep-ph/9310304,astro-ph/0104103,0711.2041,0806.2807} \\
     R548 & DA variable & 2 & 1 & $\left[0.30,\, 6.8\right]$ & Fig.~1 \cite{1211.3389}; \cite{0711.2041} \\
     L19-2 & DA variable & 2 & 1 & $<5.1$ & Fig.~5 \cite{1605.06458} \\
     PG~1351+489 & DB variable & 11 & 1 & $<3.6$ & Fig.~5 \cite{1605.07668} \\
    \bottomrule
  \end{tabular}
\end{table}

Refs~\cite{1205.6180,1211.3389,1605.06458,1605.07668} simulated the evolution of WDs with and without axions, predicting the period decrease $\dd \Pi/\dd t$ in each case. For our predictions of WD cooling rates, we interpolate these results and their stated uncertainties, using natural splines. The specific figures and objects from those papers that our implementation is based on are listed in Table~\ref{tab:coolinghints}.\footnote{For L19-2 there is also a measurement for the $k=2$, $\ell = 2$ mode, which results in a stronger preference for $\gaee \neq 0$~\cite{1605.06458}. However, we choose the $k=2$, $\ell = 1$ mode, consistent with the other DA~variable dwarfs.} Note that the plots in Refs~\cite{1205.6180,1211.3389,1605.06458,1605.07668} show the quantity~$ \mazero\cos^2(\beta^\prime)$ which, for the \dfszI axion model they consider, is proportional to the more fundamental parameters~$\caee$ or $\gaee$ (cf.\ Sec.~\ref{sec:mod:DFSZAxion}). For values of the axion-electron coupling larger than considered in the simulations (i.e.\ $\gaee/\num{e-13}>5.6$ or $8.4$), we assign the likelihood corresponding to the largest simulated coupling. This is a conservative assumption, as the disagreement between prediction and observation will in reality only worsen as the coupling increases further (until the WDs become opaque to axions -- but this would occur at couplings well beyond what we consider).

\begin{figure}
  \centering
  {
    \includegraphics[width=0.49\linewidth]{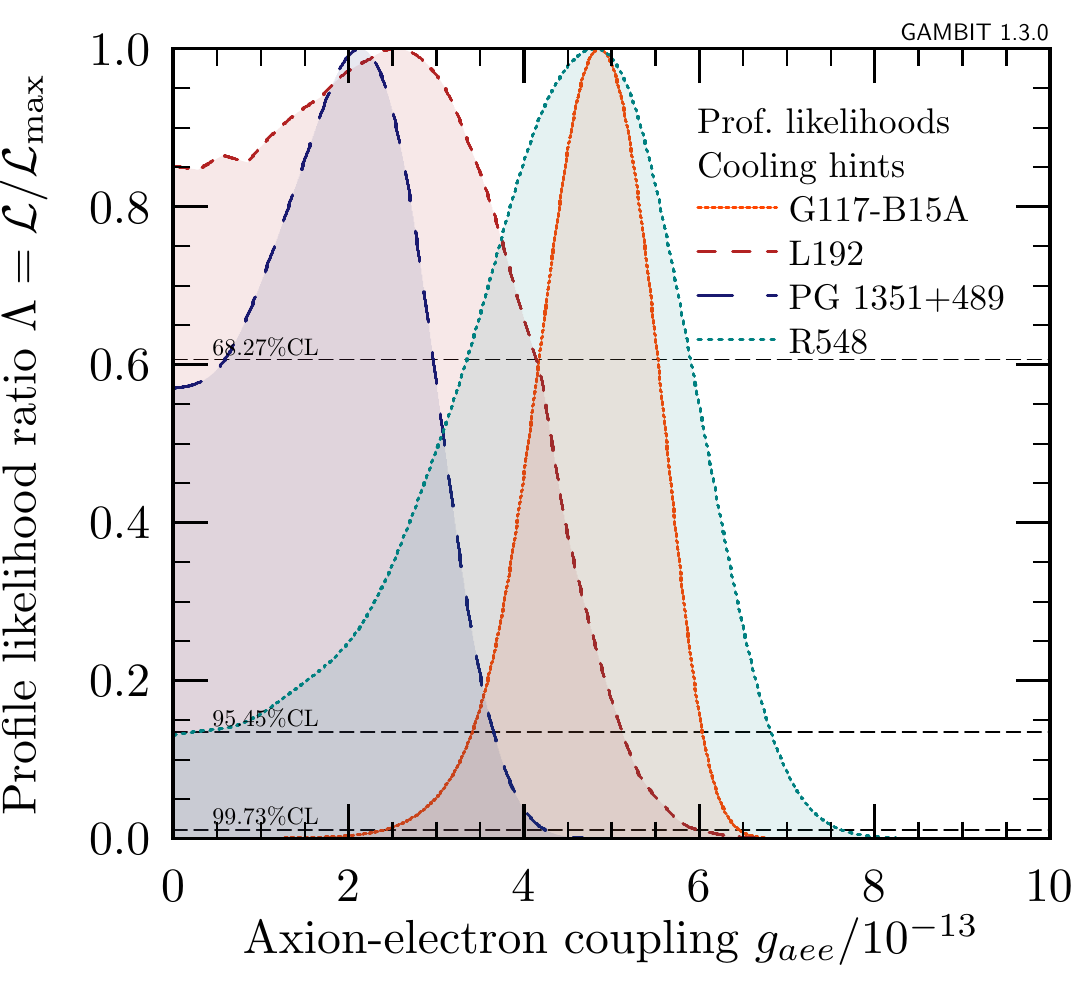}
    \hfill
    \includegraphics[width=0.49\linewidth]{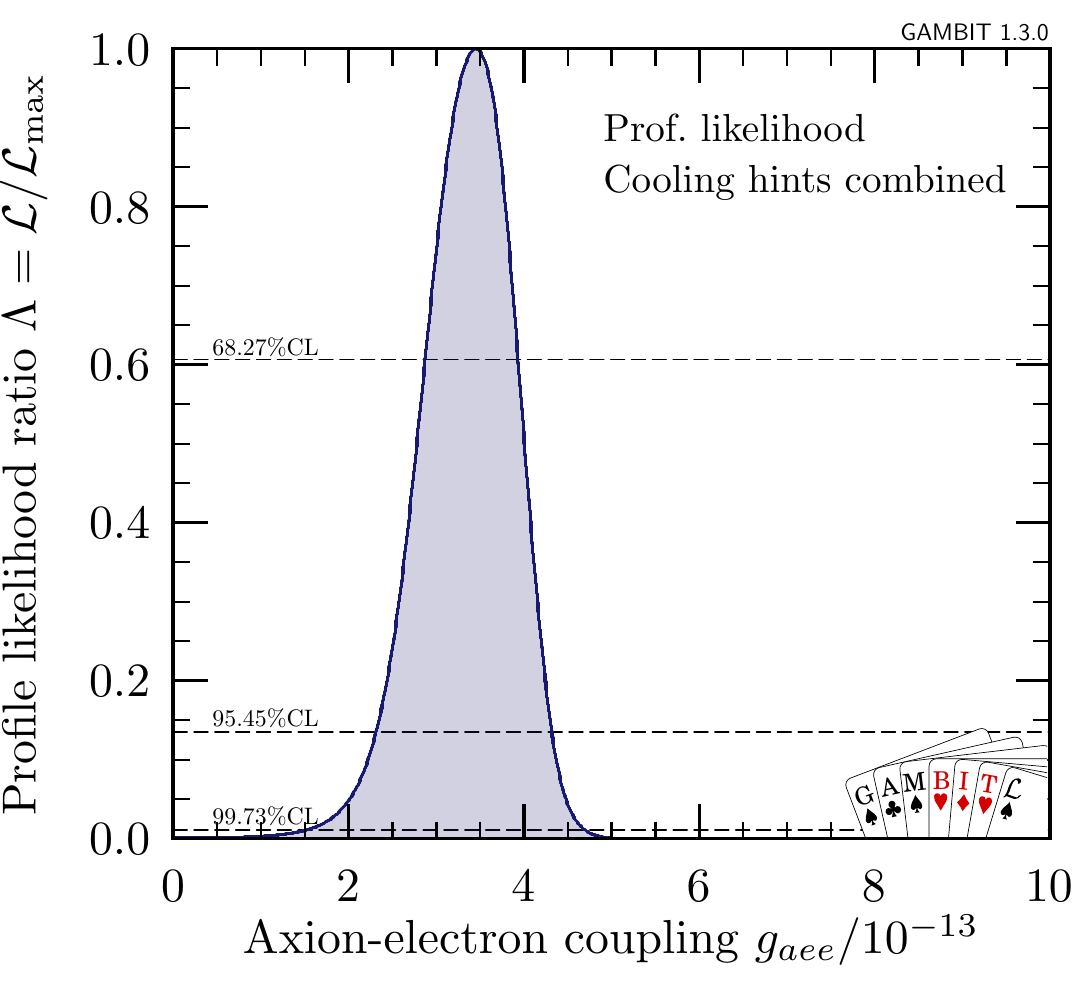}
  }
  \caption{Overview of the WD likelihoods available in \gambit. We show the separate likelihoods~(\textit{left}) as well as the combined result~(\textit{right}). To guide the eye, horizontal dashed lines indicate the confidence levels. Note that, taken at face value, the combined constraints have the potential to be significant evidence for an additional cooling channel in WDs. \label{fig:validation:cooling}}
\end{figure}

For each WD listed in Table~\ref{tab:coolinghints}, we use a simple Gaussian likelihood function for the observed ($X_{\text{obs}, i}$) and theoretically-expected ($X_{\text{pred}, i}$) period decrease, such that our total WD cooling likelihood is
\begin{equation}
	\lnL = - \frac{1}{2} \sum_{i=1}^4 \frac{\left(X_{\text{pred}, i}-X_{\text{obs}, i}\right)^2}{\sigma_{\text{pred}, i}^2+\sigma_{\text{obs,} i}^2} \, ,
\end{equation}
where, again, the predictions and corresponding uncertainties~$\sigma_{\text{pred}, i}$ are taken from the respective figures in the references listed in Table~\ref{tab:coolinghints} and interpolated via natural splines. The resulting individual and combined likelihoods can be found in \reffig{fig:validation:cooling}.

We emphasise that the interpretation of WD cooling is subject to a number of assumptions and caveats. Statistical and systematic uncertainties associated with the inputs and algorithms of stellar models were considered by the authors of Refs~\cite{1205.6180,1211.3389,1605.06458,1605.07668}, but a number of other potential issues remain. These include theoretical modelling of the transition from the main sequence to the WD phase, and the accuracy of the observed period decrease of PG~1351+489. Despite these problems, and in contrast to Ref.~\cite{1708.02111}, we include the PG~1351+489 system in our discussion. The authors of Ref.~\cite{1708.02111} exclude this object due to its similarity to R548, and the uncertainties associated with R548 being ``more conservative''. However, the different estimated uncertainties in these two systems are in fact due to a real physical effect, namely the difference in the influence of trapped vs non-trapped oscillation modes in the two systems. While the latter might give rise to concerns regarding the understanding of different modes (cf.\ Table~\ref{tab:coolinghints}), we do not conclude that the arguments in favour of excluding PG~1351+489 are strong enough to do so.

\section{Results and discussion}\label{sec:results}
In this section, we present the central findings from our global fits of various axion models, identifying the most promising regions in parameter space and comparing the various models. We present frequentist and Bayesian results side-by-side, after discussing our choice of priors for the model parameters. Unless stated otherwise, all C.L.s and C.R.s (Bayesian credible regions) are $1\sigma/2\sigma/3\sigma$ ($68.27\%/95.45\%/99.73\%$), and all C.L.s are two-sided for two degrees of freedom (d.o.f.). Secs.~\ref{sec:results:GeneralALP}--\ref{sec:results:dfszvsksvz} do not include WD cooling hints; these are the subject of a dedicated analysis in Sec.~\ref{sec:results:coolinghints}.

\subsection{Sampling algorithms and settings}
We use the differential evolution sampler \diver~\cite{ScannerBit} to sample the composite likelihood function and \twalk~\cite{ScannerBit} to sample the posterior distributions. We employ \multinest~\cite{0704.3704,0809.3437,1306.2144} primarily to compute Bayesian evidences.

We use the sampler settings established in an earlier study by the GAMBIT Collaboration as starting points~\cite{ScannerBit}. For \diver, we generally use a population size (\texttt{NP}) of \num{2e4} and a tolerance (\texttt{convthresh}) of \num{e-4}, and turn off the $\texttt{lambdajDE}$ optimisation, preferring to use regular $\texttt{jDE}$ for its slightly less aggressive optimisation. In addition to combining samples from various runs, where necessary to resolve fine-tuned regions we increase $\texttt{NP}$ to~\num{3e4} or~\num{5e4} and/or reduce $\texttt{convthresh}$ to~$\num{e-5}$. For \twalk, we use the default settings for 340 or 544~MPI processes, until reaching a tolerance ($\texttt{sqrtR}-1$) of~\num{0.01} or~\num{0.005}. All initialisation \YAML files that we use in this study are available on \textsf{Zenodo}~\cite{Zenodo_axions}; the exact scans for which we use the different settings can be ascertained by examining the input files. Because we use \multinest primarily for estimating Bayesian evidences, we set the sampling efficiency (\texttt{efr}) to~0.3, as recommend for this task~\cite{0809.3437}, and use \num{2e4}~live points (\texttt{nlive}) with a tolerance (\texttt{tol}) of~\num{e-4}.

\subsection{General ALP models}\label{sec:results:GeneralALP}
Starting at the top of the model hierarchy (see \reffig{fig:AxionModelTree}), we first consider the \genalp model. This is a phenomenological model; parameter combinations in this model need not correspond to physical models, as their couplings do not depend on the inverse of~$\fa$. The main purpose of the \genalp is to provide a straightforward, universal connection to observables and to compare to results in the literature.

\begin{table}
	\caption{Parameter ranges and scaling for \genalp models.\label{tab:priors:GeneralALP}}
	\footnotesize
	\centering
	\begin{tabular}{@{}lccc}
		\toprule
		\textbf{Model} & \multicolumn{2}{c}{\textbf{Parameter range/value}} & \textbf{Scale}\\
		\midrule
		\genalp & \iuo{\fa}{\GeV} & \prrange{e6}{e16} & log  \\
		& \iuo{\mazero}{\eV} & \prrange{e-10}{1} & log \\
		& \iuo{\gagg}{\GeV^{-1}} & \prrange{e-16}{e-8} & log \\
		& $\gaee$ & \prrange{e-22}{e-10} & log\\
		& $\thetai$ & \prrange{-3.14159}{3.14159} & flat\\
		& $\beta$ & \prrange{0}{16} & flat \\
		& \iuo{\Tcrit}{\MeV}& \prrange{e-2}{e6} & log\\
		\midrule
		Local DM density & \iuo{\rho_0}{\GeV\per\centi\metre^3} & \prrange{0.2}{0.8} & flat\\
		\bottomrule
	\end{tabular}
\end{table}

\paragraph*{Parameter ranges.}
The parameter ranges and scales that we use for {\genalp} models are given in Table~\ref{tab:priors:GeneralALP}. Because the axion potential is periodic, all normalised field values are equivalent to a value in the interval $(-\otherpi,\,\otherpi]$. For~$\gagg$, $\gaee$, and $\mazero$, there is no obvious range to choose; we adopt parameter ranges encompassing values informed by previous studies and phenomenology. Recall that $\gagg$~could be negative but the likelihood functions in the present work only depend its absolute value. We therefore scan only over positive values of $\gagg$, but label plots with $|\gagg|$, to make it explicit that the results are equally valid for the corresponding negative values. For the local DM density~$\rho_0$, we adopt the same range as in earlier GAMBIT studies~\cite{MSSM,SSDM,CMSSM,SSDM2} (and implement the same likelihood function).

The appropriateness of different ranges on $\fa$, $\beta$, and $\Tcrit$ depends on the fundamental properties of the symmetries and scales of the underlying ALP model. Although there are theoretical arguments for the existence of ALPs from e.g.\ string theory~\cite{hep-th/0605206,1201.5902}, they do not provide any quantitative guidance.\footnote{See Ref.~\cite{1809.06382} for a recent study, preforming a Bayesian analysis of (string-theory-inspired) axion models, similar to our \genalp model with $\beta = 0$ and not including coupling strengths as model parameters.} Apart from the likely case that our calculations become meaningless for $\fa \gsim \mpl$, we can only impose $\beta > 0$, as the axion mass should become smaller as the underlying symmetry is restored at higher temperatures. The ranges in Table~\ref{tab:priors:GeneralALP} are therefore an attempt to include a variety of cases around the values known or preferred for the QCD~axion.

We do not produce Bayesian results for the \genalp model, as no strong physical arguments exist for any particular choice of prior on most of its parameters. The only exception is the initial misalignment angle $\thetai$, due to the causal structure of the early-Universe cosmology mentioned in Sec.~\ref{sec:realignment}.

\begin{figure}
	\centering
	{
		\includegraphics[width=0.618\linewidth]{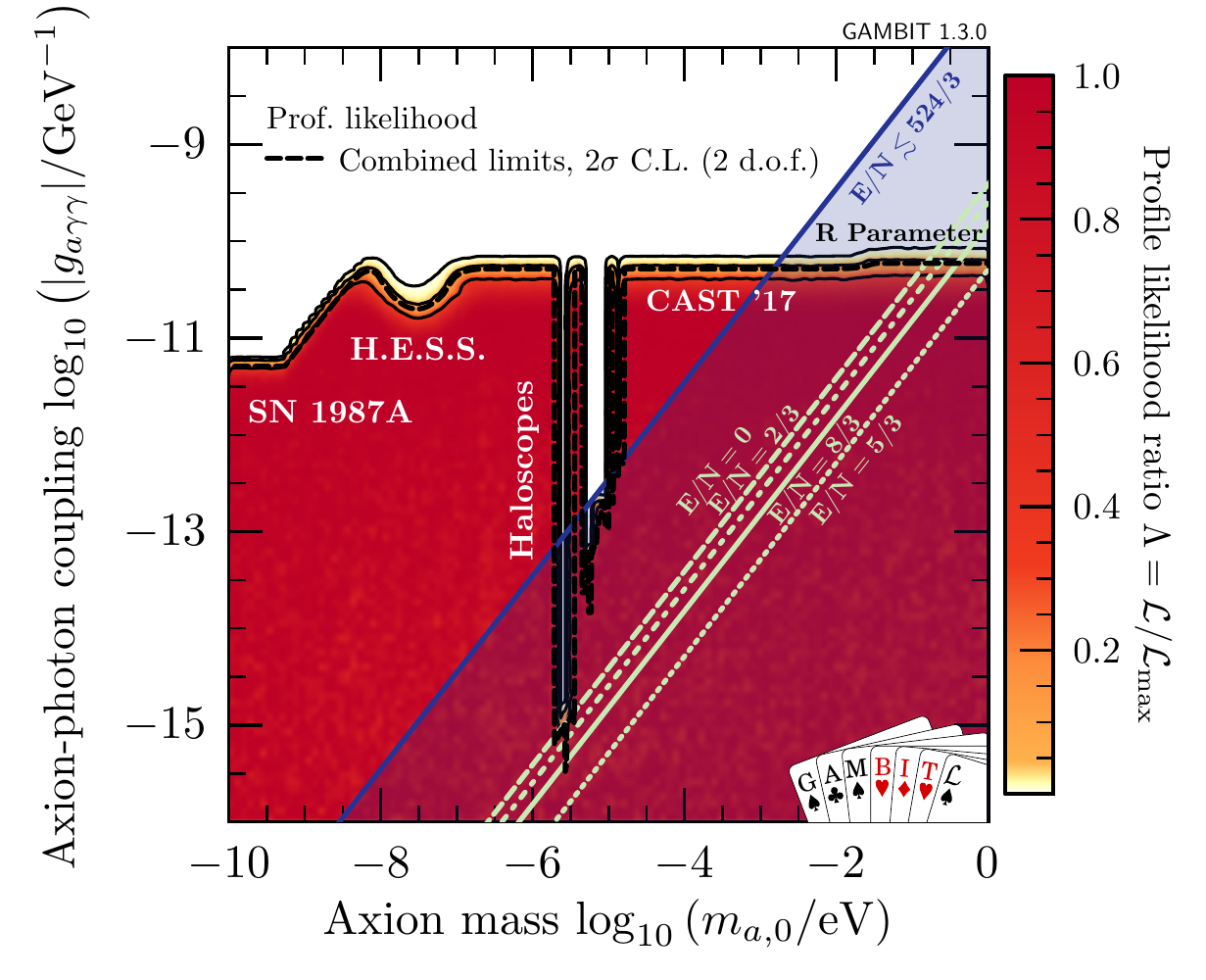}
	}
	\caption{Profile likelihood for \genalp models assuming ALPs to be all of dark matter. The constraints that dominate the exclusion contours are CAST, haloscopes (UF, RBF and ADMX), H.E.S.S., the $R$~parameter, and SN~1987A. We used \diver to sample the profile likelihood (interpolated density plot) and a root-finding algorithm with a local optimisation routine for profiling to determine the $2\sigma$ C.L. (dashed line). For comparison, we also show the band of QCD~axion models that we consider in this paper (blue shaded region; cf.\ Sec.~\ref{sec:results:QCDAxions}) and the discrete choices for $E/N$ that we use for the DFSZ- and KSVZ-type models (yellow lines; dealt with in detail in Secs.~\ref{sec:results:dfszvsksvz} and~\ref{sec:cooling:dfszvsksvz}). \label{fig:GeneralALP:overview}}
\end{figure}

\paragraph*{Frequentist results.}
We first scan the \genalp model assuming ALPs to be all of DM. The resulting limits on the axion-photon coupling (\reffig{fig:GeneralALP:overview}) are comparable to summary plots elsewhere in the literature \cite[e.g.][]{PDG17}. However, we would like to stress that unlike overplotted exclusion limits, the exclusion curve in \reffig{fig:GeneralALP:overview} arises from a composite likelihood, and profiling takes into account uncertainties in the local DM density (the only relevant nuisance parameter here).

\begin{figure}
	\centering
	{
		\includegraphics[width=0.49\linewidth]{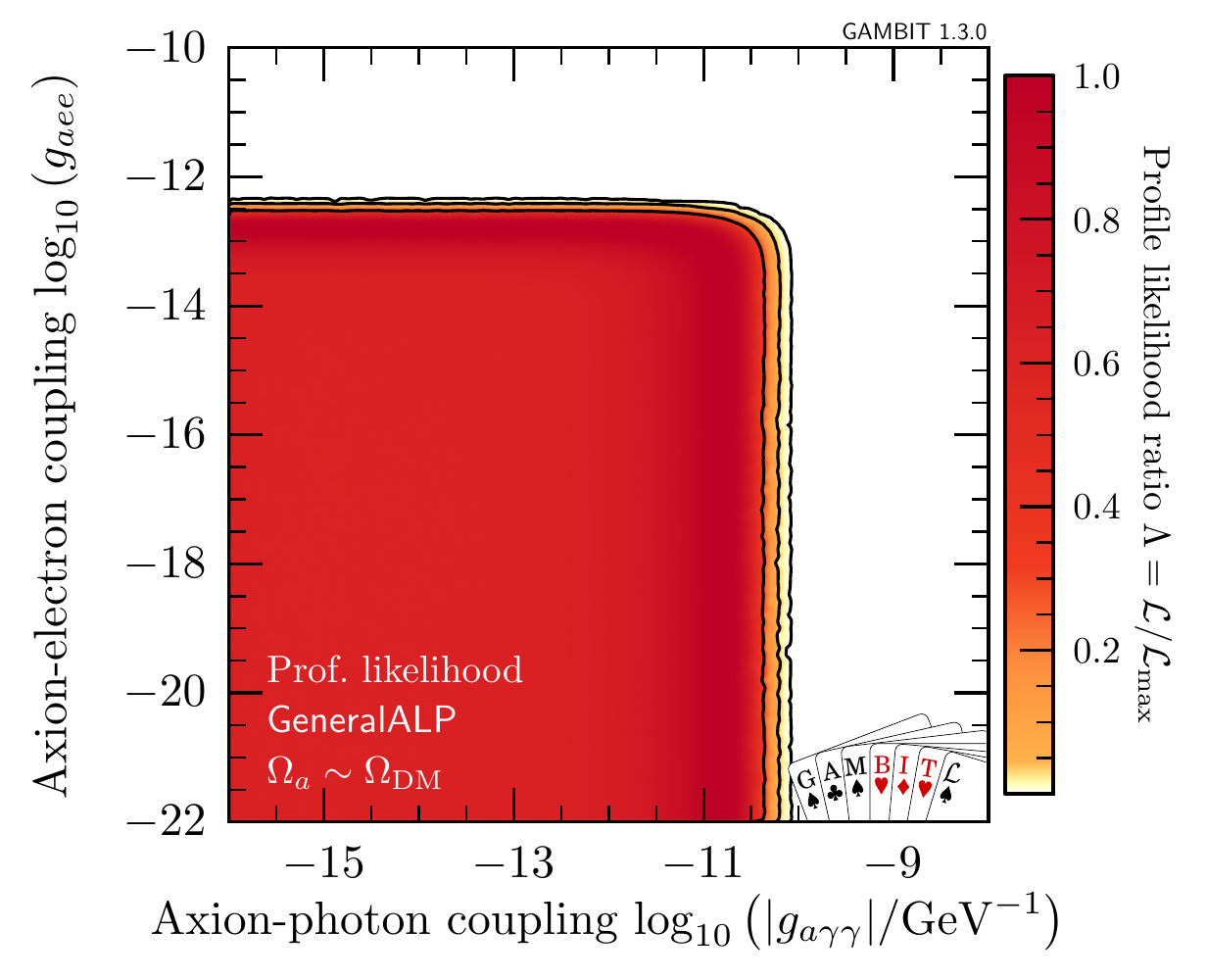}
		\hfill
		\includegraphics[width=0.49\linewidth]{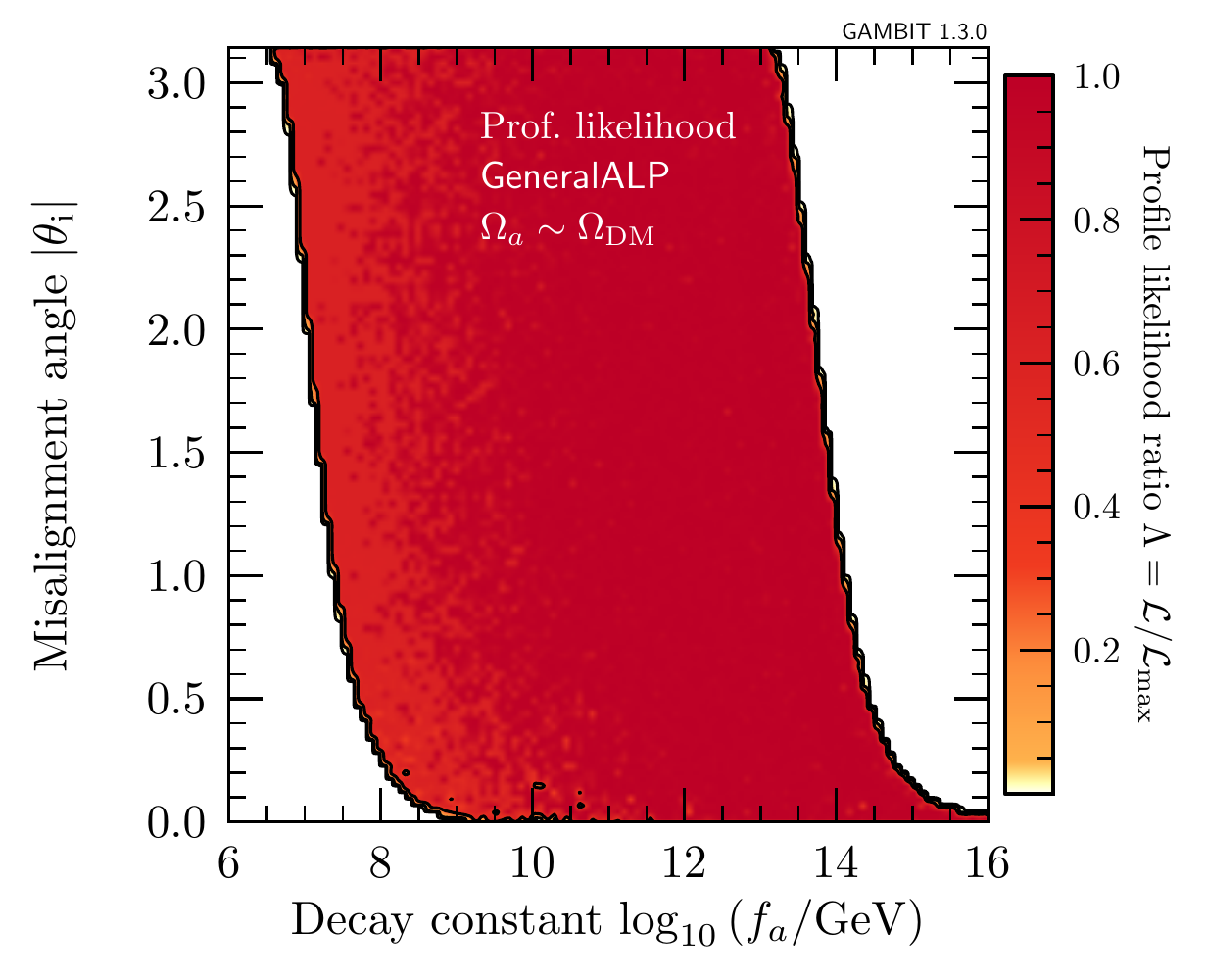}
	}
	\caption{Profile likelihoods (from \diver) for \genalp models, assuming they provide all of the dark matter in the Universe. We show limits for the axion-photon and axion-electron couplings (\textit{left}; essentially dominated by the $R$~parameter likelihood, cf.\ \reffig{fig:validation:rparameter}) and for the absolute value of the initial misalignment angle vs~$\fa$ (\textit{right}).\label{fig:GeneralALP:globalfit}}
\end{figure}

The left panel of Fig.~\ref{fig:GeneralALP:globalfit} shows that the joint constraints on the two coupling parameters are essentially dictated by the constraint on the $R$~parameter (\reffig{fig:validation:rparameter}). As a consequence, the axion-electron equivalent of \reffig{fig:GeneralALP:overview} would show that values of $\gaee \gsim \text{few} \times \num{e-13}$ are excluded across the entire mass range. The right panel of \reffig{fig:GeneralALP:globalfit} shows the possible combinations of~$\fa$ and~$|\thetai|$ that allow the \genalp to be all of DM. The extent and shape of this region is mostly due to the limited ranges of~$\mazero$, $\beta$, and $\Tcrit$. The axion potential, and therefore the initial energy density in axions, is proportional to~$\fa\,\mazero$. The observed DM abundance can only be achieved if $\fa\,\mazero$ is large enough, because $|\thetai| < \otherpi$. On the other hand, the axion starts to oscillate when $H\sim \ma$ (Sec.~\ref{sec:realignment}). The associated temperature scale depends on $\mazero$, $\beta$, and $\Tcrit$ and sets the amount by which the axion energy density is red-shifted up to the present day. To obtain the correct abundance in axions today while e.g.\ going to lower values of~$\fa$, the values of $\mazero$, $\beta$, or $\thetai$ must be increased or the value of~$\Tcrit$ decreased.

\begin{figure}
	\centering
	{
		\includegraphics[width=0.49\linewidth]{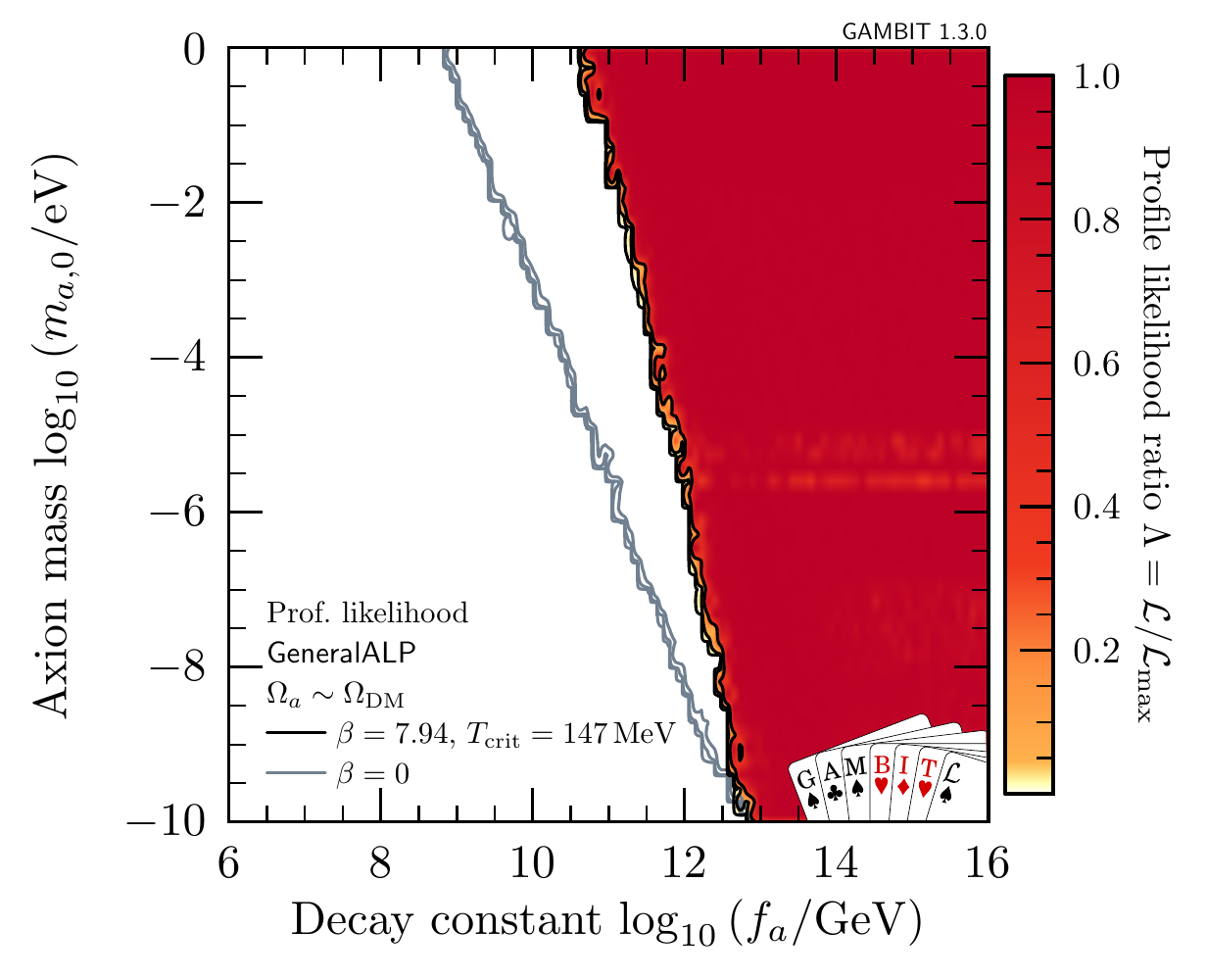}
		\hfill
		\includegraphics[width=0.49\linewidth]{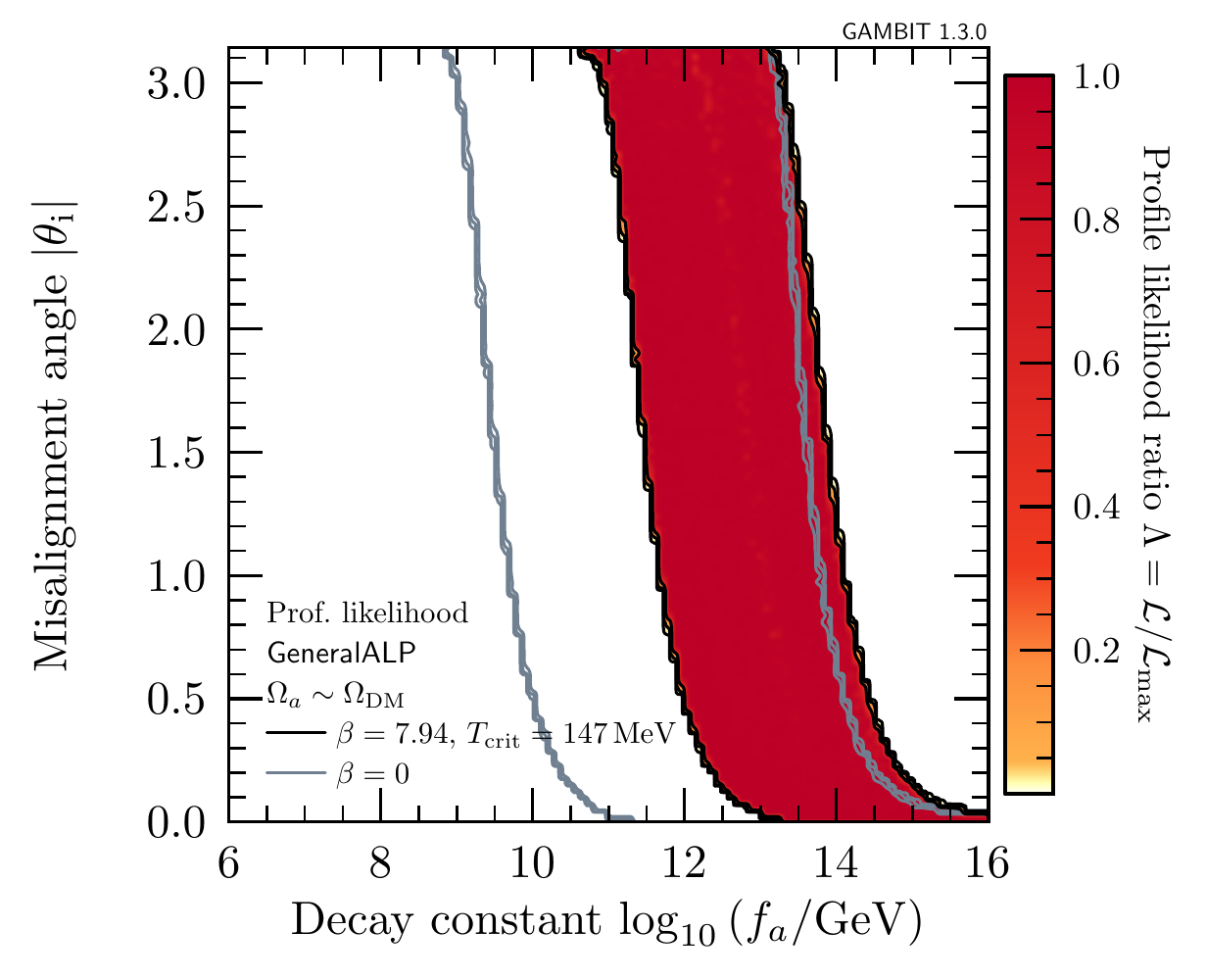}
	}
	\caption{Profile likelihoods (from \diver) for specific \genalp models that constitute all of the dark matter in the Universe. We compare the allowed regions between QCD-like ALPs with $\beta = 7.94$ and $\Tcrit = \SI{147}{\mega\electronvolt}$ (density plot and black contours) and simpler \genalp models with $\beta = 0$ (grey contours). Note that in the figure in the left panel, the region to the right of the grey contours is allowed for this model.\label{fig:GeneralALP:specialcases}}
\end{figure}

Within our selected parameter ranges, the profile likelihood does not identify preferred regions for $\Tcrit$ and~$\beta$. However, as \reffig{fig:GeneralALP:specialcases} shows, different choices of $\Tcrit$ and~$\beta$ can lead to different behaviours in the profile likelihoods of other quantities. Due to the temperature dependence of the axion energy density~\refeq{eq:axionscaling}. For example, in \reffig{fig:GeneralALP:specialcases}, we choose a generalised \qcdaxion-like model (with $\beta = 7.94$ and $\Tcrit = \SI{147}{\mega\electronvolt}$) and compare it to a \simpalp-like model ($\beta = 0$). The different slope of the exclusion region boundary in the left panel of \reffig{fig:GeneralALP:specialcases} is a consequence of how the energy density scales right after the axion field begins to oscillate. For the allowed range of values for~$\mazero$, the corresponding band of possible~$\fa$ -- given a value for the initial misalignment angle -- is also different.

\subsection{QCD axions}\label{sec:results:QCDAxions}
QCD~axions are the most well-studied type of axions to date. Unlike in previous studies, here we take into account the uncertainties due to nuisance parameters (see Sec.~\ref{sec:mod:QCDAxion}) in every part of the analysis. We also consistently scale the local DM density in axions according to their cosmological abundance. This affects the limits on the axion-photon interaction from haloscope experiments such as ADMX, as the detector signal in~\refeq{eq:genhaloscopesignal} is proportional to $\rho_{a,\text{local}}^{\phantom{2}} \, \gagg^2$. The limits on $\gagg$ therefore scale with~$1/\sqrt{\rho_{a,\,\text{local}}}$. We again cap the local axion abundance at 100\% of the local DM abundance, and penalise models that predict too much DM via the \textit{Planck} likelihood for $\OmegaCDM h^2$ (Sec.~\ref{sec:cdmconstraints}).

\paragraph*{Prior choices.}
The priors that we apply to the \qcdaxion parameters can be found in Table~\ref{tab:priors:QCDAxion}. This model imposes a number of relations between the phenomenological parameters of its parent \genalp model, which depend on nuisance parameters, i.e.\ quantities determined by simulations, theory or experiments, which are only known within an appreciable uncertainty. While we are generally not interested in inference on such parameters, their uncertainties can affect results for the actual parameters of interest. The additional nuisance parameters for \qcdaxion models are $\caggtilde$, $\LambdaQCD$, $\beta$, and $\Tcrit$. For $\caggtilde$ and $\LambdaQCD$, the nuisance likelihood is given by a 1D Gaussian for each parameter, whereas the likelihood for $\beta$ and $\Tcrit$ takes into account correlations between the two parameters (Sec.\ \ref{sec:mod:QCDAxion}). We choose flat priors from about $-5\sigma$ to $+5\sigma$ around the respective central values for all four nuisance parameters.

\begin{table}
	\caption{Prior choices for \qcdaxion models.\label{tab:priors:QCDAxion}}
	\footnotesize
	\centering
	\begin{tabular}{@{}lccc}
		\toprule
		\textbf{Model} & \multicolumn{2}{l}{\textbf{Parameter range/value}} & \textbf{Prior type} \\
		\midrule
		\qcdaxion & \iuo{\fa}{\GeV} & \prrange{e6}{e16} & log \\
		& \iuo{\LambdaQCD}{\MeV}& \prrange{73}{78} & flat \\
		& $\caggtilde$ & \prrange{1.72}{2.12} & flat \\
		& $E/N$ & \prrange{-1.33333}{174.667} & flat \\
		& $\caee$ & \prrange{e-4}{e4} & log \\
		& $\thetai$ & \prrange{-3.14159}{3.14159} & flat \\
		& $\beta$ & \prrange{7.7}{8.2} & flat \\
		& \iuo{\Tcrit}{\MeV}& \prrange{143}{151} & flat \\
		\midrule
		Local DM density & \iuo{\rho_0}{\GeV\per\centi\metre^3} & \prrange{0.2}{0.8} & flat \\
		\bottomrule
	\end{tabular}
\end{table}

The range of values that we choose for the anomaly ratio, $E/N$, is inspired by the selection criteria and range established in phenomenological studies of axion models (cf.\ Sec.~\ref{sec:mod:QCDAxion}). While the different preferred models presented in Ref.~\cite{1705.05370} form a discrete set, we assume that there is a continuous band of possible axion models, spanning a range from the lowest ($E/N = -4/3$) to the highest ($E/N = 524/3$) possible value of the anomaly ratio. Given that the number of possibilities grows very quickly if we allow for an arbitrary number of new heavy quarks in KSVZ-type models (where, however, $E/N \leq 170/3$), it is not inconceivable that such a band exists. We treat each value inside the band as equally probable before contact with data, employing a flat prior for~$E/N$.\footnote{The assignment of weights to the different discrete values or to the different parts of the band is not trivial; it becomes complicated quickly if we consider the general \qcdaxion family instead of specific DFSZ-type and KSVZ-type models, because the number of additional components (Higgs doublets or heavy quarks) is not fixed. Although it could be argued that all possible values of~$E/N$ are equally likely within each class of model, models with more additional particles might be considered more ``contrived'', and hence less probable \textit{a priori}. This is relevant because it particularly affects the higher values of~$E/N$, which cannot be achieved in the simpler models with only one new quark or two Higgs doublets. Creating a probability density function based on how often the values of $E/N$ occur for all the different cases might hence not reflect the \textit{a priori} probability for each version of the \qcdaxion model. There are also significant practical challenges to computing all possible values of~$E/N$ when the number of quarks becomes very large, as well as in the most general versions of DFSZ-type models.}  Note that this necessarily encompasses negative values for $\gagg$, as discussed in Sec.~\ref{sec:generalalpmodel}; whilst this does not impact our likelihoods (which depend only on $|\gagg|$), it does imply an asymmetric effective prior on the two signs of $\gagg$.

Assigning priors to $\fa$ and $\caee$ is more difficult. For~$\fa$, we choose a range that corresponds to our region of interest in mass: from the largest masses allowed by bounds on hot DM, to highly fine-tuned regions with very small masses and $\fa$ somewhat below the Planck scale. The logarithmic prior reflects our ignorance about the scale of new physics, given that the ability of the original QCD axion to solve the Strong~CP problem does not depend on the value of~$\fa$. We choose a generous range for $\caee$, taking a logarithmic prior around values of order unity, which may be considered the most natural value for $\caee$. Note that the lower end roughly corresponds to the minimum value that can be constrained by the $R$~parameter for the highest \qcdaxion masses we consider. Values any lower will be effectively indistinguishable.

Our choice of priors on~$\rho_0$ and $\thetai$ follow the logic from the previous section on \genalp models. Because $\rho_0$ is rather well constrained by data, the choice of log or flat prior has little impact on the final results. For $\thetai$, the causal structure of the early-Universe cosmology mentioned in Sec.~\ref{sec:realignment} means that all values of the initial misalignment angle are equally likely, so a flat prior is most appropriate.

\paragraph*{Frequentist results.}
\begin{figure}
	\centering
	{
		\includegraphics[width=0.49\linewidth]{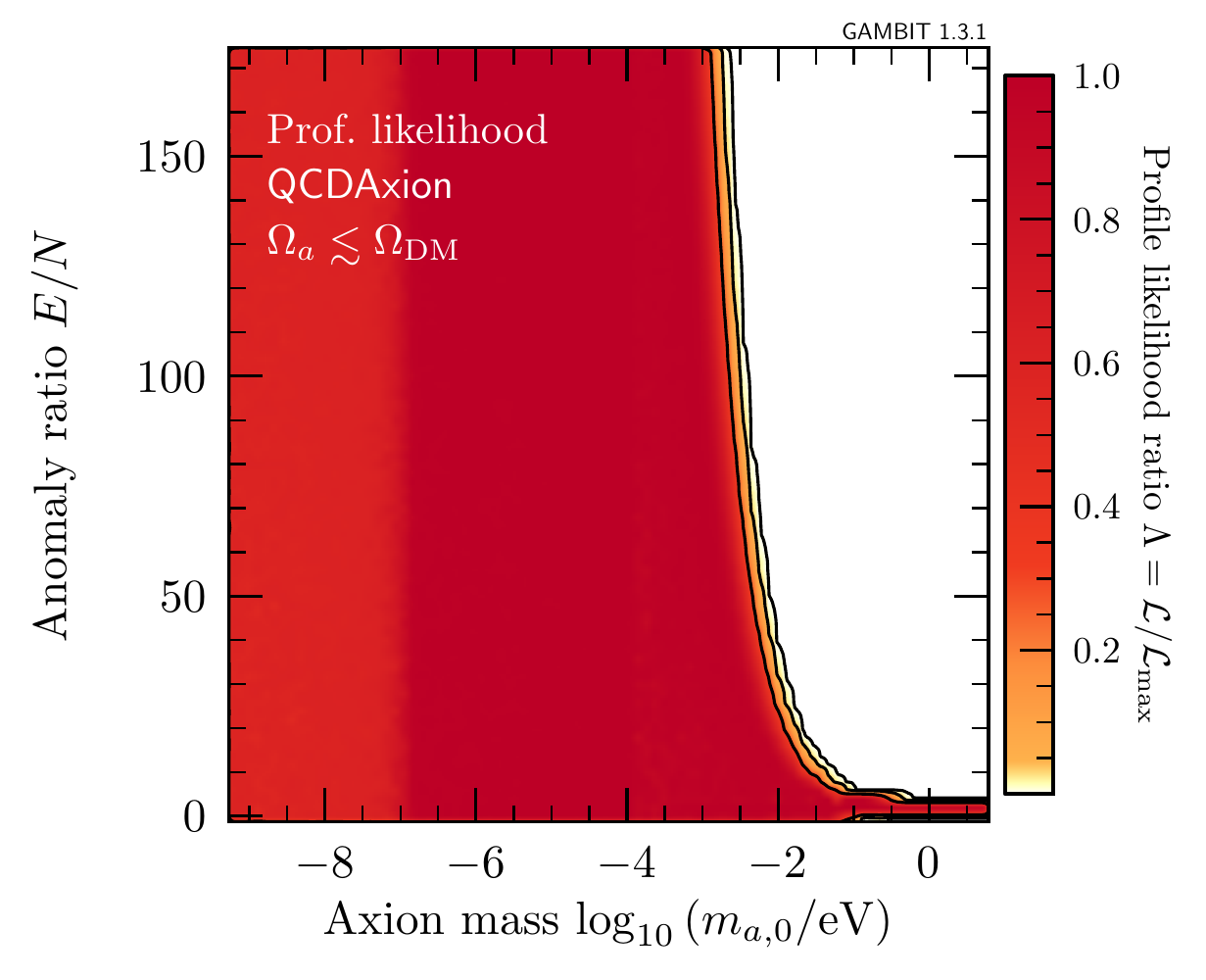}
		\hfill
		\includegraphics[width=0.49\linewidth]{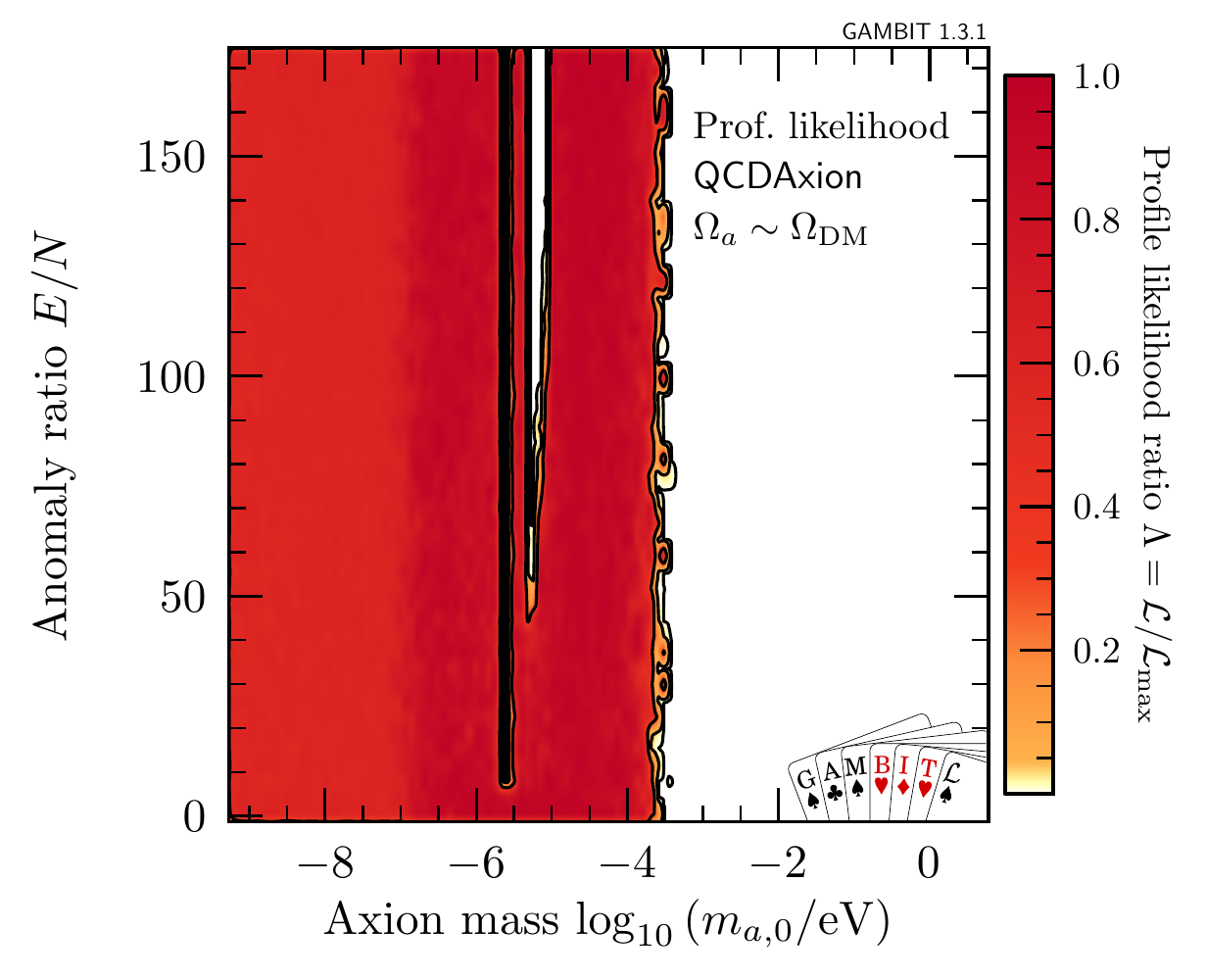}
	}
	{
		\includegraphics[width=0.49\linewidth]{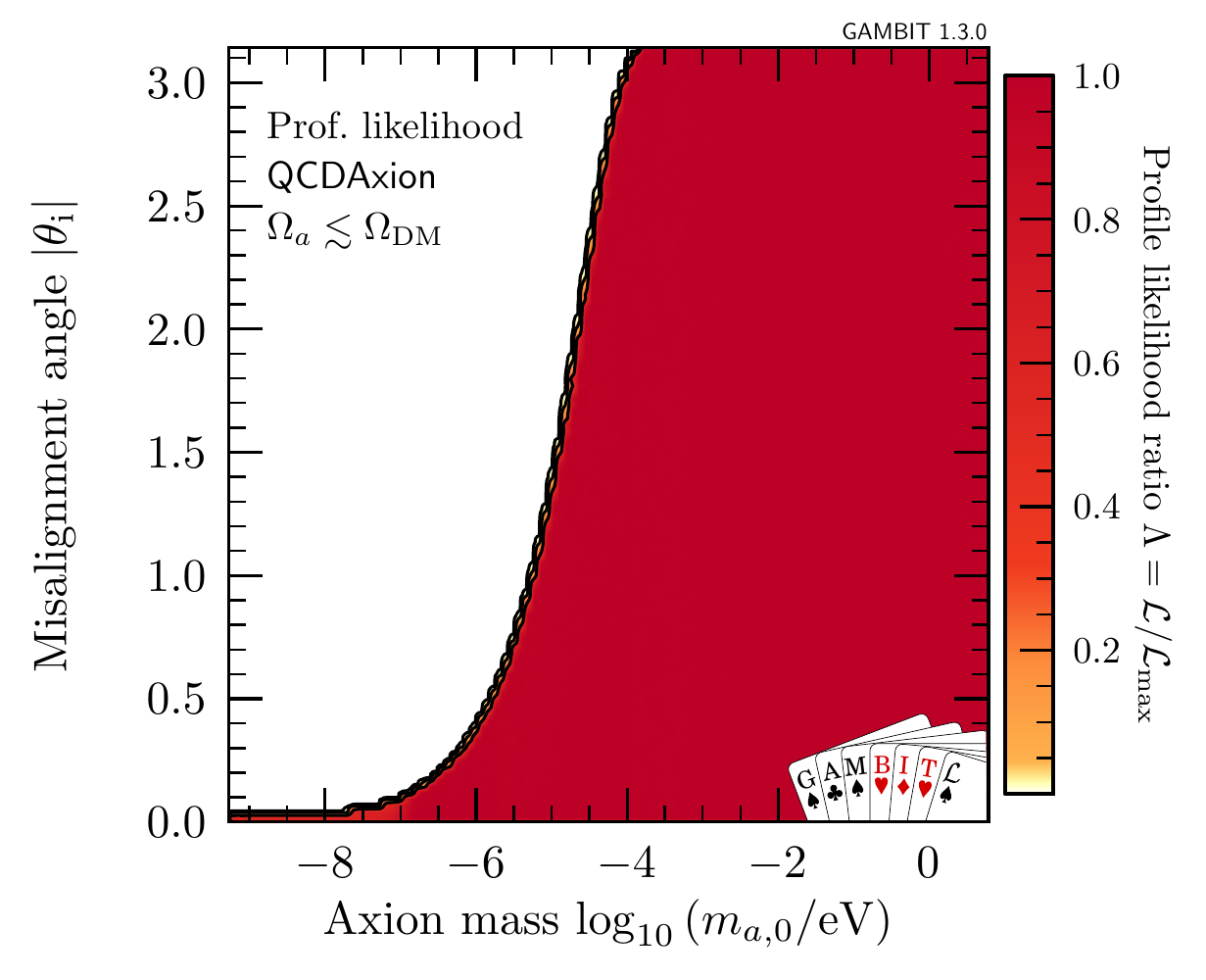}
		\hfill
		\includegraphics[width=0.49\linewidth]{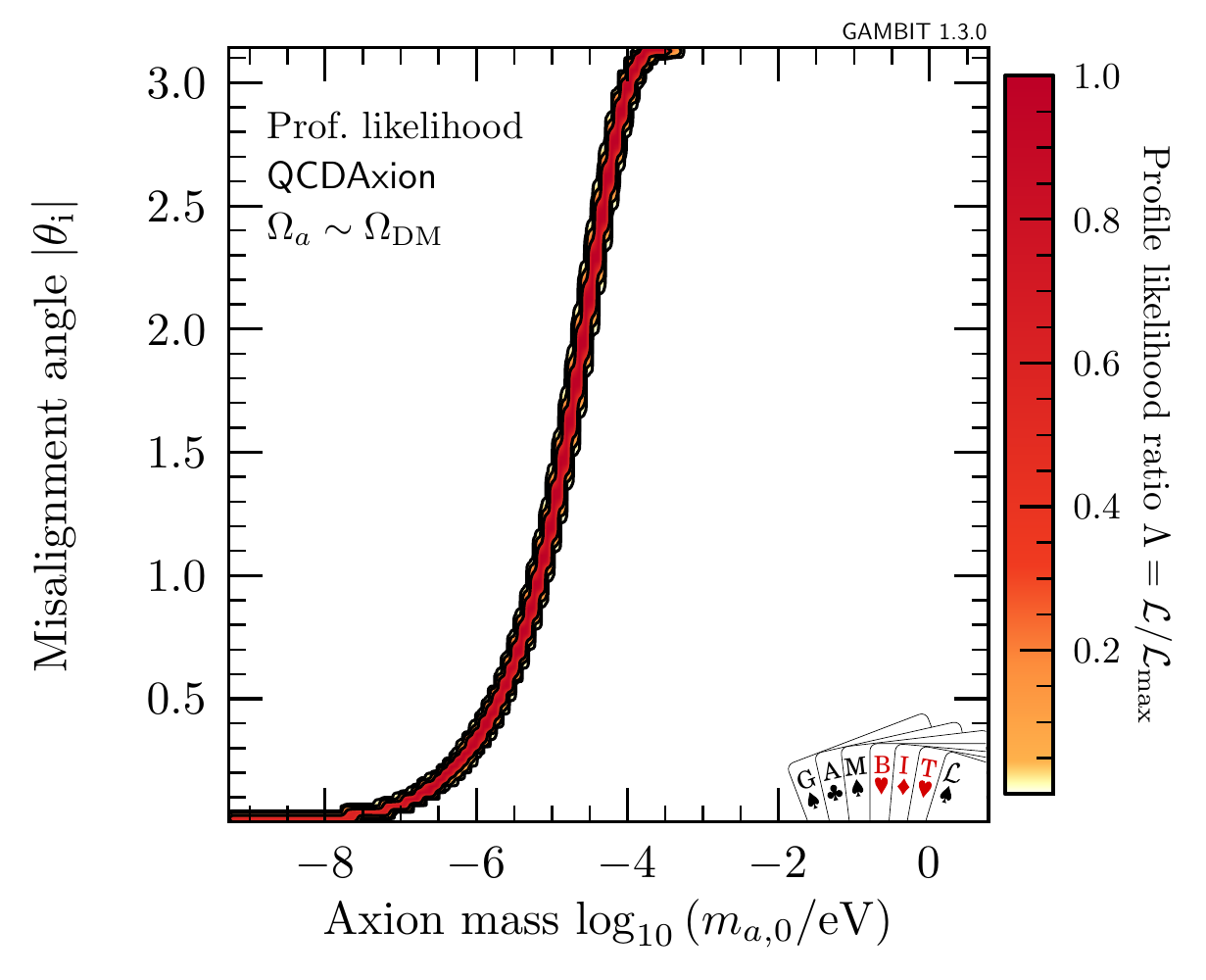}
	}
	\caption{Profile likelihoods~(from \diver) for \qcdaxion models with upper limits~(\textit{left}) and matching condition~(\textit{right}) for the observed DM relic density. The upper and lower panels show the constraints on the anomaly ratio, $E/N$, and the absolute value of the initial misalignment angle, $|\thetai|$, respectively.\label{fig:QCDAxion:frequentist}}
\end{figure}
First, let us consider statistical inference on the axion coupling strengths. There are essentially only upper limits on the axion couplings or the associated model parameters. We begin by focusing on the axion-photon interactions, as determined by the anomaly ratio $E/N$, shown in the upper row of \reffig{fig:QCDAxion:frequentist}. In the left panel we impose the relic density constraint as an upper limit, while in the right panel we demand that axions be all of DM. A notable difference between these two assumptions is that haloscopes (UF, RBF and ADMX) only provide strong limits in the latter case.

If axions are not required to be all of DM, the high-mass (low-$\fa$) region is excluded by the $R$~parameter and CAST likelihoods (cf. \reffig{fig:GeneralALP:overview}) except at very low values of $E/N$. If axions constitute all of the DM in the Universe, these constraints are not relevant because the realignment mechanism cannot produce enough DM when $|\thetai| \leq 3.14159$ and $\mazero \gsim \SI{1}{\meV}$ (cf.\ right panel of \reffig{fig:validation:realignment}), so the high-mass region is excluded.

We also see slightly lower profile likelihood values for masses below about \SI{0.1}{\micro\eV}. This is due to the role of the axion-electron coupling in the $R$~parameter likelihood: while our updated value for the helium abundance reduces the tension between theory and observations, there is still a slight preference for $\gaee \neq 0$. For small masses, however, the maximum allowed value for the axion-electron coupling, $\caee \leq \num{e4}$, is still not large enough to satisfy this small preference.

In the bottom row of \reffig{fig:QCDAxion:frequentist}, we show the allowed values for the magnitude of the initial misalignment angle, with and without the assumption that axions constitute all of DM. Due to the influence of the various nuisance parameters and the relic density likelihood, the allowed region in the right panel is not simply a line, but a band of parameter combinations that reproduce the observed DM density within the allowed uncertainties. This panel also illustrates the well-known result that the initial misalignment angle needs to be fine-tuned, i.e.\ $\left|\thetai\right| \ll 1$, for QCD axion masses of $\mazero \lsim \SI{0.1}{\micro\eV}$.\footnote{Figure~\ref{fig:QCDAxion:frequentist} contains combined results from multiple \diver runs, designed to properly sample the most fine-tuned regions of the parameter space at low $|\thetai|$ and $\mazero$.} We will investigate this issue in more detail below using a Bayesian analysis.

\paragraph*{Bayesian results.}
Breaking the PQ symmetry before inflation effectively results in a single, homogeneous value for the misalignment angle in the entire observable Universe. This gives a physical motivation for choosing a flat prior on~$\thetai$. Parameter regions in which $\thetai$ must be very small to avoid axion overproduction are hence theoretically less appealing. In a Bayesian analysis, we can see and quantify these fine-tuning issues in the (marginalised) posterior distributions, which quantify the degree of belief in certain values of the parameters given data and prior information.

\begin{figure}
	\centering
	{
		\includegraphics[width=0.49\linewidth]{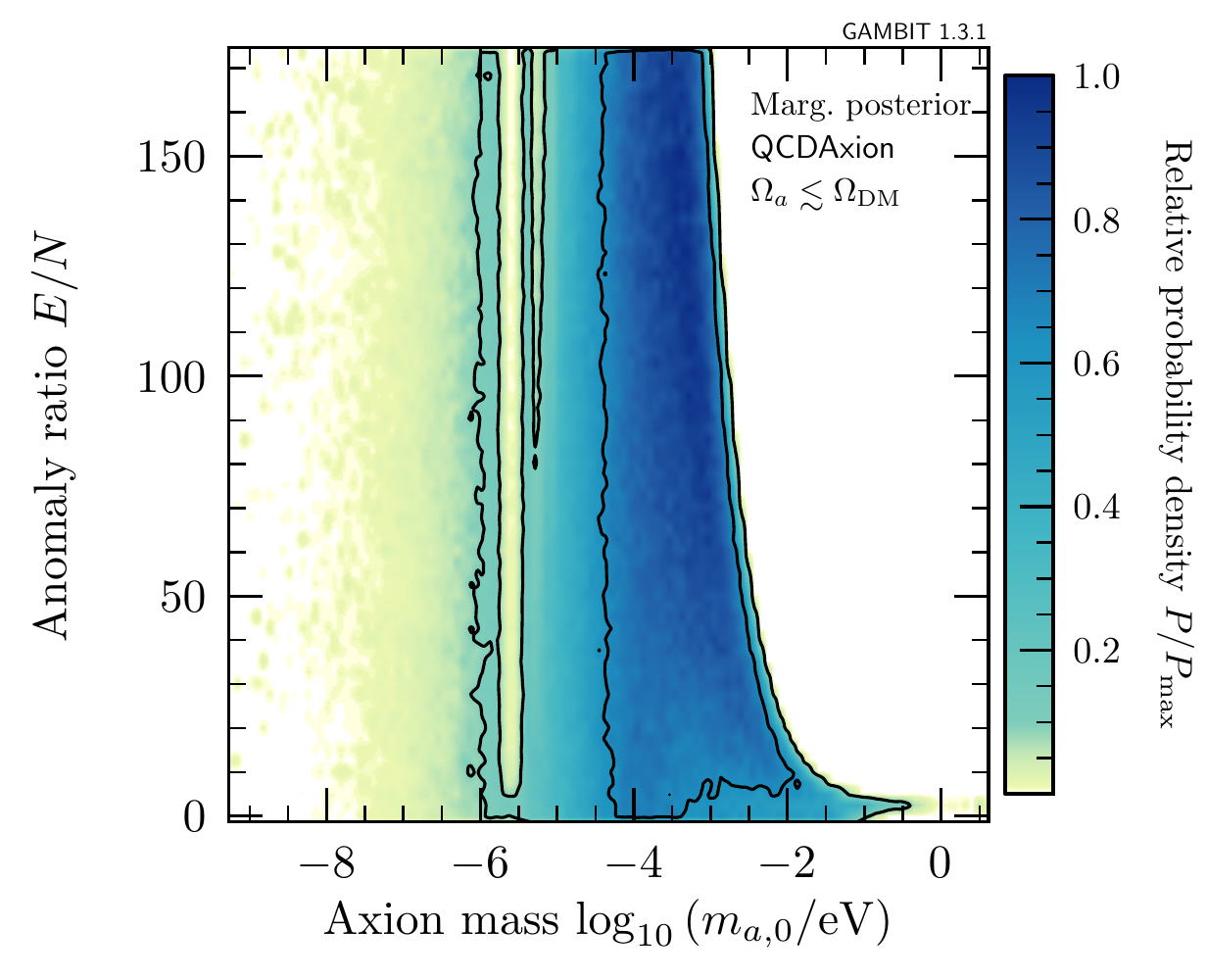}
		\hfill
		\includegraphics[width=0.49\linewidth]{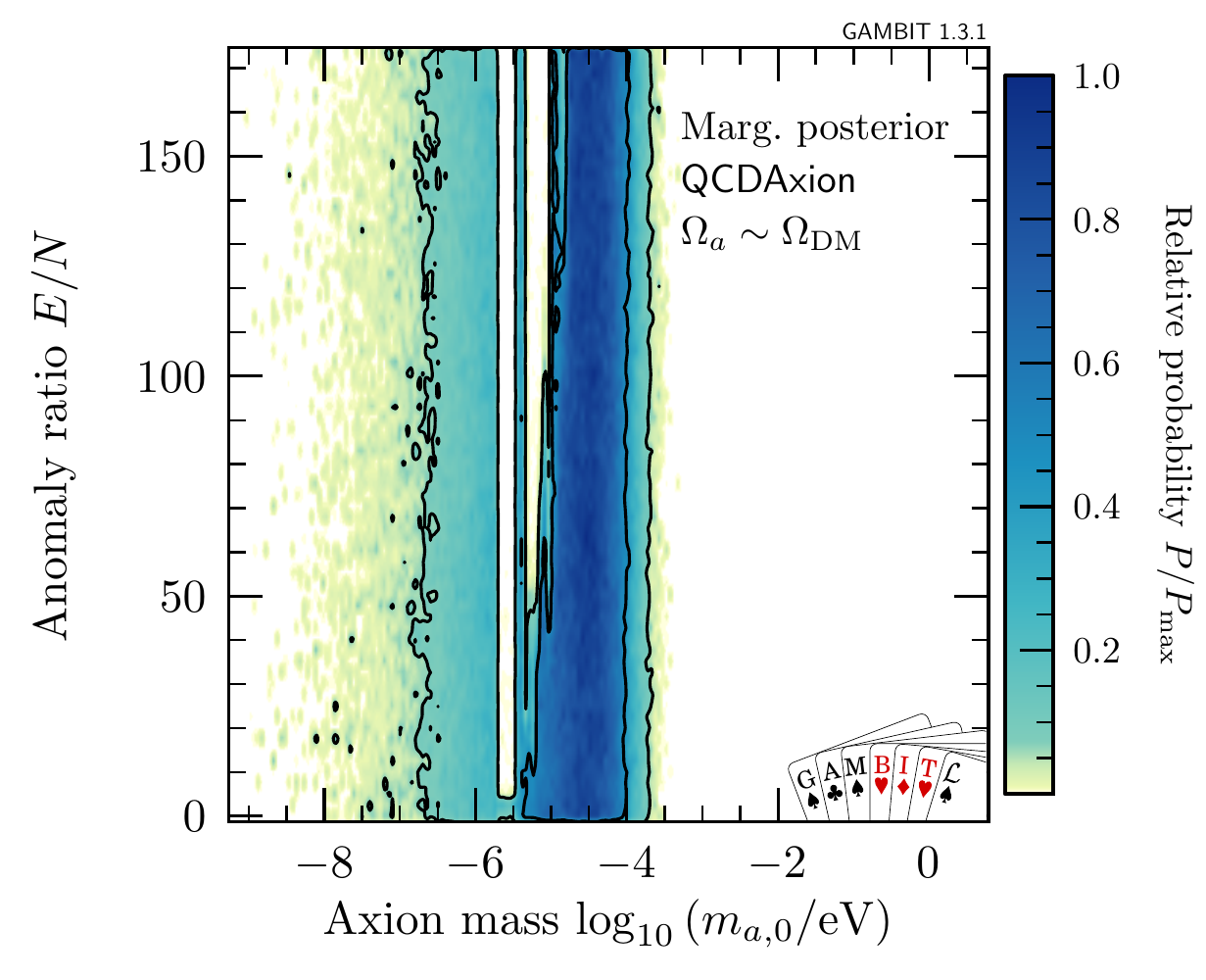}
	}
	{
		\includegraphics[width=0.49\linewidth]{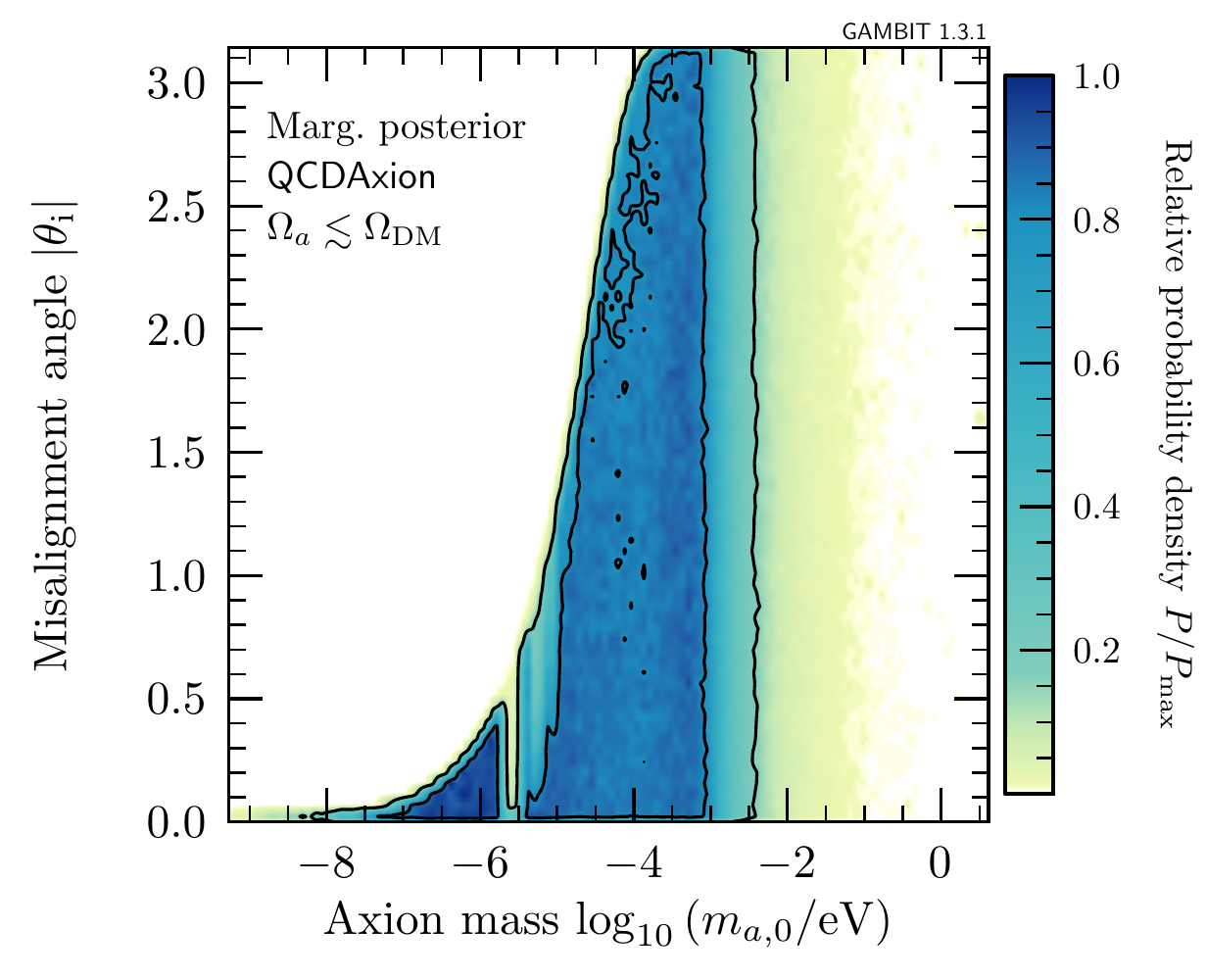}
		\hfill
		\includegraphics[width=0.49\linewidth]{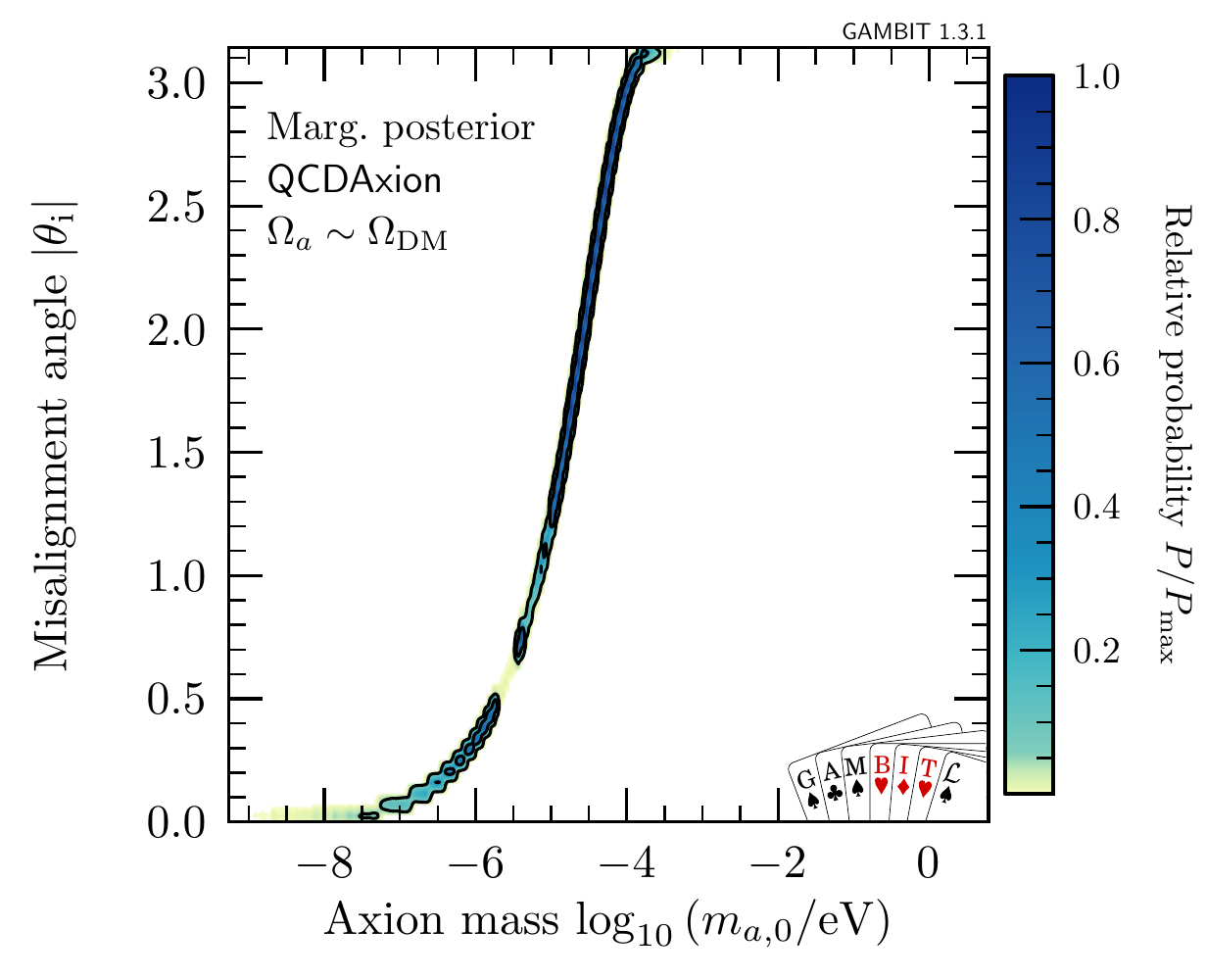}
	}
	\caption{Marginalised posteriors~(from \twalk) for \qcdaxion models with upper limits~(\textit{left}) and matching condition~(\textit{right}) for the observed DM relic density. The upper and lower panels show the constraints on the anomaly ratio, $E/N$, and the absolute value of the initial misalignment angle, $|\thetai|$, respectively.\label{fig:QCDAxion:Bayesian}}
\end{figure}

We show marginalised posteriors for the \qcdaxion model in \reffig{fig:QCDAxion:Bayesian}, once again without~(\textit{left}) and with~(\textit{right}) the requirement that QCD axions are all of DM. As a consequence of fine-tuning in~$\thetai$, the low-mass (high-$\fa$) region of the parameter space in \reffig{fig:QCDAxion:Bayesian} is disfavoured, even when taking the DM relic density as an upper limit only. This is because in the low-mass region, large absolute values of the initial misalignment angles have a small likelihood. An~$\order (1)$ value for the magnitude of the initial misalignment angle is \textit{a priori} more probable than finding a value close to zero, due to the flat prior. This conflict leads to fine-tuning becoming increasingly necessary as the axion mass decreases, which is penalised in the Bayesian analysis. Although such parameter combinations might still give valid solutions that evade all constraints, they are not as probable as others.

A similar logic applies to large axion-photon coupling, i.e.\ large~$E/N$. Due to the fine-tuning in $E/N$ necessary to evade the helioscope and $R$~parameter constraints at large axion mass (cf.\ the corresponding profile likelihood in the top left panel of \reffig{fig:QCDAxion:frequentist}), the large-$\mazero$ (low-$\fa$) region in the top left panel of \reffig{fig:QCDAxion:Bayesian} is disfavoured in the Bayesian posterior.

If we demand that axions explain all of DM, the consequences are even more dramatic. The most probable axion models are confined to the narrow band in~$\mazero$, visible in the upper right panel of \reffig{fig:QCDAxion:Bayesian}. This mass range presents a feasible target for haloscope searches, and ADMX in particular is already beginning to cut into these models from the left (low-mass end). This also explains why the band of $\mazero$-$\thetai$ values in the bottom right panel of \reffig{fig:QCDAxion:Bayesian} is not continuous, but disrupted around two points. These correspond to the ADMX and RBF/UF haloscope searches, respectively. While the RBF and UF~haloscopes cannot reach as far down into the coupling space as ADMX, they do still constrain a significant fraction of the coupling range.

\begin{figure}
	\centering
	{
		\includegraphics[width=0.4312\linewidth]{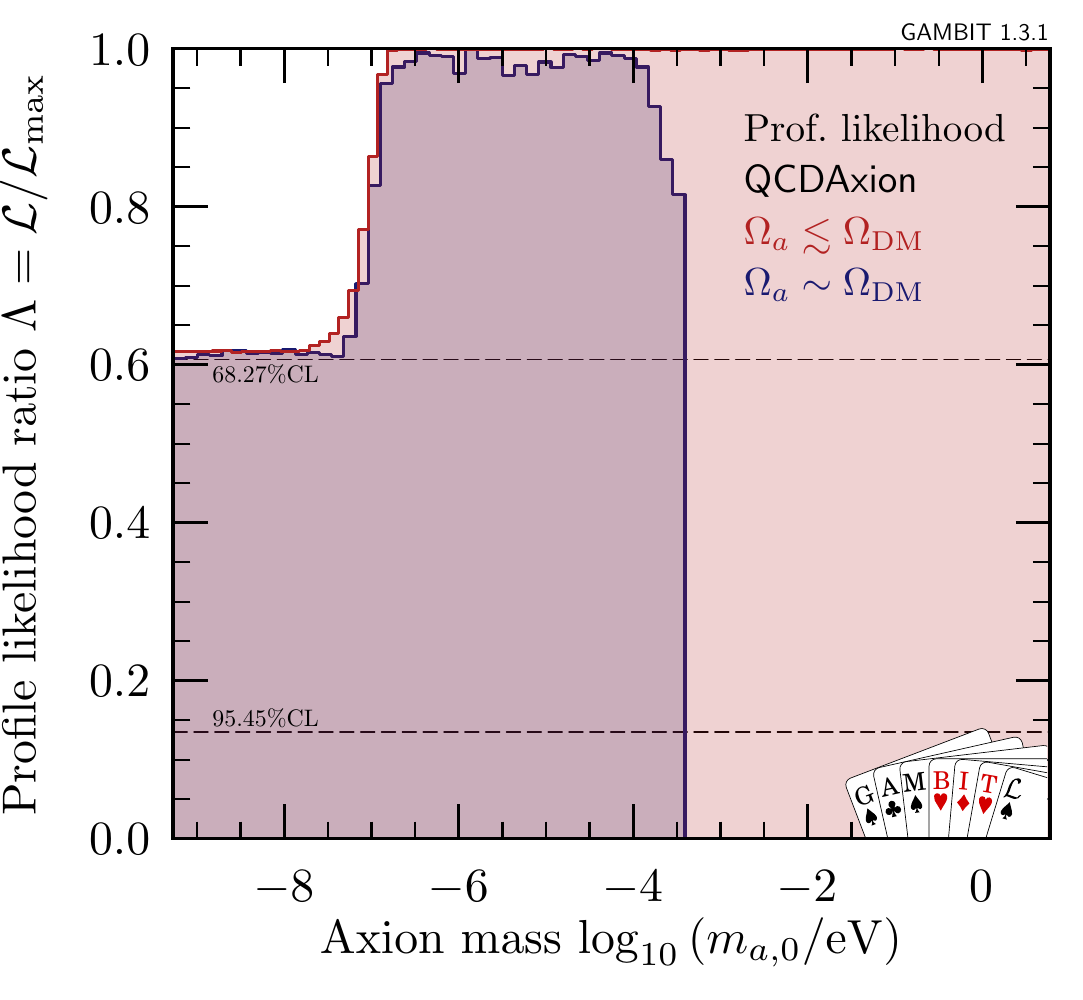}
		\hfill
		\includegraphics[width=0.4312\linewidth]{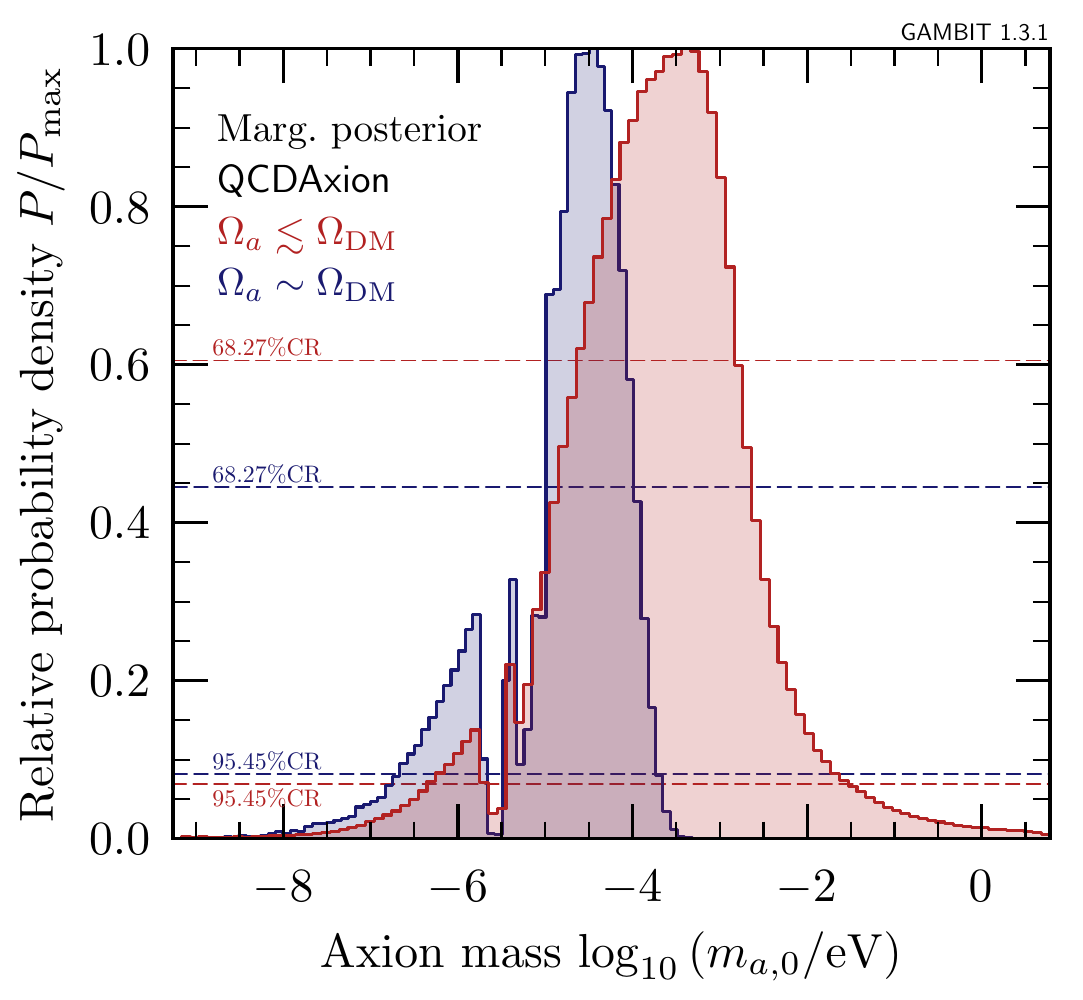}
		\hfill
	}
	\caption{Profile likelihoods~(from \diver, \textit{left}) and marginalised posteriors~(from \twalk, \textit{right}) for the mass in \qcdaxion models with upper limits~(red shading) and matching condition~(blue shading) for the DM relic density. The prior-dependence of the marginalised posteriors is investigated in Appendix~\ref{app:prior}.\label{fig:QCDAxion:mass}}
\end{figure}

The fact that the Bayesian analysis singles out a well-defined range for the QCD axion mass becomes even more apparent in \reffig{fig:QCDAxion:mass}, where we compare one-dimensional profile likelihoods and marginalised posteriors for the axion mass. The frequentist approach does not yield a clear preference for any mass range, but the posterior distributions are strongly peaked around \confirmed{$\mazero \sim \SI{100}{\micro\eV}$}. Clearly, such a result is not completely prior-independent, and we discuss the impact of adopting different priors in Appendix~\ref{app:prior}. Nevertheless, it is appealing that a Bayesian analysis can identify a preferred region of $\mazero$ which, intriguingly, falls into the range that can be covered by experimental searches. Indeed, the impact of ADMX and other haloscopes already manifests itself as dips in the right panel of \reffig{fig:QCDAxion:mass}.

The marginalised posterior in \reffig{fig:QCDAxion:mass} allows us to infer a preferred \qcdaxion mass range. When demanding that axions explain all of DM, we find that the 95\% equal-tailed credible interval for the axion mass is \updated{$\SI{0.12}{\micro\eV} \le \mazero \le \SI{0.15}{\milli\eV}$}{$\SI{0.10}{\micro\eV} \le \mazero \le \SI{0.16}{\milli\eV}$}; allowing them to constitute a fraction of DM, this becomes \updated{$\SI{0.48}{\micro\eV} \le \mazero \le \SI{3.8}{\milli\eV}$}{$\SI{0.60}{\micro\eV} \le \mazero \le \SI{7.1}{\milli\eV}$}. These numbers have minimal dependence on the adopted prior for $E/N$, but a stronger dependence on the choice of priors for $\caee$ and $\fa$ (Appendix \ref{app:prior}).

We also note that if the PQ symmetry is broken after inflation, the preferred axion mass range will generally shift to larger values due to averaging of the energy density and inclusion of topological defects (cf. Sec.~\ref{sec:axioncreation}). The lower bounds on~$\mazero$ that we quote can therefore be viewed as robust against changes of assumptions about inflation.

\begin{figure}
	\centering
	{
		\includegraphics[width=0.4312\linewidth]{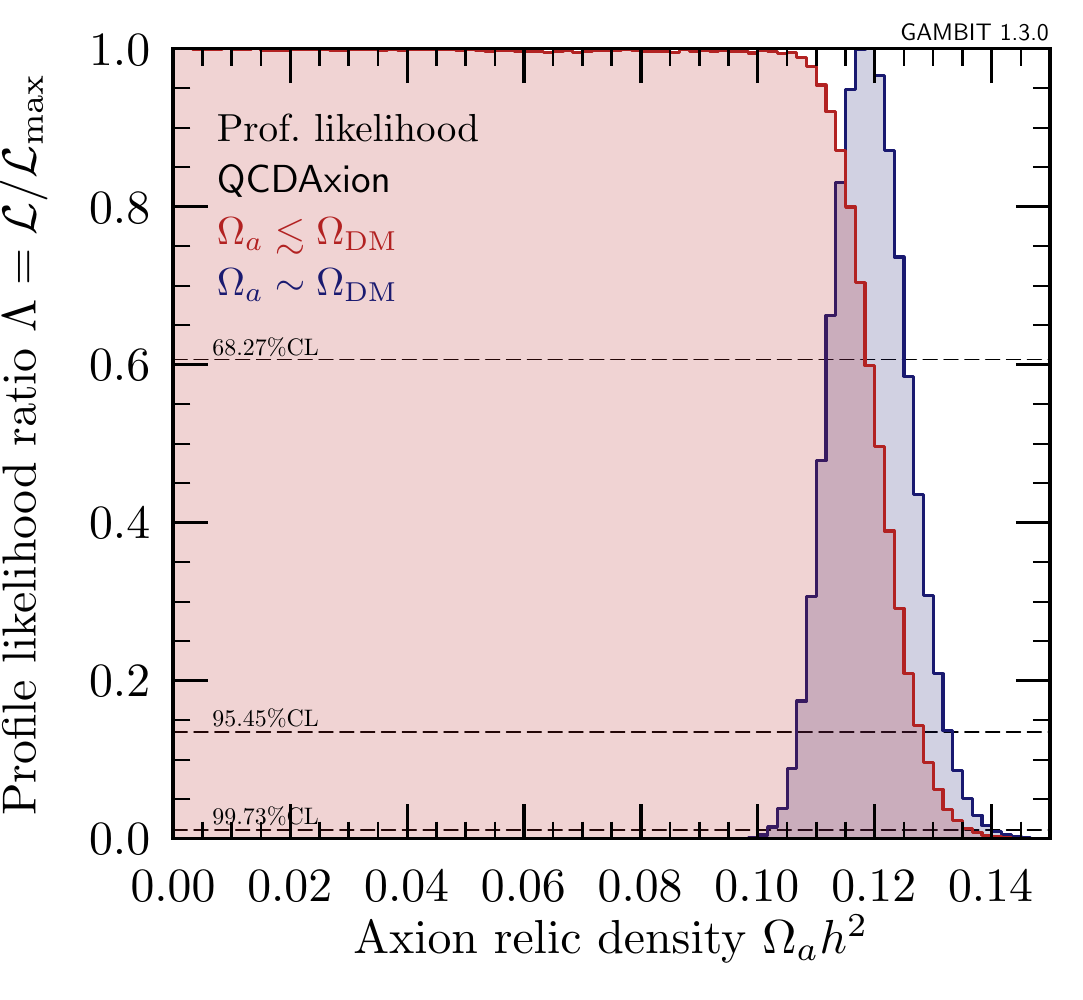}
		\hfill
		\includegraphics[width=0.4312\linewidth]{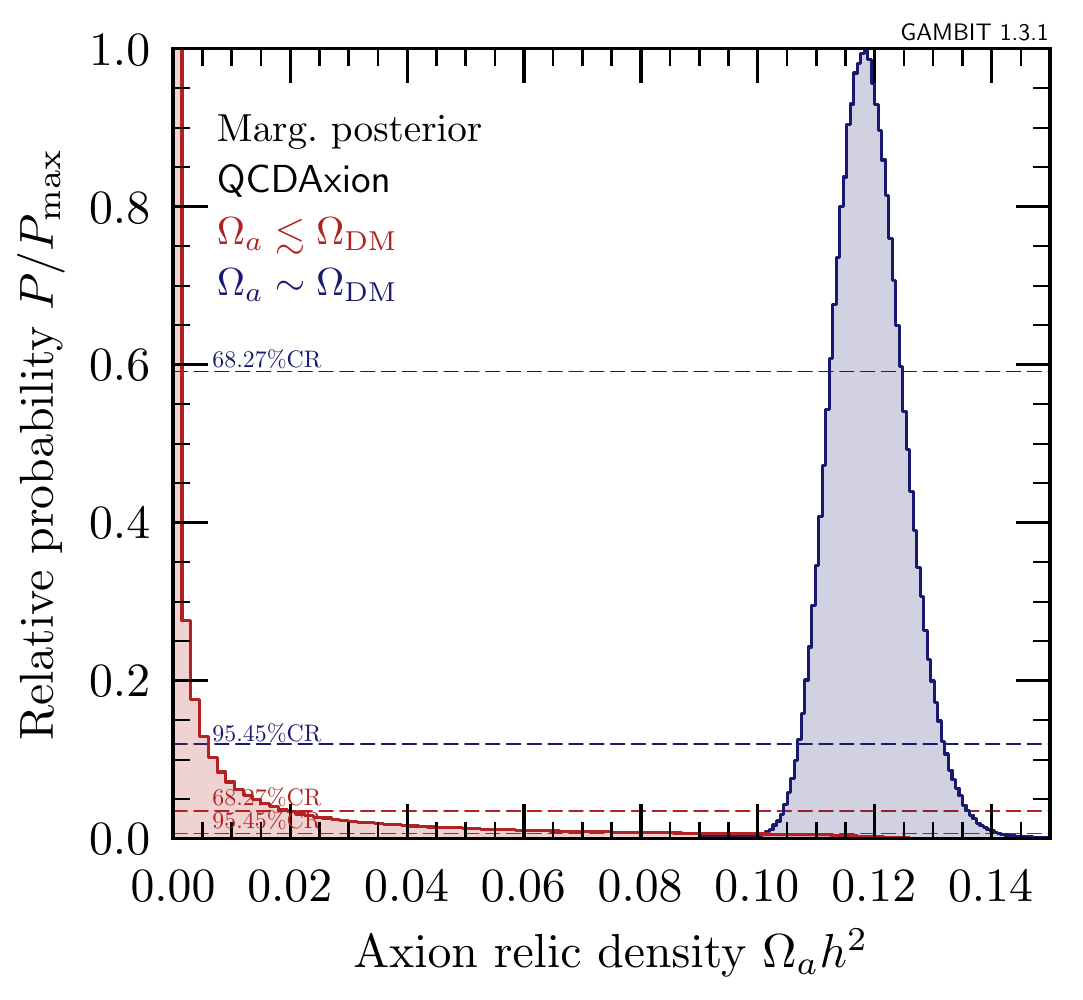}
		\hfill
	}
	\caption{Profile likelihoods~(from \diver, \textit{left}) and marginalised posteriors~(from \twalk, \textit{right}) for $\OmegaA h^2$ in \qcdaxion models with upper limits~(red shading) and matching condition~(blue shading) for the DM relic density.\label{fig:QCDAxion:omegaah2}}
\end{figure}

Finally, we also show the one-dimensional profile likelihoods and marginalised posteriors for the \qcdaxion relic density in \reffig{fig:QCDAxion:omegaah2}. Demanding that axions be all of the DM effectively dominates the outcome of this analysis. Using the DM relic density as an upper limit causes the profile likelihood to essentially follow the relic density likelihood function (left panel). In a Bayesian analysis, however, we immediately see that {\qcdaxion}s are not expected to generally provide all of the DM in the Universe, given our definition of the parameter space and priors. Imposing the DM relic density as an upper limit, the median axion relic density is \updated{\num{6.5e-3}}{\num{4.2e-3}}, or about \updated{5\%}{3.6\%} of the observed DM abundance. The 95\% credibility equal-tailed preferred range is \updated{$\num{6.8e-6} \le \OmegaA h^2 \le \num{0.10}$}{$\num{3.7e-6} \le \OmegaA h^2 \le \num{0.10}$}, which corresponds to between about \updated{0.006\% and 90\%}{0.003\% and 84\%} of the cosmological density of DM. This demonstrates that in the pre-inflationary PQ~symmetry-breaking scenario, although {\qcdaxion}s can provide a sizeable contribution to the DM density of the Universe, they probably do not contribute all of DM. Again, we stress that these statements are sensitive to the adopted prior on $\caee$ and $\fa$ (Appendix~\ref{app:prior}).

\subsection{DFSZ- and KSVZ-type models}\label{sec:results:dfszvsksvz}
The DFSZ-type (\dfszI, \dfszII) and KSVZ-type  (\ksvz) models differ from their parent model, the \qcdaxion, in that they specify the axion-photon and axion-electron coupling strengths, or at least limit them to a well-defined range for a given axion mass. They are but a few of the many possible phenomenologically-inspired models, but they serve as interesting archetypes of their respective subclasses to compare with more general {\qcdaxion} models.

\paragraph*{Prior choices.}
\begin{table}
	\caption{Prior choices for \dfszI, \dfszII and \ksvz models. Note that the priors listed in the first section of the table apply to all three models.\label{tab:priors:DFSZvsKSVZ}}
	\footnotesize
	\centering
	\begin{tabularx}{0.87\textwidth}{@{}lcccX@{}}
		\toprule
		\textbf{Model} & \multicolumn{2}{l}{\textbf{Parameter range/value}} & \textbf{Prior type} & \textbf{Comments}\\
		\midrule
		& \iuo{\fa}{\GeV} & \prrange{e6}{e16} & log & Applies to all\\
		& \iuo{\LambdaQCD}{\MeV}& \prrange{73}{78} & flat & Applies to all\\
		& $\caggtilde$ & \prrange{1.72}{2.12} & flat & Applies to all\\
		& $\thetai$ & \prrange{-3.14159}{3.14159} & flat & Applies to all\\
		& $\beta$ & \prrange{7.7}{8.2} & flat & Applies to all\\
		& \iuo{\Tcrit}{\MeV}& \prrange{143}{151} & flat & Applies to all\\
		\midrule
		\dfszI & $E/N$ & $8/3$ & delta & \\
		& $\tan(\beta')$ & \prrange{0.28}{140.0} & log & \\
		\dfszII & $E/N$ & $2/3$ & delta & \\
		& $\tan(\beta^\prime)$ & \prrange{0.28}{140.0} & log & \\
		\ksvz & $E/N$ & $0$, $2/3$, $5/3$, $8/3$ & delta & Various discrete choices\\
		\midrule
		Local DM density & \iuo{\rho_0}{\GeV\per\centi\metre^3} & \prrange{0.2}{0.8} & flat \\
		\bottomrule
	\end{tabularx}
\end{table}
Our prior choices for the DFSZ and KSVZ model can be found in Table~\ref{tab:priors:DFSZvsKSVZ}. For most of them, the rationale is the same as for {\qcdaxion}s presented in Sec.~\ref{sec:results:QCDAxions}. The only differences are in the parameters related to couplings. We fix $E/N$ to some typical values considered previously in the literature \cite[e.g.][]{1708.02111}. The range that we choose for $\tan(\beta^\prime)$ in DFSZ-type models reflects the values allowed by perturbativity bounds~\cite{1301.0309}. Our choice of a log prior for $\tan(\beta^\prime)$ reflects the assumption that each possible Higgs vacuum expectation value is equally likely; indeed, any sensible prior choice for this parameter should reflect the fact that the two Higgs doublets may be interchanged, and the prior should be invariant under inversion of the ratio of vacuum expectation values.

\paragraph*{Frequentist results.}
The DFSZ- and KSVZ-type models are essentially restrictions of the allowed QCD axion couplings. We therefore only expect to see qualitative differences in the results from the different models where the DFSZ and KSVZ interaction strengths cannot be tuned sufficiently to evade constraints from haloscopes (if axions make up all of DM) and the $R$~parameter.

\begin{figure}
	\centering
	\includegraphics[width=0.618\linewidth]{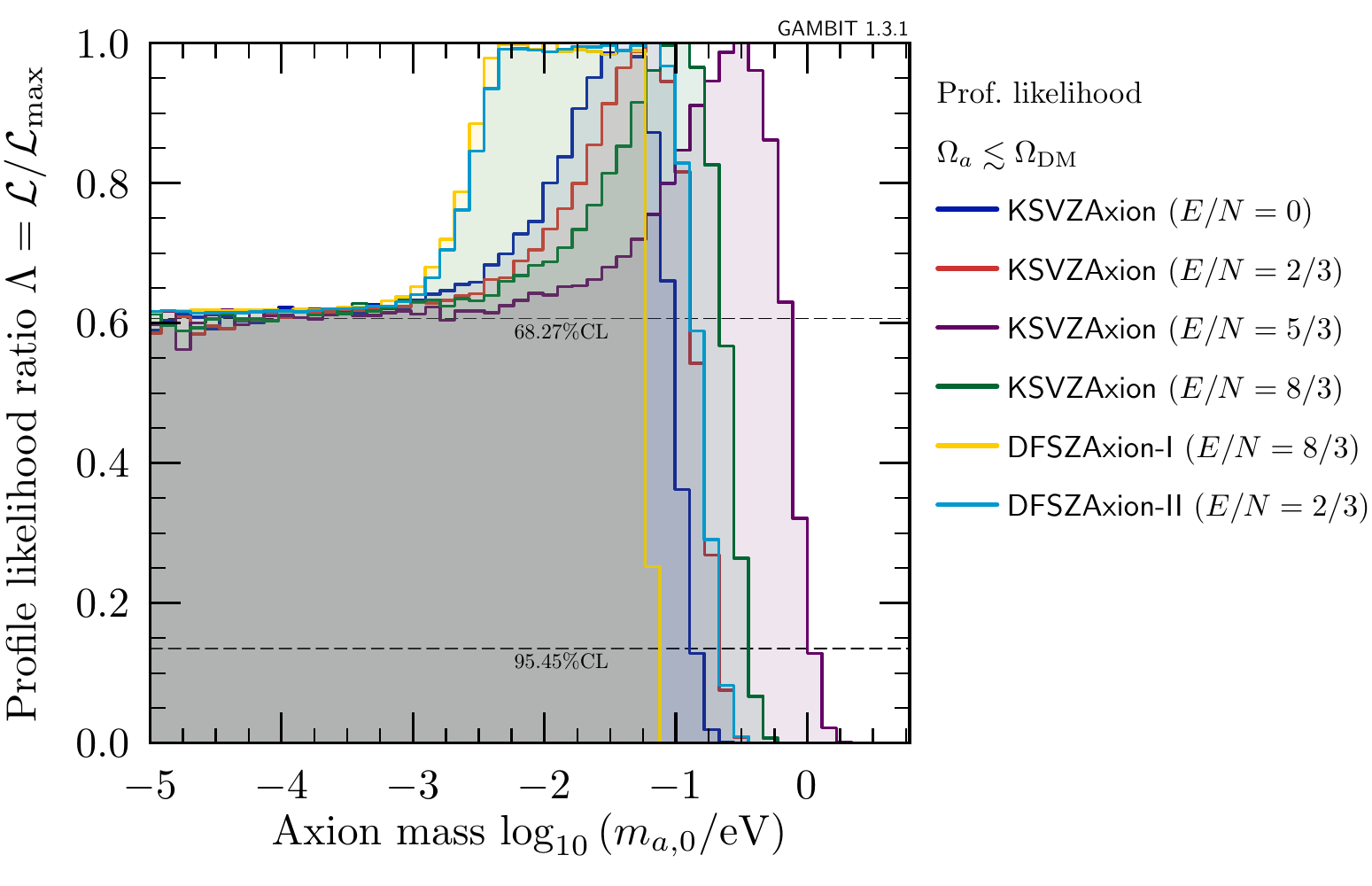}
	\caption{Profile likelihood~(from \diver) for the axion mass in the different DFSZ- and KSVZ-type models, imposing the DM relic density as an upper limit.\label{fig:dfszvsksvz:mass:oneD}}
\end{figure}
Figure~\ref{fig:dfszvsksvz:mass:oneD} shows the profile likelihood constraints on various axion models, imposing the DM density as an upper limit. We can see that the upper limit on the axion mass depends on the value of~$E/N$ in a given model (cf. \reffig{fig:GeneralALP:overview}), giving \ksvz models with $E/N = 5/3$ the largest allowed parameter space out of all the models compared here. The different values for~$E/N$ are also the reason for the different positions of the peaks in $\mazero$; the slight preference for non-zero couplings in the $R$~parameter likelihood requires slightly different axion masses for different~$E/N$.

Demanding that axions explain all of DM, \reffig{fig:dfszvsksvz:mass:oneD} would change slightly. All models with $\mazero \gsim \SI{0.1}{\meV}$ would be ruled out (not being able to provide all of the DM through the realignment mechanism), and ADMX would partially constrain all models except those with $E/N = 5/3$ (cf. \reffig{fig:GeneralALP:overview}).

\begin{figure}
	\centering
	{
		\includegraphics[width=0.49\linewidth]{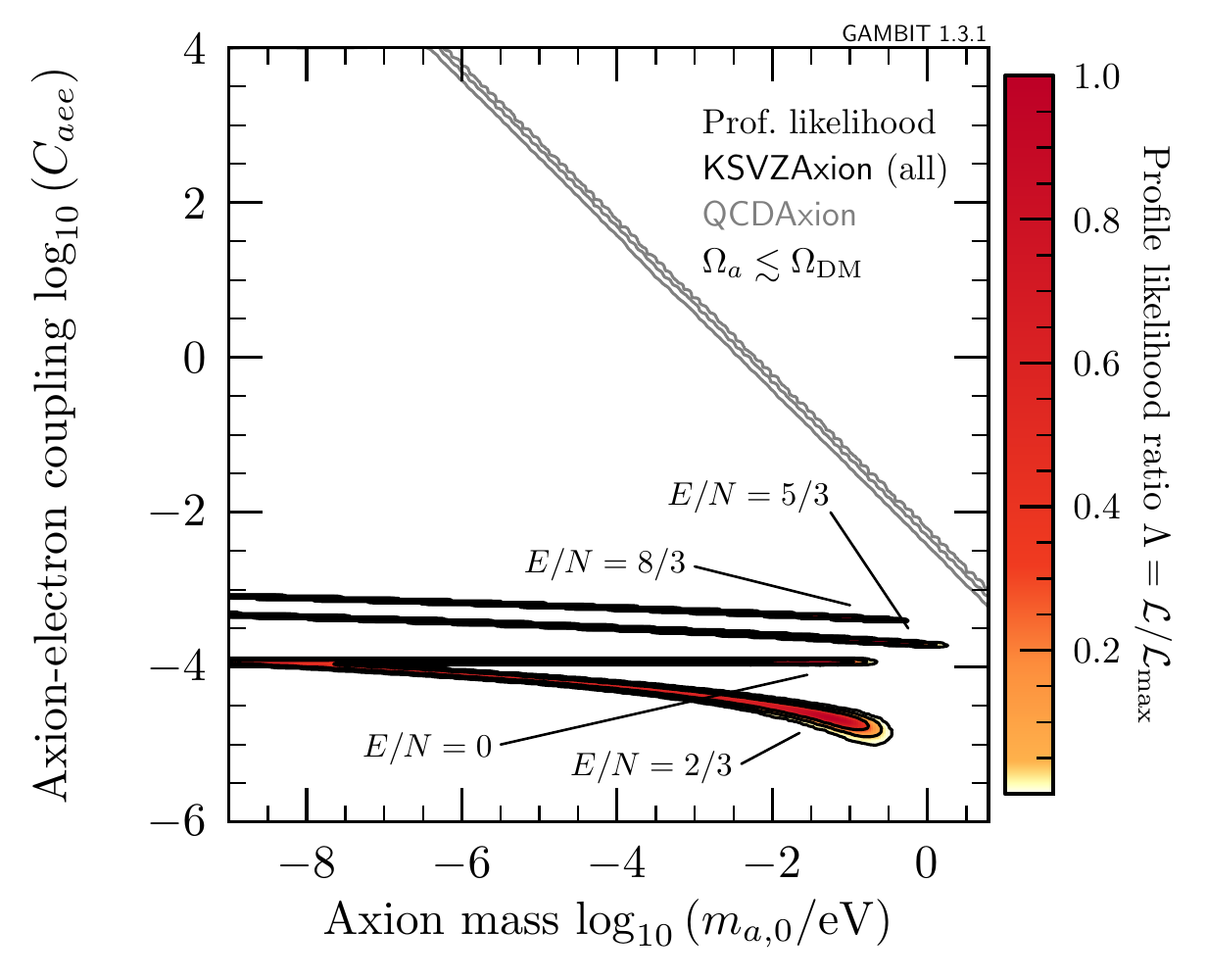}
		\hfill
		\includegraphics[width=0.49\linewidth]{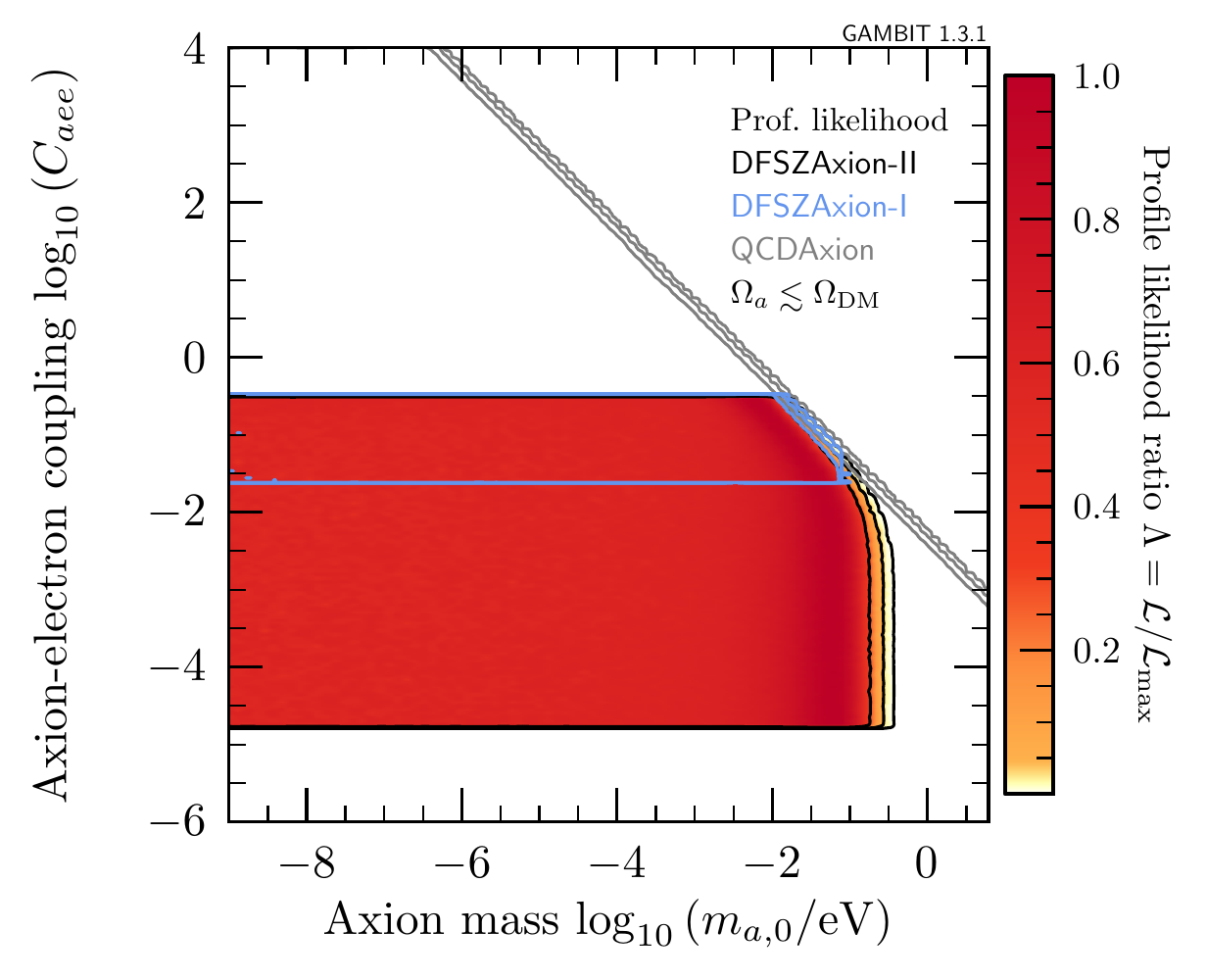}
	}
	\caption{Profile likelihood~(from \diver) for \ksvz (\textit{left}), \dfszI (\textit{right}; blue contours), and \dfszII models (\textit{right}; coloured region and black contours), compared to the profile likelihood for the \qcdaxion model (grey lines). All results use the observed relic density of DM as an upper limit on the relic axion abundance.\label{fig:dfszvsksvz:caee}}
\end{figure}

The relation between the DFSZ- and KSVZ-type models and their parent \qcdaxion model determines their allowed axion-electron couplings. Figure~\ref{fig:dfszvsksvz:caee} shows how the allowed parameter space in the $\mazero$-$\caee$ parameter plane of the \qcdaxion model is constrained further by imposing additional relations between the different model parameters in the KSVZ- and DFSZ-type models.

This is most striking in the case of the KSVZ-type models, for which $\caee$ is only induced at the loop level and depends directly on $\mazero$ \refeq{eq:ksvzgaee}. Note that the ordering of the \ksvz regions is also non-monotonic in $E/N$ due to the difference term in \refeq{eq:ksvzgaee}. The finite sizes of the allowed parameter regions are simply a result of the nuisance parameters included in the relation between $\caee$ and $\mazero$. For DFSZ-type models, $\caee$ depends on the additional parameter $\tan (\beta^\prime)$, which makes it possible to accommodate a wide range of axion-electron couplings. However, the parameter space is also more constrained in this case, as very large values of $\caee$ cannot be realised given other constraints on $\tan (\beta^\prime)$. Also note that, due to the different coupling structure in \dfszI and \dfszII models~\refeq{eq:dfsz:caee}, the same range for $\tan (\beta^\prime)$ translates into a lower minimal value of~$\caee$ in \dfszII models than in \dfszI models. The resulting possible range for~$\caee$ in \dfszII models (and \ksvz models with $E/N=5/3$) also extends to slightly lower values than the prior box that we chose for {\qcdaxion}s.

\paragraph*{Bayesian results.}
We use the nested sampling package \MultiNest to estimate the Bayesian evidences $\mathcal{Z}(\mathcal{M})$ for each model $\mathcal{M}$. From these we construct the Bayes factor \cite{Jeffreys:1939xee,10.2307/2291091,10.2307/4356165}
\begin{equation}
	\mathcal{B} \equiv \frac{\mathcal{Z}(\mathcal{M}_1)}{\mathcal{Z}(\mathcal{M}_2)} \equiv \frac{\int \!  \mathcal{L}\left(\text{data}\left| \right. \boldsymbol{\theta_1} \right) \mathcal{P}_1(\boldsymbol{\theta_1}) \, \dd \boldsymbol{\theta_1}}{\int \!  \mathcal{L}\left(\text{data}\left| \right. \boldsymbol{\theta_2} \right) \mathcal{P}_2(\boldsymbol{\theta_1}) \, \dd \boldsymbol{\theta_2}} \, ,
\end{equation}
where $\mathcal{M}_1$ and $\mathcal{M}_2$ are the two models under investigation, $\boldsymbol{\theta_1}$ and~$\boldsymbol{\theta_2}$ are their parameters, $\mathcal{P}_1$ and $\mathcal{P}_2$ are their priors and $\mathcal{L}$ is the likelihood. The Bayes factor is connected to the odds, i.e.\ the ratio of posterior probabilities, of the models being correct:
\begin{equation}
	\frac{\mathcal{P}\left(\mathcal{M}_1 \left| \text{data} \right.\right)}{\mathcal{P}\left(\mathcal{M}_2 \left| \text{data} \right.\right)} = \mathcal{B} \; \frac{\mathcal{P}(\mathcal{M}_1)}{\mathcal{P}(\mathcal{M}_2)} \, .
\end{equation}
In this paper, we assign equal prior probabilities to both models being correct, i.e.\ choose $\mathcal{P}(\mathcal{M}_1)/\mathcal{P}(\mathcal{M}_2) = 1$, so the Bayes factor is the same as the posterior odds ratio.

Using {\MultiNest}'s nested sampling (as opposed to importance nested sampling) estimates for evidences, we calculated the odds in favour of KSVZ- and DFSZ-type axion models over the \qcdaxion model. In terms of the commonly used scale for Bayes factors~\cite{Jeffreys:1939xee,10.2307/2291091}, we find that there is generally no noticeable evidence for or against any of these models, which would require an odds ratio of more than~3:1 (or less than~1:3).

Imposing the relic DM density as an upper limit, the odds in favour of any KSVZ- or DFSZ-type axion models, compared to {\qcdaxion}s, are~\confirmed{2:1}. If we demand that axions constitute all of DM, the odds reduce to~\confirmed{1:1}.

The outcome of the model comparison is not surprising, as we have not included any positive evidence for axions at this stage. We will discuss in the following section how these conclusions change when including WD cooling hints.

\subsection{Cooling hints}\label{sec:results:coolinghints}
Observables related to stellar cooling offer a unique opportunity to constrain the axion parameter space. If future observations confirm the need for additional cooling channels to explain the observed decreases in WD pulsation periods, we may be able to use WDs to measure the axion mass and coupling strengths. In this section we add the likelihoods related to the WD cooling hints to our analysis, emphasising once again the caveats and difficulties associated with assigning uncertainties to the model predictions (cf.\ Sec.~\ref{sec:coolinghints}). Here our prior choices for each model are the same as in the preceding sections. A detailed numerical comparison of our results to previous works~\cite{1512.08108,1708.02111} is not meaningful due to differences in the choice of WD likelihood function, but the findings are qualitatively similar.

\subsubsection{QCD axions}\label{sec:results:cooling:QCDAxions}
Previous studies have mostly considered the phenomenological couplings~$\gagg$ and~$\gaee$ or specific QCD~axion models with fixed~$E/N$. Here, we investigate which parts of the broader \qcdaxion parameter space can explain the cooling hints.

\paragraph*{Frequentist results.}
\begin{figure}
	\centering
	{
		\includegraphics[width=0.49\linewidth]{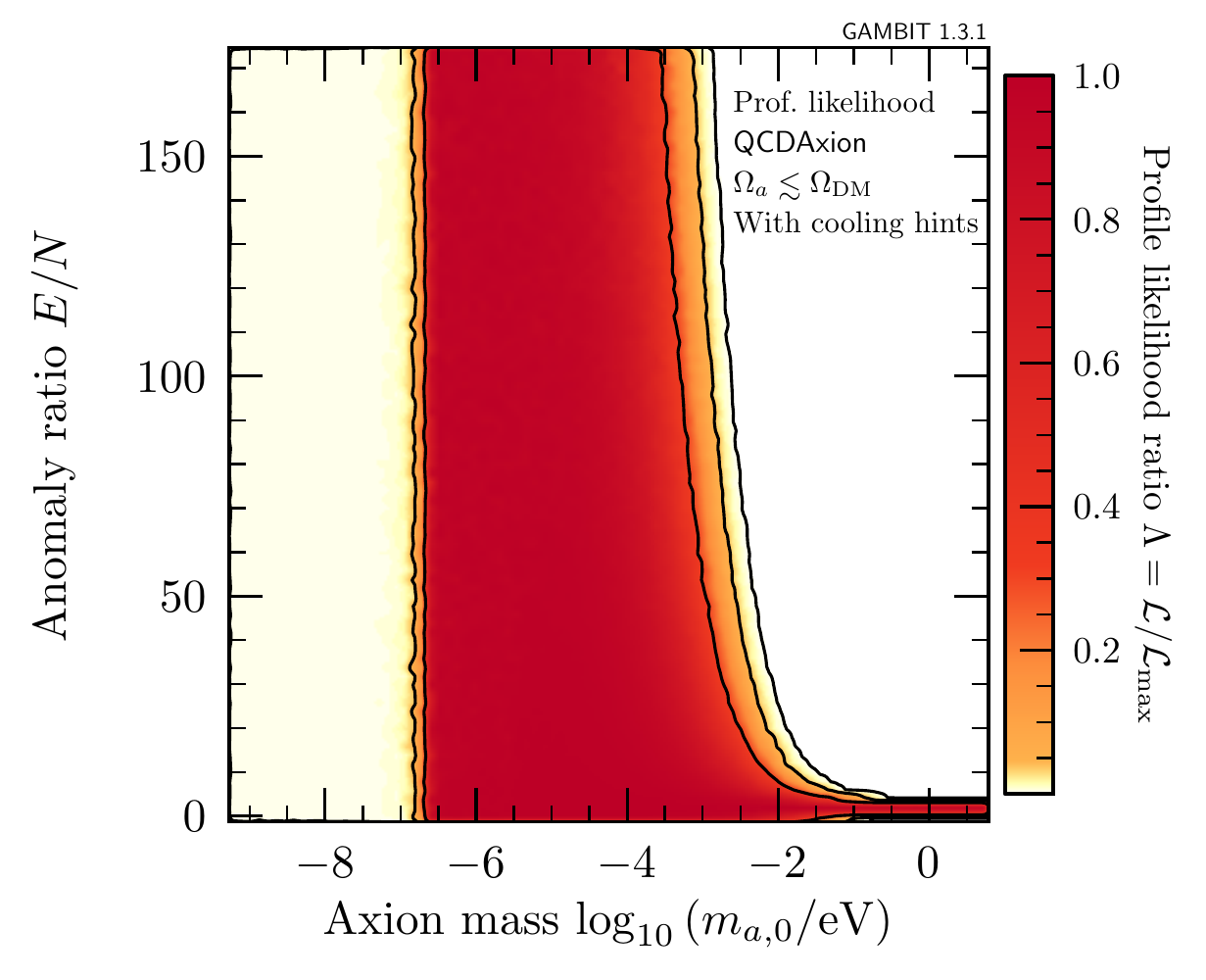}
		\hfill
		\includegraphics[width=0.49\linewidth]{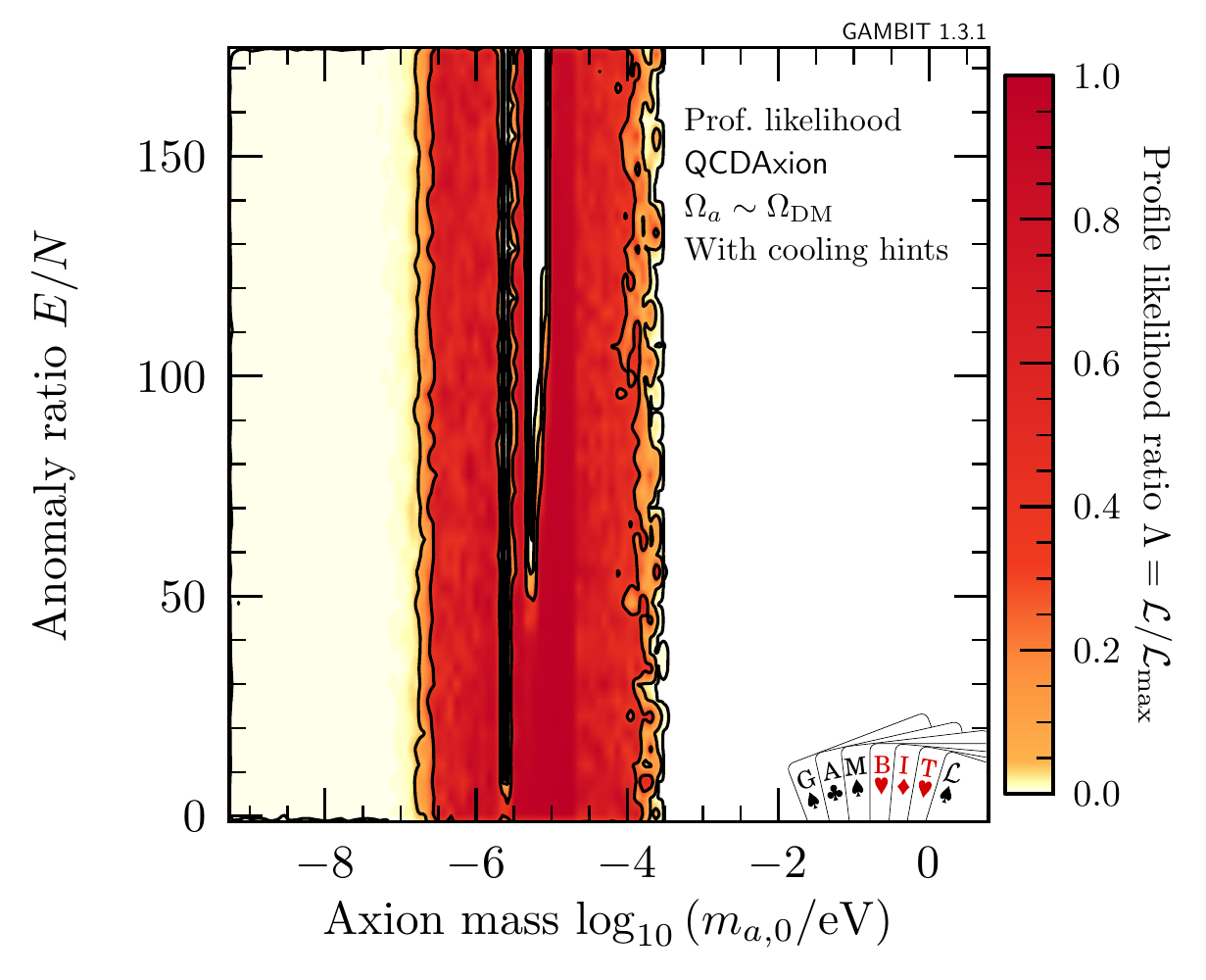}
	}
	{
		\includegraphics[width=0.49\linewidth]{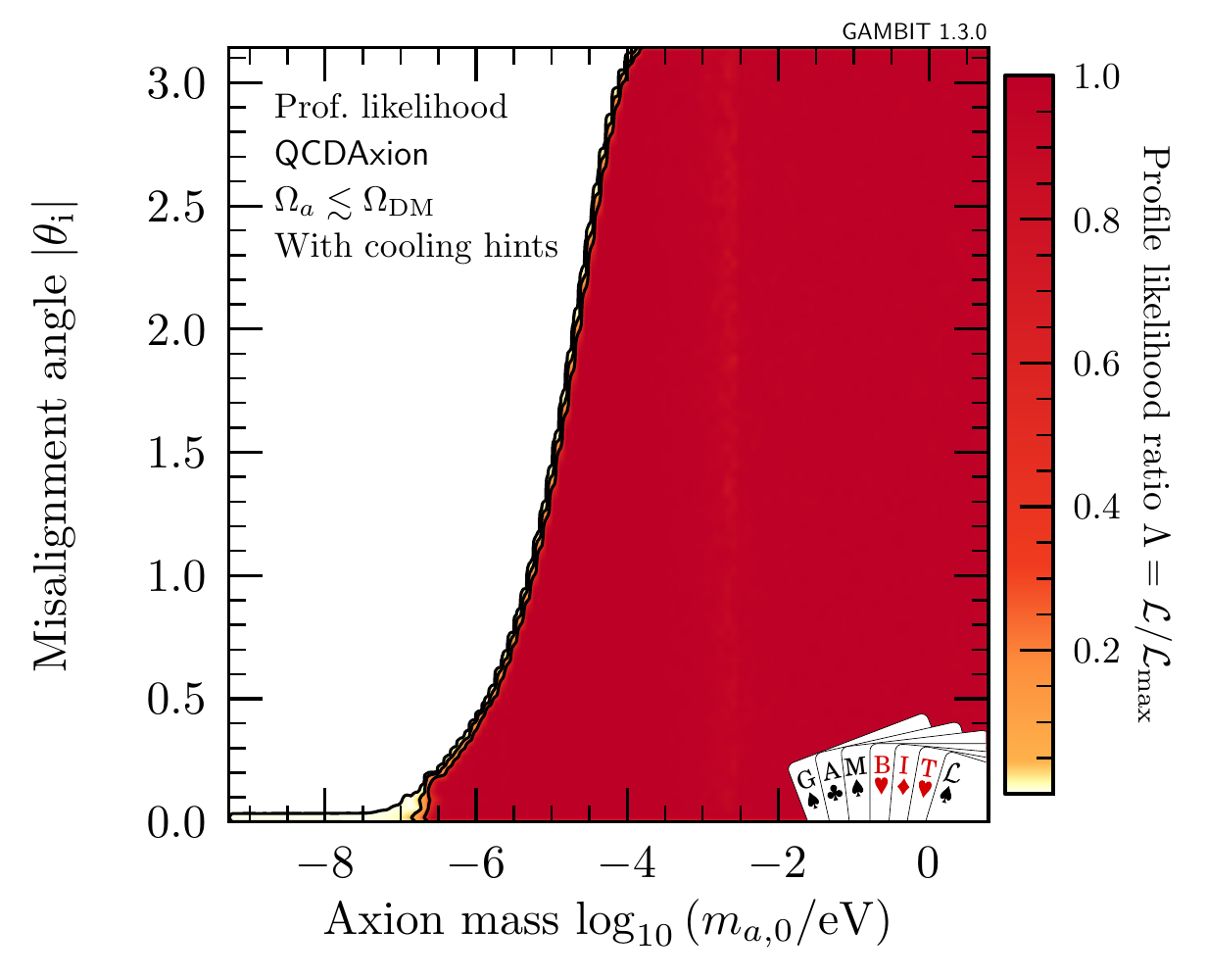}
		\hfill
		\includegraphics[width=0.49\linewidth]{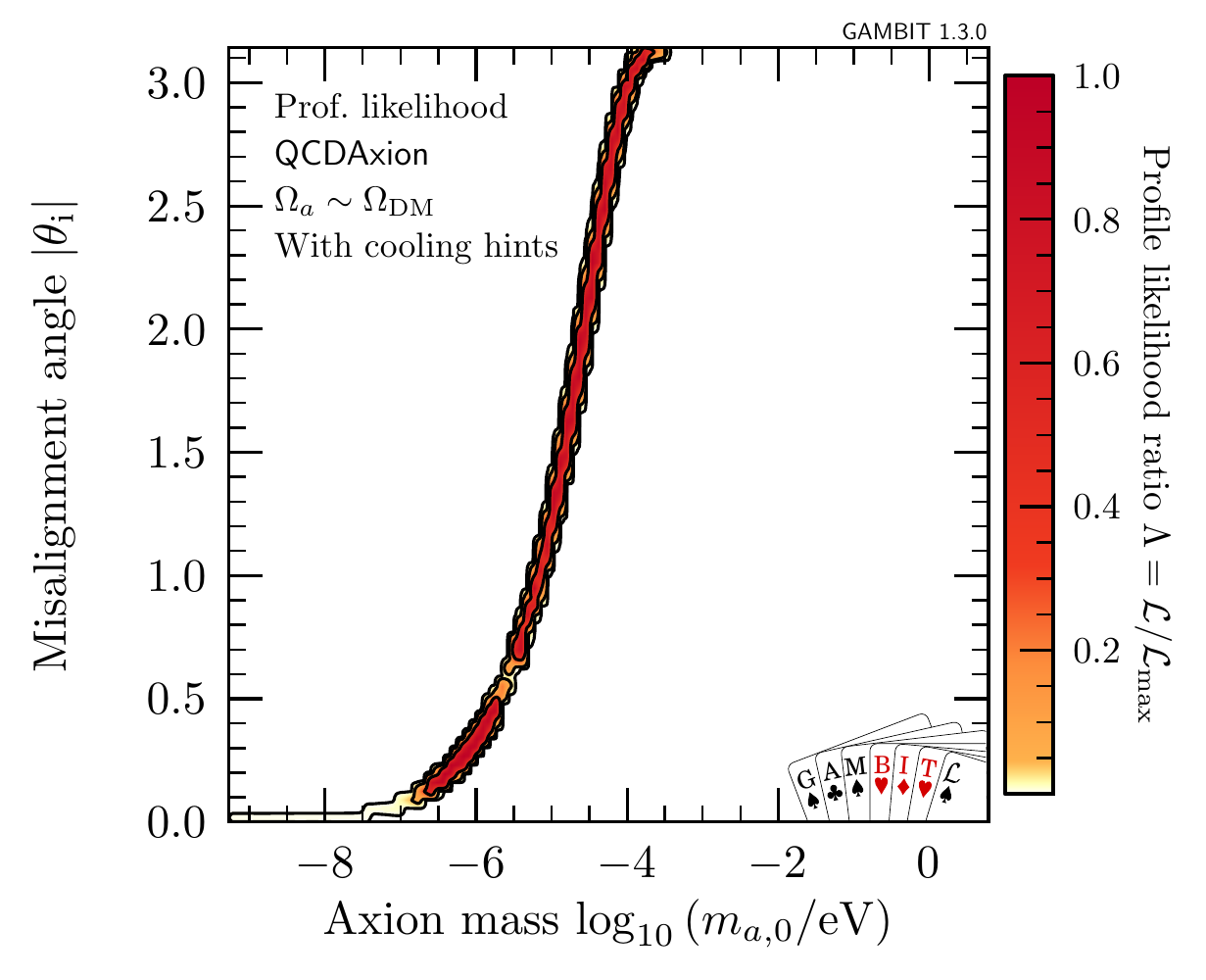}
	}
	\caption{Profile likelihoods~(from \diver) for \qcdaxion models with upper limits~(\textit{left}) and matching condition~(\textit{right}) for the observed DM relic density and including cooling hints. The upper and lower panels show the constraints on the anomaly ratio, $E/N$, and for the absolute value of the initial misalignment angle, $|\thetai|$, respectively.\label{fig:cooling:QCDAxion:frequentist}}
\end{figure}

Figure~\ref{fig:cooling:QCDAxion:frequentist} is the cooling-hint equivalent of \reffig{fig:QCDAxion:frequentist}, summarising the allowed anomaly ratio and magnitude of the initial misalignment angle. The only notable difference is at $\mazero \lsim \SI{0.1}{\micro\eV}$ ($\fa \gsim \SI{3e13}{\GeV}$), where none of the possible values for $\caee$ under consideration is large enough to fully account for the anomalous cooling. Recall that $\gaee \propto \caee \, \mazero$~\refeq{eq:qcdaxioncouplings1} and that the cooling hints point towards a relatively narrow range of couplings $\gaee$ (\reffig{fig:validation:cooling}). The overall effect of the cooling hints is therefore to disfavour lower masses. Had we chosen the range of possible values for~ $\caee$ to be smaller, these constraints would extend to even larger values of~$\mazero$ (and vice versa if we had permitted even larger values of $\caee$).

The right panels of \reffig{fig:cooling:QCDAxion:frequentist} show that \qcdaxion models can satisfy the cooling hints \emph{and} be all of the DM in the interval $0.1 \lsim \mazero/\si{\micro\eV} \lsim 300$. The lower bound on this mass region depends on the largest allowed value for $\caee$.

It is interesting to consider how good the fit of the \qcdaxion model is in an absolute sense. Most constraints are easily satisfied by the best-fit point, such that the corresponding partial likelihoods give $p$-values of order one, which we will not discuss further.\footnote{The exact numerical values depend on how many data and model d.o.f.\ one takes into account, which is often ambiguous.} One exception is the fit to the temperature dependence of the QCD~axion mass, which gives a $p$-value of order $10^{-5}$~(see Sec.~\ref{sec:mod:QCDAxion}). Ignoring this likelihood (and the two model parameters constrained by it) we are left with 7~model d.o.f.\ and 48~data d.o.f.\ when including the WD cooling hints; without the cooling hints, the data d.o.f. is 43. The corresponding $p$-value is \num{0.30} with cooling hints included, and \num{0.60} without. The decrease in $p$-value when including the WD cooling hints results from the slight discrepancies between the cooling hints themselves (cf. \reffig{fig:validation:cooling}) and their slight tension with the $R$ parameter likelihood.

\paragraph*{Bayesian results.}
\begin{figure}
	\centering
	{
		\includegraphics[width=0.49\linewidth]{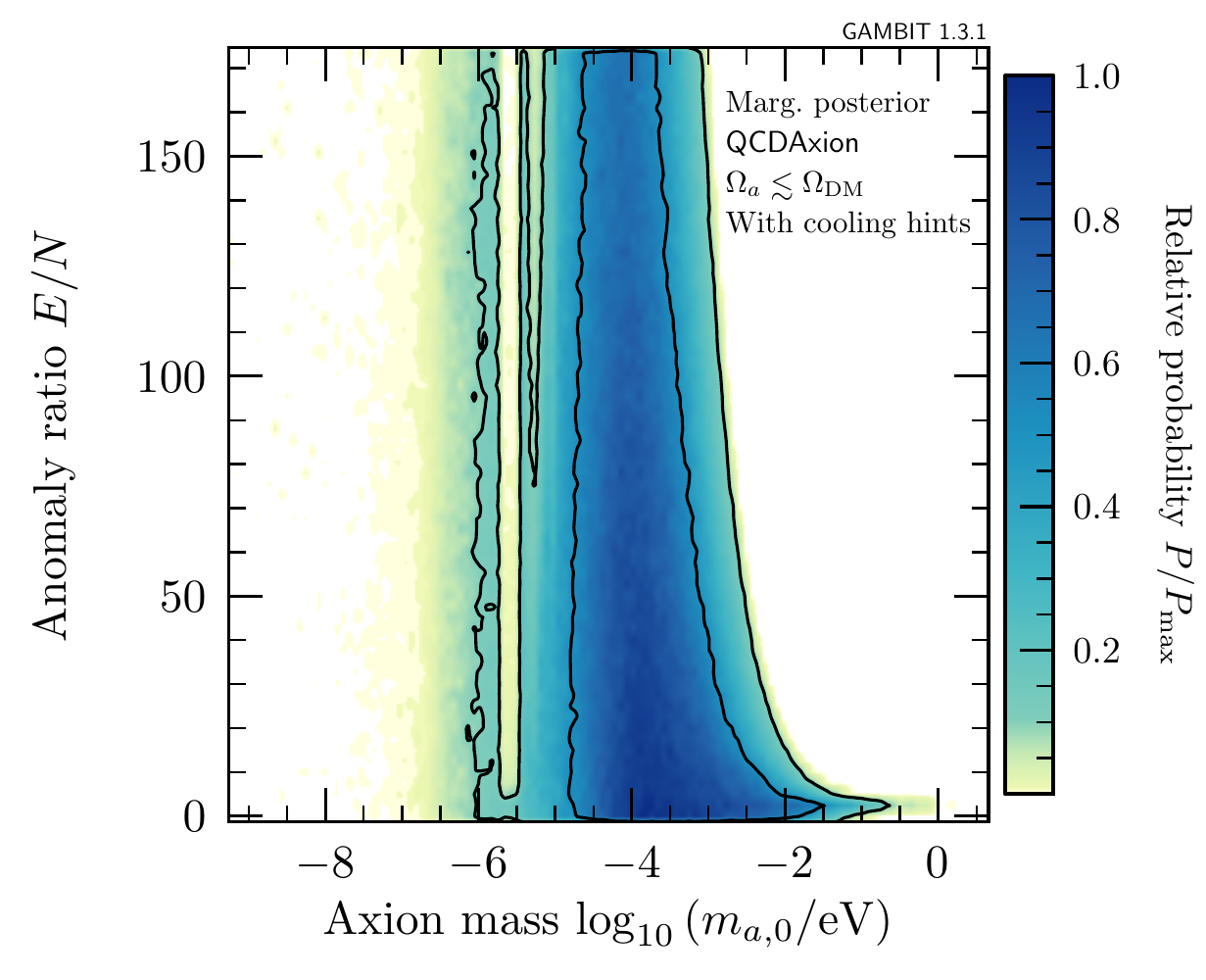}
		\hfill
		\includegraphics[width=0.49\linewidth]{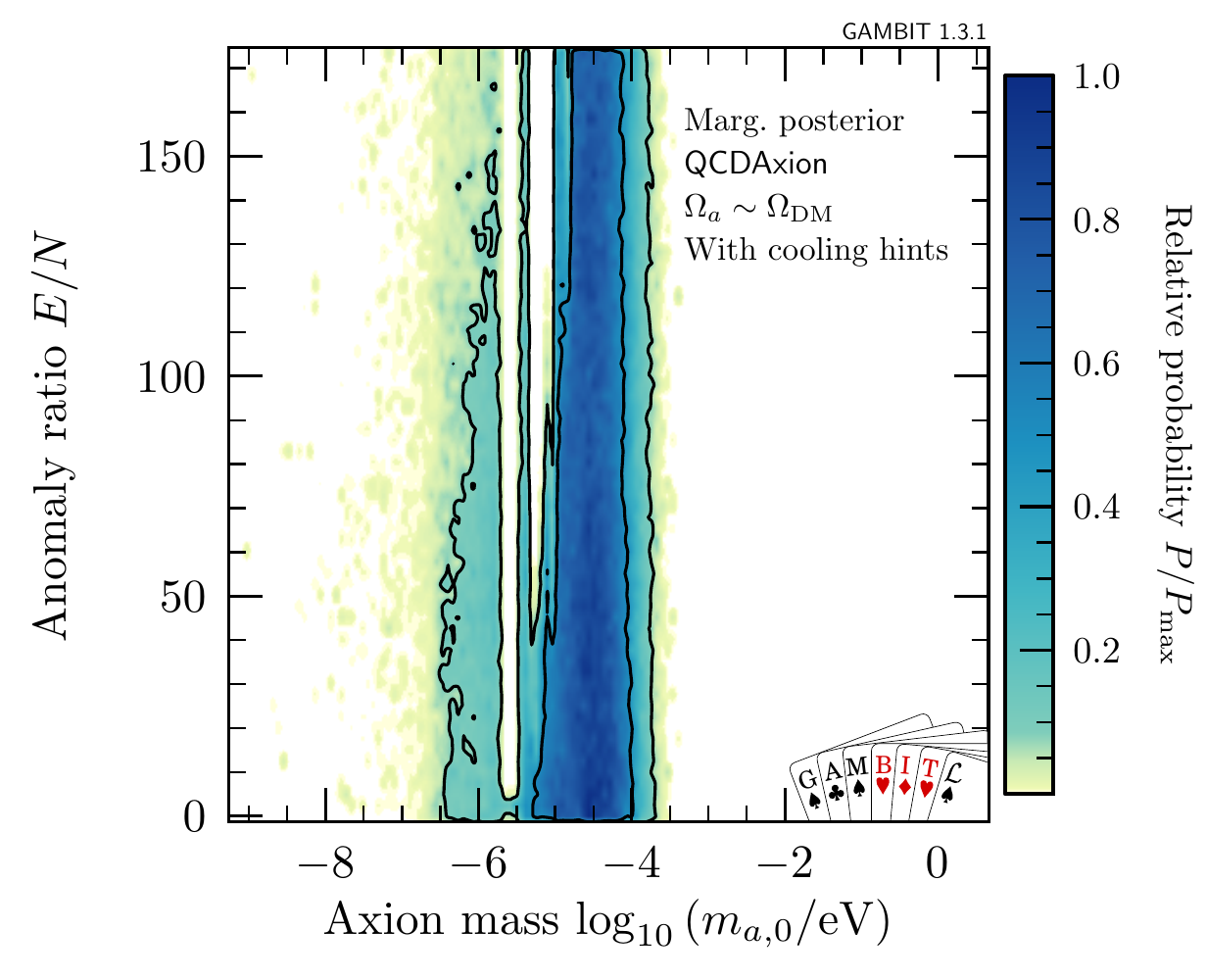}
	}
	{
		\includegraphics[width=0.49\linewidth]{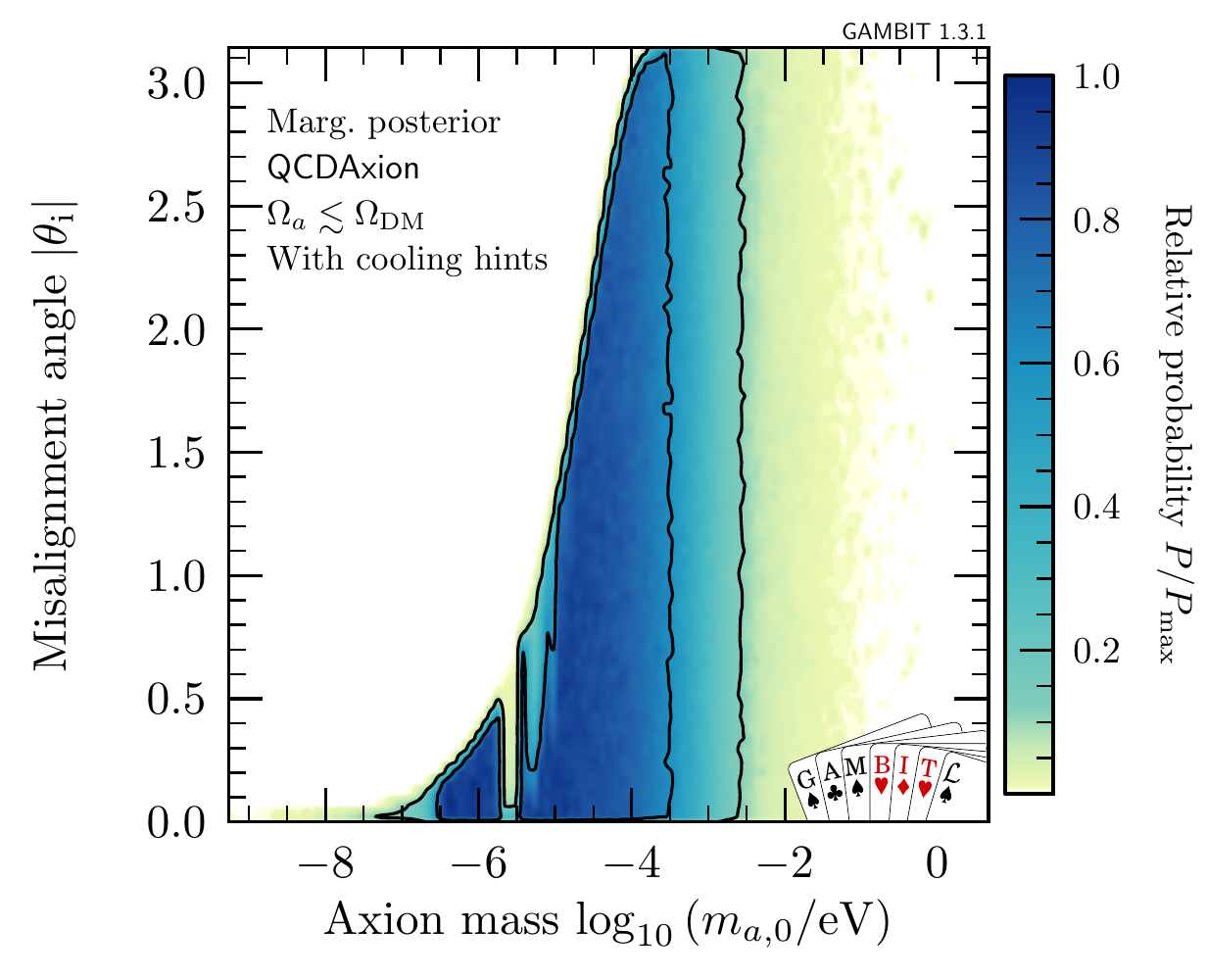}
		\hfill
		\includegraphics[width=0.49\linewidth]{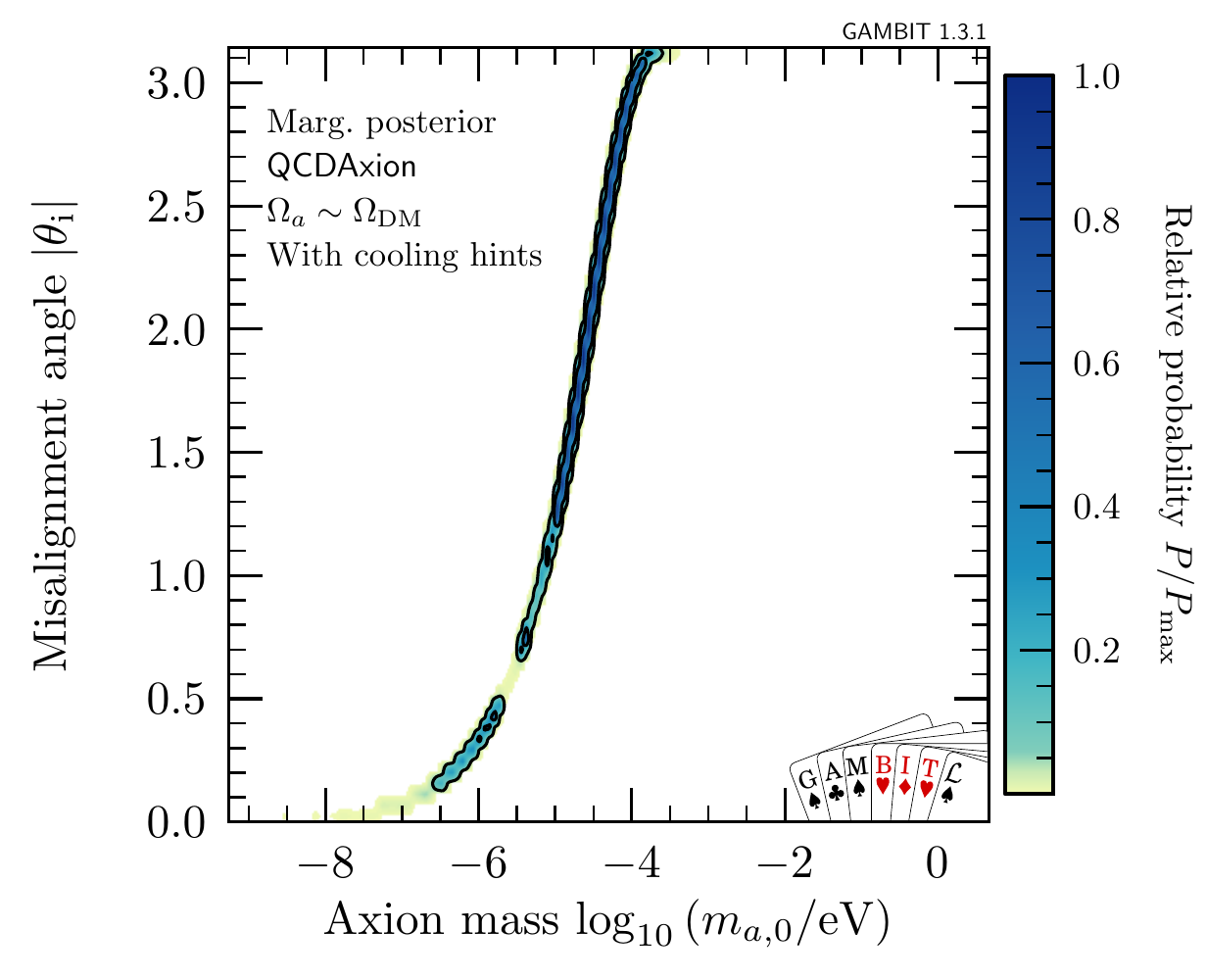}
	}
	\caption{Marginalised posteriors~(from \twalk) for \qcdaxion models with upper limits~(\textit{left}) and matching condition~(\textit{right}) for the observed DM relic density and including cooling hints. The upper and lower panels show the constraints on the anomaly ratio, $E/N$, and for the absolute value of the initial misalignment angle, $|\thetai|$, respectively.\label{fig:cooling:QCDAxion:Bayesian}}
\end{figure}
Selected results from the Bayesian analysis of QCD axions in combination with cooling hints can be found in \reffig{fig:cooling:QCDAxion:Bayesian}. Compared to the Bayesian results without cooling hints in \reffig{fig:QCDAxion:Bayesian}, we can see that the preferred mass regions in the $\mazero$-$\thetai$ plane get narrowed down slightly when we impose the DM relic density constraints as an upper limit (bottom left plot). However, for the anomaly ratio, $E/N$, this is not the case (top left plot). Generally speaking, these results identify the most credible regions for a compromise between QCD axions fitting the cooling hints and ``naturally'' not overproducing DM (which prefers masses of $\order(\text{\SIrange{1}{100}{\micro\eV}})$, cf.\ \reffig{fig:QCDAxion:Bayesian}). Despite the slight differences, which might also depend on the adopted priors, the overall results with and without cooling hints are remarkably similar. This is mainly due to the influence of the $R$~parameter likelihood included in both cases and its slight preference for non-zero couplings.

\begin{figure}
	\centering
	{
		\includegraphics[width=0.49\linewidth]{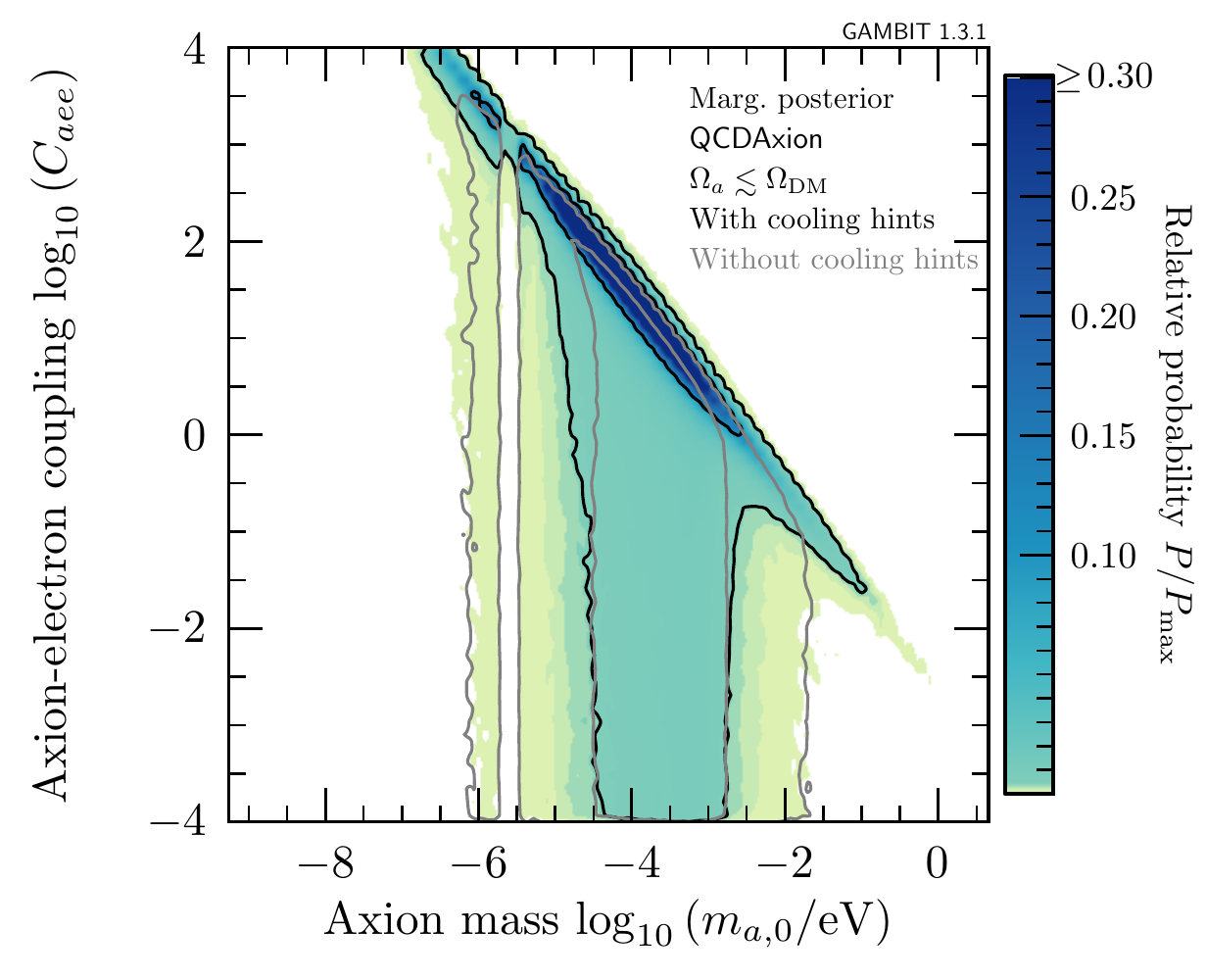}
		\hfill
		\includegraphics[width=0.49\linewidth]{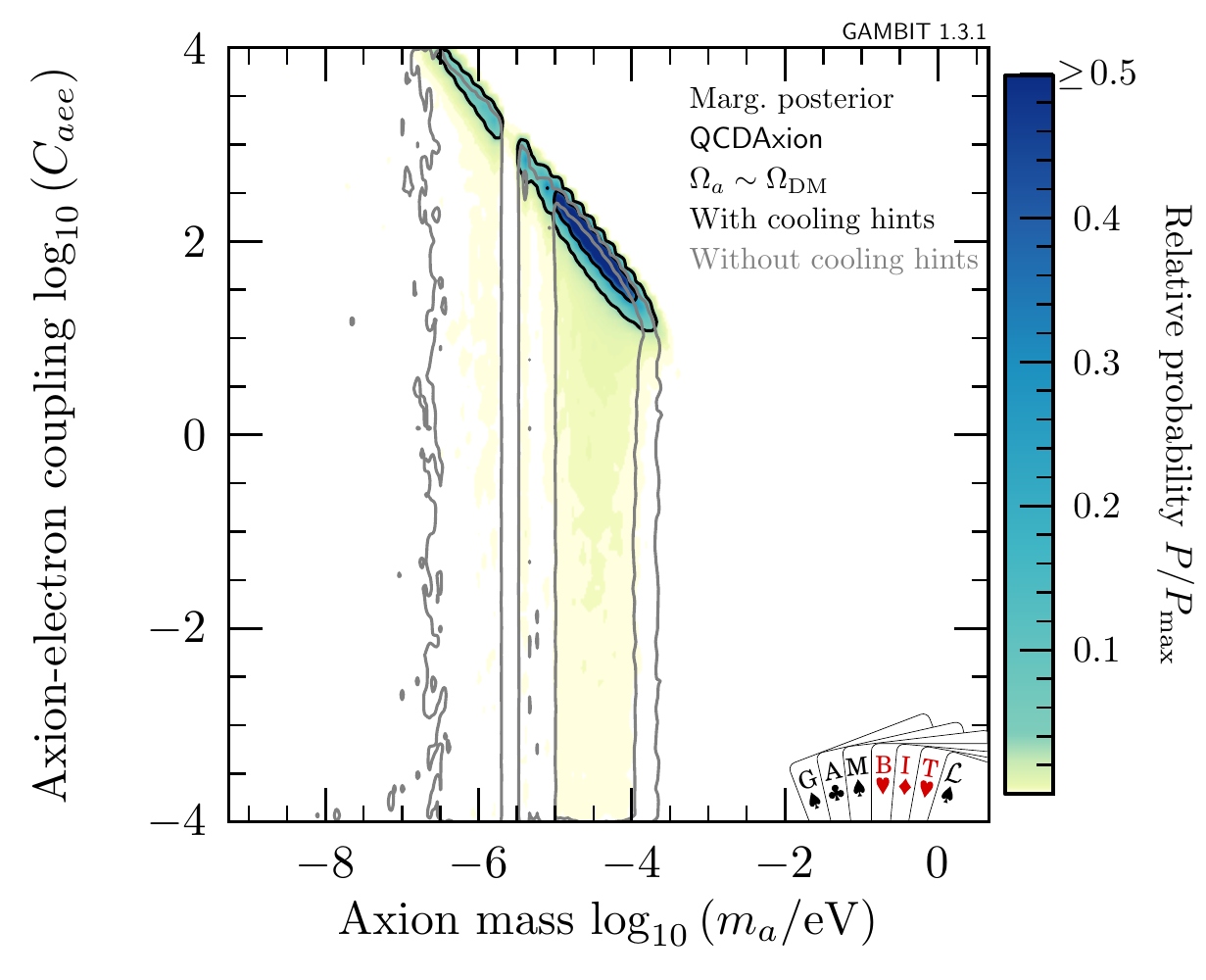}
	}
	\caption{Marginalised posteriors~(from \twalk) for \qcdaxion models with upper limits~(\textit{left}) and matching condition~(\textit{right}) for the observed DM relic density. We show the constraints on the axion-electron coupling, $\caee$, without (grey lines) and with (black lines and coloured regions) the inclusion of cooling hints.\label{fig:cooling:QCDAxion:Bayesian:Caee}}
\end{figure}

The influence of the cooling hints is illustrated further in \reffig{fig:cooling:QCDAxion:Bayesian:Caee}, which shows the regions of highest posterior probability in the $\mazero$-$\caee$ parameter plane with and without the inclusion of cooling hints. Since the cooling hints strongly require $\gaee \sim \num{3e-13}$ (cf.\ right panel of \reffig{fig:validation:cooling}), we find the highest posterior probabilities along a line of constant $\caee\,\mazero$. The chosen range of $\caee$ then implies that the cooling hints can only be explained for $\mazero \gsim \SI{0.3}{\micro\eV}$. For values of $\caee \sim 1$, the cooling hints would point towards \si{\meV}-scale axions, which is incompatible with the requirement $\Omega_a \sim \Omega_\text{DM}$. This regime is therefore disfavoured in the right panel of \reffig{fig:cooling:QCDAxion:Bayesian:Caee}. Not including the cooling hints essentially only results in a upper limit in the most credible mass regions close to where the line of constant~$\gaee$ was (grey contours in \reffig{fig:cooling:QCDAxion:Bayesian:Caee}), due to the $R$~parameter likelihood.

\begin{figure}
	\centering
	{
		\includegraphics[width=0.49\linewidth]{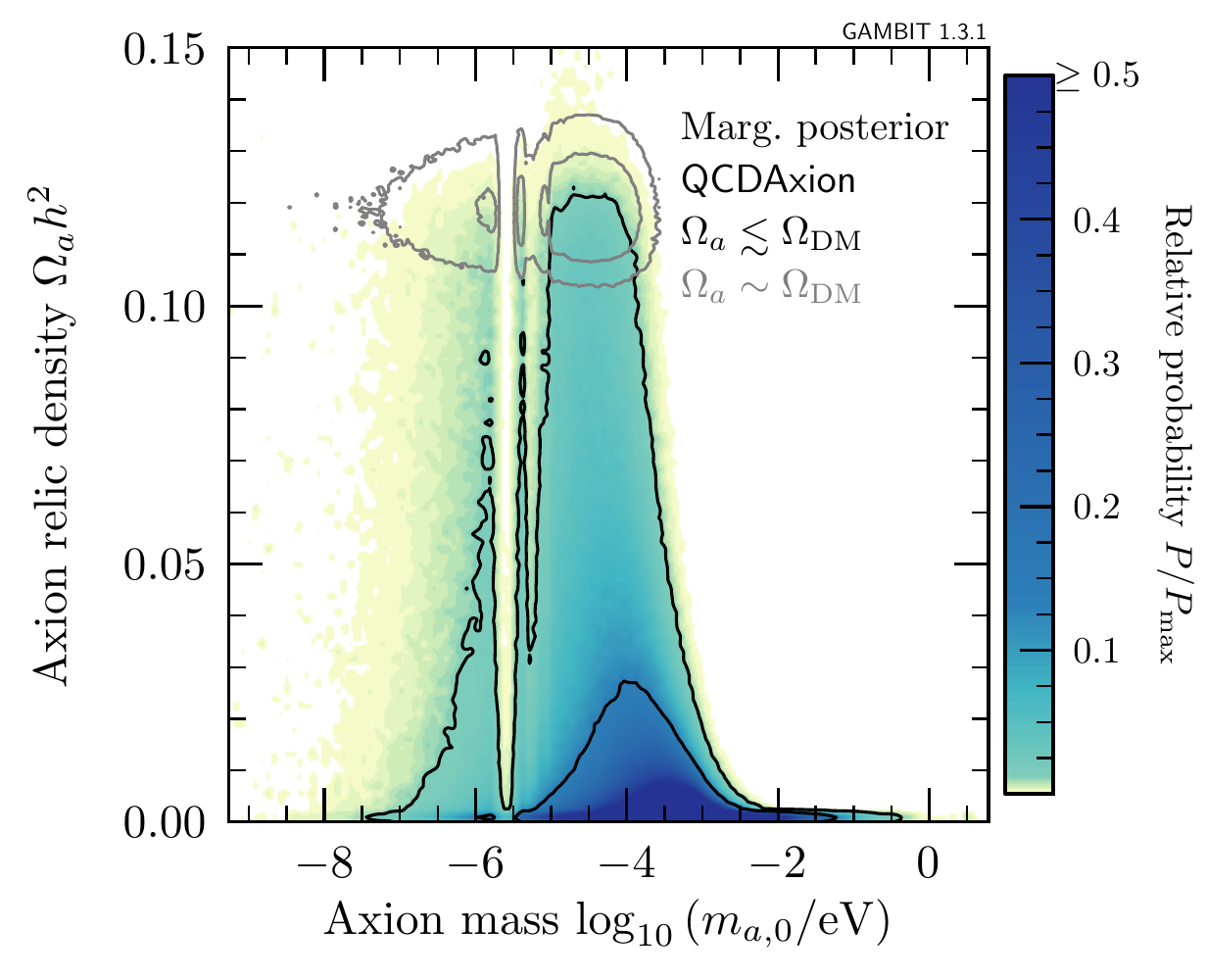}
		\hfill
		\includegraphics[width=0.49\linewidth]{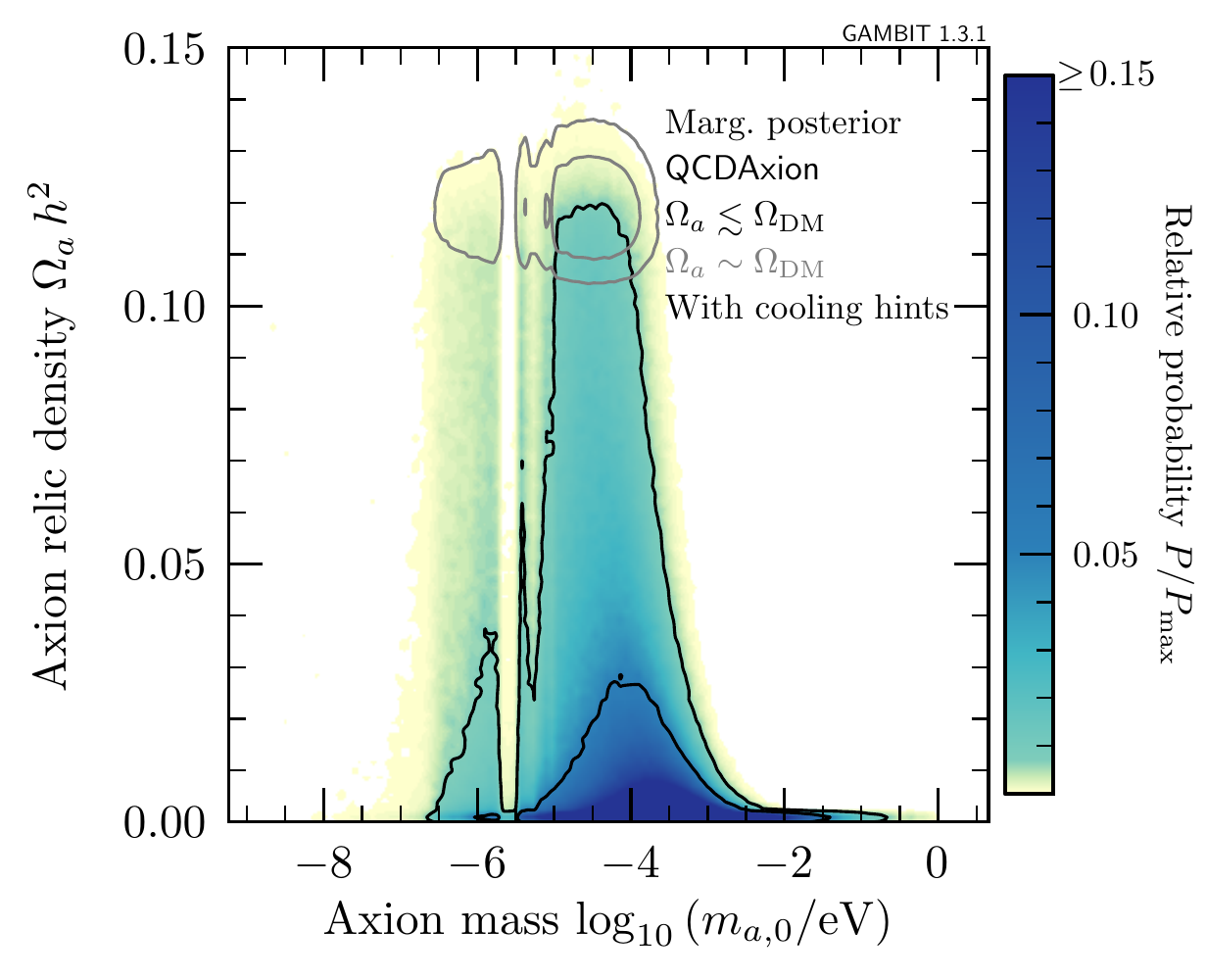}
	}
	\caption{Marginalised posteriors~(from \twalk) for \qcdaxion models with upper limits~(density plots and black contour lines) and matching condition~(grey contour lines) for the observed DM relic density. We show the constraints on the energy density in axions today, $\OmegaA h^2$, without (\textit{left}) and with (\textit{right}) the inclusion of cooling hints.\label{fig:cooling:QCDAxion:RelicDens:twoD}}
\end{figure}
As mentioned before, QCD axions can account for both the cooling hints and all of the DM in the Universe (cf.\ Figs~\ref{fig:cooling:QCDAxion:frequentist} and~\ref{fig:cooling:QCDAxion:Bayesian}). However, because the posterior probability in \reffig{fig:cooling:QCDAxion:Bayesian} is normalised, one cannot infer from this plot if these solutions occur naturally or if considerable fine-tuning is required. Figure~\ref{fig:cooling:QCDAxion:RelicDens:twoD} gives an idea of the ``naturalness'' of \qcdaxion DM by showing the marginalised posterior as a function of $\mazero$ and $\OmegaA h^2$ (with and without the WD cooling hints). In the colour density plots of both panels in \reffig{fig:cooling:QCDAxion:RelicDens:twoD}, we can see that the scan finds credible parts of the parameter space where axions account for a sizeable fraction of the DM while being consistent with all experiments and observations. The differences between including and not including the cooling hints regarding the preferred regions of $\OmegaA h^2$ are rather small, consistent with the other plots.

Similar to the discussion at the end of Sec.~\ref{sec:results:QCDAxions}, we can infer the most credible regions for the relic abundance of axions as well as for~$\mazero$. The preferred axion mass is very similar with or without the inclusion of the cooling hints. Imposing the DM relic density as an upper limit, we find \updated{$\SI{0.70}{\micro\eV} \le \mazero \le \SI{2.8}{\milli\eV}$}{$\SI{0.73}{\micro\eV} \le \mazero \le \SI{6.1}{\milli\eV}$} at 95\% credibility (equal-tailed interval); demanding that axions be all of the DM, this becomes \updated{$\SI{0.41}{\micro\eV} \le \mazero \le \SI{0.14}{\milli\eV}$}{$\SI{0.53}{\micro\eV} \le \mazero \le \SI{0.13}{\milli\eV}$}. Including the cooling hints slightly modifies the preferred range for $\OmegaA h^2$. At 95\% credibility (equal tails), \updated{$\num{1.0e-5} \le \OmegaA h^2 \le \num{0.10}$}{$\num{5.2e-6} \le \OmegaA h^2 \le \num{0.11}$}, corresponding to \updated{0.009--90\%}{0.004--94\%} of DM. The median value is \updated{$\OmegaA h^2=\num{9.3e-3}$}{$\OmegaA h^2=\num{7.1e-3}$}, or about \updated{8\%}{6\%} of the observed DM.

\begin{figure}
	\centering
	{
		\includegraphics[width=0.49\linewidth]{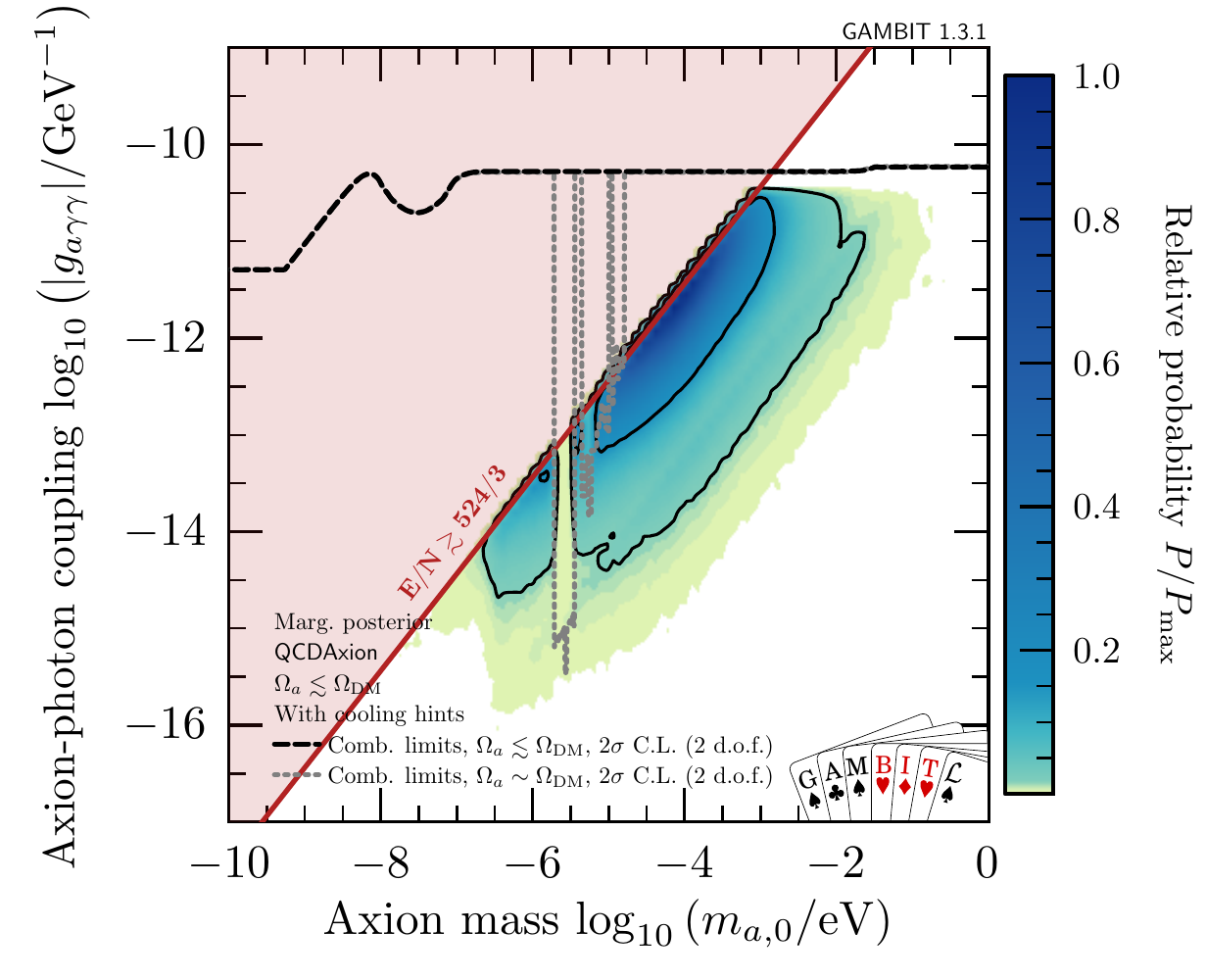}
		\hfill
		\includegraphics[width=0.49\linewidth]{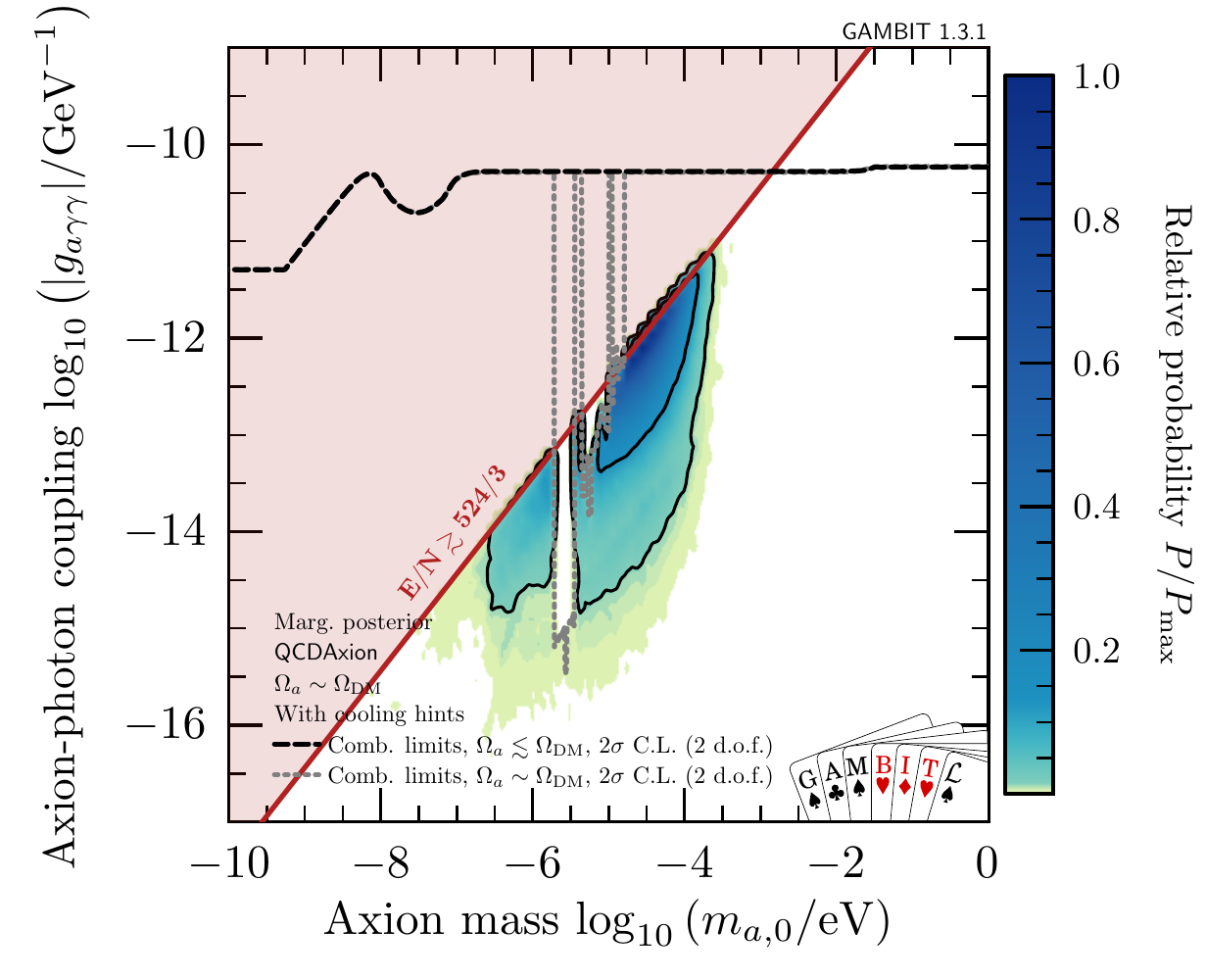}
	}
	\caption{Marginalised posteriors~(from \twalk) for \qcdaxion models with upper limits for the observed DM relic density~(\textit{left}) and matching condition~(\textit{right}). We show the constraints on the absolute value of the axion-photon coupling, $|\gagg|$, together with models \emph{outside} of the band of axion models (red line and shading) and the (frequentist) $2\sigma$ C.L. (dashed lines) for comparison. The prior-dependence of these results is investigated in Appendix~\ref{app:prior}.\label{fig:cooling:QCDAxion:Regions}}
\end{figure}

Finally, let us return to the $\mazero$--$\gagg$ parameter plane, as discussed in the \genalp model without cooling hints (see \reffig{fig:GeneralALP:overview}). In \reffig{fig:cooling:QCDAxion:Regions}, we contrast the {na\"ive} bounds on the parameter space (from the phenomenological constraints on \genalp models and the maximum possible value of~$E/N$) with the regions preferred by a Bayesian analysis. These regions are not only determined by the constraints from data (satisfying the cooling hints in both panels and matching the DM density in the right panel), but also by the fine-tuning in some parts of the parameter space. Fine tuning is necessary for avoiding overproduction of DM at small~$\mazero$, and for achieving low values of~$\gagg$ through cancellations between~$E/N$ and~$\caggtilde$ at large~$\mazero$ (cf. Eq.~\ref{eq:qcdaxioncouplings2}). For our adopted priors, the most credible parameter regions correspond to a few orders of magnitude around \updated{$\mazero \sim \SI{10}{\micro\eV}$}{$\mazero \sim \SI{100}{\micro\eV}$} and \confirmed{$\gagg \sim \SI{e-12}{\GeV^{-1}}$}. In Appendix~\ref{app:prior} we discuss how choosing different priors may affect these conclusions.

\subsubsection{DFSZ- and KSVZ-type axions}\label{sec:cooling:dfszvsksvz}
DFSZ-type models have intrinsically larger coupling to electrons than KSVZ-type models, which only obtain their interactions with electrons at loop level. DFSZ models are therefore the natural choice to account for the potential WD cooling anomalies \cite{1708.02111}, by way of an axion-electron coupling of $\gaee \sim \num{3e-13}$.

\paragraph*{Frequentist results.}
\begin{figure}
	\centering
	{
		\includegraphics[width=0.49\linewidth]{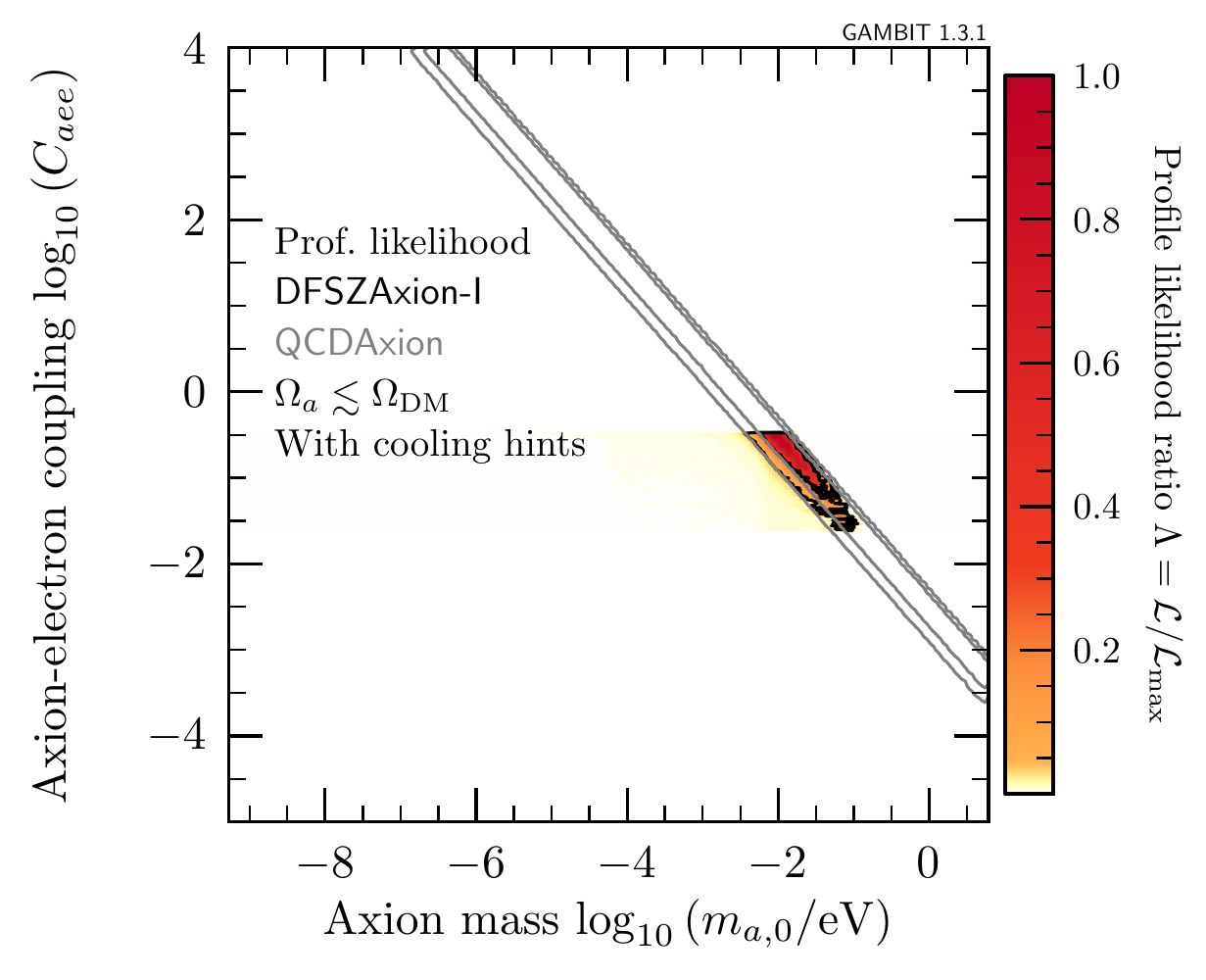}
		\hfill
		\includegraphics[width=0.49\linewidth]{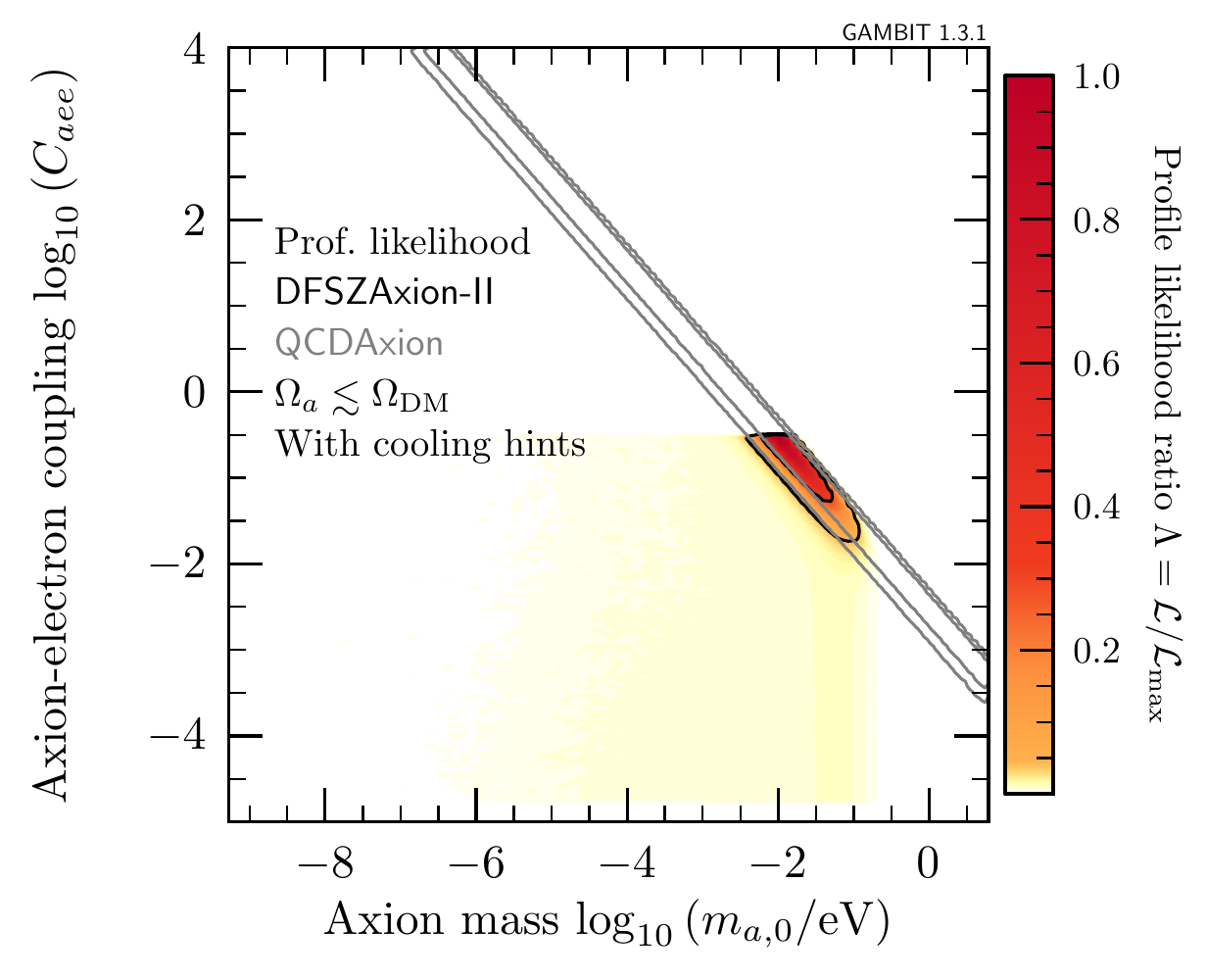}
	}
	\caption{Profile likelihoods (from \diver) for \dfszI (\textit{left}) and \dfszII (\textit{right}) models, compared to the profile likelihood for the \qcdaxion model (grey lines), taking the observed DM abundance as an upper limit on the number of axions. Contours show the $1\sigma$ and $2\sigma$ confidence regions.\label{fig:cooling:DFSZ}}
\end{figure}

\reffig{fig:cooling:DFSZ} shows the profile likelihood for the \dfszI and \dfszII models, compared to the band that maximises the profile likelihood for the \qcdaxion model. Clearly both DFSZ-type models can accommodate the cooling hints with large axion-electron couplings. We already saw in \reffig{fig:dfszvsksvz:caee} that even without the cooling hint likelihood, the \ksvz fails to achieve large $\caee$ values; with cooling hints included, the highest likelihood regions for \ksvz models are therefore essentially the same as in~\reffig{fig:dfszvsksvz:caee}.

\begin{table}
	\renewcommand{\arraystretch}{1.15}
	\caption{Best-fit values for \dfszI, \dfszII, and \ksvz models when imposing an upper limit on the DM relic density, $\OmegaA \lsim \OmegaCDM$. In the final column, we compare the likelihood of the respective best-fit points to \qcdaxion models. Note that the \qcdaxion model has two more degrees of freedom than the KSVZ models, and one more than the DFSZ-type models.\label{tab:cooling:bestfit:upperlimit}}
	\centering
	\small
\begin{tabular}{@{}lccccccrr@{}}
		\toprule
		\textbf{Model} & \mc{\textbf{$E/N$}} & \mc{\textbf{$\hat{m}_a$}} & \mc{\textbf{$\hat{C}_{a\gamma\gamma}$}} & \mc{\textbf{$\hat{g}_{a\gamma\gamma}$}} & \mc{\textbf{$\hat{C}_{aee}$}} & \mc{\textbf{$\hat{g}_{aee}$}} & \boldmath{$-2\Delta\maxlnL$} \\
		& & \mc{\si{\milli\electronvolt}} & & \mc{\SI{e-12}{\GeV^{-1}}} & \mc{\num{e-3}} & \mc{\num{e-15}} & & \\
		\midrule
		\dfszI & 8/3 & 8.90 & 0.746 & 1.35 & 333 & 266 & 0.28 \\
		\dfszII & 2/3 & 9.62 & 1.25 & 2.46 & 307 & 265 & 0.52 \\
		\midrule
		\ksvz & 0 & 39.5 & 1.92 & 15.5 & 0.118 & 0.417 & 9.57 \\
		& 2/3 & 59.4 & 1.25 & 15.2 & 0.0223 & 0.119 & 9.57 \\
		& 5/3 & 316 & 0.252 & 16.2 & 0.210 & 5.94 & 9.49 \\
		& 8/3  & 105 & 0.747 & 15.9 & 0.430 & 4.03 & 9.52 \\
	\bottomrule
\end{tabular}
\end{table}

Table~\ref{tab:cooling:bestfit:upperlimit} gives the best-fit values for the six classic axion models that we consider in this section, under the requirement that they do not exceed the observed abundance of DM. We do not report best-fit values for $|\thetai|$, as even with the maximum value included in our scans, axions only account for a few percent of the observed DM abundance. For each model, we calculate $\Delta\maxlnL$, the logarithm of the ratio of the best-fit likelihood relative to the \qcdaxion model. As anticipated, DFSZ-type models perform much better than {\ksvz}s (at the expense of having one additional degree of freedom). This is because DFSZ-type models can easily reach the required axion-electron coupling to fit the cooling hints with masses of order $\mazero \sim \SI{10}{\meV}$ (as noted previously \cite{1708.02111}). For KSVZ-type axions, the masses required to naturally fit the cooling hints are about three to four orders of magnitude larger (see e.g.\ \ref{eq:ksvzgaee}), and the associated axion-photon coupling is therefore in conflict with the $R$~parameter likelihood (as well as hot DM bounds, which are not included). Nevertheless, even for \ksvz models there is still a slight preference compared to having no axion at all, which corresponds to $-2\Delta \maxlnL = 10.54$.

For DFSZ-type models, the \dfszI model gives a better fit than the \dfszII model. This is due to the influence of the $R$~parameter likelihood, which combines with the cooling hints to force the axion-photon coupling to $\gagg \lsim \SI{2e-11}{\GeV^{-1}}$ (at 95\% C.L.). The best-fit point is therefore a balance between reaching high enough~$\gaee$ and minimising~$\gagg$. The maximum value of~$\caee$ for \dfszI models is about a factor of~1.08 larger than for \dfszII models, due to perturbativity constraints on $\tan(\beta^\prime)$, whereas the axion-photon couplings are about a factor of~0.6 lower, yielding a better fit to the $R$~parameter likelihood at any given mass.

Next, let us briefly consider the case where we demand that the classic axion models provide all of the DM. This results in much poorer maximum likelihood values of around $-2\Delta\maxlnL \approx 10.5$ compared to \qcdaxion models for \emph{all} KSVZ- and DFSZ-type models that we consider in this paper. This is not surprising because, unlike some other \qcdaxion models, they cannot account for both the cooling anomalies and DM.\footnote{Note that $\thetai \to \otherpi$ could in principle produce arbitrarily large amounts of DM, if isocurvature constraints can be avoided (at the cost of additional fine-tuning), as discussed in Ref.~\cite{1708.02111}.} The maximum-likelihood regions are also highly degenerate in this case, as none of the fits is actually ``good'' (with a maximum likelihood comparable to the case without any axion).

Finally, thanks to the model hierarchy shown in \reffig{fig:AxionModelTree}, we can perform nested hypothesis tests to determine whether or not the more constrained, specific \dfsz and \ksvz models are disfavoured compared to the broader class of \qcdaxion models. Without the cooling hints, \qcdaxion, \dfsz and \ksvz models cannot be discriminated. This situation changes if the WD~cooling hints are taken into account. Imposing the measured DM density as an upper limit only, \ksvz models (null hypothesis) can be rejected with respect to \qcdaxion models (alternative hypothesis) with a $p$-value of \updated{\num{8.3e-3}}{\num{0.008}}. \dfsz models as a null hypothesis, on the other hand, cannot be rejected (\updated{$p \geq \num{0.26}$}{$p \geq \num{0.47}$}). If we instead demand that axions be all of DM, both \dfsz and \ksvz models can be rejected with respect to the \qcdaxion model with $p$-values of \num{1.2e-3} and \num{5.3e-3}, respectively.

\paragraph*{Bayesian results.}
\begin{figure}[ht]
	\centering
	{
		\includegraphics[width=0.49\linewidth]{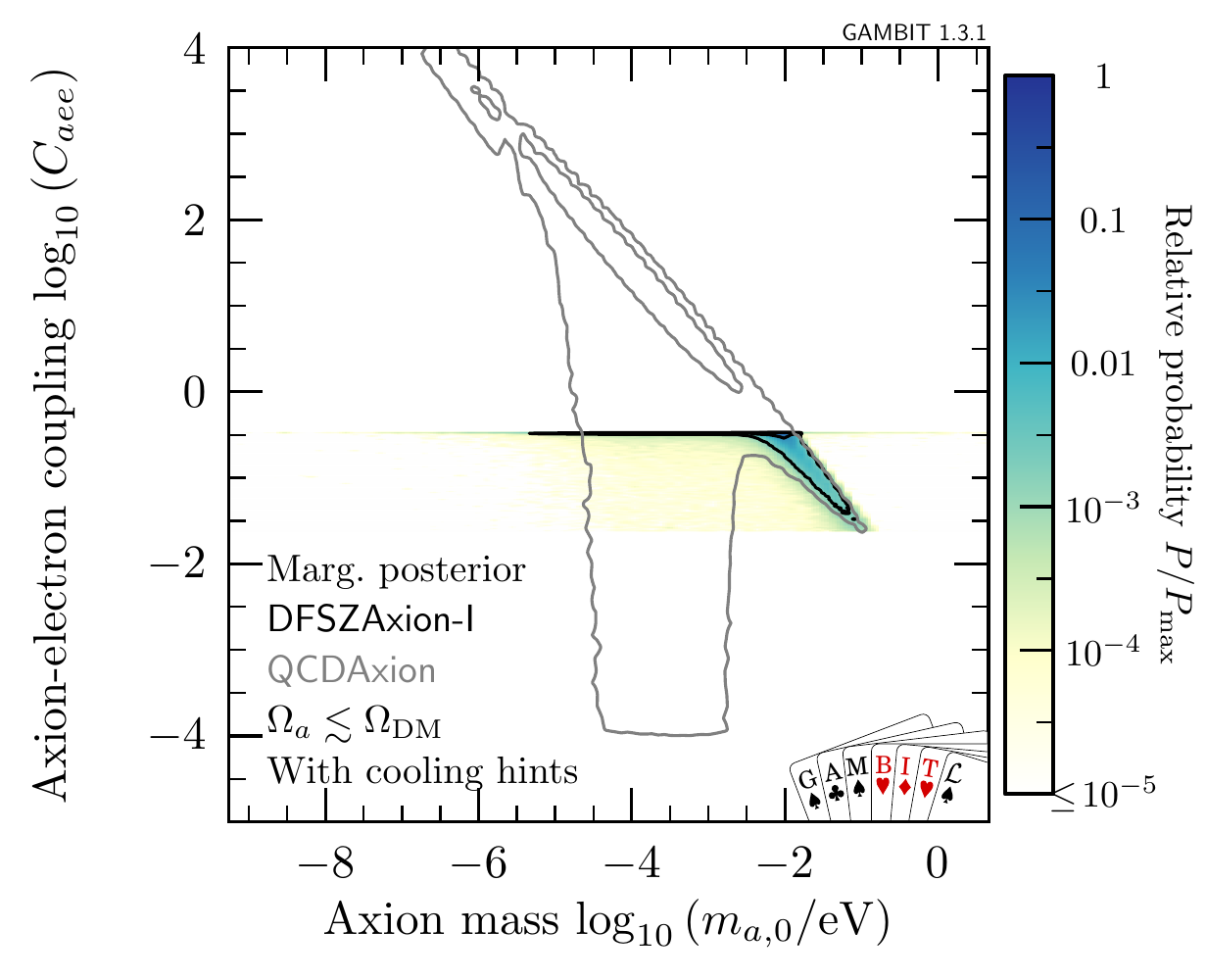}
		\hfill
		\includegraphics[width=0.49\linewidth]{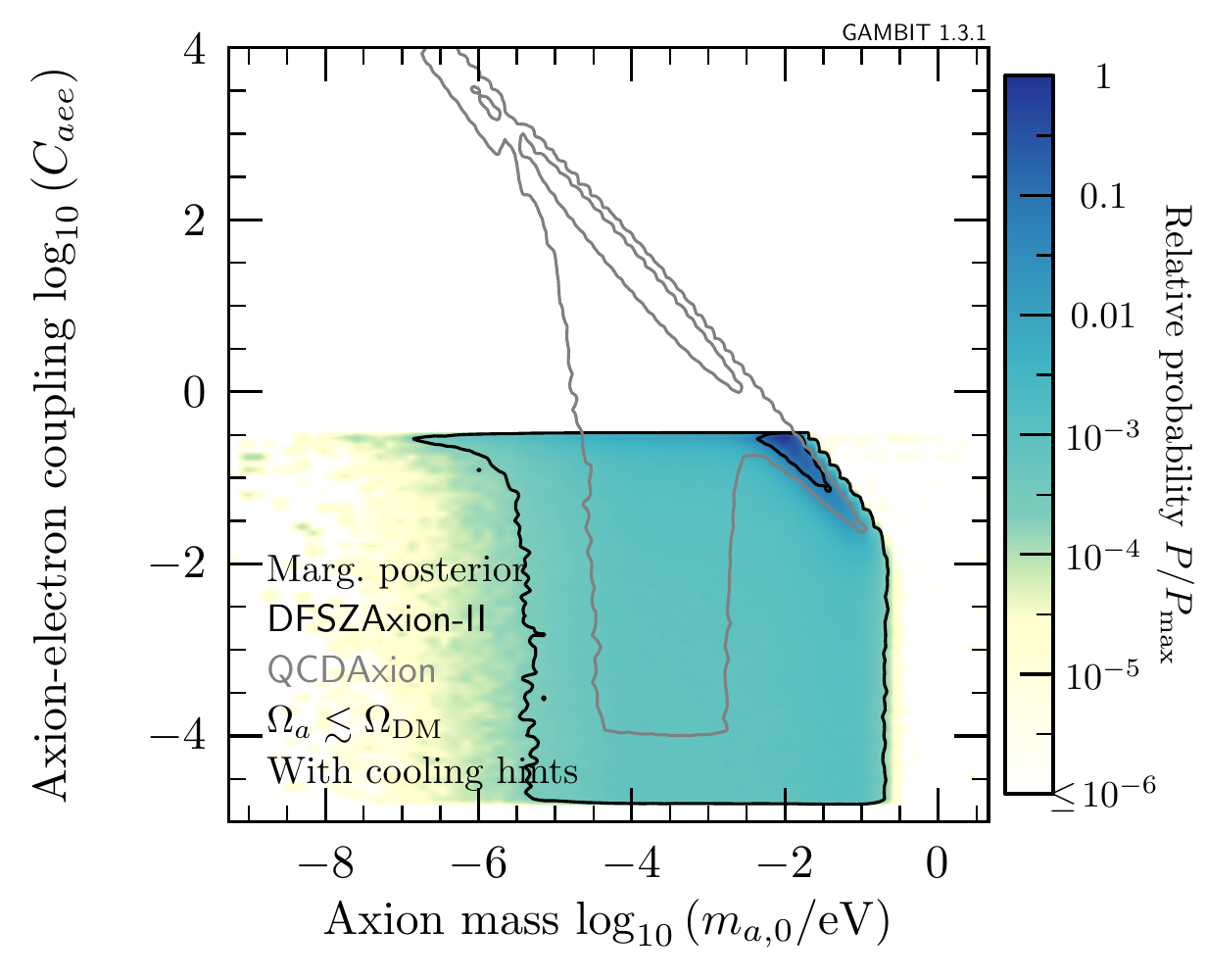}
	}
	\caption{Marginalised posteriors (from \twalk) for \dfszI (\textit{left}) and \dfszII (\textit{right}) models, compared to the marginalised posterior for the \qcdaxion model (grey lines), taking the observed DM relic density as an upper limit on the axion relic density and including cooling hints. We show only the 68.27\% and 95.45\% C.R.s for all models.\label{fig:cooling:DFSZ:post}}
\end{figure}

Figure~\ref{fig:cooling:DFSZ:post} shows posteriors for the DFSZ-type models compared to the \qcdaxion, in the $\mazero$-$\caee$ parameter plane (see also~\reffig{fig:cooling:QCDAxion:Bayesian:Caee}). These regions resemble the ones found in \reffig{fig:cooling:DFSZ}. The log prior on $\tan \beta^\prime$ enables both models to achieve sufficiently large values of $\caee$ quite naturally, despite their structural differences. At first sight, \dfszII models appear to be able to occupy a larger region of parameter space in the 68.27\% C.R. However, this is mainly a reflection of the effective prior on $\caee$, which for the \dfszI model is strongly peaked towards the upper boundary of the accessible parameter space in \reffig{fig:cooling:DFSZ:post}. In contrast, the effective prior in $\caee$ is almost flat in the \dfszII model. In terms of the fundamental parameters (i.e.\ $\tan \beta^\prime$), the credible parameter region in the \dfszI model is in fact larger, because the large values of $\caee$ needed to explain the cooling hints can be achieved more easily.

\begin{table}
	\caption{Odds ratios in favour of DFSZ- and KSVZ-type models, compared to the parent \qcdaxion, as calculated from the nested sampling evidence estimates in \multinest, and including the WD cooling hints. We either impose the DM relic density as an upper limit ($\OmegaA \lsim \OmegaCDM$) or demand that axions be all of DM ($\OmegaA \sim \OmegaCDM$). Note that the estimated uncertainties on the evidence values are small enough that the corresponding uncertainties on the odds ratios are negligible.\label{tab:odds:coolinghints:dfszvsksvz}}
	\small
	\centering
	\begin{tabularx}{0.81\textwidth}{>{\bfseries}X*{6}{c}}
		\toprule
		Model & \dfszI & \dfszII & \multicolumn{4}{c}{\ksvz} \\
		\cmidrule{4-7}
		$E/N$ & $8/3$ & $2/3$ & $0$ & $2/3$ & $5/3$ & $8/3$ \\
		\midrule
		Odds ($\OmegaA \lsim \OmegaCDM$) & 3:1 & 1:1 & \updated{1:2}{1:3} & 1:2 & 1:2 & 1:2 \\
		Odds ($\OmegaA \sim \OmegaCDM$) & 1:5 & \updated{1:6}{1:5} & 1:6 & 1:6 & 1:6 & 1:6 \\
		\bottomrule
	\end{tabularx}
\end{table}

\pagebreak

To investigate the consequences of fine-tuning in more detail, we consider all models in a Bayesian model comparison. The resulting odds ratios can be found in Table~\ref{tab:odds:coolinghints:dfszvsksvz}. If we impose the DM relic density as an upper limit, the odds ratios are still mostly inconclusive. However, for the \dfszII and \ksvz models, the trend swings in favour of the broader class of \qcdaxion models when cooling hints are added to the analysis. In contrast, the \dfszI model fares better than the \qcdaxion, with an odds ratio of 3:1. If we combine this 3:1 odds ratio with the 2:1 preference for the \qcdaxion model \updated{over \ksvz models}{over most \ksvz models}, \updated{there is a 6:1 positive preference}{there is at least a 6:1 positive preference} for the \dfszI model over all \ksvz models. This preference is not surprising, given the differences in the natural axion-electron coupling strength in these models.

If we demand that axions solve the cooling hints and constitute all of DM, model comparison confirms quantitatively (with ratios of $\leq \text{1:5}$) that there is a positive preference for the \qcdaxion over the DFSZ- and KSVZ-type models. This is because we allowed for much larger values of~$\caee$ with \qcdaxion models than with DFSZ- and KSVZ-type models. In fact, the one-dimensional marginal posterior for the {\qcdaxion} electron coupling peaks at $\caee \sim 100$ ($\caee \sim 50$ if we impose the DM relic density as an upper limit), whereas the DFSZ and KSVZ models are limited to couplings of less than one. However, it should be noted that \qcdaxion models with such a large coupling are not expected from traditional axion models, and may therefore pose a challenge for model building.

\section{Conclusions}\label{sec:outlook}
In this study we presented the first global fits of axion models in the pre-inflationary PQ~symmetry-breaking scenario, using frequentist and Bayesian methods. We identified the most viable regions of parameter space for these models and discussed the effect of adding cooling hints seen in white dwarfs. We not only considered the phenomenological parameter space, but also the underlying parameters in a generic QCD~axion model and six specific DFSZ- and KSVZ-type models. We extended previous results in the literature by including various nuisance parameters and by quantitatively considering the fine-tuning of the initial misalignment angle.

We found, in agreement with previous work, that QCD axion models are viable as explanations for both the observed cold dark matter and the white dwarf cooling hints. We showed that, for a broader class of \qcdaxion models than what is traditionally considered, these can be achieved simultaneously. We also quantitatively confirmed that this is not possible for six specific DFSZ- and KSVZ-type models. However, if the condition of being all the dark matter in the Universe is relaxed, we found that \dfszI models are positively preferred over \dfszII models, and over all four \ksvz models that we investigated.

We determined the most credible predicted ranges for the QCD axion mass and its cosmological abundance in the Bayesian statistical framework. Although these results are somewhat prior-dependent, {\qcdaxion}s appear likely to be a cosmologically relevant (but probably not dominant) component of dark matter. Moreover, the most credible axion mass range is within reach of current and planned haloscope experiments.

Global fits of QCD~axions and axion-like particles have the potential to confirm and refine previously known phenomenological statements about the relevant parameter spaces (e.g.\ exclusion limits, existence of fine-tuning, compatibility with cooling hints and dark matter). In other cases, they offer new and more rigorous insights (e.g.\ model comparison, most credible parameter regions). This is true in particular for the Bayesian analyses that we performed in this paper, because they inherently take fine-tuning into account. Due to the orthogonality of constraints and the insufficient sensitivity of most experiments, it is not (yet) possible to decisively probe the axion parameter space in the pre-inflationary PQ~symmetry-breaking scenario, but it is realistically possible to target the most likely versions of QCD~axion and axion-like particle models in the near future (see Ref.~\cite{1801.08127} for a recent review of upcoming axion searches).

\SH{The axion routines that we developed for this paper are publicly available in \darkbit within \gambitverupdate.} Statistical samples and input files from this study are freely available from \textsf{Zenodo}~\cite{Zenodo_axions}.

\acknowledgments

{
\footnotesize
We are grateful to the UK Materials and Molecular Modelling Hub for computational resources, which is partially funded by EPSRC (EP/P020194/1), PRACE for awarding us access to Marconi at CINECA, Italy, and the Imperial College Research Computing Service as well as to all those individuals and experimental collaborations who kindly provided their data and results. We thank A.~Serenelli for helpful comments on solar models, and acknowledge further useful discussions and communication with T.~Battich, P.~Brun, A.~H.~{C\'orsico}, I.~Dominguez, M.~Giannotti, I.~G.~Irastorza, J.~Jaeckel, A.~Lindner, L.~di~Luzio, D.~J.~E.~Marsh, J.~Redondo, A.~Ringwald, O.~Straniero, K.~K.~Szabo, J.~Vogel, L.~Rosenberg and G.~Rybka. We also thank our colleagues within the GAMBIT community, in particular S.~Bloor, G.~Martinez and J.~McKay. For all digitisation procedures, we acknowledge usage of the tool \textsf{WebPlotDigitizer} by A.~Rohatgi, available at \url{https://automeris.io/WebPlotDigitizer/}. SH gratefully acknowledges funding by the Imperial College President's PhD Scholarship scheme. FK is supported by the DFG Emmy Noether Grant No.\ KA 4662/1-1. PS is supported by STFC (ST/K00414X/1, ST/N000838/1, ST/P000762/1). MW is supported by the ARC Future Fellowship FT140100244.
}

\appendix

\section{Integrating solar models for the signal prediction in CAST}\label{app:solarmodelintegration}
To obtain the axion flux at Earth from axion-photon interactions, we need four inputs from a solar model: the solar radius~$R_\sol$, the temperature~$T(r)$, the plasma frequency $\mrm{\omega}{pl}(r)$ and the screening scale $\mrm{\kappa}{s}(r)$ (see Eqs.~\ref{eq:ggA}--\ref{eq:solardiscflux}). While $T(r)$ can be obtained directly from solar model files, $\mrm{\omega}{pl}(r)$ and $\mrm{\kappa}{s}(r)$ have to be calculated using the mass density~$\rho(r)$ and the mass fractions~$X_i(r)$ for each ion/atom with label~$i$ using~\refeq{eq:screeningscale}. Assuming that the plasma is fully ionised, we can recast the equations into the following form~\cite{book_raffelt_laboratories}:
\begin{align}
  \mrm{\kappa}{s}^2(r) &= \frac{4\otherpi\mrm{\alpha}{EM}}{T(r)} \frac{\rho (r)}{\mrm{m}{u}} \sum_{i} \frac{X_i(r)Z_i}{A_i}\left(1 + Z_i\right) \, ,\\
  \text{and} \quad \mrm{\omega}{pl}^2(r) &= \frac{4\otherpi\mrm{\alpha}{EM}}{m_e} \frac{\rho (r)}{\mrm{m}{u}} \sum_{i} \frac{X_i(r)Z_i}{A_i} \, ,
\end{align}
where $\mrm{m}{u}$ is the atomic mass unit, $r$ is the distance from the centre of the Sun in units of~$R_\sol$, and $Z_i$ and $A_i$ are the charge and atomic weight of the $i$th element. Note that the approximation of full ionisation is justified almost everywhere inside the Sun, i.e.\ $r \lesssim 0.95$.\footnote{Aldo Serenelli, private communication.} As the largest contribution to the axion flux comes from the innermost region ($r\lsim 0.2$~\cite{hep-ex/0702006}), we can safely employ this assumption. For elements tracked by the solar model in bulk (i.e.\ without isotopic information), we calculate the mean atomic weights $A_i$ using the isotopic composition of Ref.~\cite{AGSS}, with values based on the terrestial composition updated to use more recent data \cite{10.1515/pac-2015-0305}.

The expected number of photons in the energy range $[E_j,E_{j+1}]$ is
\begin{equation}
  s_j = \int_{E_j}^{E_{j+1}} \! \mathcal{E} (E) \; \frac{\dd\Phi(E)}{\dd E} \; \dd E \, , \label{app:eq:energyintegration}
\end{equation}
where $\mathcal{E}$ is the effective exposure and $\dd\Phi/\dd E$ is the combined photon spectrum from axion-electron and axion-photon contributions. We neglect the energy dispersion of the CAST detector, as it is always less than about 0.2\,keV~\cite{hep-ex/0702006}, and therefore smaller than or comparable to the bin width of the CAST analyses (0.3\,keV and 0.5\,keV respectively for the 2007 and 2017 analyses). We provide tabulated data within \darkbit for the effective exposure $\mathcal{E}$;\footnote{I.~Irastorza and J.~Vogel, private communication.} this can be found in \texttt{DarkBit/data} as \texttt{\metavar{dataset}\_EffectiveExposure.dat}, where \metavar{dataset} is either \texttt{CAST2007} or \texttt{CAST2017\_\metavar{X}}, where \texttt{\metavar{X}} corresponds to the data sets~\texttt{A} to \texttt{L} in Ref.~\cite{1705.02290}. All other files that we mention in the following can be found in the same folder.

Note that performing the energy integration in \refeq{app:eq:energyintegration} after the density integral in \refeq{eq:solardiscflux} is only possible because all contributions to the integral are for energies greater than $\mrm{\omega}{pl}(0)\approx\SI{0.3}{\kilo\electronvolt}$. For energies lower than $\mrm{\omega}{pl}$, axions cannot be produced from the plasma and the square root in~\refeq{eq:solardiscflux} becomes ill-defined.

To obtain the contribution to $\dd\Phi/\dd E$ from axion-photon interactions, we integrate \refeq{eq:diffsolarphotonflux} and \refeq{eq:solardiscflux} over $r$ and $\rho$, using the adaptive 51~point Gauss-Kronrod rule \texttt{gsl\_integration\_qag}, from the \textsf{gsl} library. We obtain quantities necessary for these integrations (temperature, etc.) by interpolating the solar model linearly in radius. For the contribution to $\dd\Phi/\dd E$ from axion-electron interactions, we use the spectrum published in Ref.~\cite{1310.0823} and redistribute it as \texttt{Axion\_Spectrum\_AGSS09met\_old\_gaee.dat}.

The peaks in the spectrum from axion-electron interactions specifically require using an algorithm that takes them into account as singularities. We therefore compute the contribution from axion-electron interactions to the signals~$s_j$ \refeq{app:eq:energyintegration} using the \texttt{gsl\_integration\_qagp} integrator of the \textsf{gsl} library.\footnote{Note that the \textsf{gsl} library is already required by \gambit.} For the axion-photon contribution to the signal, we again use the 51-point Gauss-Konrod rule. We perform all signal calculations to a relative accuracy of $10^{-6}$.

We compute signals at reference values of~$\gagg = \SI{e-10}{\GeV^{-1}}$ and $\gaee = \num{e-13}$, at 183~mass values ranging from \SIrange{e-3}{e2}{\electronvolt}. We use unequally spaced mass values, as the density of points needs to be higher in certain regions to obtain good interpolating functions. We provide these pre-calculated signal count files as \texttt{\metavar{dataset}\_ReferenceCounts\_\metavar{solarmodel}\_\metavar{coupling}.dat}, where \metavar{dataset} refers to \texttt{CAST2007} or \texttt{CAST2017\_\metavar{X}} results, \metavar{solarmodel} is the name of the solar model, and \metavar{coupling} is either the contribution from axion-photon interactions (\texttt{gagg}) or axion-electron interactions (\texttt{gaee}).

For the axion-photon contribution ($\metavar{coupling}=\texttt{gagg}$), we provide these files for the \texttt{GS98}~\cite{10.1023/A:1005161325181,1611.09867}, \texttt{AGSS09ph}~\cite{0909.2668}, \texttt{AGSS09met\_old}~\cite{0909.2668} and \texttt{AGSS09met}~\cite{1611.09867} models. For the axion-electron contribution ($\metavar{coupling}=\texttt{gaee}$), where a direct calculation of $\dd\Phi/\dd E$ is not easily performed, we provide a single file based on the earlier iteration of the AGSS09met model (\texttt{AGSS09met\_old}); this corresponds to the model used in Ref.~\cite{1310.0823} to compute the spectrum that we ship in \darkbit.

The user may choose which solar model to employ from the \yaml{Rules} section of the \YAML file, using the keys \yaml{solar\_model\_gagg} and \yaml{solar\_model\_gaee}, with \metavar{dataset} being \texttt{CAST2007} or \texttt{CAST2017} (no suffix in this case, as the CAST~2017 likelihood combines all the results):
\begin{lstyaml}
  Rules:
    - capability: @\metavar{dataset}@_signal_vac
      options:
        solar_model_gagg: @\metavar{solarmodel}@
        solar_model_gaee: @\metavar{solarmodel}@
\end{lstyaml}
If these options are left unspecified, the defaults will be chosen. For the axion-photon coupling, this is \texttt{AGSS09met}~\cite{1611.09867}; for the axion-electron coupling, the default is \texttt{AGSS09met\_old}~\cite{0909.2668}. If the corresponding \texttt{\metavar{dataset}\_ReferenceCounts\_\metavar{solarmodel}\_\metavar{coupling}.dat} file cannot be found, \darkbit will attempt to create such a file by solving~\refeq{app:eq:energyintegration}. For the axion-photon coupling, it will attempt to do this on the basis of a provided solar model file, which should be called \texttt{SolarModel\_\metavar{solarmodel}.dat}. We have included the solar model file \texttt{SolarModel\_AGSS09met.dat} from Ref.~\cite{1611.09867} as an example of the expected formatting. For the axion-electron coupling, \darkbit will attempt to generate the relevant \texttt{\metavar{dataset}\_ReferenceCounts\_\metavar{solarmodel}\_gaee.dat} file on the basis of a provided spectrum file \texttt{Axion\_Spectrum\_\metavar{solarmodel}\_gaee.dat}. If the relevant solar model or spectrum file cannot be located, or has the wrong format, \gambit will terminate. Note that this will currently happen for any choice of \yaml{solar\_model\_gaee} except for the default value. However, if the user can provide the spectrum file for any other model, the calculation will go ahead.

\section{Numerical implementation of the solution to the axion field equation}\label{app:axioneom}
The most general equation of motion for a scalar field~$\phi(t,\vc{x})$ in the background of a Friedmann-Robertson-Walker-Lema\^itre universe with Hubble parameter $H\equiv\dot{a}/a$ and potential~$V$ is given by~\cite[app.~B.12]{weinberg_cosmology}
\begin{equation}
\ddot{\phi} + 3H(t)\dot{\phi} + \frac{\Delta\phi}{a^2(t)} + V'[\phi] = 0 \, ,
\end{equation}
while the energy density associated with~$\phi$ can be calculated using:
\begin{equation}
\ede = \frac{1}{2}  \dot{\phi}^2 - \frac{1}{2a^2(t)} \left(\nabla\phi\right)^2 + V[\phi]  \, . \label{app:axionede}
\end{equation}
To obtain the energy density in axions today, it is sufficient to solve the \emph{homogeneous} field equation for $\theta\equiv a/\fa$, the normalised axion field:
\begin{equation}
\ddot{\theta} + 3H\dot{\theta} + \frac{V'[\theta]}{\fa^2} = 0 \, . \label{eq:homaxioneq}
\end{equation}
We assume the potential
\begin{equation}
V\left[\theta\right]=\fa^2 \, \ma^2 \left[1-\cos \left(\theta\right) \right] \, ,
\end{equation}
for QCD~axions and ALPs alike. For QCD~axions, this is a reasonable approximation to the full potential \cite{DiVecchia:1980yfw}. The axion energy density~$\varepsilon_a\simeq\rho_a$ from~(\ref{app:axionede}) is therefore given by
\begin{equation}
\rho_a = \frac{1}{2}\fa^2 \, \dot{\theta}^2 + \fa^2 \, \ma^2 \left[1-\cos \left(\theta\right) \right] \, .
\end{equation}
The initial conditions for~(\ref{eq:homaxioneq}) are some initial misalignment angle $\theta(0)\equiv\thetai$ and vanishing initial derivatives $\dot{\theta}(0)\equiv 0$.\footnote{The initial value for the derivative can be taken as zero assuming that $|\dot{\theta}/H| \ll 1$ at early times~\cite[ch.~10.3.2]{book_kolbturner}. The consequence of that assumption is essentially independent of the cosmological history and equation of state~\cite{1709.01090}.} For early times, the solution of~(\ref{eq:homaxioneq}) is a constant value, $\theta(t)=\thetai$. The behaviour only changes around the oscillation time $\mrm{T}{osc}$, defined as the point when $3H(\mrm{T}{osc})=\ma(\mrm{T}{osc})$.\footnote{We will solve the field equation numerically around this point and the exact definition of the onset of oscillation is therefore irrelevant. See e.g.\ Ref.\ \cite{1510.07633} for a more detailed discussion of the general relevance of this definition.}

Assuming that the scale factor~$a$ is a monotonically increasing function of cosmic time~$t$, and that the temperature~$T$ is a monotically decreasing function of $t$, we may re-write~(\ref{eq:homaxioneq}) as
\begin{equation}
\mathrm{D}_X\theta+\frac{V'\left[\theta\right]}{\fa^2} = 0 \, , \label{eq:axioneom_tau}
\end{equation}
where $X=t, \, \alpha, \, \tau$ with $\alpha\equiv a/\mrm{a}{osc}$, $\tau\equiv  T/\mrm{T}{osc}$ and
\begin{align}
  \mathrm{D}_t &= \frac{\dd^2}{\dd t^2} + 3H\frac{\dd}{\dd t} \, ,\\
  \mathrm{D}_\alpha &= H^2\alpha^2\left[\frac{\dd^2}{\dd \alpha^2} + \left(\frac{1}{\alpha} + \frac{1}{H} \frac{\dd H}{\dd\alpha}\right) \frac{\dd}{\dd \alpha}\right] \, , \\
  \mathrm{D}_\tau &= \left(\frac{H\alpha}{\alpha^\times}\right)^2\left[\frac{\dd^2}{\dd \tau^2} + \left(\frac{H^\times}{H} + \frac{\alpha^\times}{\alpha} - \frac{\alpha^{\times\times}}{\alpha^\times}\right)\frac{\dd}{\dd \tau}\right] \, , \label{eq:tauderivative}
\end{align}
where $^\times$ denotes derivatives w.r.t.~$\tau$. Conservation of entropy can subsequently be used to relate $\alpha$ and temperature:
\begin{equation}
\alpha = \left(\frac{g_S(T)}{g_S(\osc{T})}\right)^{-1/3}\tau^{-1} \equiv \gamma(\tau)/\tau \, .
\end{equation}
Equation~(\ref{eq:tauderivative}) has the advantage that an explicitly temperature-dependent axion mass or the effective relativistic degrees of freedom, $g$ and~$g_S$, can be easily incorporated. The various terms can be further simplified as follows:
\begin{align}
  \frac{\alpha^\times}{\alpha} &= \frac{\gamma^\times}{\gamma} - \frac{1}{\tau} \, , \\
  \frac{\alpha^{\times\times}}{\alpha^\times} &= \frac{\alpha^{\times\times}}{\alpha} \, \frac{\alpha}{\alpha^\times} = \frac{\tau \gamma^{\times\times}}{\tau \gamma^\times - \gamma} - \frac{2}{\tau} \, .
\end{align}
Rescaling $\vartheta\equiv\theta/\mrm{\theta}{i}$, we may use (\ref{eq:tauderivative}) to rewrite~(\ref{eq:axioneom_tau}) as
\begin{equation}
\vartheta^{\times\times} + \frac{F_2(\tau)}{\tau} \, \vartheta^\times + \left(\frac{F_1(\tau)\ma(\tau)}{\tau H(\tau)}\right)^2 \, \frac{\sin\left(\thetai\vartheta\right)}{\thetai} = 0 \, , \label{eq:axioneom_finalform}
\end{equation}
where the auxiliary functions $F_i$ for $i=1,2$ are given by
\begin{align}
  F_1(\tau) &= - 1 + \frac{\tau\gamma^\times}{\gamma}  = -\left(1 + \frac{\tau }{3}\frac{g_S^\times}{g_S}\right)\, , \label{app:F1}\\
  F_2 (\tau)& = \phantom{-} 2 + \frac{\tau H^\times}{H} - \frac{\tau^2 g_S^{\times\times}+4 \tau g_S^\times}{\tau g_S^\times + 3 g_S} \, , \\
  & = \phantom{-0 +} \frac{\tau}{2} \frac{g^\times}{g} - \frac{\tau^2 g_S^{\times\times}+4 \tau g_S^\times}{\tau g_S^\times + 3 g_S} \, , \label{app:F2}
\end{align}
where we assume a radiation-dominated universe, $H \propto \sqrt{g(T)} \, T^2$, for the last line. We use interpolated values for~(\ref{app:F1}) and~(\ref{app:F2}) based on the analytic forms for the effective degrees of freedom in Ref.~\cite{0910.1066}. The tables used for this procedure are included in \gambit.

Note that if the changes in the effective degrees of freedom are not important, $F_1(\tau) \approx -1$, $F_2(\tau) \approx 0$, and the field equation~(\ref{eq:axioneom_finalform}) reads in the harmonic limit of $\left|\thetai \right| \ll 1$:
\begin{equation}
\theta^{\times\times} \approx - \left(\frac{\ma(\tau)}{H\tau}\right)^2 \, \theta \, ,
\end{equation}
i.e.\ we obtain the approximate behaviour of a (damped) harmonic oscillator. In that limit, the comoving axion number density $n_a\equiv\rho_a a^3/\ma$ is conserved on average. We can therefore stop integrating the differential equation once specified conditions for $\ma/3H$ and $\theta$ are met. Because the solution $\theta(t)$ oscillates as a function of time or temperature, we define a peak amplitude $\hat{\theta}$ in order to check that $\left|\theta(t_\star)\right|\ll 1$ from some time $t_\star$:
\begin{equation}
\hat{\theta}(t) \equiv \frac{\sqrt{2\rho_a(t)}}{\fa\ma(t)} \, .
\label{eq:thetahat}
\end{equation}
In the harmonic limits, \refeq{eq:thetahat} indeed coincides with the amplitude of oscillation when all the energy of the field is stored in the potential part of the energy density.\footnote{Note that due to the inequality $1-\cos(x)\leq x^2/2$, we can apply this procedure even if $\theta$ is not much smaller than unity. In fact, it should be valid even if the true axion potential is not cosine-shaped, as long as the true potential is also harmonic for small values of $\theta$.} Once the axion field starts to oscillate, fulfilment of the condition $\hat{\theta}\ll 1$ indicates that the harmonic limit has been achieved. This provides much clearer limiting behaviour than e.g. $|\theta|$.

We use the \textsf{gsl} implementation of Brent's method to solve for~$\mrm{T}{osc}$ to a precision of \num{e-6}. We then impose the initial conditions at some relatively high temperature~$\mrm{\tau}{start}=\num{e5}$, and evolve the differential equation with \texttt{gsl\_odeiv2\_step\_bsimp} until some relative temperature~$\mrm{\tau}{stop}$ for which $\ma/3H>\num{e3}$ and $\hat{\theta}<\num{e-2}$. We then calculate the energy density from realignment axions today using the conservation of comoving number density
\begin{equation}
	\rho_a = \frac{ \mazero}{m_{a,\star}} \, \frac{g_S(\mrm{T}{CMB})}{g_S(T_\star)} \, \left(\frac{\mrm{T}{CMB}}{T_\star}\right)^3 \, \rho_{a,\star} \, ,
\end{equation}
where $\mrm{T}{CMB}$ is the CMB temperature.

\section{H.E.S.S. likelihood implementation details}\label{app:hessdetails}
\begin{figure}
  \centering
  \includegraphics[width=0.618\linewidth]{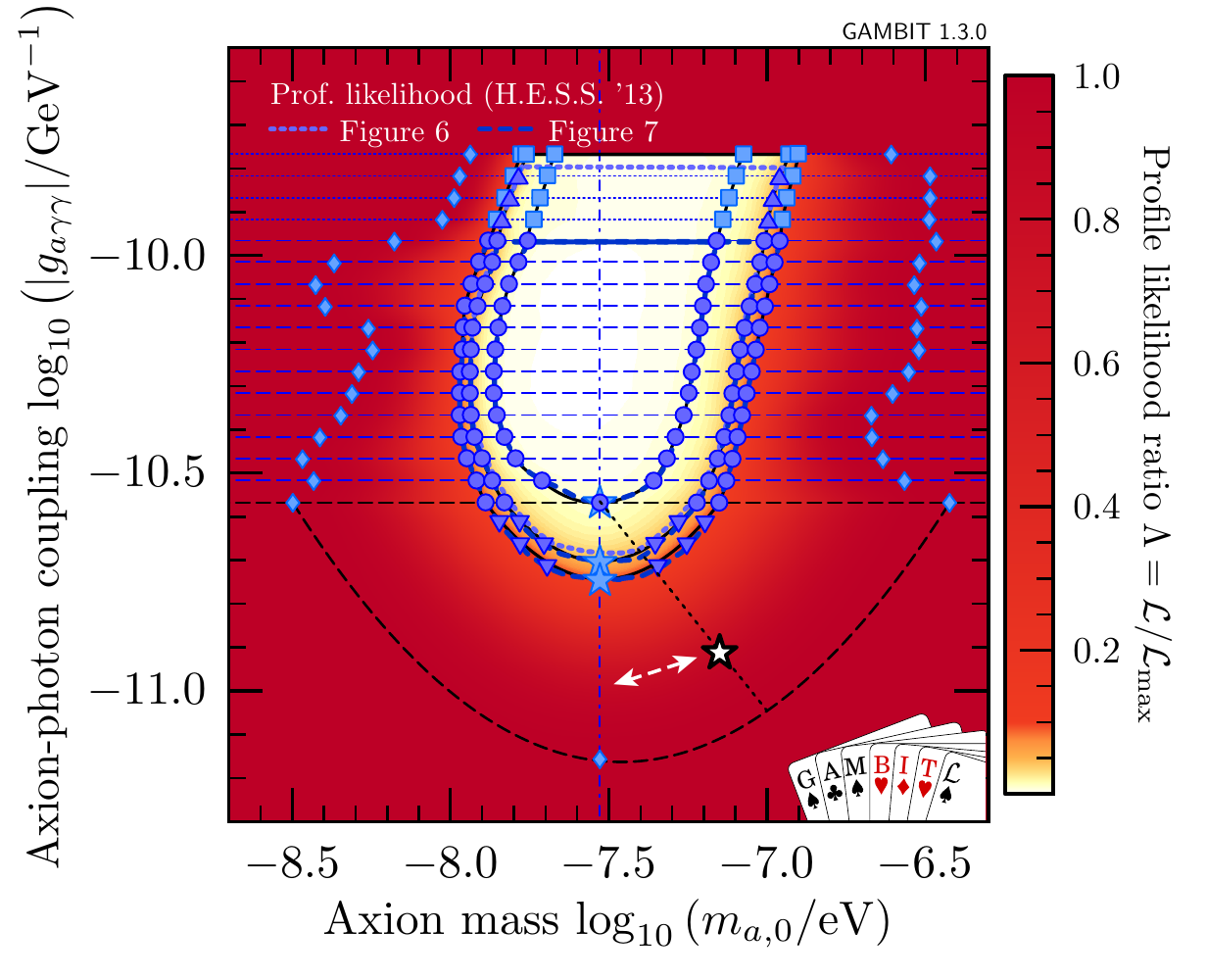}
  \caption{Detailed description of our implementation of the H.E.S.S. exclusion limits from gamma-ray spectral distortions of PKS 2155-304 (\reffig{fig:validation:h.e.s.s.}). We show the points that form the basis of our interpolation: points directly from the exclusion curves (all blue triangles, circles and stars), estimated points (squares) as well as points directly inferred from the interpolation scheme (diamonds). See the main text for further details.\label{fig:explanation:h.e.s.s.}}
\end{figure}

Our approximation of the H.E.S.S. likelihood is based on the information provided in Figs~6 and 7 of Ref.~\cite{1311.3148} on limits on axion-photon conversion in galactic cluster magnetic fields. We choose 13~lines of constant axion-photon coupling (corresponding to specific values of the axion decay rate~$\Gamma$ in their figure) that intersect with the exclusion curves at least five times (dashed black and blue horizontal lines and blue circles in \reffig{fig:explanation:h.e.s.s.}).

We subsequently define a family of one-dimensional natural cubic splines along each dashed line by including two additional points \emph{outside} of the known data. We optimise each of those splines by iteratively modifying the locations of the additional points along the dashed lines, until the values and first derivatives of the splines at the additional points go to zero (blue diamonds in \reffig{fig:explanation:h.e.s.s.}). This forcibly prevents ringing, ensuring that the interpolated log-likelihood is negative everywhere. To evaluate the log-likelihood at a given axion mass on one of the horizontal dashed lines, we use the resultant interpolating cubic spline from the fitting procedure. If a coupling value of interest sits between two horizontal dashed lines, we interpolate linearly between the values at a given mass on the adjacent dashed lines. We assign zero log-likelihood to all points outside of the area defined by the exterior points (blue diamonds).

This leaves two more regions in \reffig{fig:explanation:h.e.s.s.}. The first is at higher couplings, where Fig.~6 of Ref.~\cite{1311.3148} does not show any limits, but Fig.~7 does (dotted horizontal lines and blue triangles in \reffig	{fig:explanation:h.e.s.s.}). To complete the likelihood curve families in this region, we infer additional points (by eye; light blue squares in \reffig{fig:explanation:h.e.s.s.}) and follow the procedure described in the previous paragraph. We assign zero log-likelihood to all couplings above the uppermost horizontal line in \reffig{fig:explanation:h.e.s.s.}.

At low couplings, less than five contours are available for interpolating (inverted triangles in \reffig{fig:explanation:h.e.s.s.}). Here, we construct a single cubic spline in the vertical direction (blue stars and vertical dot-dash line in \reffig{fig:explanation:h.e.s.s.}), in order to determine an exterior point at low coupling (the diamond at the bottom of the plot). We define three parabolae, based on nine points: the set of two diamonds and four outer circles on the dashed black line around $\gagg = \SI{e-10.6}{\GeV^{-1}}$, the two stars below this line, and the one diamond at the bottom of the plot. We designate the lower parabola, passing through the exterior point (bottommost diamond) at low coupling and drawn as a dashed black curve, as the zero-log-likelihood contour. We use the middle parabola to model the 99\% C.L. exclusion curve, and the upper one to model the 95\% C.L. curve. These three parabolae allow us to map any point (represented by the white star) inside the lower parabola to a likelihood, by interpolating between the three parabolae. We do this using the one-dimensional spline constructed along the central vertical axis, using as input to the interpolation the position of the point in question between the central star and the zero-log-likelihood parabola. We carry out all interpolations in log space for both the axion mass and the axion-photon coupling.

\section{Prior-dependence of results}\label{app:prior}
For QCD~axions some prior choices are relatively straightforward (such as the use of a flat prior for~$\thetai$) or do not have a significant impact on the results (such as the prior assignments for the well-constrained nuisance parameters). In general, however, the prior beliefs can have a significant impact on the posteriors and Bayes factors, especially in the absence of constraining data. In this appendix, we investigate different prior choices for the parameters $\fa$, $\caee$, and $E/N$ in order to assess the robustness of our results and to understand how our conclusions may be affected by a variation of priors. All the new results that we present in this appendix are based on \multinest runs only, as we need the evidence values for examining the impacts of priors on model comparison, but are not sufficiently interested in the finer details of posterior maps to warrant also running \twalk.

\begin{figure}
	\centering
	{
		\includegraphics[width=0.46340504409369\linewidth]{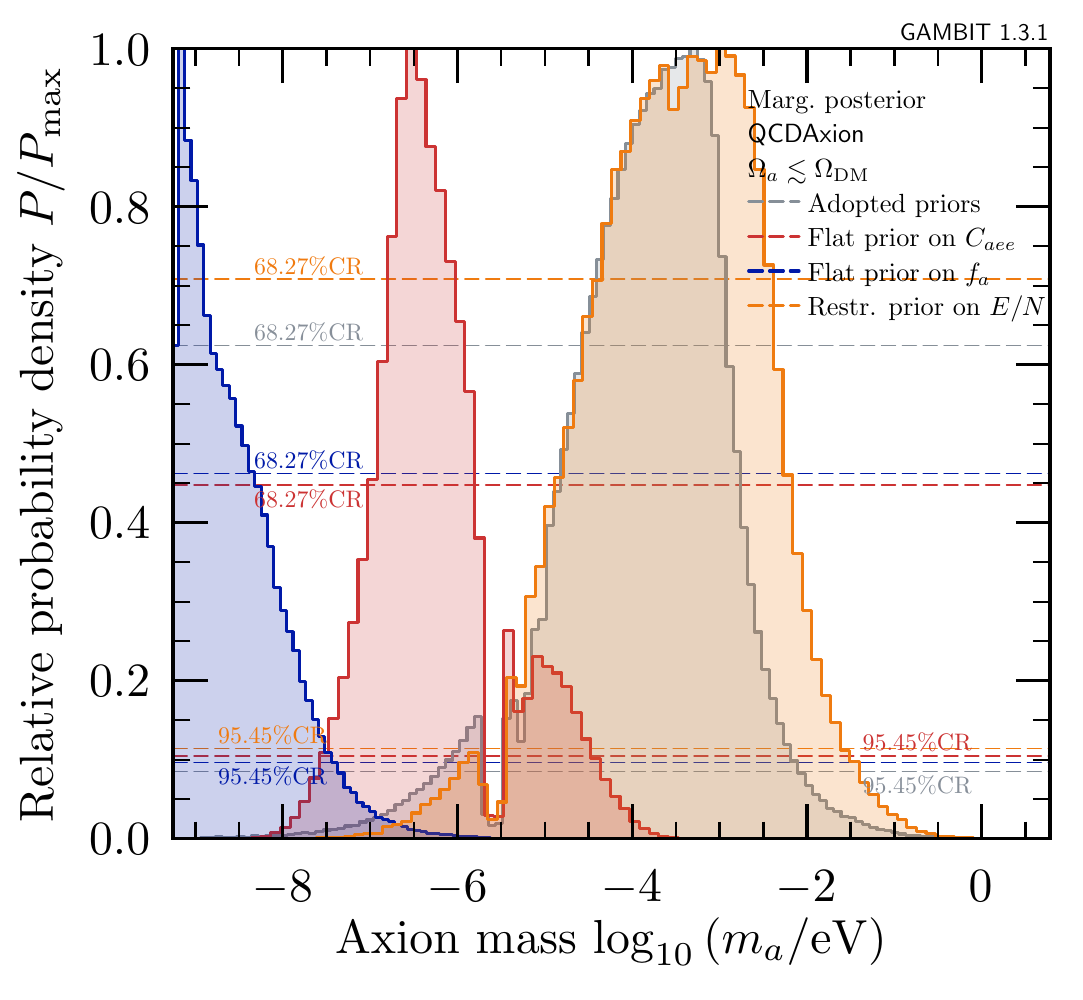}
		\hfill
		\includegraphics[width=0.52659495590631\linewidth]{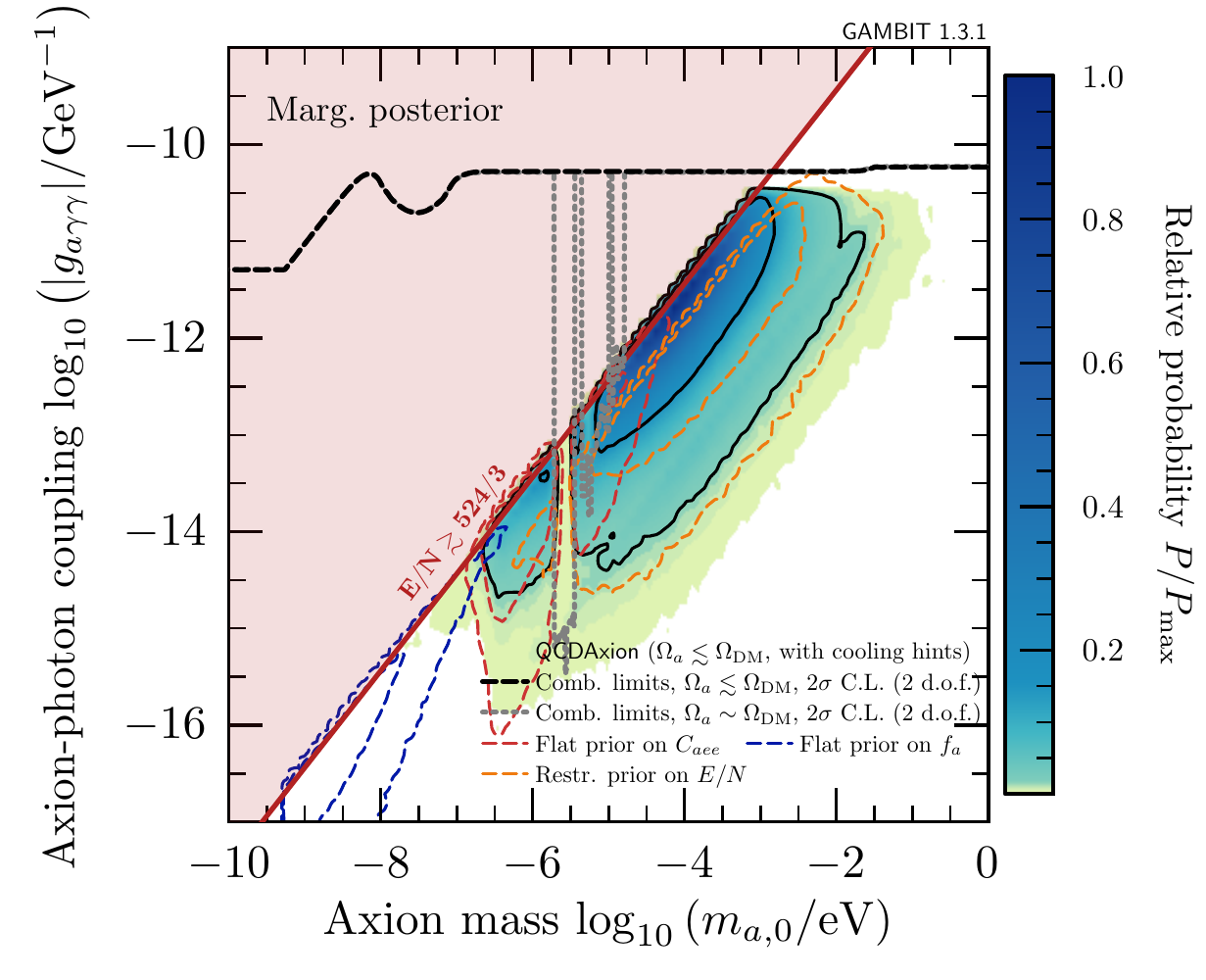}
		\hfill
	}
	\caption{Marginalised posteriors for \qcdaxion models with three alternative priors. Here we impose the observed DM relic density as an upper limit. We show the one-dimensional posterior for the axion mass without the cooling hints (\textit{left}; to be compared to the right panel of \reffig{fig:QCDAxion:mass}) and the two-dimensional posterior for $\mazero$ vs $\gagg$, including the cooling hints (\textit{right}; to be compared with \reffig{fig:cooling:QCDAxion:Regions}). The priors adopted in the main part of this study are marked in grey in the left panel, and in the colour scale and black contours in the right panel.\label{fig:app:priordep}}
\end{figure}

First, let us revisit the determination of the most credible regions for the axion mass in \qcdaxion models from Sec.~\ref{sec:results:QCDAxions}. The left panel of \reffig{fig:app:priordep} shows the marginalised posterior distribution for the axion mass, $\mazero$, for a number of alternative priors. The grey line and shading correspond to the prior choice made in the main text (right panel of \reffig{fig:QCDAxion:mass}). We can see that choosing a flat prior on either~$\caee$ or~$\fa$ shifts the most credible region for the axion mass to lower masses. The effect is more extreme for a flat prior on~$\fa$ than for a flat prior on~$\caee$. On the other hand, restricting the prior of the anomaly ratio to $E/N \leq 170/3$ (the highest values achievable in KSVZ-type models) does not have a significant impact on the result.

The different prior choices also alter the most credible regions of the axion-photon couplings (cf. \reffig{fig:cooling:QCDAxion:Regions}). The resulting posterior distributions are shown in the right panel of \reffig{fig:app:priordep}. As expected, priors that imply a preference for smaller axion masses also lead to a preference for smaller values of $\gagg$, while changes in prior that do not significantly affect the most probable region for $\mazero$ also do not impact the findings for $\gagg$ very much. In particular, for the alternative priors on~$E/N$ and~$\caee$, the regions of highest posterior density still overlap with the region that we obtained with our adopted prior, whereas the regions resulting from a flat prior on~$\fa$ are almost completely disjoint from the others.

The preference for very small axion masses for the case of a flat prior on $\fa$ is a direct consequence of the relation between $\fa$ and $\mazero$, which implies that the prior probability for $\mazero$ is strongly peaked at the lowest possible axion masses. This prior then overwhelms the preference for large~$\mazero$ from the combination of tuning in the initial misalignment angle and the relic density requirement. In other words, the preferred mass range in this case is dictated almost entirely by the assumed prior. A log prior on $\fa$, on the other hand, essentially corresponds to a log prior on $\mazero$, which has the important advantage that any preference in the axion mass range is the result of phenomenological requirements on the model. A logarithmic prior for $\fa$ is therefore less informative than a flat one, which favours a particular scale of new physics near the upper end of the prior range.

However, it is also important to determine how changing the range of the logarithmic prior impacts our results. Taking the example of the grey curve in the left panel of \reffig{fig:app:priordep} (canonical priors, no cooling hints, $\OmegaA \lsim \OmegaCDM$), we find that restricting the log prior to  $\fa \in [10^7,\, 10^{15}]\,\si{\GeV}$ results in a 95\% equal-tailed credible interval for the axion mass of \updated{$\SI{0.50}{\micro\eV} \le \mazero \le \SI{3.8}{\milli\eV}$}{$\SI{0.64}{\micro\eV} \le \mazero \le \SI{6.8}{\milli\eV}$}, compared to \updated{$\SI{0.48}{\micro\eV} \le \mazero \le \SI{3.8}{\milli\eV}$}{$\SI{0.60}{\micro\eV} \le \mazero \le \SI{7.1}{\milli\eV}$} for the prior range that we adopted in the main paper ($\fa \in [10^6,\, 10^{16}]\,\si{\GeV}$). Narrowing this range further to $\fa \in [10^8,\, 10^{14}]\,\si{\GeV}$ \updated{only raises the lower edge of the interval slightly:}{also compresses the interval slightly more:} \updated{$\SI{0.63}{\micro\eV} \le \mazero \le \SI{3.6}{\milli\eV}$}{$\SI{0.81}{\micro\eV} \le \mazero \le \SI{6.1}{\milli\eV}$}. Clearly the range of the logarithmic prior on $\fa$ has little impact on our results. The median values and credible intervals of~$\OmegaA$ are even more stable, with the lower bound of the 95\% equal-tailed credible interval slightly increasing from~\updated{$\num{6.8e-6}$}{$\num{3.7e-6}$} to~\updated{$\num{7.3e-6}$}{$\num{4.3e-6}$} for $\fa \in [10^8,\, 10^{14}]\,\si{\GeV}$ (corresponding to about \updated{0.006\%}{0.004\%} of the cosmological density of DM).

\begin{table}
	\caption{Odds ratios in favour of DFSZ- and KSVZ-type models, compared to the parent \qcdaxion model, as calculated from nested sampling evidence estimates by \multinest. We impose the DM relic density as an upper limit ($\OmegaA \lsim \OmegaCDM$). Note that the estimated uncertainties on the evidence values are small enough that the corresponding uncertainties on the odds ratios are negligible.\label{tab:app:priordep:odds}}
	\small
	\centering
	\begin{tabularx}{0.95\textwidth}{>{\bfseries}X*{6}{c}}
		\toprule
		Model & \dfszI & \dfszII & \multicolumn{4}{c}{\ksvz} \\
		\cmidrule{4-7}
		$E/N$ & $8/3$ & $2/3$ & $0$ & $2/3$ & $5/3$ & $8/3$ \\
		\midrule
		\multicolumn{7}{l}{Priors adopted in this study}\\[0.5em]
		Odds (w/o WD cooling) & 2:1 & 2:1 & 2:1 & 2:1 & 2:1 & 2:1\\
		Odds (with WD cooling) &  3:1 & 1:1 & \updated{1:2}{1:3} & 1:2 & 1:2 & 1:2 \\
		\midrule
		\multicolumn{7}{l}{Reduced prior range on~$E/N$ in \qcdaxion models}\\[0.5em]
		Odds (w/o WD cooling) & 1:1 & 2:1 & 2:1 & 2:1 & 2:1 & 2:1\\
		Odds (with WD cooling) &  2:1 & \updated{1:2}{1:1} & 1:4 & 1:4 & 1:3 & 1:4\\
		\midrule
		\multicolumn{7}{l}{Flat prior on~$\caee$ in \qcdaxion models}\\[0.5em]
		Odds (w/o WD cooling) & \updated{23:1}{24:1} & \updated{27:1}{29:1} & \updated{25:1}{29:1} & \updated{26:1}{31:1} & \updated{31:1}{34:1} & \updated{28:1}{32:1} \\
		Odds (with WD cooling) &  \updated{3:1}{4:1} & \updated{1:1}{2:1} & 1:2 & 1:2 & 1:2 & 1:2 \\
		\midrule
		\multicolumn{7}{l}{Flat prior on~$\fa$ in \emph{all} models}\\[0.5em]
		Odds (w/o WD cooling) & 1:1 & 1:1 & 1:1 & 1:1 & 1:1 & 1:1 \\
		Odds (with WD cooling) & 1:1 & 1:1 & 1:1 & 1:1 & 1:1 & 1:1\\
		\bottomrule
	\end{tabularx}
\end{table}

We also calculate the Bayes factors for KSVZ- and DFSZ-type models compared to \qcdaxion models, adopting the alternative priors for the appropriate models. The resulting odds ratios are given in Table~\ref{tab:app:priordep:odds}, in addition to the values for the priors that we adopt in Sec.\ \ref{sec:results}.

As in \reffig{fig:app:priordep}, adopting a reduced range for the anomaly ratio $E/N$ has little impact, slightly increasing the odds in favour of KSVZ-type models. Choosing the alternative prior on the axion-electron coupling constant, $\caee$, the odds in favour of KSVZ- and DFSZ-type models increase by an order of magnitude (if cooling hints are not included). This is because the flat prior on~$\caee$ causes a preference for large~$\fa$ (smaller~$\mazero$). This, in turn, implies a considerable fine-tuning in~$\thetai$ in order to avoid overproducing DM. After including the cooling hints, axion-electron interactions are much more tightly constrained by data, and the prior-dependence of the odds ratios is reduced.

Finally, a flat prior on~$\fa$ favours the lowest allowed axion masses in all models. This results in even more fine-tuning of the initial misalignment angle than using a flat prior on~$\caee$ for \qcdaxion models. However, although the Bayesian evidence in \qcdaxion models is drastically reduced, the odds ratios in favour of KSVZ- and DFSZ-type models turn out to be 1:1. This is because the alternative prior is applied to \emph{all} models: the fine-tuning required to overcome the prior pushes all evidences equally low, washing out any preference one way or another from the data.

We see that prior choice can indeed significantly influence the results of our analysis. This is true in particular for a flat prior on~$\fa$, where the prior-dependence dominates the results. On the other hand, we also saw that data with a strong preference for certain model parameter values, such as the cooling hints, can reduce the prior-dependence. In the face of more such data, the impact of prior choices tends to become less pronounced, ideally leaving only the physically-motivated prior on the initial misalignment angle able to influence final results.

\section{Overview of new capabilities}\label{app:capabilities}
\begin{sidewaystable}[p]
	\caption{New, axion-related capabilities for the \darkbit module in \gambit.\label{tab:app:capabilities}}
	\tiny
	\renewcommand{\arraystretch}{1.0}
	\centering
	\begin{threeparttable}
		\begin{tabularx}{0.95\textwidth}{@{} l X l @{}}
			\toprule
			\textbf{Capability} & \mr{\textbf{Function (Return Type):}\\ \textbf{Brief Description}} & \textbf{Dependencies}\\
			\midrule
			\cpp{QCDAxion\_ZeroTemperatureMass} & \mr{\cpp{QCDAxion\_ZeroTemperatureMass\_Nuisance\_lnL (double):}\\ Nuisance likelihood for the QCD-related energy scale $\LambdaQCD$, which defines the low-energy QCD~axion mass.} & \\
			\cpp{QCDAxion\_TemperatureDependence} & \mr{\cpp{QCDAxion\_TemperatureDependence\_Nuisance\_lnL (double):}\\ Nuisance likelihood for the parameters $\beta$ and $\Tcrit$, which define the temperature\\ dependence of the QCD~axion mass for high temperatures.} & \\
			\cpp{QCDAxion\_AxionPhotonConstant} & \mr{\cpp{QCDAxion\_AxionPhotonConstant\_Nuisance\_lnL (double):} \\ Nuisance likelihood for the contribution of axion-pion interactions in the chiral Lagrangian\\ to the axion-photon coupling.} & \\
			\midrule
			\cpp{ALPS1\_signal\_vac} & \mr{\cpp{calc\_ALPS1\_signal\_vac (double):}\\ Calculates the signal prediction for one data frame of the ALPS 1 experiment (evacuated setup).} & \\
			\cpp{ALPS1\_signal\_gas} & \mr{\cpp{calc\_ALPS1\_signal\_gas (double):}\\ Calculates the signal prediction for one data frame of the ALPS 1 experiment (setup filled with Argon gas).} & \\
			\cpp{lnL\_ALPS1} & \mr{\cpp{calc\_lnL\_ALPS1 (double):}\\ Log-likelihood for all runs of the ALPS 1 experiment.} & \mr{\cpp{ALPS1\_signal\_vac},\\ \cpp{ALPS1\_signal\_gas}}\\
			\midrule
			\cpp{CAST2007\_signal\_vac}\tnote{*} & \mr{\cpp{calc\_CAST2007\_signal\_vac (std::vector<double>):}\\ Signal prediction for a 2004 vacuum run of the CAST experiment.} & \\
			\cpp{lnL\_CAST2007} & \mr{\cpp{calc\_lnL\_CAST2007 (double):}\\ Log-likelihood for CAST 2007 experimental results.} & \cpp{CAST2007\_signal\_vac} \\
			\cpp{CAST2017\_signal\_vac}\tnote{*} & \mr{\cpp{calc\_CAST2017\_signal\_vac (std::vector<std::vector<double>>):}\\ Signal prediction for the combined 2013, 2014 and 2015 vacuum runs of the CAST experiment.} & \\
			\cpp{lnL\_CAST2017} & \mr{\cpp{calc\_lnL\_CAST2017 (double):}\\ Log-likelihood for CAST 2017 experimental results.} & \cpp{CAST2017\_signal\_vac} \\
			\midrule
			\cpp{Haloscope\_signal} & \mr{\cpp{calc\_Haloscope\_signal (double):}\\ Calculates the relative expected signal for a haloscope ignoring experimental parameters\\ and mass dependence.} & \\
			\cpp{lnL\_Haloscope\_ADMX1} & \mr{\cpp{calc\_lnL\_Haloscope\_ADMX1 (double):}\\ Log-likelihood for the combined ADMX 1998 to 2011 experimental results.} & \cpp{Haloscope\_signal} \\
			\cpp{lnL\_Haloscope\_ADMX2} & \mr{\cpp{calc\_lnL\_Haloscope\_ADMX2 (double):}\\ Log-likelihood for the ADMX 2018 experimental result.} & \cpp{Haloscope\_signal} \\
			\cpp{lnL\_Haloscope\_RBF} & \mr{\cpp{calc\_lnL\_Haloscope\_RBF (double):}\\ Log-likelihood for the RBF experiment.} & \cpp{Haloscope\_signal} \\
			\cpp{lnL\_Haloscope\_UF} & \mr{\cpp{calc\_lnL\_Haloscope\_UF (double):}\\ Log-likelihood for the UF experiment.} & \cpp{Haloscope\_signal} \\
			\midrule
			\cpp{AxionOscillationTemperature} & \mr{\cpp{calc\_AxionOscillationTemperature (double):}\\ Calculates the temperature scale at which the axion field starts to oscillate.} & \\
			\cpp{RD\_oh2}\tnote{\textdagger} & \mr{\cpp{RD\_oh2\_Axions (double):}\\ The axion dark matter relic density.} & \\
			\midrule
			\cpp{RParameter} & \mr{\cpp{calc\_RParameter (double):}\\ Calculates the expected R parameter for a system with axions.} & \\
			\cpp{lnL\_RParameter} & \mr{\cpp{calc\_lnL\_RParameter (double):}\\  Log-likelihood for the observed R parameter.} & \cpp{RParameter} \\
			\midrule
			\cpp{lnL\_HESS\_GCMF} & \mr{\cpp{ calc\_lnL\_HESS\_GCMF (double):}\\ Log-likelihood for H.E.S.S. constraints on axions (galactic cluster magnetic field result).} & \\
			\midrule
			\cpp{lnL\_SN1987A} & \mr{\cpp{calc\_Haloscope\_signal (double):}\\ Log-likelihood for an axion signal from SN1987A.} & \\
			\midrule
			\cpp{lnL\_WDVar\_G117B15A} & \mr{\cpp{calc\_lnL\_WDVar\_G117B15A (double):}\\ LogLikelihood for the white dwarf G117-B15A (cooling hint).} & \\
			\cpp{lnL\_WDVar\_R548} & \mr{\cpp{calc\_lnL\_WDVar\_R548 (double):}\\ Log-likelihood for the white dwarf R548 (cooling hint).} & \\
			\cpp{lnL\_WDVar\_PG1351489} & \mr{\cpp{lnL\_WDVar\_PG1351489 (double):}\\ Log-likelihood for the white dwarf PG 1351+489 (cooling hint).} & \\
			\cpp{lnL\_WDVar\_L192} & \mr{\cpp{lnL\_WDVar\_L192 (double):}\\ Log-likelihood for the white dwarf L19-2 (cooling hint).} & \\
			\bottomrule
		\end{tabularx}
		\begin{tablenotes}
			\item[*] Options explained in Appendix~\ref{app:solarmodelintegration}
			\item[\textdagger] New module function able to deliver a pre-existing capability
		\end{tablenotes}
	\end{threeparttable}
\end{sidewaystable}
For reference, in Table~\ref{tab:app:capabilities} we provide a complete list of the new capabilities, dependencies and options that we have added to \darkbit whilst preparing this paper.

\FloatBarrier

\pagestyle{empty}
\bibliography{R1.5.bib}
\end{document}